\documentclass[b5paper,10pt]{book}

\usepackage{phdthesis}
\usepackage[
  paperwidth=17cm
  ,paperheight=24cm
  ,inner=2.6cm
  ,outer=2cm
  ,top=2.3cm
  ,bottom=2.3cm
]{geometry}

\usepackage[hyphens]{url}
\usepackage[strict]{changepage}
\usepackage{graphicx,times}
\usepackage{tabularx,amsmath}
\usepackage{multirow}
\usepackage{booktabs}
\usepackage{amssymb}
\usepackage[square,comma,numbers,sort&compress,sectionbib]{natbib}
\usepackage{wrapfig}
\usepackage{enumerate}
\usepackage{supertabular}
\usepackage{multicol}
\usepackage{setspace}
\usepackage{import}
\usepackage{booktabs} 
\usepackage{graphicx}
\usepackage{algpseudocode}
\usepackage{pdfpages}
\usepackage{times}
\usepackage{microtype}
\usepackage{tabularx,amsmath}
\usepackage{amssymb}
\usepackage{multirow}
\usepackage{color}
\usepackage[usenames,dvipsnames,table]{xcolor}
\usepackage[normalem]{ulem}
\usepackage[english]{babel}
\usepackage{algorithm}
\usepackage{paralist}
\usepackage{flushend}
\usepackage{enumerate}
\usepackage{enumitem}
\usepackage{setspace}
\usepackage{textcomp}
\usepackage{listings}
\usepackage{keyval}
\usepackage{rotating}
\usepackage{acronym}
\usepackage{nicefrac}
\usepackage{csquotes}
\usepackage{afterpage}
\newcommand\blankpage{
    \null
    \thispagestyle{empty}
    \addtocounter{page}{-1}
    \newpage
    }
\usepackage{tikz}

\acrodef{rq:ecir}[\ref{rq:ecir}]{Can we leverage prior knowledge of entities of interest to bootstrap the discovery of new entities of interest?} 
\acrodef{rq:plos}[\ref{rq:plos}]{Are there common temporal patterns in how entities of interest emerge in online text streams?} 
\acrodef{rq:wsdm}[\ref{rq:wsdm}]{Can we leverage collective intelligence to construct entity representations for increased retrieval effectiveness of entities of interest?} 
\acrodef{rq:sigir}[\ref{rq:sigir}]{Can we predict email communication through modeling email content and communication graph properties?} 
\acrodef{rq:umap}[\ref{rq:umap}]{Can we identify patterns in the times at which people create reminders, and, via notification times, when the associated tasks are to be executed?} 

\acrodef{pt:1}{Analyzing, Predicting, and Retrieving Emerging Entities} 
\acrodef{ch:plos}{Analyzing Emerging Entities} 
\acrodef{ch:ecir}{Predicting Emerging Entities} 
\acrodef{ch:wsdm}{Retrieving Emerging Entities} 

\acrodef{pt:2}{Analyzing and Predicting Activity from Digital Traces} 
\acrodef{ch:sigir}{Analyzing and Predicting Email Communication} 
\acrodef{ch:umap}{Analyzing and Predicting Task Reminders}

\newcommand{\downdown}{$^{\blacktriangledown}$}
\newcommand{\down}{$^{\triangledown}$}
\newcommand{\upup}{$^{\blacktriangle}$}
\newcommand{\up}{$^{\vartriangle}$}

\newcommand{\dubbelop}{$^{\blacktriangle}$}

\newcommand{\dubbelneer}{$^{\blacktriangledown}$}

\usepackage[backref=page]{hyperref}
\hypersetup{
	pdfborder = {0 0 0},
	pdftitle = {Entities of Interest},
	pdfsubject = {PhD Thesis},
	pdfkeywords = {digital traces, discovery, emerging entities, entity linking, entity retrieval, recipient recommendation, reminders},
	pdfauthor = {David Graus}
	}

\usepackage{bookmark}
\renewcommand*{\backref}[1]{} 
\renewcommand*{\backrefalt}[4]{
\ifcase #1
\or (Cited on page~#2.)  
\else 
(Cited on pages~#2.)  
\fi
}

\begin{document}

\includepdf[pages=-]{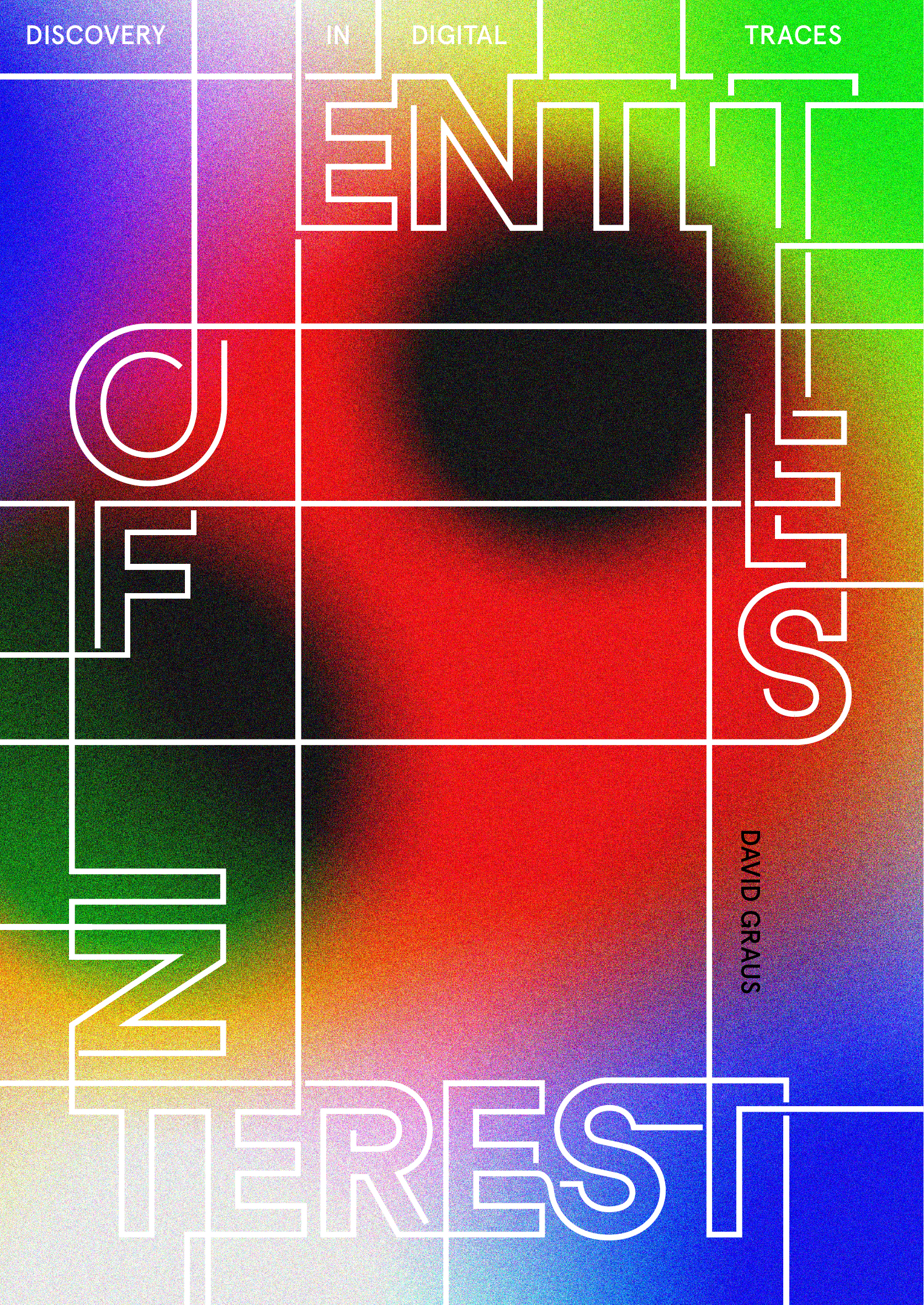}

\frontmatter
\clearpage{}

{\pagestyle{empty}
\newcommand{\printtitle}{
{\Huge\bf Entities of Interest \\[0.8cm]
\Large\bf Discovery in Digital Traces
}}

\begin{titlepage}
\par\vskip 2cm
\begin{center}
\printtitle
\vfill
{\LARGE\bf David Graus}
\vskip 2cm
\end{center}
\end{titlepage}

\mbox{}\newpage
\setcounter{page}{1}

\clearpage
\par\vskip 2cm
\begin{center}
\printtitle
\par\vspace {4cm}
{\large \sc Academisch Proefschrift}
\par\vspace {1cm}
{\large ter verkrijging van de graad van doctor aan de \\
Universiteit van Amsterdam\\
op gezag van de Rector Magnificus\\
prof. dr. ir. K.I.J. Maex\\
ten overstaan van een door het College voor Promoties ingestelde \\
commissie, in het openbaar te verdedigen in \\
de Agnietenkapel\\
op vrijdag 16 juni 2017, te 10:00 uur} \\ 
\par\vspace {1cm} {\large door}
\par \vspace {1cm}
{\Large David Paul Graus}
\par\vspace {1cm}
{\large geboren te Beuningen} 
\end{center}

\clearpage
\noindent
\textbf{Promotiecommissie} \\\\
\begin{tabular}{@{}l l l}
Promotor: \\
& prof.\ dr.\ M.\ de Rijke & Universiteit van Amsterdam \\  
Co-promotor: \\
& dr.\ E. Tsagkias & 904Labs \\
Overige leden: \\
& prof.\ dr.\ A.P.J.\ van den Bosch & Radboud Universiteit Nijmegen \\
& prof.\ dr.\ ing.\ Z.J.M.H.\ Geradts & Universiteit van Amsterdam \\
& dr.\ ir.\ J.\ Kamps & Universiteit van Amsterdam \\
& dr.\ E.\ Kanoulas & Universiteit van Amsterdam \\
& prof.\ dr.\ D.W.\ Oard & University of Maryland \\
\end{tabular}

\bigskip\noindent
Faculteit der Natuurwetenschappen, Wiskunde en Informatica\\

\vfill

\noindent
\begin{figure}[!h]
   \begin{minipage}[t][1cm]{0.3\textwidth}
      \vspace{0pt}
      \includegraphics[width=\linewidth]{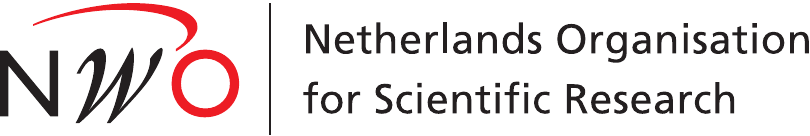}
   \end{minipage}
   \hfill
   \begin{minipage}[t][1cm]{0.65\textwidth}
      \vspace{0pt}
The research was supported by the Netherlands Organization for Scientific Research (NWO) under project number 727.011.005. \\
   \end{minipage}
\end{figure}

\begin{figure}[!h]
   \begin{minipage}[t][1.5cm]{0.3\textwidth}
      \vspace{0pt}
      \includegraphics[width=\linewidth]{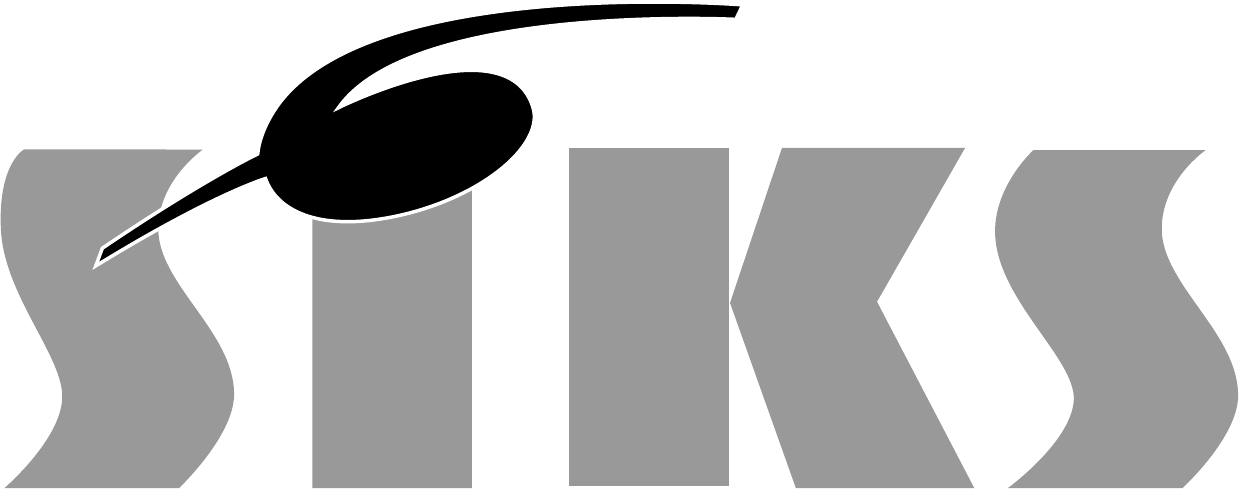}
   \end{minipage}
   \hfill
   \begin{minipage}[t][1.5cm]{0.65\textwidth}
      \vspace{0pt}
    SIKS Dissertation Series No.~2017-23 \\
    The research reported in this thesis has been carried out under the
    auspices of SIKS, the Dutch Research School for Information and
    Knowledge Systems.
   \end{minipage}
\end{figure}

\noindent
\begin{figure}[!h]
   \begin{minipage}[t][2cm]{0.28\textwidth}
   	  \centering
      \vspace{0pt}
      \includegraphics[width=\linewidth]{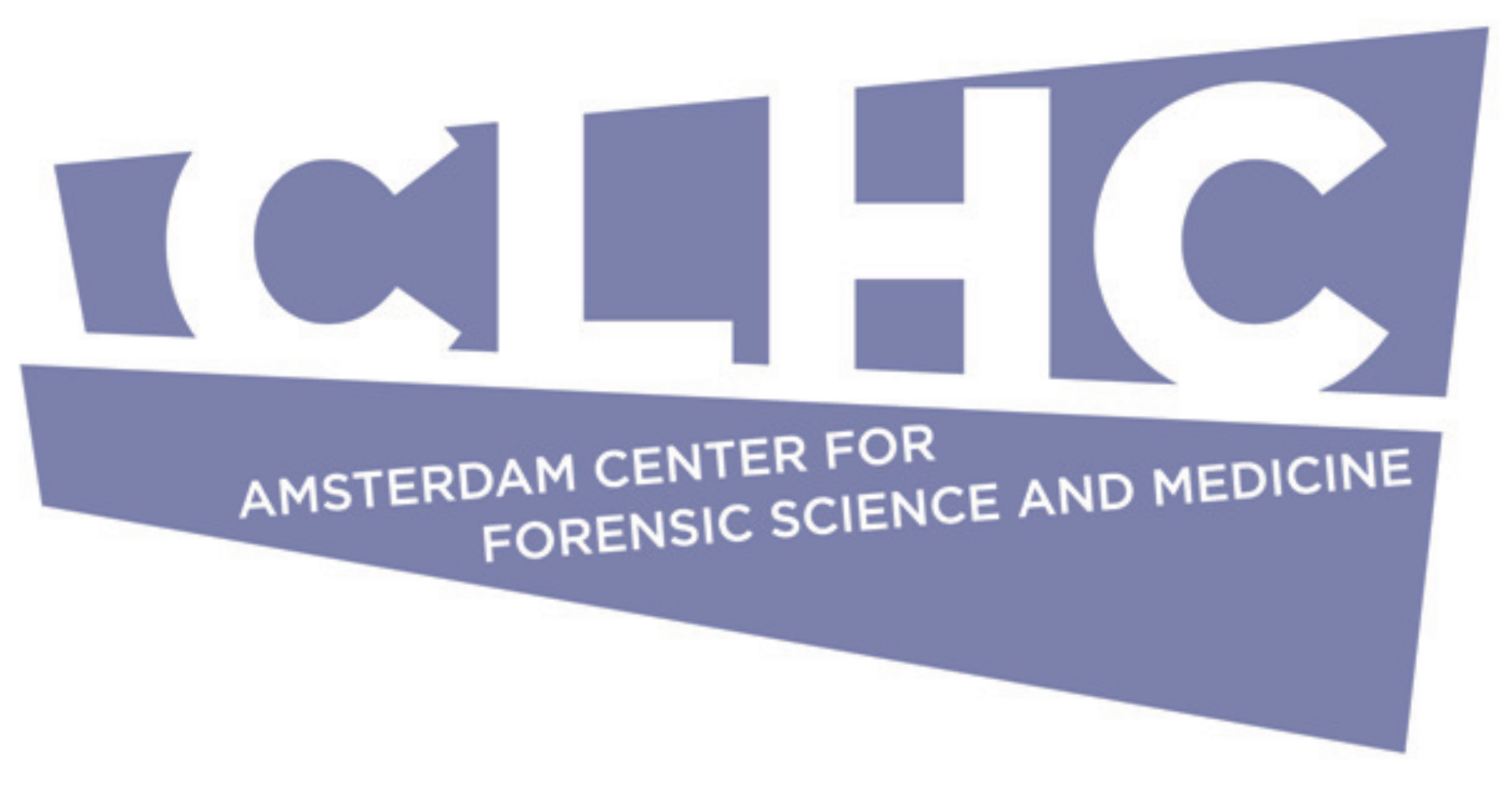}
   \end{minipage}
   \hfill
   \begin{minipage}[t][2cm]{0.65\textwidth}
      \vspace{0pt}
The printing of this thesis was financially supported by the Co van Ledden Hulsebosch Center, Amsterdam Center for Forensic Science and Medicine. \\
   \end{minipage}
\end{figure}

\noindent
Copyright \copyright~2017 David Graus, Amsterdam, The Netherlands\\
Cover by Rutger de Vries/perongeluk.com\\
Printed by Off Page, Amsterdam\\
\\
ISBN: 978-94-6182-800-2\\

\clearpage
}
\clearpage{}

\clearpage{}
{\thispagestyle{empty}

\vspace*{-1cm}
{
\begin{center}
\noindent
\textbf{Acknowledgements}
\end{center}
}

\noindent
I started the process that culminated in the little booklet you are now reading in 2012.
At that time, 
I had never touched a unix, 
I didn't know what an ssh was, 
and I hardly knew how to write a line of code. 
I hate to be overly dramatic,\footnote{I really don't\ldots} but I have grown, not only as a researcher, also as a person, and this growth is not exclusively to be attributed to the passing of time. 
Over the course of my PhD I have become more confident, about myself and my skills, and I have learned I can pick up anything I put my mind (and time) to. 
The latter is by far the greatest life's lesson my PhD has brought me. 

For all of this, I have to thank some people. 
First, P\&C, for setting up the prior by raising me in a home where I naturally got exposed to computers, technology, and (overly) critical thinking. 
P, for putting me and Mark in front of a Sinclair ZX81 at young age, traumatizing us with the scariest ASCII-rendered games imaginable. 
It shaped me into the tough trooper I am today. 
C, for putting up with a grumpy household when our attempts at assembling computers from parts scrambled from all over the place invariably failed. 

Next, Maarten, for taking a ``chance'' with me. 
I was quite wet behind the ears, as one would say, before I started my PhD. 
I am glad and thankful that this did not hold Maarten back for giving me the opportunity to see how I would pan out as a computer scientist. 
I was a coin-flip. Thanks for flipping! 

Next next, ILPS. 
First, the senior population at ILPS. 
In particular, Edgar, who at the time understood better than me what I was doing, and who saw that I could do and might enjoy this PhD. 
Manos and Wouter, who guided, supervised, co-authored, and co-lived this period with me. 
My peers at ILPS, of which I will name a few because these are my acknowledgements and I get to decide whether I single people out or not: Daan, Anne, Tom, and Marlies. 
I enjoyed interfering with your day-to-day activities, sharing misery, disagreeing (hi Anne), coffee breaks, bootcamps and runs, and everything else. 

Then, my co-authors, for partaking in that thing we do in science, which conveniently provided the basis of this dissertation. 
Paul, Ryen, and Eric for mentoring me during my internship at Microsoft Research in the summer of 2015. Best summer ever. 

Finally, my brothers from other mothers, Rutger and Marijn, for putting up with me being right all the time, slightly cocky, and (at times) tending towards the far-right end of the spectrum of self-confidence. I'm sorry. 

Moving on to the nonliving things: Polder, Oerknal, Maslow, and Joost (in order of appearance) for providing shelter for unwinding, escaping, brainstorming, philosophizing and reflecting on life, academia, and beyond. 
Oh, and also for drinking beers. 
Twitter, for giving me a place where I could write down stuff when Christophe would stop listening to the monologues I typically directed at my computer screen (but were intended for all to hear). 
Spotify and Bose, for sheltering me from the real world during extended periods of (coding) cocooning. 

Okay, this is not an exhaustive list, but I have to both start and stop somewhere. 
Before we part ways and move onto the researchy stuff, I will leave you with one big fat clich\'e---which I know loses some of its expressive power by virtue of being a clich\'e (but which I hope to counteract a bit by being explicit about my awareness of it being a clich\'e):
\begin{center}
\emph{It was a great journey!} 
\end{center}

\begin{flushright}
David
\end{flushright}

}\clearpage{}

\tableofcontents

\makeatletter
\mainmatter
\clearpage{}

\chapter{Introduction}
\label{ch:introduction}

\begin{flushright}
\rightskip=1.8cm``Dis-moi ce que tu manges, je te dirai ce que tu es.'' \\
\vspace{.2em}
\rightskip=.8cm---\textit{Jean Anthelme Brillat-Savarin}
\end{flushright}
\vspace{1em}

\noindent
From the things you ``like'' on Facebook, algorithms can infer many personal and demographic traits with surprisingly high accuracy. 
For example, political preferences, religious beliefs, but even more obscure factoids, such as whether someone's parents were together during their childhood~\cite{Youyou27012015}. 
The predictive power of these algorithms relies on the availability of \emph{digital traces} of a large number of people. 
In the era of \emph{Big Data}, digital traces are available in abundance. 
We willingly share, post, like, link, play, and query, all the while leaving behind rich digital traces. 
Traces that, when aggregated, can provide meaningful insights into behavior, preferences, and profiles of people. 

We distinguish between two types of digital traces: 
\emph{Active digital traces} that people leave behind deliberately, e.g., blogposts, tweets, or forum posts,
and \emph{passive digital traces}, that people may leave behind unknowingly, e.g., by clicking on links on websites, querying web search engines, or by simply visiting websites~\cite{4653470}. 
The quality of and our reliance on online services has enabled wide availability of these digital traces. 
Combined with rapid developments in computer hardware, software, and algorithms, the availability of our digital traces has given rise to (yet) a(nother) resurge of Artificial Intelligence (AI). 
With it, a new economy is on the rise that revolves around applied \emph{machine learning} for understanding users, e.g., inferring their traits, preferences, and behavior, for applications such as targeted advertising, personalizing content, and more generally, improving online services. 

People have voiced concerns around this rise of the ``algorithm,'' 
hailing it as the demise of our privacy and freedom~\cite{davidgraus:bigdata-privacy}, 
the death of politics~\cite{morozov},  
and in general harmful~\cite{zeynep}. 
But indisputably, mining, collecting, analyzing, and leveraging our digital traces has brought many advances. 
At its most visible level, it has permeated into our day-to-day lives, and has enabled our access to exceeding and unforeseen amounts of data, information, and knowledge. 
While this information access may seem self-evident to us, it is not.  
Enabling effective access to these huge amounts of information relies to a large degree on understanding people's behavior, preferences, and traits. 
The ``relevance'' of information is highly user-specific, it depends on the context of the user (e.g., the time of day, geographical location, or a task the user may be executing), and more personal aspects such as taste, preference, or background~\cite{Mizzaro:1997:RWH:262192.262203}. 

\noindent
We would be unable to find what we wanted if it were not for 
the search engines that learn from their users~\cite{schuth-phd-thesis-2016,Hofmann:2013:RHI:2433396.2433419}, 
the filtering algorithms that find and prioritize content relevant to us~\cite{Eslami:2015:IAA:2702123.2702556,Das:2007:GNP:1242572.1242610}, 
and the recommendation algorithms that allow services such as Netflix and Spotify to serve a wider variety of content, by no longer being constrained to catering to the masses like traditional media. 

Next to these everyday-life, practical applications, analyzing and mining digital traces has brought advances for (social) good, too. 
In the health domain, search engine logs have led to discoveries of new and previously unreported side effects of prescription drugs~\cite{pmid23467469}, and adverse drug reactions~\cite{yomtov2013}. 
Social media posts have allowed us to develop methods for identifying people who suffer from depression~\cite{ICWSM136124}. 
And more generally, the research fields of sociology, anthropology, (social) psychology, and (digital) humanities have benefited from the possibilities brought by large-scale analysis of (digital) traces~\cite{manovich2011}. 

And this is just the beginning. 
More and more areas are starting to reap the benefits of advances in artificial intelligence and the ability to extract information from large collections of digital data. 

\subsection*{Discovery in Digital Traces}
The Enron scandal~\cite{wiki:enron}, 
the Hillary Clinton email controversy~\cite{wiki:clinton}, 
the Panama papers~\cite{wiki:panama}, 
or any data-release by WikiLeaks~\cite{wiki:wikileaks}, 
are all examples of cases in which large amounts of (digital) traces needed to be investigated, explored, and turned upside down, to gain insights and discover ``evidence.'' 
This \emph{E-Discovery} task is at its core \emph{``finding evidence of activity in the real world''} in ``Electronically Stored Information'' (ESI)~\cite{INR-025}. 
Discovery in digital data finds applications in many domains and scenarios, e.g., 
in (investigative) journalism~\cite{doi:10.1080/21670811.2014.976418}, 
digital forensics~\cite{Marchionini:2006:ESF:1121949.1121979,Ieong200629}, 
and litigation, where data may be requested by a plaintiff with the aim of gaining insights and finding evidence for a legal case~\cite{INR-025}. 

Whether in litigation, journalism, or digital forensics, in this discovery task, users set out to browse and explore large collections of data (usually text), with the aim of discovering new information, or of finding answers to any questions they may have.

Gaining insights and making sense of our digital traces may seem trivial. 
Often in written form, digital traces hold the promise of being explicit, structured, unambiguous, and readily interpretable. 
However, even written language is not as structured as it may seem. 
First, language is noisy. 
Ambiguity is common across languages, the meaning of words may change over time, and without understanding the continuously changing context in which language is created, it is often impossible to understand its intended meaning~\cite{firth57}.
Second, the amount of digital data we produce on a daily basis is enormous and ever-growing.
We are exposed to and produce an ever-increasing volume of data, both online on e.g., social media, collaboration platforms, and email, and offline on e.g., laptops, mobile phones, computers and storage media~\cite{howmuchinfo2003}.
To address the above challenges, the need arises for automated methods for \emph{sense-making} of digital traces for discovery, with the goal of both understanding their \emph{content} and the \emph{contexts} in which they are created. 

In this thesis, we draw inspiration from this discovery scenario, and propose methods that aim to support sense-making of digital traces. 
We focus on textual traces, e.g., emails, social media streams, and user interaction logs. 
In two parts in this thesis, we address two different aspects that are central to sense-making  from digital traces. 

In the first part of the thesis, we study methods for sense-making of the \emph{content} of (textual) digital traces. 
More specifically, our objects of study are the \emph{real-world entities} that are referenced in digital traces. 
Knowing, e.g., 
which people are mentioned in email communication between employees of a company, 
which companies are discussed by stockbrokers on Twitter, 
or which people and companies are mentioned in the Panama papers, 
is central to the exploratory and investigative search process that is inherent to sense-making~\cite{Ahn2010383}. 
In this part, our \emph{entities of interest} are real-world entities: \blockquote{things with distinct and independent existence}~\cite{define:entity} that exist~\emph{in the real world}.
We thus study the occurrence of concrete entities, e.g., companies, organizations, locations, and products in digital traces such as email, social media, and blogs.

In the second part of the thesis, we focus on the \emph{context} of digital traces.  
Here, our \emph{entities of interest} are the producers of digital traces, i.e., the people that leave behind the traces. 
Uncovering real-world activity is a central task in the E-Discovery sense-making process~\cite{INR-025}, thus, we focus on methods for uncovering this (evidence of) real-world activity from digital traces.
Our aim is to understand real-world behavior from digital traces. 
We present two case studies where we analyze and leverage patterns in behavior to predict activity of the people who leave behind digital traces.

\section{Research Outline and Questions}
\label{section:introduction:rqs}

As outlined above, we distinguish two research themes on automated methods for sense-making of digital traces: 

In the first part, our \emph{entities of interest} are the real-world entities that are referenced in digital traces. 
Identifying the real-world entities that appear in ESI such as emails, social media, or forums and blogs, supports the exploratory and complex search process that underlies the discovery process. 
In the discovery scenario, the \emph{entities of interest} may not be known in advance, i.e., they may not be well-known or established entities. 
For this reason, we focus on so-called \emph{emerging} real-world entities, i.e., entities that are not described (yet) in publicly available knowledge bases such as Wikipedia. 
We address three tasks, (i)~analyzing, (ii)~predicting, and (iii)~retrieving newly emerging real-world entities. 

In the second part of this thesis, our \emph{entities of interest} are the producers of the digital traces. 
We aim to 
(i)~understand their behavior and real-world activity by analyzing digital traces, and 
(ii)~predict their (future) real-world activity, leveraging the analyses and uncovered patterns. 
In this part, we present two case studies where we (i)~analyze and predict email communication by studying the communication network and email content, and (ii)~analyze interaction data with a mobile intelligent assistant to predict when a user will perform an activity.

\subsection*{Part~\ref{pt:1}.~\acl{pt:1}} 
The discovery task is a lot like looking for needles in haystacks, without knowing what the needles look like. 
Users of search systems in the discovery context (e.g., forensic analysts, lawyers, or journalists) may be interested in finding evidence of real-world activity, without knowing beforehand what type of activity nor whose activity they are looking for. 
Traditional IR methods fall short in supporting this type of exploratory search~\cite{Stasko:2008:JSI:1466620.1466622}. 
They largely rely on \emph{lexical matching}, i.e., matching words in a search query to words in a document, to allow users to retrieve documents from a collection. 
However, this type of search is not suitable for the typical discovery scenario, where the aim is to understand the 5 W's: 
\emph{Who was involved? What happened? Where, when and why did it happen?}~\cite{INR-025}. 
For answering these questions, it is typically difficult to formulate an exhaustive set of search queries, which means the keyword-based lexical matching paradigm does not suffice. 

Semantic Search, i.e., ``search with meaning''~\cite{INR-032}, is an alternate search paradigm that aims to move beyond lexical matching, and improve document retrieval by incorporating additional (external) knowledge in the search process. 
This additional knowledge can be in the form of 
the discussion structure of email threads, 
the topical structure of documents, or 
information on the entities and their relations that are mentioned in documents~\cite{vandijk2011}. 

In the first part of this thesis we focus on real-world entities that appear, i.e., are referenced, in digital traces. 
Understanding which real-world entities appear and are discussed in digital traces is of central importance in exploratory search~\cite{Khalid:2008:INE:1793274.1793371}, and in answering the 5 W's. 

A central task in semantic search is EL\cite{INR-032}. 
EL revolves around linking the mentions of real-world entities in text to their representations in an external KB. 
Linking mentions of real-world entities in text to their KB representations effectively boils down to disambiguation, and enables the enrichment of text with additional information and structure. 
Knowing the unambiguous real-world entities that occur in text enables better support for searching through large text collections~\cite{bontcheva2012making} and the complex information seeking behavior that is inherent to the discovery process~\cite{Ahn2010383}, through, e.g., enabling filtering for specific entities (e.g., a person) or entity types (e.g., companies), and by providing insights into how different entities co-occur. 

EL relies on external knowledge from a reference KB that contains (descriptions of) real-world entities~\cite{Rao2013}. 
An example and often-used instantiation of such a reference KB is Wikipedia, the world's largest encyclopedia. 
Wikipedia is online, collectively built, and known to democratize information~\cite{10.2307/20864471}. 
At the time of writing, the English Wikipedia contains over 5 million articles that represent both abstract and real-world entities, written by over 29 million users~\cite{wikipediacount2016}. 
With this broad coverage, it spans the majority of entities that emerge in the media, and play a role in our public discourse, it has been dubbed a ``global memory place,'' where collective memories are built~\cite{Pentzold01052009}.  
In addition, the rich metadata of Wikipedia, e.g., its hyperlink structure, category structure, and infoboxes, can be effectively employed as additional signals for disambiguating and improving EL, which makes Wikipedia an attractive standard reference KB for EL systems. 

Due to the ambiguous nature of language, knowing the words that refer to entities (i.e., \emph{entity mentions}) alone is unlikely to be sufficient for answering the 5 W's. 
To answer \emph{who}, \emph{where}, or \emph{what} the mention refers to, the most important step of EL is not to identify an entity mention, but to correctly \emph{link} it to its referent KB entry. 
The main challenge in EL is ambiguity: 
a single entity mention (e.g., \emph{``Michael''}) can refer to 
multiple real-world entities in the KB (e.g., \emph{Michael Jordan}, \emph{Michael Jackson}, or even any other real-word entity that may not be described in the KB). 
At the same time, a single KB entity may be referred to by multiple entity mentions (e.g., \emph{Michael Jackson} may be referred to with \emph{``the king of pop,''} \emph{``MJ,''} \emph{``Mr. Jackson,''} or simply \emph{``Michael''}). 

EL methods using Wikipedia have been effectively applied to many challenging domains where context is sparse, and language may be noisy, e.g., 
Twitter~\cite{meij2012} and 
television subtitles~\cite{Odijk:2013aa}. 
However, the E-Discovery scenario provides an additional challenge over noisy and large volumes of data: the \emph{entities of interest} may not be included in the reference KB. 

In the first part of this thesis our \emph{entities of interest} are these entities that are absent from the reference KB. 
Entities about which we may have little prior knowledge. 
We call these entities \emph{emerging entities}, as they are initially absent from the KB, but of interest to an end-user, i.e., worthy of being incorporated into the KB. 

We address three tasks that relate to emerging entities: 
The first is largely observational in nature, here, we study how entities emerge in online text streams, by analyzing the temporal patterns of their mentions before they are added to a KB. 
Next, we bootstrap the discovery of emerging entities in social media streams, by generating pseudo-ground truth to learn to predict which entity mentions represent emerging KB entities. 
And finally, we collect external and user-generated additional entity descriptions to construct and enrich entity representations to allow searchers to retrieve them more effectively.

\subsubsection*{Chapter~3.~\acl{ch:plos}}

We start our exploration of real-world entities in digital traces with a large-scale analysis of how entities emerge. 
The digital traces we study are online text streams (e.g., news articles and social media posts), and the entities of interest are emerging entities. 
Understanding the temporal patterns under which entities emerge in online text streams, before they are added to a KB may be helpful in predicting newly emerging entities of interest, or in distinguishing between different types of entities of interest. 
In Chapter~3 
we study the temporal patterns of emerging entities in online text streams, and answer the following research question: 

\begin{enumerate}[label=\textbf{RQ\arabic*},ref={RQ\arabic*},resume]
\item \acl{rq:plos}\label{rq:plos}
\end{enumerate}

\noindent
To answer this question, we analyze a large collection of entity time series, i.e., time series of entity mentions in the timespan between the first mention of an entity in online text streams and its subsequent incorporation into the KB. 
Our collection contains over 70,000 entities that emerge in a timespan of over 18 months. 
We find that entities emerge in broadly similar patterns. 
We distinguish between entities that emerge gradually after being introduced in public discourse, and entities that emerge more abruptly. 
Furthermore, we study the differences and similarities between how entities emerge in different types of online text streams, and the differences and similarities in the emergence of different types of emerging entities.

\subsubsection*{Chapter~\ref{ch:ecir}.~\acl{ch:ecir}}
Motivated by the findings of the dynamic nature of knowledge bases, and continuous stream of newly emerging entities in Chapter~\ref{ch:plos}, 
in Chapter~\ref{ch:ecir} we address the task of predicting newly emerging entities. 
The digital traces we study are social media streams (i.e., Twitter), and our entities of interest are entities that are similar to a set of seed entities, i.e., entities that are described in a reference KB. 
In the discovery process a searcher may have some prior knowledge of the entities of interest, e.g., she may have a reference list of entities with the goal of exploring a document collection in search for similar entities. 
In such a scenario, where the entities of interest are absent from the KB, traditional EL methods fall short. 
In Chapter~\ref{ch:ecir} we propose a method that leverages the prior knowledge that is encoded in a reference KB, and existing methods for recognizing and linking KB entities, to recognize newly emerging entities that are not (yet) part of the KB. 
In this chapter, we answer the following research question: 

\begin{enumerate}[label=\textbf{RQ\arabic*},ref={RQ\arabic*},resume]
\item \acl{rq:ecir}\label{rq:ecir}
\end{enumerate}

\noindent
We propose an unsupervised method, where we use an EL method and an incomplete reference KB to generate pseudo-ground truth to train a named-entity recognizer to detect mentions of emerging entities. 
Mentions of emerging entities are entity mentions that were not linked by the EL method, but that do occur in similar contexts as those that are in the reference KB. 
We compare the effect of different sampling strategies on the pseudo-ground truth, and the resulting predictions of emerging entities. 
We show that sampling based on textual quality and the confidence score of the EL method are effective methods for increasing the effectiveness of discovering absent entities. 
Furthermore, we show that with a small amount of prior knowledge, our method is able to cope with missing labels and incomplete data, justifying the approach of generating pseudo-ground truth. 
Finally, the method we propose is domain and/or language independent, as it does not rely on language or domain-specific features, making it particularly suitable for the discovery scenario.

\subsubsection*{Chapter~\ref{ch:wsdm}.~\acl{ch:wsdm}}
Finally, in Chapter~\ref{ch:wsdm}, the last chapter of Part~\ref{pt:1}, we address the task of \emph{retrieving} entities of interest. 
As seen in Chapter~\ref{ch:plos}, entities may suddenly appear in bursts, as events unfold in the real world.
In order to capture the changing contexts in which entities appear, and improve the retrieval (i.e., search) of entities, we propose a method that dynamically enriches and expands the (textual) representations of entities. 
We enrich representations of an entity by collecting descriptions from a variety of different dynamic sources that represent the \emph{collective intelligence}. 
The different sources constitute different (user-generated) digital traces, e.g., social media, social tags, and search engine logs. 
In expanding their representations with these collective digital sources, we bridge the gap between the informal ways how users search for and refer to entities, and the more formal ways of how they are represented in the KB. 
In this chapter, we answer the following research question: 

\begin{enumerate}[label=\textbf{RQ\arabic*},ref={RQ\arabic*},resume]
\item \acl{rq:wsdm}\label{rq:wsdm}
\end{enumerate}

\noindent
We combine the collected additional descriptions from digital traces into a single entity representation, by learning to weight and combine the heterogeneous content from the different sources. 
Our method learns directly from users' past interactions (i.e., search queries and clicks), and enables the retrieval system to continuously learn to optimize the entity representations towards how people search for and refer to the entities of interest. 
We find that incorporating dynamic description sources into entity representations enables searchers to better retrieve entities.
In addition, we find that informing the ranker of the ``expansion state'' of the entity overcomes challenges related to heterogeneity in entity descriptions (i.e., popular entities may receive many descriptions, and less popular entities few), and further improves retrieval effectiveness.

\subsection*{Part~\ref{pt:2}.~\acl{pt:2}}
In the second part of this thesis we take a different view on entities of interest. 
Here, our entities of interest are the people who leave behind the digital traces, i.e., the producers of digital traces. 
Our aim is to better understand the producers' context under which they leave behind digital traces. 
This part revolves around finding evidence of activity in the real world, which is an essential task in the sense-making process of E-Discovery~\cite{INR-025}. 
We present two case studies of analyzing and predicting human activity given digital traces. 

Analyzing and predicting user behavior has a rich history in the information retrieval and the user modeling, adaptation and personalization communities~\cite{umap2009}. 
Understanding human behavior by studying historic interactions and digital traces finds many applications, e.g., 
improving personalization and recommendation systems~\cite{Liu:2010:PNR:1719970.1719976,35599}, 
improving search engines~\cite{Hofmann:2013:RHI:2433396.2433419,Baeza-yates05modelinguser,Agichtein:2006:IWS:1148170.1148177}, 
or improving information filtering~\cite{Morita:1994:IFB:188490.188583}. 

In both case studies we look at the impact of the \emph{contexts} in which digital traces are created, and their \emph{content}, on predicting activity of entities of interest. 
In our first case study, we analyze and predict email communication in an enterprise. 
In our second case study, we study interaction logs of users with an intelligent assistant on a mobile device, and aim to predict user task execution.

\subsubsection*{Chapter~\ref{ch:sigir}.~\acl{ch:sigir}}
Analyzing communication patterns can be helpful in answering questions like \emph{who was involved?}, and \emph{who knew what?}, which are central questions in the discovery process. 
Increased understanding of the aspects that guide communication, and being able to predict communication flows can be helpful in identifying atypical or unexpected communications, a valuable signal in the discovery process~\cite{disco2009}. 
In this chapter, the digital traces under study are enterprise email, and the entities of interest are emailers. 

We study the impact of contextual and content aspects of email in predicting communication.
More specifically, as context, we study the impact of leveraging the enterprise's ``communication graph'' for predicting email recipients.
The communication graph provides contextual clues of the creation of digital traces, e.g., the ``position'' of an emailer in a communication graph may implicitly capture their position in the company, and the strength of ties, or the proximity between two emailers may capture their professional or social relations. 
Next, we study the impact the content of the digital traces, i.e., we leverage the similarity between emails for predicting likely recipients.  
In this chapter, we answer the following research question: 

\begin{enumerate}[label=\textbf{RQ\arabic*},ref={RQ\arabic*},resume]
\item \acl{rq:sigir}\label{rq:sigir}
\end{enumerate}

\noindent
To answer this question, we present a hybrid model for email recipient prediction, that leverages both the information from the communication graph of the email network, and the Language Models (LM) of the emailers, estimated with the content of the emails that are sent by each user. 
We find that both the context and content of digital traces provide a strong baseline for predicting recipients, but are complementary, i.e., combining both signals achieves the highest accuracy for predicting the recipients of email. 
We obtain optimal performance when we incorporate the number of emails a recipient has received so far, and the number of emails a given sender sent to a recipient at that point in time in our model.

\subsubsection*{Chapter~\ref{ch:umap}.~\acl{ch:umap}}
Our second case study revolves around a proliferating type of digital traces; user interaction logs with mobile devices, and more specifically, user interactions with a personal intelligent assistant. 
With the rise of mobile devices in E-Discovery, an ever important task is to understand real-life behavior of people through traces from their mobile devices. 

Intelligent assistants are becoming ubiquitous; Google's Now, Apple's Siri, Amazon's Echo, Facebook's M, and Microsoft's Cortana have all been introduced in rapid succession over the last few years. 
In this chapter, we study user interaction logs with Microsoft's Cortana. 
Through their personal, human-like and conversational nature, intelligent assistants are more closely embedded in a user's daily life and activity than some of the technologies they borrow from (e.g., search engines, calendar, planning, and time management apps). 
By closing the gap between a user's offline and online world, intelligent assistants have a potentially large impact on a user's life, such as on her productivity, time management, or activity planning. 
The digital traces left behind through interacting with personal assistants may thus contain important signals related to the users' offline real-world activities. 
We focus on studying Cortana's reminder service, as reminders represent tangible traces of people's real-life (planned) activities and tasks. 
Understanding the temporal patterns related to reminder setting and task execution can help in inferring and understanding people's behavior in the real world. 

We aim to identify common categories of tasks that people remind themselves of, and study temporal patterns linked to the types of tasks people execute. 
In this chapter, we answer the following research question: 

\begin{enumerate}[label=\textbf{RQ\arabic*},ref={RQ\arabic*},resume]
\item \acl{rq:umap}\label{rq:umap}
\end{enumerate}

\noindent
More specifically, we apply a data-driven analysis to identify a body of common tasks types that give rise to the reminders across a large number of users. 
We arrange these tasks into a taxonomy, and analyze their temporal patterns. 
Furthermore, we address a prediction task, and much like the work presented in Chapter~\ref{ch:sigir}, we study the impact of both the content and contexts of the digital traces. 
We show that the time at which a user creates a reminder (context) is a strong indication of when the task is scheduled to be executed.
However, including the description of the task reminder (content) further improves prediction accuracy. 
 
\section{Main Contributions}
\label{section:introduction:contributions}
In this section, we list the theoretical, technical, and empirical contributions made in this thesis.

\subsubsection*{Analyses and Algorithms}

\begin{description}
\item[A large-scale study of emerging entities] 
In Chapter~3 we study a large dataset which contains over 70,000 entities that emerge in a timespan of over 18 months in over 579M documents, and show that entities emerge in broadly two types of patterns: with an initial burst of increased attention leading up to incorporating in Wikipedia, or a more gradual pattern, where the attention builds up over time. 
Furthermore, we identify characteristic differences between how entities emerge in news and in social media streams. 
Finally, we show that specific entity types are more strongly associated with specific emergence patterns. 
\item[Unsupervised method for generating pseudo-ground truth using EL] In Chapter~\ref{ch:ecir} we propose an unsupervised method that uses an EL method for generating training material for a named-entity recognizer to detect entities that are likely to become incorporated in a KB. 
Our method can be applied with any trainable Named-Entity Recognition (NER) model and EL method that is able to output a confidence score for a linked entity. 
Furthermore, our method is not dependent on human annotations that are necessarily limited and domain and language specific, is not restricted to a particular class of entities, and is suitable for domain and/or language adaptation as it does not rely on language specific features or sources. 
\item[Dynamic collective entity representations] 
In Chapter~\ref{ch:wsdm} we employ collective intelligence to construct ``dynamic collective entity representations,'' i.e., we dynamically create entity representations that encapsulate the different ways of how people refer to or talk about the entity. In doing so, we show we can bridge the gap between the words used in a KB entity description and the words used by people to refer to entities. 
\item[Generative hybrid model for email recipient prediction] 
In Chapter~\ref{ch:sigir} we propose a hybrid generative model aimed at calculating the probability of an email recipient, given the sender and the content of the email. 
We show how both the communication graph properties as email content properties contribute to a highly accurate prediction of email recipients. 
\item[Large-scale study of user interaction logs with a personal intelligent assistant] 
In \\
Chapter~\ref{ch:umap} we present the first large-scale study of the creation of common task reminders. 
We show that users largely remind themselves for short-term chores and tasks, such as to go somewhere, communicate, or perform daily chores. 
We develop a taxonomy of types of time-based reminders, facilitated by the data we have about the actual reminders created by a large populations of users. 
We show how reminders display different temporal patterns depending on the task type that they represent, the creation time of the reminder, and the terms in the task description. 
We study temporal patterns in reminder setting and notification, demonstrating noteworthy patterns, which we leverage to build predictive models. The models predict the desired notification time of reminders, given the reminder text and creation time. 
\end{description}

\subsubsection*{Empirical Contributions}

\begin{description}
\item[A large-scale study of emerging entities] 
In Chapter~\ref{ch:umap} we apply a hierarchical agglomerative clustering method to over 70,000 entity document-mention time series, and uncover two distinct patterns of emerging entities. 
We discover two distinct patterns of how entities emerge in public discourse. 
\item[Two sampling methods for pseudo-ground truth] Chapter~\ref{ch:ecir} presents two sampling methods for automatically generated training data.
We show that (i) sampling based on the textual quality improves performance of NER and consequently performance of predicting emerging entities, and (ii) sampling based on the confidence score of the EL method which provides the pseudo-training data labels results in fewer labels but better performance. 
\item[Effect of KB size in emerging entity prediction] In Chapter~\ref{ch:ecir} we study the effect of the size of the KB (i.e., the amount of prior knowledge) on predicting newly emerging entities, and find consistent and stable precision regardless of KB size, which justifies our emerging entity prediction method that assumes incomplete data by design. 
\item[Dynamic collective entity representations] In Chapter~\ref{ch:wsdm} we show entity representations enriched with external descriptions from various sources better capture how people search for entities than their original KB descriptions. Furthermore, we show that informing the ranker of the expansion state of the entity, i.e., the number and type of additional descriptions the entity holds, increases retrieval effectiveness.
\item[Email recipient prediction] In Chapter~\ref{ch:sigir} we show that combining communication graph with email content features achieves optimal predictive power. We show that the number of email received by a recipient is an effective method for estimating the prior probability of observing a recipient, and the number of emails sent between two users is an effective way of estimating the ``connectedness'' between these two users, and is a helpful signal in ranking recipients. 
\end{description}

\section{Thesis Overview}
\label{section:introduction:overview}
The first chapter, which is the one you are currently reading, introduces and motivates the research topic of the thesis: automated methods for sense-making of digital traces. 
Furthermore, this chapter provides an overview of the main contributions, content, and origins of the further chapters in this thesis. 
Chapter~\ref{ch:background} discusses the background and related work that serves as a base for the following chapters.
We describe 
Information Retrieval (IR), 
Semantic Search, 
Entity Linking (EL), and 
user modeling and 
analysis. 
The core of this thesis consists of two parts.

\noindent
In Part~\ref{pt:1} of this thesis, we study emerging entities and incomplete knowledge bases. 
It consists of three chapters.
Chapter~3
 presents an empirical study of temporal patterns of entities as they emerge in online text streams, before they are incorporated into Wikipedia. 
Chapter~\ref{ch:ecir} presents a method of leveraging an incomplete reference KB to generate pseudo-training data for training a system for discovering emerging entities, similar to the reference entities but absent from the KB. 
Chapter~\ref{ch:wsdm} presents a novel method of leveraging collective intelligence and learning from past behavior of users to construct dynamic entity representations, aimed at improving entity retrieval effectiveness. 

Part~\ref{pt:2} of this thesis revolves around analyzing and predicting human behavior from digital traces. 
Chapter~\ref{ch:sigir} presents a case study of enterprise email communication, and provides insights into the different aspects that guide communication between people. 
Chapter~\ref{ch:umap} presents a large-scale user log study of an intelligent personal assistant, and shows the types of task people tend to remind themselves about. 

Finally, Chapter~\ref{ch:conclusions} concludes this thesis, where we summarize the content and findings of this thesis, discuss the limitations of the presented work, and briefly reflect onto future work. 

The two parts of this thesis (Part~\ref{pt:1} and Part~\ref{pt:2}) are self-contained and form independent parts. 
For the reader's convenience, both the background chapter (Chapter~\ref{ch:background}), and the conclusions (Chapter~\ref{ch:conclusions}) follow this structure, i.e., the relevant background material, conclusions, and future work parts are organized separately for each part, so that both parts can be read independently. 
 
\section{Origins}
\label{section:introduction:origins}

This thesis is based on five publications~\cite{graus-generating-2014,Graus:2014:RRE:2600428.2609514,1701.04039,Graus:2016:APT:2930238.2930239,Graus:2016:DCE:2835776.2835819}.
In addition, it draws on ideas from four others~\cite{Graus:2014:SSE:2633211.2634354,graus2012context,graus2013semantic,graus2013yourhistory}.
This section lists for each chapter the publication it is based on, as well as the contributions of the author and co-authors. 

\paragraph{Chapter~\ref{ch:plos}}
 is based on \emph{The Birth of Collective Memories: Analyzing Emerging Entities in Text Streams}, currently under submission at the Journal of the Association for Information Science and Technology
 (JASIST).
This publication is written with Odijk and de Rijke. 
Graus wrote the clustering and analysis code. 
Odijk contributed to the data parsing pipeline (i.e., adding timestamps to entity annotations). 
Graus performed the experiments and the analysis. 
All authors contributed to the text. 

\paragraph{Chapter~\ref{ch:ecir}} is based on \emph{Generating Pseudo-ground Truth for Predicting New Concepts in Social Streams} which was published at ECIR 2014 by Graus, Tsagkias, Weerkamp, Buitinck, and de Rijke. 
Graus implemented the general pipeline and algorithms, Tsagkias and Weerkamp contributed code to the sampling methods. 
Graus performed the experiments and analyses. 
All authors contributed to the text. 

\paragraph{Chapter~\ref{ch:wsdm}} is based on \emph{Dynamic Collective Entity Representations for Entity Ranking} which was published at WSDM 2016 by Graus, Tsagkias, Weerkamp, Meij, and de Rijke. 
Graus implemented the general pipeline and algorithms, Tsagkias and Weerkamp contributed code to the sampling methods. 
Graus performed the experiments and analyses. 
All authors contributed to the text. 

\paragraph{Chapter~\ref{ch:sigir}} is based on \emph{Recipient Recommendation in Enterprises Using Communication Graphs and Email Content} which was published at SIGIR 2014 by Graus, van Dijk, Tsagkias, Weerkamp, and de Rijke. 
Graus implemented the general pipeline and algorithms, Tsagkias and Weerkamp contributed code to the sampling methods. 
Graus performed the experiments and analyses. 
All authors contributed to the text. 

\paragraph{Chapter~\ref{ch:umap}} is based on \emph{Analyzing and Predicting Task Reminders} which was published at UMAP 2016 by Graus, Bennett, White, and Horvitz.
Graus performed the log-analysis, developed the task type taxonomy, implemented the algorithms for task reminder execution time prediction. 
Graus performed the experiments and analyses. 
All authors contributed to the text. 
This work was done while on an internship at Microsoft Research. \\

\noindent
Finally, this thesis also indirectly builds on publications on 
entity linking for generating user profiles for personalized recommendation of events~\cite{graus2013yourhistory}, 
a context-based entity linking method~\cite{graus2012context}, 
entity linking of search engine queries~\cite{Graus:2014:SSE:2633211.2634354}, and 
tweets~\cite{microposts2015_neel_garbacea.ea:2015}.
And publications around Semantic Search for E-Discovery~\cite{vandijkdesi6,graus2013semantic}.  \clearpage{}

\clearpage{}

\chapter{Background}
\label{ch:background}

\begin{flushright}
\rightskip=1.8cm``Nescire autem quid ante quam natus sis acciderit, \\ id est semper esse puerum.'' \\
\vspace{.2em}
\rightskip=.8cm\textit{---Cicero, De Oratore, XXXIV}
\end{flushright}
\vspace{2em}

\noindent
The work presented in this thesis sits at the crossing between the fields Information Retrieval (IR) and Natural Language Processing (NLP).
This thesis deals with enabling and improving access to, and retrieval of information, and with improving and leveraging computer systems' ability to process, understand, and extract information from natural language. 
In this chapter, we present all relevant background and related work that serves as a basis for understanding this thesis. 
We start at the very basics, with a high level explanation of Information Retrieval and the current state of affairs and challenges, in Section~\ref{sec:IR}.
Next, we move on to explain Semantic Search, Entity Linking (EL), and Knowledge Bases in Section~\ref{sec:SS}.
In Section~\ref{sec:rwpart1} we discuss work related to the first part of this thesis, on analyzing, predicting, and retrieving emerging entities. 
Finally, we discuss the related work for the second part of this thesis on analyzing and predicting user behavior from digital traces, in Section~\ref{sec:userbehavior}. 

\section{Information Retrieval}
\label{sec:IR}
IR is most prominent in our everyday lives in the search engines people consider their entry point to the internet. 
Sometimes literally: 
a large share of queries issued to search engines are so-called \emph{navigational queries} --- queries that represent the user's intent of visiting a specific website~\cite{Broder:2002:TWS:792550.792552}. 
To illustrate: the 10 most frequent queries issued to the Yahoo search engine include: \texttt{facebook}, \texttt{amazon}, and \texttt{yahoo}~\cite{Guy:2016:STA:2911451.2911525}. 
Modern-day search engines have no trouble addressing these kind of information needs, quickly serving us websites from indices of many billions of ``documents.'' 

\begin{figure}[t]
\includegraphics[width=\linewidth]{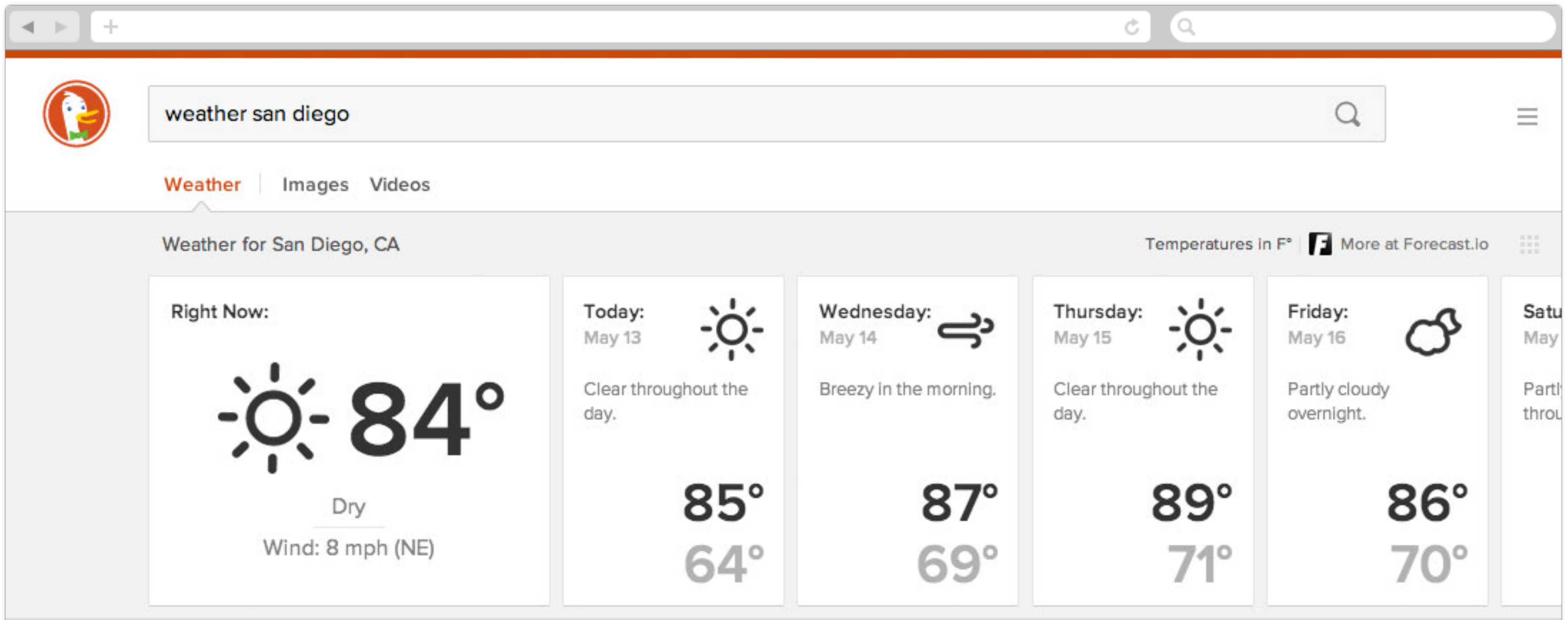}
\label{fig:ddg_weather}
\caption{An example of a \emph{knowledge card} displaying the weather forecast for San Diego on the search engine result page of DuckDuckGo.}
\end{figure}

But in web search, the landscape is rapidly changing. 
At Google, the volume of mobile search has surpassed that of desktop search~\cite{googleadwords}. 
Around 20\% of queries from mobile devices in the US are voice input~\cite{pichai2016:googleio_keynote}, and over half of the teenagers in the USA use voice input to issue their search queries. 
Mobile web search changes the game. 
Voice queries are longer on average, and use richer language than text queries~\cite{Guy:2016:STA:2911451.2911525}. 
Displaying results on a mobile device means a shift from the ``10 blue link''-paradigm towards newer, richer, more reactive, and more interactive ways of presenting the user with results.
One such example is the proliferation of \emph{knowledge cards} in many current web search engines and intelligent personal assistants such as Apple's Siri and Microsoft Cortana. 
Knowledge cards are small panels that summarize information to answer a query, e.g., the weather forecast, or central information around an entity of interest. 
See Figure~\ref{fig:ddg_weather} for an example knowledge card. 

Even outside mobile web search, in the more ``traditional'' setting of desktop web search, natural language queries (and in particular natural language questions) are on the rise~\cite{White:2015:QVQ:2740908.2742769}. 
Compared to the aforementioned navigational queries, understanding the more complex information needs that come with longer queries is a more challenging task. 
Web search engines do not adequately support more complex exploratory search scenarios~\cite{HassanAwadallah:2014:SCS:2661829.2661912}. 

Outside the domain of web search, plenty more complex information needs and search tasks exist, e.g., 
(re-)finding email messages~\cite{Carmel:2015:RTR:2806416.2806471}, 
desktop search~\cite{Dumais:2003:SIS:860435.860451}, 
researching prior art in patents~\cite{Joho:2010:SPU:1840784.1840789}, 
or finding related work for a background chapter in a PhD thesis~\cite{pertti2001}.

The shift towards more complex queries in web search and the proliferation of complex search scenarios in other domains are just two examples of where the traditional IR approach of matching words (in queries) to words (in documents), known as \emph{lexical matching}, no longer suffices. 
More complex information needs call for a better understanding of these information needs; they call for moving beyond words, towards better understanding of the intents, needs, and meaning that underlies queries and search behavior. 
\emph{Semantic search}~\cite{Li:2014:SMS:2692909.2692910} aims to fill this gap, by incorporating meaning (i.e., semantics), natural language structure, and (external) knowledge into the search process. 

\subsection{Retrieval Models} 
Before we delve into semantic search, we provide a brief overview of the traditional lexical matching-based retrieval models that we use in this thesis. 
Generally speaking, a retrieval model aims to assign a score of ``relevance'' (itself a very poor-understood notion) to documents ($d$) from an index, given a user-issued query ($q$). 
The most naive way of quantifying how ``relevant'' a document is to a query, is to count the occurrences of the query words in the document. 
This query term frequency-based method suffers one drawback: not all words are equally important in a document~\cite{ManningRaghavanSchuetze08}. 
Weighting terms according to their importance alleviates this. 

\subsubsection{TF-IDF}
One such way of term weighting is to assign higher weights to terms that occur in fewer documents: the underlying assumption is that words that occur frequently in a document, but in few documents overall, are the most representative of the document.
This assumption is modeled by combining a term's frequency with its \emph{inverse document frequency}~\cite{SparckJones:1988:SIT:106765.106782}. 
This is the Term-Frequency, Inverse Document Frequency (\emph{TF-IDF}) weighting scheme:
\begin{align}
Score(q,d) = \sum_{t \in q}tf(t,d) \cdot \textup{log}\frac{N}{df_t}.
\end{align}
\noindent
Here, $q$ is the user issued query, $d$ the document, $tf(t,d)$ the term frequency of query-term $t$ in $d$, $N$ the total number of documents (i.e., collection size), and $df_t$ the document frequency of query-term $t$ in the collection. 
This is a tried-and-tested, i.e., old, but still frequently used scoring method. 
Another scoring method (that represents the state-of-the-art in lexical matching-based retrieval), Okapi BM25~\cite{okapi-at-trec-3}, shares the same notion of term frequency and inverse document frequency for estimating the relevance of a document to a query, with more elaborate techniques to account for query and document length.

\subsubsection{Language Modeling and the Query-Likelihood Model}
Another way to estimate the relevance of a document to a query is to use language modeling. 
A statistical language model (LM) is a probability distribution over words~\cite{rosenfeld2000}, which we can use to estimate the probability (i.e., how likely it is) that a query is generated by a particular LM (i.e., particular distribution). 
If we compute for each document in a collection its own LM, we can rank the documents based on the probability of generating query $q$. 
This ranking method is known as the \emph{query likelihood model}, which postulates that documents with higher probabilities to generate the query should rank higher:
\begin{align}
	Score(q,d) = P(q \mid d) = \prod_{t \in q} P(t \mid d)^{n(t,q)},
\end{align}
\noindent
where the document LMs are derived from word occurrences in the documents, i.e., the probability of observing a word given a document ($P(w \mid d)$) is typically defined as:
\begin{align}
	P(w_i \mid d_j) = \frac{n(w_i, d_j)}{\sum_N n(w_i, d_j)},
\end{align}
\noindent
where $n(w_i, d_j)$ is the frequency of term $w_i$ in document $d_j$. 
Using these simple wordcounts yields a unigram LM, also known as the bag-of-words model, where the order of words (in both the query and the document) does not affect the score. 
Extending this model to incorporate word order yields $n$-gram models.

While these models are simple and computationally cheap, they suffer one major drawback. 
Both TF-IDF and the query likelihood model, but also more advanced models, e.g., BM25, rely on lexical matching, i.e., words from the query need to appear in the documents to get non-zero relevance scores.

\subsection{Evaluation} 
A fundamental aspect in IR is evaluation. 
In order to assess the performance and usefulness of novel retrieval models, methods, and algorithms, it is central to be able to compare the output of different methods (typically rankings), to measure differences and improvements. 
These evaluation metrics rely on the availability of \emph{ground truth}, i.e., a sets of assessments (e.g., relevance scores for documents per query), usually collected beforehand, which can serve as a golden standard to compare against. 

The topic of evaluation in IR is huge, and to address everything is beyond the scope of this thesis. 
Here, we restrict to listing and describing the evaluation metrics we employ in this thesis, which we break down into set-based, rank-based, and classification evaluation metrics. 
We point the interested reader to~\cite{eval-chapter-manning} for a more comprehensive overview. 

\subsubsection{Set-based Metrics}

\textbf{Precision} corresponds to the fraction of retrieved items (e.g., documents) in the system's output that are relevant (i.e., the items that are in the \emph{ground truth} set). 
It is computed by taking the fraction of True Positives (TP), i.e., the set of retrieved items that are in the \emph{ground truth}, over all retrieved items of the system, i.e., the set of True Positives and False positives (FP), where False Positives correspond to the items that are in the system's output, but not in the \emph{ground truth}, i.e., 
$\frac{TP}{TP+FP}$. 

\textbf{Recall} is a related set-based metric, which corresponds to the fraction of relevant items in the system's output. 
It is computed by taking the fraction of True Positives (TP) over the set of True Positives and False Negatives (FN), where False Negatives correspond to items that are in the ground truth, but not in the system's output, i.e., $\frac{TP}{TP+FN}$.

We employ Precision and Recall to compute our method's effectiveness of predicting emerging entities in Chapter~\ref{ch:ecir}. 
 
\textbf{P@k} is a variation of Precision that is helpful in ranking scenarios where the set of ground truth items is very large (e.g., in a web search scenario).
P@k corresponds to the Precision of a system's output up to a certain rank ($k$). 
We employ P@k in Chapter~\ref{ch:wsdm}. 

\subsubsection{Rank-based Metrics}
Precision and Recall are set-based metrics, which do not take the ordering of results into account.
To turn the set-based Precision into a metric that does take order into account, we can compute the \textbf{MAP} (Mean-average Precision). 
First, Average Precision (AP) is the average Precision over all ranks (i.e., all values of $k$), up to the point where recall is 1 (i.e., the rank at which all relevant items are retrieved). 
By subsequently averaging AP over multiple sets (e.g., over multiple queries), we yield MAP. 
We employ MAP in Chapter~\ref{ch:wsdm} and Chapter~\ref{ch:sigir}. 

\subsubsection{Classification Metrics}
\textbf{Accuracy} is a classification metric, i.e., it measures how well a system assigns class labels to a set of samples, by computing the fraction of correct class labels (w.r.t. a set of ground truth labels) over all assigned class labels. 
We employ accuracy in Chapter~\ref{ch:umap}, where we address a multiclass-classification task. 
We distinguish between micro- and macro-averaged accuracy. 
\textbf{Macro-averaged accuracy} does not take different distributions in class membership into account, and simply computes the accuracy by taking the total set of correct predictions over the total number of predictions. 
\textbf{Micro-averaged accuracy} does take different class distributions into account, by computing the accuracy per class, and averaging these class-specific accuracies over the number of classes.

\section{Semantic Search}
\label{sec:SS}
With the rise of more complex queries, information needs, search tasks, and challenging domains, the lexical matching-based retrieval models described above may not suffice. 
Semantic Search, or ``search with meaning,'' aims to move beyond the constraints of simple keyword matching, by incorporating additional ``structure'' in the search process. 
This can be in terms of semantic matching, e.g., by applying topic modeling methods to project documents and queries in some (lower dimensional) semantic space, where matching can happen without direct lexical overlap between query and document~\cite{Li:2014:SMS:2692909.2692910}. 

In this thesis, however, we focus on one particular form of additional ``structure:'' real-world knowledge. 
By incorporating knowledge from external sources, e.g., knowledge bases such as Wikipedia, one can enrich text and improve matching~\cite{Dalton:2014:EQF:2600428.2609628,Hasibi:2016:EEL:2970398.2970406}. 
But incorporating real-world knowledge also better supports the exploratory search process that is inherent to E-Discovery~\cite{bontcheva2012making,Ahn2010383}.
In particular, the real-world entities that occur in documents are central to question answering from text (the questions being, e.g., the 5 W's)~\cite{Khalid:2008:INE:1793274.1793371}.

\subsection{Knowledge Bases and Entities}
Given a KB that describes real-world entities and concepts (e.g., Wikipedia), EL addresses the task of identifying and disambiguating occurrences of entities in text. 
Or, more specifically: 

\begin{displayquote}
Matching a textual entity mention, possibly identified by a named entity recognizer, to a knowledge base entry, such as a Wikipedia page that is a canonical entry for that entity~\cite{Rao2013}.
\end{displayquote}

\noindent
EL is a key component in modern-day applications such as semantic search and advanced search interfaces, and plays a major role in accessing and populating the Web of Data~\cite{eps271285}. 
It can also help to improve NLP tasks~\cite{JAIR:2009:Gabrilovich}, 
or to ``anchor'' a piece of text in background knowledge; authors or readers may find entity links to supply useful pointers~\cite{CIKM:2011:He}. 
Another application can be found in search engines, where it is increasingly common to link queries to entities to present entity-specific overviews~\cite{ICSC:2010:Balasubramanian}, 
or to improve ad-hoc retrieval~\cite{Dalton:2014:EQF:2600428.2609628,Hasibi:2016:EEL:2970398.2970406}. 

\noindent
EL can be traced back to several related tasks that preceded it, such as record linkage~\cite{doi:10.1080/01621459.1969.10501049} in the Database community, and coreference resolution~\cite{Soon:2001:MLA:972597.972602} in the NLP community. 
However, in its current form EL has received a lot of attention from the IR community, as well as industry, since the Text Analysis Conference (TAC) Knowledge Base Population track introduced an EL task in 2009~\cite{tac2009}. 

The EL task can formally be described as follows: 
given an entity mention $m$ (a term or phrase) occurring in reference document $d$, identify the entity $e$ from a knowledge base $KB$ that is the most likely referent of $m$.
In this thesis, EL plays a central role in Part~\ref{pt:1} (i.e., in Chapters~\ref{ch:plos}, \ref{ch:ecir}, and \ref{ch:wsdm}). 
EL consists of two distinct steps: 
(i) recognizing mentions of entities in the knowledge base in text (\emph{entity mention detection}), and subsequently 
(ii) linking them to their referent knowledge base entries (\emph{entity disambiguation}). 
We briefly explain the core challenges and methods for both steps.

\paragraph{Entity mention detection.}
The first step, known as entity mention detection, aims to identify the word sequences that refer to entities (\emph{entity mentions}). 
One approach is to apply \emph{named-entity recognition} (NER). 
NER has a rich history, and has been studied since the 1990s in the Message Understanding Conferences~\cite{Grishman:1996:MUC:992628.992709}.
Years later NER gained more traction as part of the CoNLL shared tasks, where winning systems achieved around 90\% accuracy of recognizing named-entities in news articles~\cite{TjongKimSang:2002:ICS:1118853.1118877}. 
NER methods typically rely on learning patterns in sequences of words, by leveraging the structure of language, 
e.g., the functions of words and their surrounding words (entity mentions are typically proper names~\cite{TjongKimSang:2002:ICS:1118853.1118877}), 
or features relating to their surface form (e.g., in many languages capitalized words are more likely to refer to entities). 

An alternative approach for \emph{entity mention detection} is so-called lexical matching, or dictionary-based mention detection~\cite{shen2014}. 
This method leverages the rich metadata of a KB such as Wikipedia, by creating a lexicon or dictionary that maps entities (i.e., Wikipedia page IDs) to entity surface forms (i.e., entity mentions). 
These surface forms are extracted from the rich variety of different ways used to refer to Wikipedia pages. 
For example, a single entity on Wikipedia can be represented by its: 
title, 
anchor texts from hyperlinks on other Wikipedia pages (i.e., the text that is used to hyperlink from one Wikipedia page to another), 
and redirect pages (mostly manually-added links that cover (common) misspellings or alternative names of Wikipedia pages), e.g., the Wikipedia page for \texttt{Kendrick Lamar} has redirect pages such as \texttt{K-Dot} (Kendrick Lamar's former stage name), 
and also misspellings, such as \texttt{Kendrick Lamarr} and \texttt{Kendrick Llama}. 

After extracting all these candidate entity mentions and generating the mappings of surface forms to entities, statistics on their usage can be further leveraged to estimate the probability that a (sequence of) word(s) is a reference to a KB entity (i.e., an entity mention). 
One commonly used and intuitive method of estimating the probability that an $n$-gram is an entity mention is the so-called \emph{keyphraseness}. 
It boils down to the fraction of the number of times an $n$-gram is used to refer to an entity (i.e., the $n$-gram is used as an anchor text in Wikipedia), over the number of times the $n$-gram occurs in Wikipedia (including the number of times it is not part of a hyperlink)~\cite{Mihalcea:2007:WLD:1321440.1321475}. 
To illustrate, the phrase \emph{Kendrick} appears 5,037 times in Wikipedia articles,\footnote{Wikipedia dump dated June 2014} of which it is part of a hyperlink (only) 24 times, resulting in \emph{Kendrick's} keyphraseness score of $\frac{24}{5,037}=0.005$. 
In contrast, the phrase \emph{Kendrick Lamar} appears 698 times, of which 501 times as a hyperlink, yielding a keyphraseness score, i.e., the prior probability of the phrase \emph{Kendrick Lamar} being an entity mention, of $\frac{501}{698}=0.718$. 

Lexical matching-based EL approaches have shown to be successful in the challenging domain of social streams~\cite{meij2012}, and have shown to be suitable for adaptation to new genres and languages~\cite{Odijk:2013aa}. 
As an added advantage, these methods are independent of language-dependent linguistic annotation pipelines, which are prone to cascading errors~\cite{Finkel:2006:SPC:1610075.1610162}. 
Their drawback, however, is that they are unable to identify entity mentions that are not in the lexicon, e.g., misspellings not covered by Wikipedia's redirect pages.

\paragraph{Entity disambiguation.}
Given the detection of (candidate) entity mentions, the next step is to assign the referent entities from the KB to the mentions, or alternatively, determine that the mention refers to an entity that is not in the KB. 
Also known as \emph{entity disambiguation}, this task is commonly approached by one of the following two approaches: 
(i) \emph{local} and 
(ii) \emph{global} entity disambiguation methods. 

Local methods attempt to resolve a mention (i.e., assign the correct entity to a mention) by only considering properties of a (candidate) entity mention and the candidate KB entities. 
A common local method of disambiguating entity mentions is by employing the \emph{commonness}-score~\cite{Mihalcea:2007:WLD:1321440.1321475}.
Commonness is a simple fraction of the number of times the entity mention is used as an \emph{anchor} for a specific entity $e$, over the number of times it is used as an anchor to any entity in the KB~\cite{medelyan2008}. 
Continuing with the previous example, the entity mention \emph{``Kendrick''} is used as an anchor text of a hyperlink 24 times, of which in 8 cases, the hyperlink points to the Wikipedia page \texttt{Kendrick, Idaho}, i.e., the probability that the mention \emph{``Kendrick''} refers to \texttt{Kendrick, Idaho} is $\frac{8}{24}=0.333$, whereas the probability that the mention refers to \texttt{Kendrick\_Lamar} is $\frac{2}{24}=0.083$. 

Commonness thus favors the most common target entity for a mention, which makes it work well in practice, particularly in domains that refer to popular entities by nature (e.g., news streams). 
Because of the emphasis on single occurrences of entity mentions, local EL methods have been shown to be highly effective in domains in which context is limited and/or noisy, e.g., microblog posts~\cite{meij2012}. 
Additional methods have been proposed to expand context when it is limited~\cite{bron:link11,cassidy2012analysis}. 

There is an increasing interest in approaches that link multiple entities in a document simultaneously~\cite{ji2011kbp}. 
These so-called \emph{global} methods are based on a notion of ``coherence'' in the set of entities in a document. 
There are several methods for defining the ``coherence'' of a set of entities. 
One is to rely on the structural information of the KB, e.g., the link or category structure~\cite{milne08:learn}, leveraging categories and contexts \cite{silviu-disambne:2007}, or by using graph-based metrics~\cite{hit}. 
The TAC Knowledge Base Population (KBP) track has seen many systems that incorporate global approaches to EL~\cite{ji2011kbp,Cucerzan2011,Cassidy2010,Fernandez2010,Radford2011}. 
Also, several publicly available EL frameworks that aim to resolve each entity in a document apply global methods, e.g., `DBpedia Spotlight'~\cite{isem2011mendesetal} and the `GLOW' framework \cite{Ratinov2011}.
The most common global approach uses entity ``relatedness''~\cite{milne08:learn}. 
Relatedness is based on the overlap of the sets of ``related'' entities of a pair of entities. 
Related entities can be, e.g., all entities that are linked to or from the Wikipedia page that represents the entity of interest. 

A known challenge of global methods is that entity mentions in the document can themselves be ambiguous, i.e., the number of pairwise comparisons between candidate entities for multiple mentions is exponential in their number. 
The problem of optimizing the global coherence function is NP-hard \cite{silviu-disambne:2007}.
A common strategy to deal with this problem is to decrease the search space, e.g., by only considering non-ambiguous entity mentions (i.e., mentions that are only associated with a single target entity)~\cite{milne08:learn}. 
While this evidently reduces complexity, the presence of such entities cannot always be assumed. 
Another approach is to resolve the ambiguous entity mentions to their most popular candidate~\citep{dai}. 
\citet{silviu-disambne:2007} introduces a method for joint disambiguation that does not consider all combinations of assigning one disambiguation per surface form at a time. 
Instead, all possible disambiguations of a surface form (candidates) are considered simultaneously. 
Another method is to set a threshold on the number of entities to consider per mention (e.g., at most 20 candidate entities~\cite{Ratinov2011}). 
\citet{Ratinov2011} train a local approach to resolve individual entity mentions, before applying the global approach for coherence. 

\paragraph{Entity linking in social streams.}
The domain of social streams brings additional challenges to the EL task.
Whereas typical NER approaches perform in the upper 90\% range in terms of accuracy on ``clean,'' grammatically correct and correctly spelled news articles, the accuracy drops considerably when performing NER on misspelled, short, and noisy text, such as social media posts. 
The noisy character and unreliable capitalization of social streams degrades the effectiveness of NER methods~\cite{Derczynski:2013:MNI:2481492.2481495,Finkel:2006:SPC:1610075.1610162,Derczynski201532}.
One approach to addressing this is to tailor the linguistic annotation pipeline to Twitter, which relies heavily on large amounts of training data~\cite{Ritter12,Bontcheva:2013aa}. 
Another approach to combat the lack of context, proposed by \citet{cassidy2012analysis}, is to expand a social media post with additional posts of the author, and additional posts found by a clustering algorithm, before optimizing the EL system's output for global coherence.
Finally, incorporating spatiotemporal features~\cite{TACL323} has been shown to improve EL effectiveness in social streams. 
The method we follow in this thesis however, embraces the little context and ignores the notion of global coherence, opting instead for machine learning methods with features that focus on learning mappings between individual $n$-gram and entities~\cite{meij2012}. 

\section{Emerging Entities}
\label{sec:rwpart1}
In Part~\ref{pt:1} of this thesis, we focus on analyzing, predicting, and retrieving ``emerging entities,'' 
i.e., entities that are not (yet) part of the KB. 
This falls into the domain of KB construction (KBC) or (cold start) knowledge base population (KBP), which encompasses the ambitious goal of creating a KB from raw input data (e.g., a collection of documents). 
KBP and KBC span many smaller subtasks, e.g., (new) entity mention detection, entity clustering, relation extraction, and slot filling~\cite{Niu:2012:ELK:2607592.2607595,Dong:2014:KVW:2623330.2623623}. 

We narrow our focus, and address the analysis, prediction, and representation of emerging entities in three successive chapters.
More specifically, in Chapter~\ref{ch:plos} we study whether there exist common temporal patterns in how entities emerge in online text streams, 
in Chapter~\ref{ch:ecir} we propose a method that leverages EL to generate training data for predicting new mentions of emerging entities in social streams, 
and finally, in Chapter~\ref{ch:wsdm} we study a method for improving entity ranking by mining additional entity descriptions from different external sources, e.g., social media and search engine logs. 

\subsection{Analyzing Emerging Entities} 
In the first chapter, we analyze entities as they ``emerge;'' i.e., we study the time series of entity mentions in online text streams in the time span between their first mention (i.e., the entity surfaces) and the moment at which the entity is incorporated into the KB.

Previous work on studying the expansion of Wikipedia through the addition of new pages studies the phenomenon from the perspective of Wikipedia itself, e.g., by analyzing how newly created articles fit in Wikipedia's semantic network, studying the relation between activity on talk pages and the addition of new content to articles, or by studying controversy and disagreement on new content through ``edit wars''~\cite{10.1371/journal.pone.0141892,Keegan01052013,10.1371/journal.pone.0038869,10.1371/journal.pone.0030091}. 

Studying emerging entities has received considerable attention from the natural language processing and information retrieval communities. 
Most notably, different methods and systems for identifying and linking unknown or emerging entities have been proposed~
\cite{NakasholeTW13,Hoffart:2014:DEE:2566486.2568003,Lin:2012:NNP:2390948.2391045,voskarides-query-dependent-2014}.
More recently, \citet{Farber2016} formalize and analyze the specific challenges and aspects that come with linking emerging entities, while \citet{reinanda-document-2016} study the problem of identifying relevant documents for known and emerging as new information comes in.

\subsection{Predicting Emerging Entities} 
Addressing the task of detecting entities that are not in the KB, i.e., identifying emerging entities, can be traced back to the introduction of the ``NIL clustering'' subtask in the TAC KBP Entity Linking track in 2011~\cite{ji2011kbp}. 
The task was a straight-forward EL task: participants were given a reference KB (derived from Wikipedia), a collection of documents (mostly news articles), and offsets for entity mentions that occur in the documents. 
The task was to decide for each mention whether it refers to an entity that exists in the reference KB (and if yes, which), or whether it refers to a ``NIL entity:'' an entity that is not in the KB. 
In the latter case, the system needs to assign a new ID (NIL ID) to the mention, and assign other mentions that refer to the same underlying real-world entity to the same NIL ID (i.e., cluster mentions to NIL IDs). 
Since common EL methods rely on ranking candidate entities to mentions, the inclusion of mentions for out-of-KB entities is typically approached by learning a threshold (for the EL system's confidence or retrieval score) to determine whether or not the mention refers to an entity already described in the KB. 
This approach was chosen, e.g., by \cite{Bunescu06usingencyclopedic,Kulkarni:2009:CAW:1557019.1557073,NakasholeTW13}. 

In the TAC KBP Entity Linking track the entity mention detection step, shown to be the bottleneck in state-of-the-art EL systems~\cite{hachey2013}, was not part of the task. 
However, in a realistic scenario, when entity mentions are not given, Wikipedia lexicons and/or lexical matching-based approaches cannot detect unseen entity mentions of new entities, which means different approaches need to be leveraged, e.g., named-entity recognition (NER). 
The method for detecting emerging entities with ambiguous names, presented by \citet{Hoffart:2014:DEE:2566486.2568003}, leverages lexical matching methods for detecting entity mentions. 
However, in this approach, only ambiguously named emerging entities can be detected, making it complementary to the method we present in Chapter~\ref{ch:ecir}, which focuses on detecting new (i.e., unseen) mentions of emerging entities. 
\citet{Lin:2012:NNP:2390948.2391045} learn to identify entity mentions of absent entities by leveraging Google Books' n-gram statistics. 
Their method relies on part-of-speech tagging to identify noun phrases in documents, and deciding for each noun phrase that cannot be linked to a Wikipedia article whether or not it refers to an (absent) entity. 
Another way to model the task of detecting emerging entities is \emph{entity set extraction}, where given a seed list of entity mentions, the task is to retrieve more entities that are similar to the seed set~\cite{Pantel:2009:WDS:1699571.1699635}. 
However, the task differs as it does not rely on a KB, and consequently, the rich structure and metadata that comes with it. 

\paragraph{Automatically generated pseudo-ground truth.}
The approach we propose in Chapter~\ref{ch:ecir} revolves around automatically generating training data using an EL system. 
Automatically generating training data for machine learning methods has been a longstanding and recurring theme in machine learning, dating back to the 1980s~\cite{tague:80}. 

The idea of using EL systems for generating pseudo-ground truth, or \emph{silver standard} training data, is not new either. 
\citet{guo-chang-kiciman:2013:NAACL-HLT} used an EL system to collect additional context words for entities, to aid the disambiguation step. 
Their mention detection step relies on the Wikipedia lexicon, meaning their system is unable to detect unseen entity mentions. 
Similarly, \citet{Zhang:2010:ELL:1873781.1873926} apply EL to improve the disambiguation step in EL, by replacing unambiguous mentions in entity linked documents with ambiguous ones to generate artificial but realistic training data. 
\citet{Zhou:2010:RSF:1873781.1873931} generate training data from Wikipedia to improve the disambiguation step as well.
More specifically, they consider Wikipedia links as positive examples, and generate additional negative samples by taking each other entity that may be referred to by the same anchor text.

Attempts have been made to simulate human annotations for generating training data for named-entity recognition, too. 
\citet{Kozareva:2006:BNE:1609039.1609041} uses regular expressions for generating training data for NERC. 
\citet{Nothman2013151} leverage the anchor text of links in between two same articles in different Wikipedia translations for training a multilingual NERC system. 
\citet{Wu:2009:DAB:1699648.1699699} investigate generating training material for NERC from one domain and testing on another.
\citet{Becker05optimisingselective} study the effects of sampling automatically generated data for training NER. 

What is new to the method for predicting mentions of emerging entities we propose in Chapter~\ref{ch:ecir}, is that it directly optimizes for the task at hand, i.e., we generate training data for a named-entity recognizer to detect a specific type of entity, i.e., entities that are worthy of being added to the KB.

\subsection{Constructing Entity Representations for Effective Retrieval} 
A task closely related to EL is entity ranking: in the end, most EL systems approach the disambiguation problem as an entity ranking problem. 
However, there is a huge difference between ranking entities for a candidate entity mention, and ranking entities for a user-issued query. 
In particular, there may be a vocabulary gap between how users refer to entities, and how entities are described in knowledge bases.
The task of retrieving the correct entity for user-issued queries is of central importance in the general web search domain~\cite{ERD2014}, but also in a discovery or exploratory search setting, where analysts may be interested in retrieving entities with partial names, or nicknames. 

As with many NLP and information extraction tasks, research into entity ranking too has flourished in recent years, largely driven by benchmarking campaigns, such as
the INEX Entity Ranking track~\cite{inex2007,inex2008,inex2009}, and 
the TREC Entity track~\cite{trec2010,trec2011}. 
Entity ranking is commonly addressed --- much like EL --- by exploiting content and structure of knowledge bases, for example by 
including anchor texts and inter-entity links~\cite{shen2014}, 
category structure~\cite{scholer2004,balog2010}, 
entity types~\cite{Kaptein:2010:ERU:1871437.1871451}, or 
internal link structure~\cite{Vercoustre:2008:ERW:1363686.1363943}. 

More recently, researchers have also started to focus on using ``external'' information for entity ranking, such as query logs, i.e., users' past search behavior. 
\citet{billerbeck2010} use query logs of a web search engine to build click and session graphs, and walk this graph to answer entity queries. 
\citet{Mottin:2013:ERU:2595588.2595595} use query logs to train an entity ranking system using different sets of features. 
\citet{hong2014} follow a similar line of thought and enrich their knowledge base using linked web pages and queries from a query log. 

The entity ranking task is usually approached in a static setting, in which web pages and queries are added to the knowledge base before any ranking is performed. 
One of the few initial attempts to bring in time in entity ranking is a position paper by \citet{Balog:2012:USK:2390148.2390150}, who propose temporally-aware entity retrieval, in which temporal information from knowledge bases is required. 
Finally, closing the loop, EL can be applied to queries to improve entity ranking~\cite{Hasibi:2016:EEL:2970398.2970406}. 
Entity ranking is inherently difficult due to the potential mismatch between the entity's description in a knowledge base and the way people refer to the same entity when searching for it. 

\paragraph{Document expansion.}
One way to bridge this gap is to expand (textual) entity representations with terms that people use to refer to entities. 
Collecting terms associated with documents, and adding them to the document (document expansion) is in itself not a novel idea. 
Previous work has shown that it improves retrieval effectiveness in a variety of retrieval tasks. 
\citet{singhal1999} are one of the first to use document expansion in a (speech) retrieval setting, motivated by the vocabulary mismatch introduced by errors made by automatic speech transcription. 
Ever since, using external sources for expanding document representations has been a popular approach to improve retrieval effectiveness. 

In particular, it was shown that anchors can improve ad-hoc web search~\cite{westerveld2001,wu2012,metzler2009,eiron2003}.
\citet{kemp2002} state that document transformation using search history, i.e., adding queries that lead to the document to be clicked, brings documents closer to queries and hence improves retrieval effectiveness. 
Similarly, \citet{xue2004} study the use of click-through data by adding queries to clicked document representations. 
In this case, the click-through data includes a score that is derived from the number of clicks the query yields for a single document. 
\citet{gao2009} follow a similar approach, but add smoothing to click-through data to counter sparsity issues. 
\citet{amitay2005} study the effectiveness of query reformulations for document expansion by appending all queries in a reformulation session to the top-$k$ returned documents for the last query. 
\citet{scholer2004} propose a method to either add additional terms from associated queries to documents or replace the original content with these associated queries, all with the goal of providing more accurate document representations.

Looking at other sources for expansion, \citet{bao2007} improve web search using social annotations (tags). 
They use the annotations both as additional content and as a popularity measure. 
\citet{lee2013} explore the use of social anchors (i.e., content of social media posts linking to a document) to improve ad hoc search. 
\citet{noll2008} investigate a variety of ``metadata'' sources, including anchors, social annotations, and search queries. 
They show that social annotations are concise references to entities and outperform anchors in several retrieval tasks. 
\citet{efron2012} show that document expansion can be beneficial when searching for very short documents (tweets). 

As an alternative to expanding entity representations, more recently, advances in neural nets and deep learning have given rise to (continuous) entity representations, i.e., entity embeddings~\cite{Bordes_learningstructured,Lin:2015:LER:2886521.2886624,NIPS2013_5028}. 
One of the drawback of these methods is that it is not trivial to employ them in an online, dynamic scenario, where representations may continually change.

\paragraph{Fielded retrieval.}
A common approach to incorporate different document expansions into a single document representation is to create fielded documents~\cite{robertson2004,zaragoza2004}. 
Based on fielded documents, a variety of retrieval methods have been proposed. 
\citet{robertson2004} introduce BM25F, the fielded version of BM25, which linearly combines query term frequencies over different field types. 
\citet{broder2010} propose an extension to BM25F, taking into account term dependencies. 
\citet{svore2009} use a machine learning approach for learning BM25 over multiple fields, the original document fields (e.g., title and body) and so-called ``popularity'' fields (e.g., anchors, query-clicks). 
\citet{macdonald2013} compare the linear combination of term frequencies before computing retrieval scores to directly using retrieval scores in the learning to rank setting and show that it is hard to determine a clear winner. 
\citet{perezaguera2010} note how one of the challenges in structured IR is the fact that field importance differs among collections and that different collections means field importance needs to be optimized per collection.

\section{Analyzing and Predicting Activity}
\label{sec:userbehavior}
In Part~\ref{pt:2} of this thesis, 
we study the link between their digital traces and activities. 
Mining digital traces has a rich history in exploratory analyses and improving search engines and online products, as discussed in Chapter~\ref{ch:introduction}. 
More specifically, large-scale user logs from many users have been used for a range of purposes to improve online services and advance our understanding of how people use systems. 
Search engine queries and search-result clicks have been used 
to understand how people seek information online~\cite{White:2007:IBV:1242572.1242576}, 
train search engine ranking algorithms to better serve user needs~\cite{Agichtein:2006:IWS:1148170.1148177,Joachims:2002:OSE:775047.775067}, and more generally, 
teach us about how humans behave in the world~\cite{Richardson:2008:LWT:1409220.1409224}. 
Although large-scale log analysis of online behavior has typically focused on search and browsing activity, recent work has targeted the large-scale usage of communication tools too, such as 
instant messaging~\cite{Leskovec:2008:PVL:1367497.1367620} and email~\cite{Koren:2011:ATE:2020408.2020560}. 

In this thesis, we focus on mining digital traces with the aim to understand the underlying behavior that gives rise to the traces.
We address an email prediction task in Chapter~\ref{ch:sigir}  
and aim to gain insights into usage patterns with the reminder service of an intelligent personal assistant in Chapter~\ref{ch:umap}. 

\subsection{Mining Email}
Email has a rich history as an object of study in computer science, once more spurred by the availability of data, specifically, the acquisition and subsequent release of the ``Enron corpus,'' by computer scientist Andrew McCallumn~\cite{klimt2004enron} for a reported sum of \$10,000~\cite{nyt-enron}. 

\noindent
Research initially focused on information extraction tasks such as 
recognizing speech acts~\cite{Cohen04learningto,Carvalho:2005:CCE:1076034.1076094}, 
people-related tasks, such as 
name recognition and entity resolution~\cite{diehl06nrr,minkov05extractingpersonal,minkov06email,elsayed-oard-namata:2008:ACLMain}, 
contact information extraction~\cite{balog06expertmail,culotta04extractingsocial}, 
identity modeling and resolution~\cite{elsayed06identity}, 
discovery of peoples' roles~\cite{leuski04email}, 
and finally email-related tasks, such as
automated classification of emails into folders~\cite{bekkerman2004}, 
e.g., by leveraging the social network features of the email network~\cite{Yelupula:2008:SNA:1593105.1593229}. 
The Enron email collection was also used very directly in an E-Discovery task in the TREC Legal~\cite{treclegal2009,treclegal2010} tracks, which revolved around finding responsive documents for a given production request (in litigation). 

Studying email collections as communication graphs (or social networks) has yielded insights into 
how communication patterns change and relate to real-world events (e.g., how interpersonal communication intensified during Enron's crisis period~\cite{Diesner2005}), 
how important nodes (= email addresses) or influential members of the network can be discovered by looking at network properties~\cite{Shetty:2005:DIN:1134271.1134282}, 
and how one can discover shared interests among members of an email network~\cite{schwartz93graph}. 

More recently, research interest in email has been rekindled. 
Spurred by publications from industry, where access to huge amounts of real email data and traffic has resulted in work on 
automated classification of email into folders~\cite{Grbovic:2014:MFY:2661829.2662018}, 
predicting recipients' follow-up actions for emails~\cite{DiCastro:2016:YGM:2835776.2835811,Dabbish05understandingemail},
or predicting an email's (personalized) ``importance''~\cite{Aberdeen10thelearning}, and 
inferring the activities that govern email communication~\cite{activity-modeling-email}. 

Finally, work on recipient prediction (the task we address in Chapter~\ref{ch:sigir}) generally included the additional constraint that one or more seed recipient is given, a task also known as CC-prediction \cite{pal2006cc,6273570}.  
Previous attempts at the type of recipient prediction we address leverages either 
the communication network, constructed from previously sent emails~\cite{Roth:2010:SFU:1835804.1835836}, 
or the content of the current email~\cite{Carvalho:2008:RUI:1793274.1793314}. 
Finally, motivated by privacy concerns previous work typically addresses recipient prediction by restricting to a sender's ego network for prediction (i.e., only considering the local network of the sender, as opposed to the global communication network). 
In Chapter~\ref{ch:sigir} we ignore these constraints, motivated by our offline, batch, discovery scenario, and rather attempt to gain insights into enterprise-wide communication patterns, and in particular the importance of communication graph properties and email content features. 

\subsection{User Interaction Logs}
As mentioned in Chapter~\ref{ch:introduction}, interaction logs with personal intelligent assistants may provide rich digital traces for inferring people's real life activities. 
In the case of intelligent assistants, analyzing user logs may support inferring the users' intents~\cite{Armentano:2009:RUI:1611644.1611664} or current activities~\cite{Partridge2009}. 
Where previous work addressed modeling long-term goals~\cite{Barua2014}, the Cortana reminder logs we study in Chapter~\ref{ch:umap} may help us in understanding short term goals. 
To study the relation between a user's reminders and real life activities and tasks, several areas of research are relevant. 
We focus largely on research on memory and (completing planned) tasks, and 
review research in the following areas: 
(i) reminders, 
(ii) memory aids, and
(iii) prospective memory. 

\paragraph{Reminders.}
A number of systems have been developed to help remind people of future actions~\cite{DeVaul00thememory,Dey:2000:CCS:647986.757284,jogger-models-for-context-sensitive-reminding,lamming1995,df8972d307b64a36ae5b2e2a3f526694}, many of which leverage contextual signals for more accurate reminding. These systems can help generate reminders associated with a range of future actions, including location, events, activities, people, and time. Two of the most commonly supported types of reminders are location- and time-based (and combinations thereof~\cite{DeVaul00thememory,Marmasse:2000:LID:647986.741313}). Location-based reminders fire when people are at or near locations of interest~\cite{Ludford:2006:ICM:1124772.1124903,doi:10.1080/10447318.2013.796440}. Time-based reminders are set and triggered based on time~\cite{Fertig:1996:LRR:249170.249187,Hicks2005430}], including those based on elapsed time-on-task~\cite{doi:10.1108/IJPCC-07-2014-0042}. While time-based reminders can provide value to many users, particular groups may especially benefit from time-based reminders. These include the elderly~\cite{McGee-Lennon:2011:UMR:1978942.1979248}, those with memory impairments~\cite{doi:10.1080/09602010343000138}, and people seeking to comply with prescribed medications~\cite{pmid19848576}. In Chapter~\ref{ch:umap} we study time-based reminders in the Cortana reminder service, and omit location and person-based reminders, which are less common in our data, and more challenging to study across users as they rely heavily on personal context and relationships between the user and the locations and persons that trigger the reminders.

\paragraph{Memory aids.}
Memory aids help people to remember past events and information. Studies have shown that people leverage both their own memories via recall strategies and the use of external memory aids to increase the likelihood of recall~\cite{intonspeterson1986}. Aids can assume different forms, including paper~\cite{Malone:1983:POD:357423.357430} to electronic alternatives~\cite{Barreau:1995:FRF:221296.221307,Fertig:1996:LAD:257089.257404,Rhodes1996}. One example of a computer-based memory aid is the Remembrance Agent~\cite{Rhodes1996}, which uses words typed into the text processor to retrieve similar documents. People also leverage standard computer facilities to support future reminding (e.g., positioning documents in noticeable places on the computer desktop)~\cite{Barreau:1995:FRF:221296.221307}, which is inadequate for a number of reasons, including that the reminder is not pushed to the user~\cite{Fertig:1996:LRR:249170.249187}. Other work has focused on the use of machine learning to predict forgetting, and the need for reminding about events~\cite{jogger-models-for-context-sensitive-reminding}. Cortana is an example of an interactive and intelligent external memory aid. Studying usage patterns and user behavior lets us better understand users' needs, developing improved methods for system-user interaction and collaboration, and more generally, enhance our understanding of the types of tasks where memory aids are necessary.

\paragraph{Prospective memory.}
Prospective memory (PM) refers to the ability to remember actions to be performed at a future time~\cite{prospectivememory1996,pmid2142956}. Beyond simply remembering, successful prospective memory requires recall at the appropriate moment. PM failures have been an area of study~\cite{ACP:ACP767,pmid9282220}. Studies have shown that failures can be linked to external factors such as interruptions~\cite{Czerwinski:2004:DST:985692.985715,OConaill:1995:TWD:223355.223665}. Prospective tasks are usually divided into time-based tasks and event-based tasks~\cite{pmid2142956}. Time-based tasks are tasks targeted for execution at a specific future time, while event-based tasks are performed when a particular situation or event occurs, triggered by external cues, e.g., a person or a location~\cite{prospectivememory1996}. Laboratory studies of PM have largely focused on retention and retrieval performance of event-based PM, as this is straightforward to operationalize in an experimental setting. Time-based PM is a largely overlooked type in PM studies~\cite{Dismukes01082012}, as this type of ``self-generated'' PM is difficult to model in a lab setting. The Cortana reminder logs we study in Chapter~\ref{ch:umap} represent a real-life collection of time-based PM instances. They provide insights in the type and nature of tasks users are likely to forget to execute.

\section{What's next}

The rest of this thesis is structured in two self-contained parts that can be read independently. 

In Part~\ref{pt:1} of this thesis, our \emph{entities of interest} are emerging real-world entities, i.e., entities that are not described (yet) in publicly available KBs such as Wikipedia. 
First, we study how entities emerge in online text streams in Chapter~\ref{ch:plos}. 
Next, we set out to predicting newly emerging entities in social streams in Chapter~\ref{ch:ecir}. 
Finally, in Chapter~\ref{ch:wsdm} we address the dynamic nature in which entities of interests may appear in online text streams, and leverage collective intelligence to improve the retrieval effectiveness of real-world entities by capturing the changing contexts in which they may appear. 

Next, in Part~\ref{pt:2} we switch our focus, and our \emph{entities of interest} are the producers of the digital traces. 
Here, we present two case studies. 
First, we analyze and predict email communication by studying the communication network and email content in Chapter~\ref{ch:sigir}, and then we analyze interaction data of users with an intelligent assistant, to predict when a user will perform an activity in Chapter~\ref{ch:umap}. 

Finally, we summarize our findings, formulate implications and limitations of our work, and highlight some areas for potential future work in Chapter~\ref{ch:conclusions}.

\clearpage{}

\part{\acl{pt:1}}
\label{pt:1}

\clearpage{}
\chapter{\acl{ch:plos}}
\label{ch:plos}

\acrodef{rq:clusters}[\ref{rq:clusters}]{Can we discover distinct groups of differently emerging entities by clustering their emergence time series?} 
\acrodef{rq:substreams}[\ref{rq:substreams}]{Do news and social media text streams exhibit different emergence patterns?} 
\acrodef{rq:entities}[\ref{rq:entities}]{Do different types of entities exhibit different emergence patterns?} 
\acrodef{CM}{collective memory}
\acrodef{EB}{early bursting}
\acrodef{LB}{late bursting}

\begin{flushright}
\rightskip=1.8cm``The two most important days in your life are the day \\ you are born and the day you find out why.'' \\
\vspace{.2em}
\rightskip=.8cm\textit{---Mark Twain}
\end{flushright}
\vspace{2em}

\section{Introduction}
\label{sec:introduction}

In the first part of this thesis we focus on \emph{emerging} real-world entities, i.e., entities that are not described (yet) in a knowledge base. 
In this first chapter, we take a closer look at how these entities emerge. 
This chapter serves as an introductory study into emerging entities, our object of study, which we study in subsequent chapters in the contexts of predicting them in social streams (Chapter~\ref{ch:ecir}), and enabling their effective retrieval (Chapter~\ref{ch:wsdm}). 

We study entities that emerge in online text streams, and are subsequently added to Wikipedia, the largest publicly available and collectively maintained KB.
Wikipedia has been dubbed a global memory place~\cite{Pentzold01052009}, where collective memories are stored~\cite{halbwachs1997memoire}. 
The moment when an entity is incorporated into Wikipedia, i.e., stored in collective memory, can thus be considered parallel to the collective's \emph{discovery} of an entity. 
Studying emerging entities, hence, provides insights into what the collective considers \emph{entities of interest}. 

We take a macro view and aim to understand the properties of and circumstances under which entities of interest emerge. 
We answer the following research question: \acf{rq:plos}.

Studying online text streams and Wikipedia allows us to make observations and gain insights into the process of emerging entities at a large scale. 
Every day, new content is being added to Wikipedia, with the knowledge base receiving over 6 million monthly edits at its peak~\cite{Suh:2009:SNS:1641309.1641322}. 
These edits may appear under different circumstances. 
Domain experts may find information missing on Wikipedia, and take up the task of contributing this new information. Alternatively, as events unfold in the real world, new, previously unknown, and unheard-of entities (people, organizations, products, etc.) emerge into public discourse. 
These new entities emerge online in news articles and social media postings that may describe or comment on events, e.g., 
the Olympics may introduce new athletes onto the world stage, 
a newly announced smartphone or video game console may generate a wave of activity on social media, 
or the opening of a new restaurant may be reported in local news media and pop up in social media. 

In this chapter, we analyze a sample of online social media and news text streams, spanning over 18 months and comprising over 579 million documents. 
We focus on the \emph{emergence patterns} of these entities, i.e., how a new entity's exposure develops and evolves in the timespan between the entity's first mention in online text streams, and when an article devoted to the entity is subsequently added to Wikipedia. 
More precisely, we define an entity's \emph{emergence pattern} to be its ``document mention time series,'' i.e., the time series that represents the number of documents that mention a specific entity over time, starting at the moment it is first mentioned, up to the moment it is incorporated into Wikipedia. 

\begin{figure}[t]
\centering
\includegraphics[width=.95\textwidth]{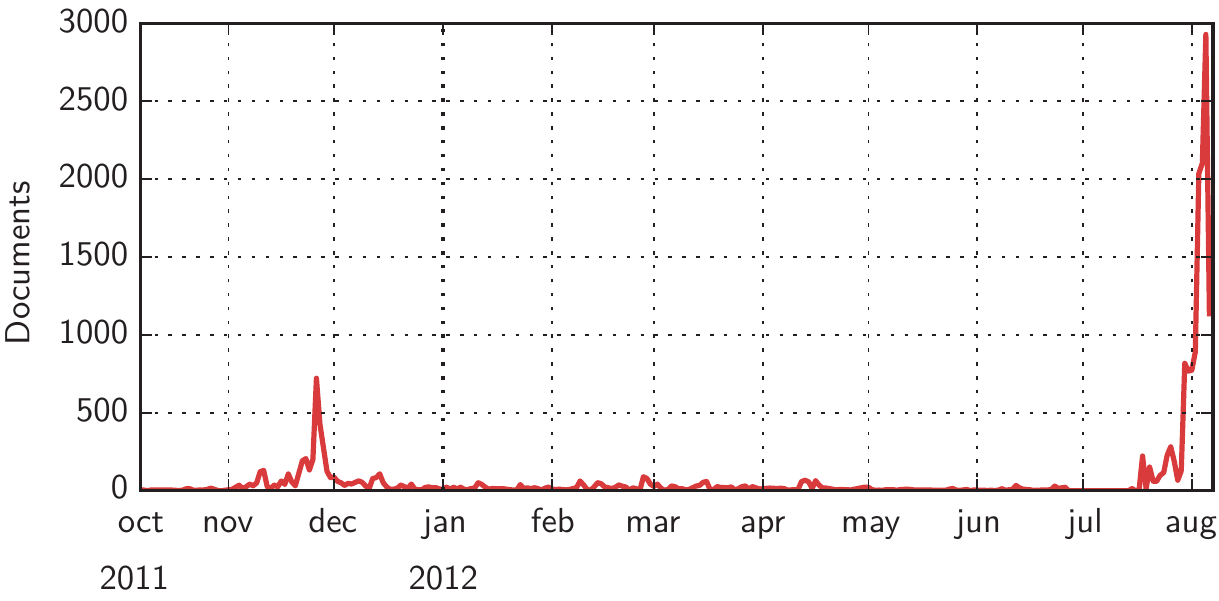}
\caption{Emergence pattern of the entity \texttt{Curiosity (Rover)}, first mentioned in our text stream in October 2011. 
The Wikipedia page for Curiosity was created nine months later, on August 6, 2012. 
There are two distinct \emph{bursts}, one late November 2011, the second shortly before the entity is added to Wikipedia. 
The two document bursts correspond to the Mars rover's launch date (November 26, 2011) and its subsequent Mars landing (August 6, 2012). 
The time series shows us that while the launch did generate publicity, yielding a burst of documents (and signaling increased attention for the entity), at that point it was not deemed important enough to be added to Wikipedia.}
\label{fig:curiosity}
\end{figure}

An example of one such emergence pattern is shown in Figure~\ref{fig:curiosity}. 
The time series shows the number of documents that mention a given entity per day on the $y$-axis (i.e., the \emph{emergence volume}), with the $x$-axis representing the time between the first mention of the entity in the online text stream and the day it is added to Wikipedia (i.e., the \emph{emergence duration}).\footnote{This is the English version of Wikipedia; see Section~\ref{sec:methods-and-data} for details on the data used in this chapter.}
Since emergence durations naturally differ between entities, our time series are of variable length. 
As we are interested in broad attention patterns, we study document mention time series, as opposed to total mention volume. And
because we want to study global, broad, and long-term patterns, our time series are at a granularity of days, not hours.

The main findings of this chapter are as follows. 
By clustering time series of mentions of entities as these entities emerge in news and social media streams, we find broadly two different emergence patterns: entities that show a \emph{strong initial burst} around the time of their introduction into public discourse, and \emph{late bursting} entities that exhibit a more gradual emergence pattern. 
Furthermore, we find meaningful differences between how entities emerge in social media and news streams: entities that emerge in social media streams exhibit slower emergences than those that emerge in news streams.  
Finally, we show how different types of real-world entities exhibit different emergence patterns; we find that the entities that emerge fastest are entity types that know shorter life-cycles such as devices (e.g., smartphones), and ``cultural artifacts'' (e.g., movies and music albums). 

We proceed by formulating three sub research questions in Section~\ref{sec:researchquestions}.
The data that we use in our study is described in Section~\ref{sec:methods-and-data}. In Section~\ref{section:methods} we detail the methods used to analyze the data. Results of our analysis are presented in Section~\ref{sec:plos-results}. We conclude in Section~\ref{sec:plos-conclusion}.
 
\section{Research Questions} 
\label{sec:researchquestions}
In studying emergence patterns of entities, we apply three different methods for grouping entities, representing alternative views on emerging entities. 
In Section~\ref{sec:rq1} below, we apply a burst-based unsupervised hierarchical clustering method to cluster similar entity emergence patterns, so as to discover groups of entities with broadly common emergence patterns. 
That is, we group entity time series by similarities in the peaks of comparatively higher activity (i.e., peaks of exposure in public discourse). This type of analysis is meant to help us answer the following research question:

\begin{enumerate}[label=\textbf{RQ3.\arabic*},ref={RQ\arabic*},labelwidth=1cm,leftmargin=!,align=left]
\item \acl{rq:clusters}\label{rq:clusters}
\end{enumerate}

\noindent
We show that entities emerge in two distinct patterns: so-called \ac{EB} entities, that show a strong initial burst around the time of their introduction into public discourse, or \ac{LB} entities, that exhibit a more gradual emergence pattern. 
\ac{EB} entities are shown to exhibit shorter emergence durations than \ac{LB} entities, i.e., they are deemed entities of interest more rapidly. 

In Section~\ref{sec:rq2}, we adopt an alternative view of emerging entities, and examine their emergence patterns in different types of text stream, comparing patterns between entities that emerge in news to social media streams. 
This view is motivated by perceived differences in the nature of the professionally curated, authoritative, and high impact, ``mainstream media'' versus the user-generated, unedited, social media streams. 
We apply two grouping methods.
First, we group the time series by type of text stream, and provide their descriptive statistics, hypothesizing that news streams will exhibit shorter emergence durations than social media streams, due to their reach and impact. 
We also analyze the cross-pollination between the two types of stream, i.e., we study whether entities appear first in either of the streams, or whether they appear in both at the same time, etc. 
These analyses help us answer the following research question:

\begin{enumerate}[label=\textbf{RQ3.\arabic*},ref={RQ\arabic*},resume,labelwidth=1cm,leftmargin=!,align=left]
\item \acl{rq:substreams}\label{rq:substreams}
\end{enumerate}

\noindent
We find that news and social media streams show broadly similar emergence patterns for entities. 
However, entities that first emerge in social media streams on average are incorporated in the KB comparatively faster than entities that first emerge in news streams. 

Finally, in Section~\ref{sec:rq3}, we study the similarities and differences between different types of entities as they emerge in public discourse. 
Specifically, we leverage DBpedia, the structured counterpart of Wikipedia, to group emerging entities by their underlying entity types, such as companies, athletes, and video games. 
This allows us to gain insights into whether different types of entities exhibit different emergence patterns. In addition, by considering entity types we study whether the news stream and the social stream exhibit different focal points, i.e., where professionally curated news streams exhibit a focus on the mainstream and user-generated social media streams surface more niche entities. This analysis helps us answer the following research question:

\begin{enumerate}[label=\textbf{RQ3.\arabic*},ref={RQ\arabic*},resume,labelwidth=1cm,leftmargin=!,align=left]
\item \acl{rq:entities}\label{rq:entities}
\end{enumerate}

\noindent
We find that different entity types exhibit substantially different emergence patterns. 
Furthermore, we find that specific entity types are more or less likely to emerge in either the news or social media streams, which can be largely attributed to the different nature of both streams, and their authors. Finally, we find that entity types that have shorter emergence durations remain popular over time. The findings suggest that the entity types that are incorporated relatively fast in the KB, play a more central place in public discourse. 
 
\section{Data}
\label{sec:methods-and-data}

As described in the introduction, we study the emergence patterns of entities by looking at the lead time between the first mention of an entity, i.e., its first appearance in online text streams, up to the moment they are incorporated into Wikipedia. 
To study these emergence patterns, we generate a custom dataset of time\-stamped documents that are annotated retrospectively with links to Wikipedia, including for each link (i)~the creation date of the associated Wikipedia page, and (ii)~whether the Wikipedia page existed at the time the document was created. Our dataset spans 7.3 million documents with 36.2 million references to 79,482 unique emerging entities. 
We create this dataset by extending an existing document-stream dataset (TREC KBA's StreamCorpus 2014\footnote{\url{http://trec-kba.org/kba-stream-corpus-2014.shtml}}) with an additional set of annotations to Freebase (FAKBA1\footnote{\url{http://trec-kba.org/data/fakba1/}}). We enrich the dataset by including Wikipedia creation dates of the Freebase entities, and the relative age of the entity to the document. 
We subsequently reduce this dataset by filtering out documents that have ``future leaks,'' i.e., that contain references to entities whose Wikipedia pages were created after the dataset's timespan. 
Below, we describe our data acquisition and preparation processes in detail.

\subsection{Emerging Entities in News and Social Media}
Our dataset is derived from the \emph{TREC KBA StreamCorpus 2014}, a dataset of roughly 1.2 billion timestamped documents from various sources, e.g., blog and forum posts and newswires. It spans 572 days (or roughly 18 months) from October 7\textsuperscript{th}, 2011 to May 1\textsuperscript{st}, 2013 (line~1 in Table~\ref{table1}). 
The StreamCorpus is composed of multiple subsets, with slightly different periods and volumes (i.e., numbers of documents). 
All documents in the corpus that were automatically classified as being written in English have been automatically tagged for named entities with the Serif tagger~\cite{boschee2005automatic}, yielding roughly 580M tagged documents. This annotated subset of documents was the official collection for the TREC Knowledge Base Acceleration (KBA) task in 2014~\cite{Frank2014EvaluatingSF}. 

\citet{fakba1} further annotated these 580M documents with Freebase entities, resulting in the \emph{Freebase Annotations of TREC KBA 2014 Stream Corpus, v1 (FAKBA1) dataset}, which spans over 394M documents (line~2 in Table~\ref{table1}). 
It is essential to our study that the Freebase dump used for linking the entities is dated after the span of the StreamCorpus. Because of this, we can isolate entities that are mentioned (i.e., linked) in documents before their corresponding Wikipedia page was created. 

To extract Wikipedia page creation dates for the entities present in the TREC KBA StreamCorpus 2014, we first map the Freebase entities linked in the FAKBA1 collection to their corresponding Wikipedia pages, using the available mappings in Freebase. We extract Wikipedia page creation dates from a dump of Wikipedia with the full revision history of all pages (\texttt{enwiki-latest-page-meta-history.xml}). 
We append the Wikipedia page creation date or \emph{entity timestamp} (denoted $e_{T}$) to each entity in FAKBA1. 
In addition, we include the entity's ``age'' relative to the document timestamp ($doc_{T}$): the difference between the Wikipedia page creation date and the document timestamp, i.e., $e_{age} = e_{T} - doc_{T}$. 
The resulting dataset, FAKBA1, extended with the entity age and entity timestamp, is denoted \emph{FAKBAT} (Freebase Annotations of TREC KBA 2014 Stream Corpus with Timestamps), see line~3 in Table~\ref{table1}. 

As a next step, we filter to retain only documents that contain emerging entities. 
Emerging entities are entities with $e_{age} < 0$, i.e., entities mentioned in documents dated before the entity's Wikipedia creation date. We denote the resulting subset of FAKBAT documents with emerging entities \emph{OOKBAT} (Out of Knowledge Base Annotations (with) Timestamps).
This yields a set of nearly 24M documents (line~4 in Table~\ref{table1}). 

To be able to study an emerging entity's complete emergence pattern, we take two additional filtering steps.
First, we prune entities with creation dates later than the last document in our stream, to ensure the entities emerged in the timespan of our document stream, i.e., we remove all entities whose Wikipedia page has a creation date later than May 1\textsuperscript{st}, 2013.
Next, we prune all entities that are mentioned in fewer than 5 documents, to be able to visualize and study their time series. 
This yields our final dataset, which comprises 79,482 emerging entities (line~5 in Table~\ref{table1}). 

\begin{table}[tb]
\centering
\caption{Descriptive statistics of our dataset acquisition. Coverage over preceding dataset in brackets. 
Looking at the second and third row in the table, we note that roughly two-thirds of the FAKBA1 entities can be mapped to Wikipedia. However, this portion represents 98\% of the mentions. 
Closer inspection showed that the missing one-third, i.e., the entities we could not map to Wikipedia creation dates, were Freebase entities that were not linked to their Wikipedia counterparts, most notably, WordNet concepts and entities from the ``MusicBrainz'' knowledge base (i.e., artists, albums, and artists). 
The last two rows show that one in ten of the entities emerge during the span of the dataset, however, they constitute a mere 1\% of the mentions.}
\label{table1}
\begin{tabular}{l@{ }l @{ } r@{ }r @{ } r@{ }r @{ } r@{ }r}
\toprule
& {\bf Dataset} & \multicolumn{2}{c}{\bf \# entities} & \multicolumn{2}{c}{\bf \# mentions} & \multicolumn{2}{c}{\bf \# documents} \\ 
\midrule
1. & \emph{TREC KBA}~\cite{Frank2014EvaluatingSF} & N/A & & N/A & & 579,838,246 &  \\ 
2. & \emph{FAKBA1}~\cite{fakba1} & 3,272,980 && 9,423,901,114 && 394,051,027 & (68.0\%) \\ 
\midrule
3. & FAKBAT & 2,254,177 & (68.9\%) & 9,221,204,641 & (97.8\%) & 394,051,027 & (100\%) \\ 
4. & OOKBAT & 225,291 & (10.0\%) & 94,929,292 & (1.0\%) & 23,896,922 & (6.1\%) \\
\midrule
5. & Emerging entities & 79,482 & (35.3\%) & 36,242,096 & (38.2\%) & 7,291,700 & (30.5\%) \\ 
\bottomrule
\end{tabular}
\end{table}

\subsubsection{Entity Types}
In the analysis in Section~\ref{sec:rq3}, we leverage an entity's ``class'' from the DBpedia ontology.\footnote{\url{http://mappings.dbpedia.org/server/ontology/classes/}}
Freebase provides mappings to Wikipedia and its structured counterpart DBpedia. 
In the DBpedia ontology, an entity is assigned to one or more classes in a tree-like class structure.
We map each of our emerging entities to the classes assigned in DBpedia, e.g., the entity \texttt{Barack\_Obama} is mapped to the Person, Politician, Author, Award Winner classes.
We extract these mappings by extracting all triples that have \texttt{rdf:type} property (e.g., \texttt{Barack\_Obama} \texttt{<rdf:type>} \texttt{Person}).
Out of the 79,482 emerging entities in our dataset, we have 39,713 class-mappings (i.e., a coverage of 50.0\%). 

\subsubsection{Entity Popularity}
Finally, as a proxy for an entity's popularity, which we use to study the composition of clusters and the different substreams in Section~\ref{sec:rq3}, we extract Wikipedia pageview statistics. 
More specifically, we extract the total number of pageviews each entity received during 2015. 
We choose to use the pageview counts of a year that falls outside of the timespan of our dataset so as to minimize the effects of timeliness (i.e., we want to separate the true ``head'' entities from the ones that have a shorter lifespan). 

\subsection{Time Series of Emerging Entity Mentions}
\label{sec:timeseries}

\begin{figure}[t]
\centering
\includegraphics[width=.95\linewidth]{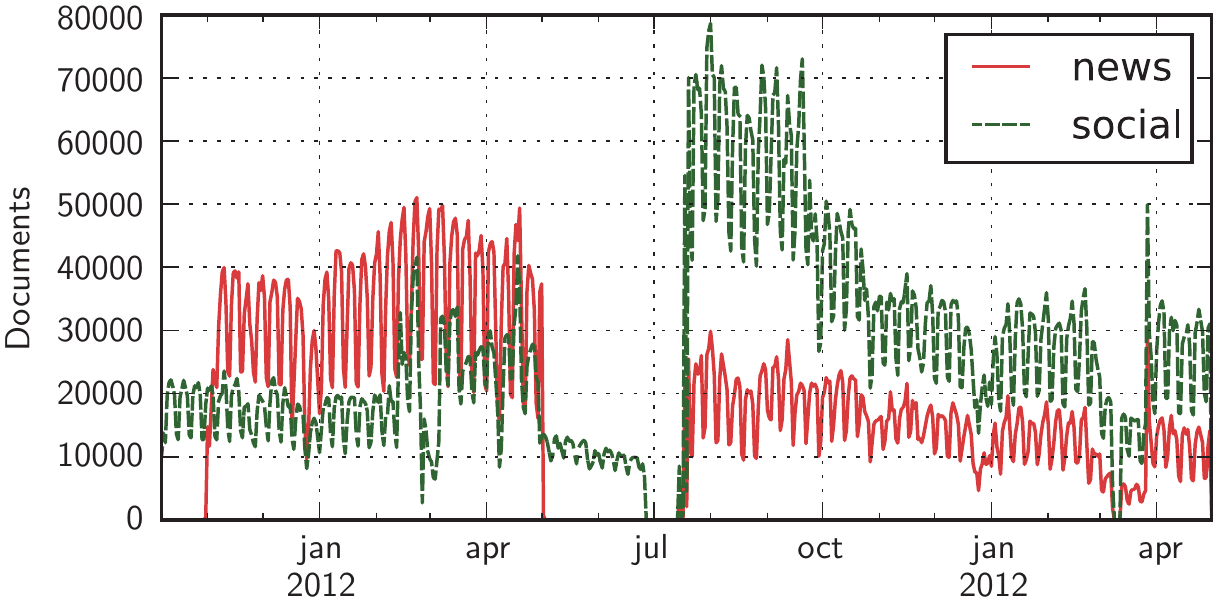}
\caption{Total document volume (i.e., count) for the news (green line) and social media (grey line) text streams over time ($n$ = 23,896,922). 
The streams are composed of multiple separate datasets, explaining the gap in May through mid-July of the news stream.
(Best viewed in color.)}
\label{fig:substreamvolume}
\end{figure}

We use the Emerging Entities dataset that we have created (line~5 in Table~\ref{table1}) to generate time series that describe their emergence. 
Figure~\ref{fig:substreamvolume} shows the document volume over time of the different streams in our dataset. It shows a larger number of news documents compared to social media documents in the first half of the data, and a larger number of social media compared to news documents in the second half. 
In total, the news stream comprises 1,836,022 documents, making it substantially smaller than the social media stream, at 5,357,014 documents. 
The news and social media streams themselves are composed of multiple datasets from different sources~\cite{ldietz:trec-kba-overview-2012}, which explains the gap seen around May 2012 up to somewhere in July 2012. 

\begin{figure}[t]
\centering
\includegraphics[width=.95\linewidth]{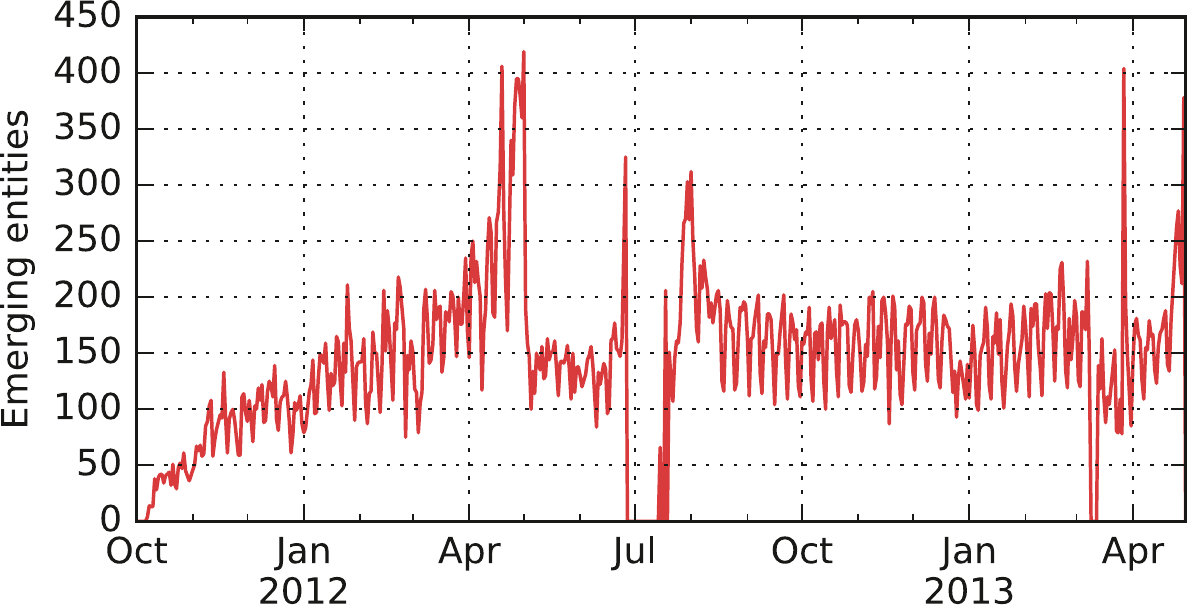}
\caption{Distribution of emerging entities in our dataset ($n$ = 79,482). 
The $x$-axis shows the timespan of our dataset in days, the $y$-axis shows the number of entities that emerge in our dataset on that day (i.e., have their Wikipedia pages created).}
\label{fig:entitycreationdates}
\end{figure}

\noindent
We note that 79,482 entities emerge during the 18+ month (572 days) timespan of the TREC KBA StreamCorpus 2014 dataset, i.e., on average over 138 new entities emerge per day. 
Looking at the distribution at which these entities emerge over time, we observe in Figure~\ref{fig:entitycreationdates} that they do not emerge uniformly. 
In particular, Figure~\ref{fig:entitycreationdates} shows a gradual increase of emerging entities between the start of our dataset and May 2012, at which it peaks.
The subsequent gap can be explained by the absence of the news stream during that time (see also Figure~\ref{fig:substreamvolume}). 

The core unit in our analysis are so-called \emph{entity document mention time series}, i.e., time series that represent the number of documents that mention an entity over time (see, e.g., \texttt{Mars Curiosity}'s document mention time series in Figure~\ref{fig:curiosity}).  These time series are characterized by several properties.
First, the time series are of variable length: 
each entity's time series starts at the first mention of the entity in our dataset, and ends at the day the Wikipedia page for that entity was created. 
For some entities, the time series may span several days, whereas others may span months. 
Second, the time series in our dataset are not temporally aligned.
They exhibit different absolute timings, where the date of the first mention (i.e., the start of the time series) and last mention (i.e., the end of the time series) varies between emerging entities. 
 
\section{Methods}
\label{section:methods}

Our first research question, \ac{rq:clusters}, revolves around discovering common emergence patterns. We apply a time series clustering method for discovering groups of entities with similar emergence patterns. 
In this section, we first describe our time series clustering method, explain and motivate the choices for representing the emerging entity's time series. After that, we describe the general time series analysis methods that we use. 

\subsection{Time Series Clustering}
\label{subsec:clustering}
Clustering time series consists of three steps: 
First, we need to normalize our time series as they might span very different periods of time. 
Next, we measure the similarity between time series by applying a similarity metric.
And third, we apply a hierarchical agglomerative clustering method to group entities in groups with similar emergence patterns. 

\subsubsection{Normalization} 
\label{subsec:norm}
As described in Section~\ref{sec:timeseries}, the time series in our dataset might span different periods and are not temporarily aligned. 
For these reasons, we cannot rely on time series analysis and modeling methods that leverage aligned time series or seasonal patterns. 
Because of variable lengths, we cannot leverage similarity methods that assume a correspondence between the data points between two time series such as, e.g., Euclidean distance. 
Furthermore, to be able to visualize clusters and groups of similar time series, we linearly interpolate the time series to have equal length~\cite{ijca2012clustering}.
Finally, as we are interested in the similarity in emergence patterns, not in individual differences between popular and long-tail entities (i.e., absolute number of mentions), we standardize all time series by subtracting the mean and dividing by standard deviation to account for the differences in volume/popularity~\cite{Vlachos:2004:ISP:1007568.1007586}. 

\subsubsection{Burst Similarity} 
Typically, time series similarity metrics rely on fixed-length time series, and leverage seasonal or repetitive patterns~\cite{WarrenLiao20051857}. 
But as noted above, our time series are of variable length, and not temporally aligned. 
For this reason, common time series similarity metrics such as Dynamic Time Warping (DTW) are not applicable~\cite{Bemdt94usingdynamic}. 
We are interested in the moments at which the attention or focus around an entity in public discourse increases, i.e., we are looking for periods with higher activity. 
These so-called time series ``bursts'' may be correlated to real-world activity and events around the entity. 

To address the nature of our time series as well as our focus on bursts, we turn to BSim~\cite{Vlachos:2004:ISP:1007568.1007586} (Burst Similarity) as the similarity metric we leverage to compare time series. 
It relies on detecting bursts, and using the overlap in bursts between time series as the notion of similarity. 

\begin{figure}[t]
\centering
\includegraphics[width=\linewidth]{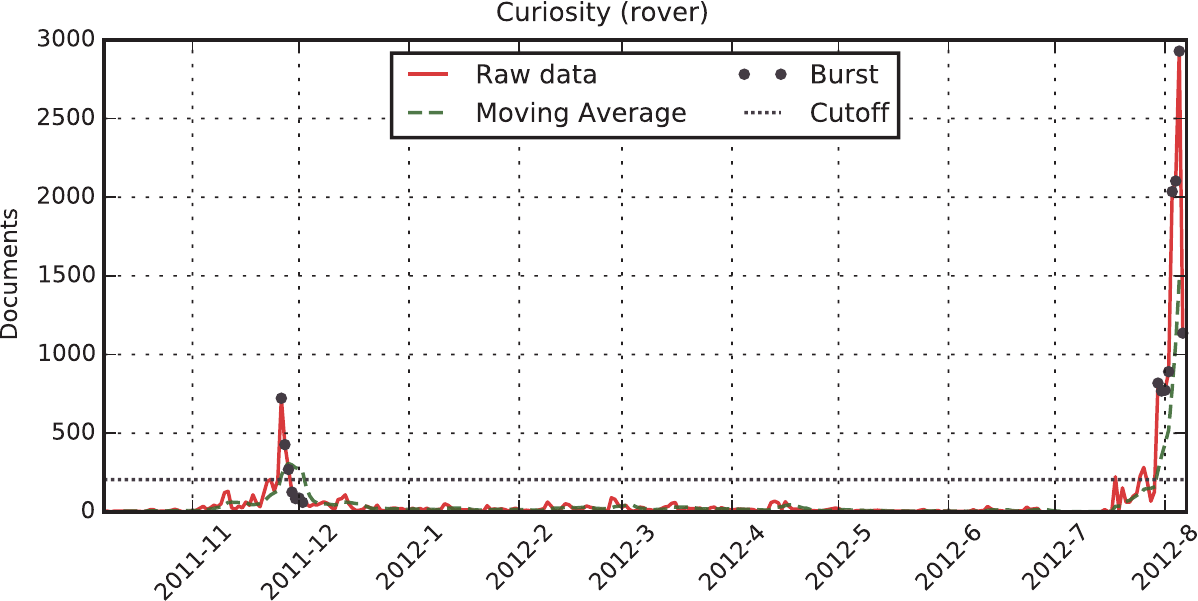}
\caption{Emergence pattern of \texttt{Curiosity (rover)}. (Best viewed in color.)}
\label{fig:curiositybursts}
\end{figure}

For burst detection, we compute a moving average for each (raw) entity document mention time series ($T_{e}$), denoted $T_{e}^{MA}$.
We set our parameter $w$, indicating the size of the rolling window to 7 days. 
Bursts are the points in $T_{e}^{MA}$ that surpass a cutoff value ($c$).
We set $c = 1.5 \cdot \sigma_{MA}$, where $\sigma_{MA}$ is the standard deviation of $MA$. 
These parameter choices for $w$ and $c$ are in line with previous work~\cite{Vlachos:2004:ISP:1007568.1007586}. 
Figure~\ref{fig:curiositybursts} shows an example time series ($T_{e}$), with the bursts detected for the emerging entity \texttt{Curiosity (rover)}.
The detected bursts correspond to the earlier described launch and landing of the Mars Rover. 

\subsubsection{Hierarchical Agglomerative Clustering} 
Now that we can measure similarity between time series, we need to identify clusters of similar time series.
To this end, we compute the Similarity Matrix ($SM$) with all pair-wise burst similarities. 
More specifically, to cluster emerging entities, we first apply $L_2$ normalization to $SM$, and then convert it to a distance matrix $DM$ ($DM = 1 - SM$). 
Next, we apply hierarchical agglomerative clustering (HAC) on $DM$ using the \texttt{fastcluster} package~\cite{JSSv053i09} to discover groups of similar time series at different levels of granularity. 
We apply \emph{Ward's method} as our linkage criterion. Ward's method is an iterative approach, where one starts with singleton clusters, and aims to merge the pair of samples that maximally decreases the within-cluster variance at each successive iteration~\cite{doi:10.1080/01621459.1963.10500845}. 

\subsection{Time Series Analysis}
\label{subsec:analysis}

Throughout this chapter, we take an exploratory approach to analyzing, visualizing, and comparing patterns of groups of time series to discover meaningful clusterings~\cite{von2012clustering}. 
We apply different grouping strategies, e.g., implicit groups, such as the clusters we find by using our clustering method, or explicit groups, e.g., entities that emerge either in social media or in news streams. 

We apply two analysis methods to study different groups of time series: 
(i)~visualization of group signatures that represent common emergence patterns within a group of time series, and 
(ii)~analysis of descriptive statistics that reflect properties of the underlying time series. 

\subsubsection{Visualization}
To compare patterns and trends across different groups of emerging entities, we visualize and compare the so-called \emph{group signatures}, i.e., the average of all time series that belong to a group.
See Figure~\ref{fig:global} below for an example of the group signature of all emerging entities in our dataset ($n$ = 79,482). 

Two challenges specific to the time series that we study in this chapter arise when visualizing them.  
First, their \emph{duration}. 
The time series we study in this chapter (may) differ in length, as emergence durations differ between entities. 
Second, their \emph{alignment}. 
The time series are not temporally aligned, as their start (i.e., $x$ = 0) is marked when the entity is first mentioned in the online text stream, and end at the time at which the entity is incorporated into Wikipedia. 
For these reasons, we linearly interpolate the time series to the (overall) highest emergence duration, effectively ``stretching'' them to have equal length.
Next, we align them in relative duration, i.e., the first and last mentions for each entity is set at the start and end of the $x$-axis, respectively.
This allows us to visualize both the clusters themselves, and the corresponding cluster signatures. 

\subsubsection{Descriptive Statistics}
While studying group signatures of time series allows us to discover similar patterns and study and compare broad patterns and trends, they do not paint the full picture. 
More fine-grained properties of emergence patterns, e.g., the average \emph{emergence duration} (the time between a new entity's first mention in the text stream and its subsequent incorporation into Wikipedia) and the \emph{emergence volume} (the total number of documents that mention the entity as it emerges) are difficult to convey through visualization alone. 
In order to study these aspects, we represent the groups of time series through different descriptive features that reflect the emergence and burst behavior of the time series that belong to a group. 
For an overview of the emergence and burst statistics that we consider, see Table~\ref{tab:tsfeatures}.

\subsubsection{Chi-square Goodness of Fit Test}
In Section~\ref{sec:rq3} we compare the distribution over entity classes per online text stream (i.e., comparing the entity classes in the social media stream to those in the news stream). 
To assess whether the differences in these class-distributions are statistically significantly different, we apply a chi-square goodness of fit test to both distributions. 
In addition, we rank the classes by their contribution to the difference, using chi-$grams$, i.e., we compute for each class: $\frac{observed - expected}{\sqrt{expected}}$, where $observed$ corresponds to the number of entities that belong to a particular class in one set (e.g., the entities that emerge in social media) and $expected$ corresponds to the number of entities that belong to a particular class in the global population (i.e., all entities that emerge in our dataset).

\begin{table}[t]
\centering
\caption{Descriptive statistics used for analyzing and comparing different groupings of emerging entities. We distinguish between \emph{time series statistics} (top three) and \emph{burst statistics} (bottom three). 
All statistics are computed for the period ranging from the emerging entity's first mention in the corpus to the creation date of the Wikipedia page devoted to it.
The burst durations and values are computed over the normalized time series (see Section~\ref{subsec:norm}).}
\label{tab:tsfeatures}
\begin{tabular}{ll}
\toprule
\textbf{Emergence volume} & Number of documents that mention the entity \\
\textbf{Emergence duration} & Number of days from first mention to incorporation \\
\textbf{Emergence velocity} & $\frac{Volume}{Duration}$ (average number of documents per day) \\
\midrule
\textbf{Burst number} & Total number of bursts \\
\textbf{Burst duration} & Normalized average durations of bursts (i.e., burst widths) \\
\textbf{Burst value} & Normalized average heights of bursts (i.e., burst heights) \\
\bottomrule
\end{tabular}
\end{table}
 
\section{Results}
\label{sec:plos-results}

In this section, we present the analyses that answer our three research questions. 
First, we study the time series clusters that result from our clustering method in Section~\ref{sec:rq1}.
Next, in Section~\ref{sec:rq2}, we study similarities and differences between emergence patterns in social media and newswire streams.
In Section~\ref{sec:rq3}, we study the underlying entity types, and their emergence patterns. 

\subsection{Emergence Patterns}
\label{sec:rq1}

The first research question we aim to answer, \acf{rq:clusters}, 
is at the core of our study into emerging entities. 
Finding similar patterns and studying how an entity appears in online text streams before it is added to Wikipedia allows us to gain insights into the circumstances in which entities emerge, i.e., in which they are deemed \emph{entities of interest}. 

\begin{sidewaysfigure}
\includegraphics[width=\linewidth]{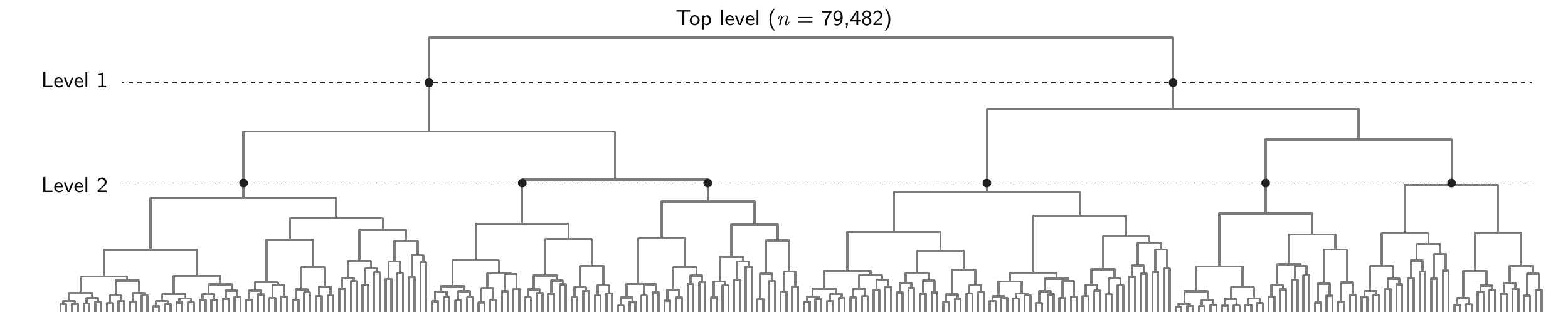}
\caption{Dendrogram resulting from applying hierarchical agglomerative clustering using BSim~\cite{Vlachos:2004:ISP:1007568.1007586} similarity, on our corpus of emerging entity time series ($n$ = 79,482). 
The three cutoff-points at which we analyze the clusters are denoted Top level, Level 1 (2 clusters), and Level 2 (6 clusters). 
For clarity, the tree is truncated by showing no more than 7 levels of the hierarchy. }
\label{fig:dendrogram}
\end{sidewaysfigure}

\noindent
Figure~\ref{fig:dendrogram} shows a cluster tree that results from applying hierarchical clustering on the BSim similarity matrix computed for the time series of all entities under consideration, as explained in Section~\ref{subsec:clustering}. 
The tree shows a clear subdivision into smaller clusters. 
At its highest level, the tree shows two distinct clusters, each of which is broken down in multiple smaller sub-clusters.  
The cluster tree in Figure~\ref{fig:dendrogram} helps us to organize our analysis; in the following section, we look at different granularities of similarly emerging entities, by exploring three levels of this cluster tree: first, we study the \emph{global emergence patterns}, by taking the time series that are at the root node of the tree (i.e., the first level of the tree, denoted \emph{Top level} in Figure~\ref{fig:dendrogram}).
Then we study the main two clusters, at the next level of the tree (denoted \emph{Level 1} in Figure~\ref{fig:dendrogram}). 
Finally, move down another level in the tree and study the set of six clusters at \emph{Level 2} in Figure~\ref{fig:dendrogram}. 

\subsubsection{Global Emergence Pattern}
\begin{figure}
\centering
\includegraphics[trim={7.5cm 0 0 0},clip,height=1.5in]{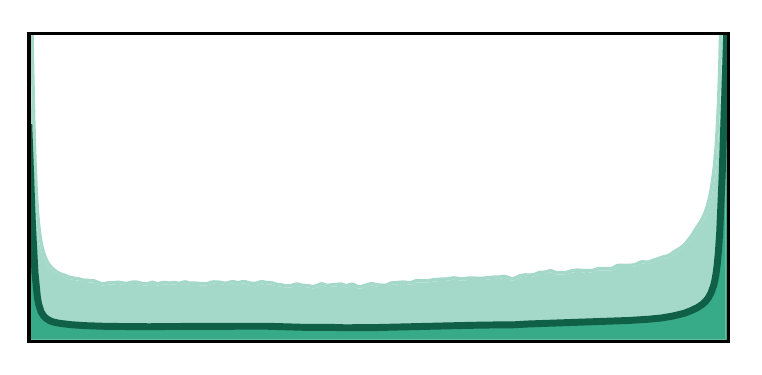}
\caption{Global cluster signature (of all emerging entities) where $n$ = 79,482. 
That is, the top level in the dendrogram in Figure~\ref{fig:dendrogram}. 
The axes are not labeled since all time series values (i.e., document counts) are standardized, and the series are linearly interpolated to have equal length. 
The solid line represents the cluster signature (i.e., the average time series), the lighter band represents standard deviation.}
\label{fig:global}
\end{figure}

First, we examine the emergence pattern of the global average, by taking the group at the root node of the cluster tree (i.e., all time series in our dataset, where $n$ = 79,482). 
Figure~\ref{fig:global} shows the global emergence signature, and Table~\ref{tab:global} shows the associated descriptive statistics. 

Figure~\ref{fig:global} shows how both the emerging entities' introduction into public discourse (the first mention at the left-most side of the plot) and subsequent incorporation into Wikipedia (the right-most side of the plot) occur in bursts of documents, i.e., overall, the largest number of documents that mention a newly emerging entity are either at the start or at the end of their time series. 
This can be explained as follows. The entrance into public discourse represents the first emergence of an entity, whereas being added to Wikipedia is likewise likely to happen in a period of increased attention, e.g., a real-world event that puts the entity in public discourse.  Between these two bursts, the number of documents that mention the entity seems to increase gradually as time progresses, suggesting that on average, the number of documents that mention a new entity, and thus the attention the entity receives in public discourse increases over time before it reaches ``critical mass,'' i.e., when the entity is deemed important enough to be incorporated into the KB. 

\begin{table}[t]
\centering
\caption{Global time series and burst descriptive statistics.}
\label{tab:global}
\begin{tabular}{r@{ $\pm$ }r r r@{ $\pm$ }r r r@{ $\pm$ }r r}
\toprule
\multicolumn{3}{c}{duration (\# days)} & \multicolumn{3}{c}{volume (\# docs)} & \multicolumn{3}{c}{velocity (docs/day)} \\
mean & std & med. & mean & std & med. & mean & std & med. \\
\midrule
245 & 153 & 221 & 183 & 1,180 & 32 & 0.87 & 5.6 & 0.19 \\
\midrule
\multicolumn{3}{c}{n\_bursts} & \multicolumn{3}{c}{bursts durations} & \multicolumn{3}{c}{bursts values} \\
\midrule
3.8 & 2.62 & 3 & 0.03 & 0.03 & 0.02 & 0.03 & 0.08 & 0.02 \\
\bottomrule
\end{tabular}
\end{table}

Next, we look at the global emergence pattern's descriptive statistics; see
Table~\ref{tab:global}.
Here, we see how on average it takes 245 days between an emerging entity's first mention and being added to the KB. 
In these 245 days, an emerging entity is mentioned in 183 documents on average. 
However, both the emergence durations and emergence volumes show large standard deviations (of 153 days and 1,180 documents, respectively), indicating that they differ substantially between entities, further motivating our clustering approach to zoom into the different underlying patterns. 
The emergence velocity shows how, on average, an entity is mentioned in less than a document a day. 

To better understand how the documents that mention an emerging entity are distributed over time, e.g., to gain insights into whether they are concentrated in a few bursts, or spread out more uniformly over the timeline, we turn to the burst statistics in Table~\ref{tab:global}. 
On average, an entity is associated with 3.8 bursts, indicating that entities are likely to resurface multiple times in public discourse before being deemed important enough to be added to the KB. 
The average burst durations span a mere 3\% of an emerging entity's time series, indicating that emerging entities spend the majority of their time ``under the radar.''
The heights of these bursts (i.e., burst values) show a comparatively large standard deviation, suggesting that the heights differ substantially between different entities and bursts. 

In summary, globally, entities experience a long time span between surfacing in online text streams and being incorporated into the KB; they are associated with multiple bursts and thus display a resurfacing behavior. Finally, the large standard deviations seen at the descriptive statistics suggests the entities show large variations in terms of their emergence patterns. 
Below, we study whether grouping the time series of new entities by similar burst patterns allows us to find groups of broadly similarly emerging entities in terms of their group signatures, and emergence and burst-features.

\subsubsection{Clusters at Level~1 in Figure~\ref{fig:dendrogram}: Early vs.\ Late Bursts}
\label{subsec:mainclusters}

\begin{figure}[h]
\centering
\includegraphics[trim={0 0 0 0},clip,width=\linewidth]{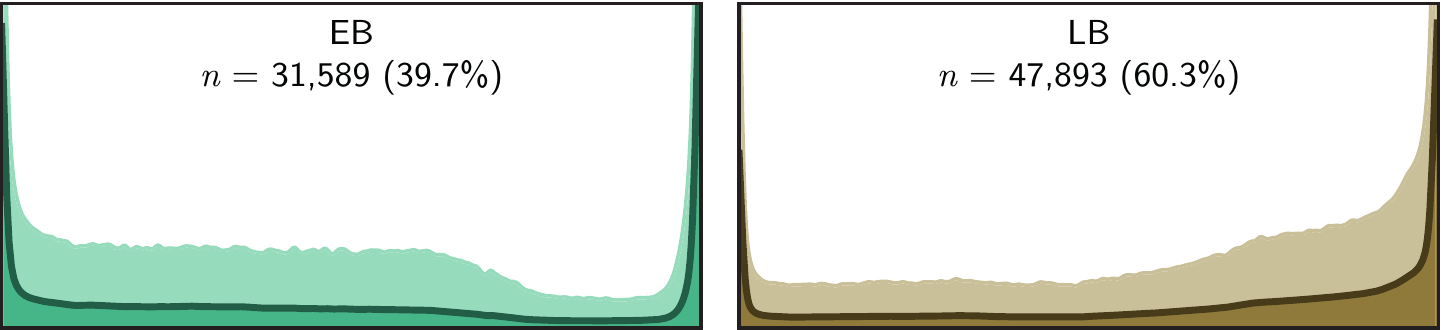}
\caption{Cluster signatures of the \acl{EB} entities (left plot) and \acl{LB} entities (right plot) clusters, denoted Level 1 in Figure~\ref{fig:dendrogram}. 
The solid lines represent the cluster signature (average), the lighter bands represent the standard deviation.} 
\label{fig:clusters}
\end{figure}

In our first attempt at uncovering distinct patterns in which entities emerge, we look at the first two main clusters that appear at Level~1 of the cluster tree in Figure~\ref{fig:dendrogram}. 
The resulting cluster signatures are shown in Figure~\ref{fig:clusters}. 

Much like the global cluster signatures we studied in the previous section, the clusters at Level~1 show two main bursts: the \emph{initial burst} around the first mention, and the \emph{final burst} around the last mention before an entity is added to Wikipedia. 
However, as we will show, the left cluster, which we call \emph{\acf{EB} entities}, is characterized by a stronger initial burst, with the majority of the documents that mention the entity concentrated at the time around the first mention when the entity surfaces in the online text streams. 
This suggests that the cluster contains new entities that suddenly emerge and experience a (brief) period of lessened attention, before being incorporated into the KB. 
The right cluster, which we denote as \emph{\acf{LB} entities}, shows a more gradual pattern in activity towards the point at which the entity is added to the KB, much like we saw in the global signature (Figure~\ref{fig:global}). 

Looking at Figure~\ref{fig:clusters}, we note two main differences between the group signatures of the \acl{EB} and \acl{LB}  entities. 
First, the way documents are distributed between the initial and final burst. 
The \ac{EB} entities cluster show a more ``abrupt'' final burst: the signature shows the majority of the documents in the wake of the initial burst, i.e., at the left-hand side of the plot. 
In the next phase, the volume of documents seems to gradually wind down into a relatively quiet period, which finally seems to abruptly transition into the final burst at the right hand-side of the plot. 
In contrast, the \ac{LB} entities cluster shows a relatively subtle initial burst, which like the \ac{EB}-cluster appears to quiet down, followed by a gradual increase of document volume that leads up to the final burst. 

A second difference is the height difference between the initial and final bursts. 
The \ac{EB} cluster shows roughly equally high initial and final bursts. 
The \ac{LB} cluster shows a substantially smaller initial burst, which suggests the introduction into public discourse is comparatively subtler than its addition, i.e., these entities may emerge ``silently,'' suggesting the entities are less central, more niche, and less widely supported. 
In summary, the first split separates the entities that are associated with a strong ``initial'' emergence in online text streams from the entities that more gradually build up to the moment at which they are incorporated into the KB. 

\begin{table}[t]
\centering
\caption{Comparison of \acl{EB} and \acl{LB} entities clusters statistics. }
\label{tab:clusterstats}
\begin{tabular}{l @{} r r@{ $\pm$ }r r r@{ $\pm$ }r r r@{ $\pm$ }r r}
\hline
& proportion & \multicolumn{3}{c}{duration (\#days)} & \multicolumn{3}{c}{volume (\#docs)} & \multicolumn{3}{c}{velocity (docs/day)} \\
& & mean & std & med. & mean & std & med. & mean & std & med. \\
\hline
\ac{EB} & 0.40 & 224 & 146 & 195 & 118 & 804 & 22 & 0.70 & 6.45 & 0.15 \\
\ac{LB} & 0.60 & 259 & 156 & 238 & 225 & 1,371 & 42 & 0.99 & 4.96 & 0.23 \\
\hline
& & \multicolumn{3}{c}{n\_bursts} & \multicolumn{3}{c}{burst durations} & \multicolumn{3}{c}{burst values} \\
\hline
\ac{EB} & 0.40 & 3.32 & 2.20 & 3  & 0.03 & 0.03 & 0.02 & 0.05 & 0.11 & 0.02 \\
\ac{LB} & 0.60 & 4.12 & 2.82 & 4 & 0.03 & 0.03 & 0.02 & 0.03 & 0.05 & 0.01 \\
\hline
\end{tabular}
\end{table}

Next, to better understand the different characteristics of the two clusters, we study the descriptive time series and burst statistics of the time series in the \acf{EB} and \acf{LB} entity clusters in Table~\ref{tab:clusterstats}. 
Before we proceed, we determine whether the differences between the statistics of the two clusters are statistically significant. 
To do so, we perform a Kruskal-Wallis one-way analysis of variance test, also known as a non-parametric ANOVA. 
Following this omnibus test, we perform a post-hoc test using Dunn's multiple comparison test (with p-values corrected for family-wise errors using Holm-Bonferroni correction). 
Comparing all descriptive statistics from the two clusters, we find that all differences are statistically significant at the $\alpha$ = 0.05 level. 

\noindent
Zooming into the statistics, we make the following observations. 
First, there are more \ac{LB} than \ac{EB} entities, at 60.3\% vs. 39.7\%. 
\ac{LB} entities emerge more slowly than \ac{EB} entities: the \ac{LB} entities exist for a longer period in public discourse (with an \emph{emergence duration} of 259 days, versus 224 for \ac{EB} entities), and receive more attention (with an \emph{emergence volume} of 225 documents, versus 118 documents for \ac{EB} entities) before being deemed important enough to be incorporated into the KB. 
The \ac{EB} entities' shorter emergence durations and lower emergence volumes suggest a higher ``urgency'' or timeliness, suggesting this cluster may contain entities that represent sudden events, e.g., large-scale natural catastrophes or societal events, that will typically be incorporated in Wikipedia soon after they first emerge in public discourse. 
The descriptive statistics of the \ac{LB} entities too may indicate they comprise less timely or urgent entities, e.g., recurring events, such as sports events that may appear in public discourse long before being part of the KB. 

This view of the slower, less timely \ac{LB} entities, and the more urgent, fast, and timely \ac{EB} entities is supported by the burst statistics. 
First, the average burst heights of \ac{EB} entities are higher (at 0.05 on average, versus 0.03 for \ac{LB} entities), suggesting \ac{LB} entities see a more evenly spread volume of documents that mention them. 
Next, \ac{EB} entities show a lower number of bursts (3.22 on average, versus 4.12 for \ac{LB} entities). Fewer and higher bursts, together with shorter emergence durations and lower emergence volumes supports the view of more urgent or timely \ac{EB} entities, i.e., those that are more quickly (in terms of time and number of documents) incorporated into the KB, while exhibiting more bursty patterns. 
In the next section, we show that entities that emerge in news likewise exhibit higher burst heights and fewer bursts on average, further exploring the notion of higher urgency or importance. 

In summary, we find that the two clusters at Level~1 in Figure~\ref{fig:dendrogram} differ substantially and significantly in terms of their emergence patterns and burst properties. 
Our visual inspection of the cluster signatures and the analysis of burst and emergence features suggests \ac{LB} entities emerge more slowly, i.e., build up attention more slowly before being added to the KB, whereas \ac{EB} entities are associated with more sudden and higher bursts of activity, prior to being incorporated into the KB.

\subsubsection{Clusters at Level~2 in Figure~\ref{fig:dendrogram}}
\begin{figure}
\centering
\includegraphics[height=.9\textheight]{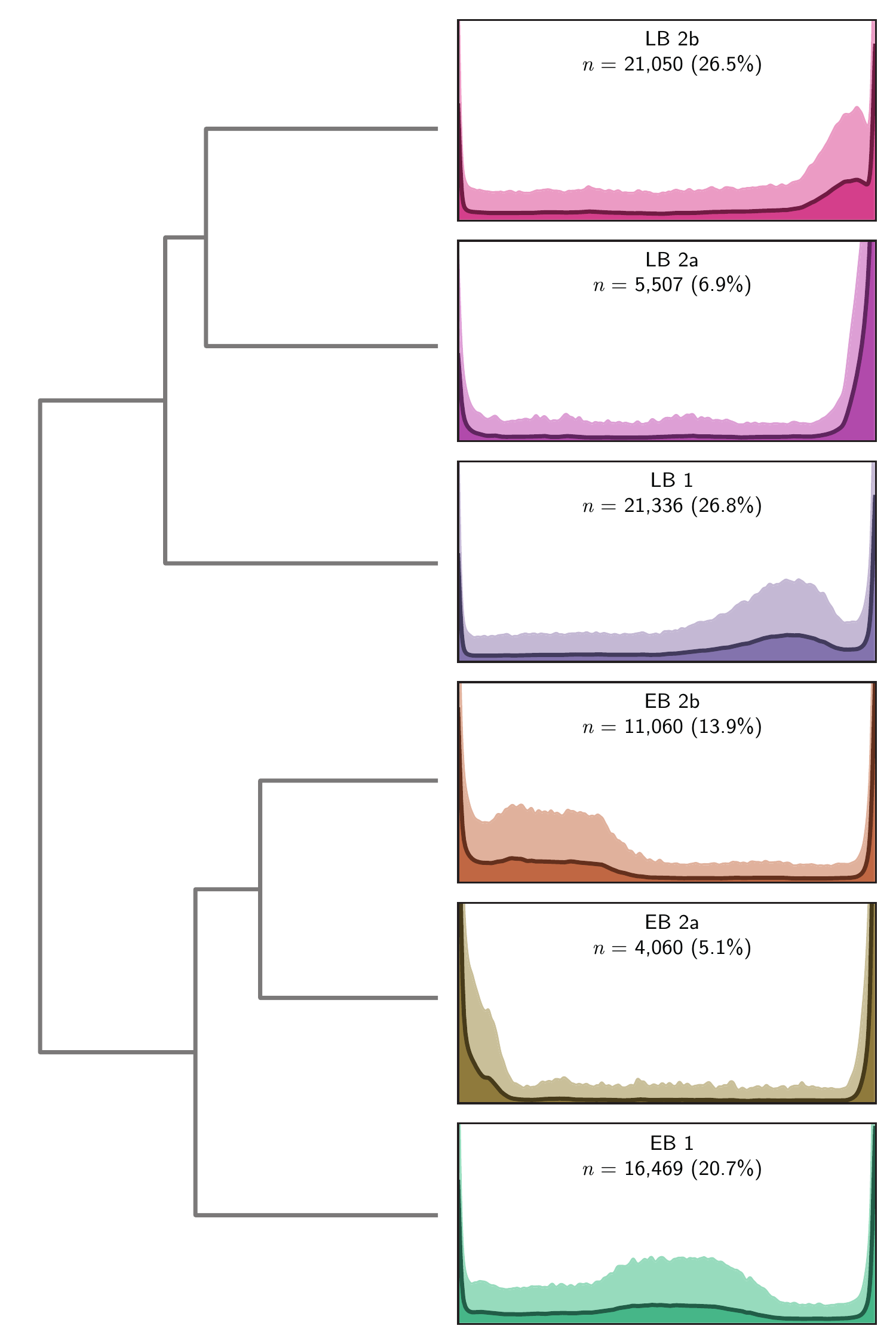}
\caption{Signatures of the clusters at Level 2 in Figure~\ref{fig:dendrogram}. 
Solid lines represent the cluster signatures (average of all samples in the cluster), lighter bands represent standard deviation.} 
\label{fig:subclusters}
\end{figure}

Next, we look at the clusters at Level~2 in Figure~\ref{fig:dendrogram}, i.e., the different clusters that make up the \ac{LB} and \ac{EB} clusters. For brevity, we refer to these clusters as \emph{subclusters}.
Figure~\ref{fig:subclusters} shows the signatures of the six subclusters. 
In general, the distinction between faster \ac{EB} entities, and comparatively slower \ac{LB} entities, remains in the subclusters. 

\noindent
Using the  procedure described in the previous section, we test whether the differences between the subclusters are statistically significant. 
Overall, the properties of time series within each subcluster different significantly, except for \acl{EB} 1 and \acl{LB} 2a, which may not differ statistically significantly in terms of emergence volume.
And in terms of burst statistics, some clusters show inconclusive differences. Specifically, \acl{EB} 1 and 2a do not differ statistically significantly in terms of burst values, and \acl{EB} 2a and \acl{LB} 2b do not differ statistically significantly in terms of burst durations. 

The cluster signatures in Figure~\ref{fig:subclusters} show that the entities that belong to the \ac{LB} entities subclusters (the top three plots) exhibit an increase in document volume in the period leading up to the final burst before entering into the KB. 
This suggests that even when the entity first appears in online text streams with a relatively big burst, its subsequent incorporation into Wikipedia is not instantaneous. 
Consider, e.g., the previously shown example of the Mars Curiosity (in Figure~\ref{fig:curiositybursts}); whereas its first burst happens around October 2011, it is added to Wikipedia after a relatively quiet period, 9 months later. 
Whereas the \ac{LB} entities subclusters both differ in the ``steepness'' of the final burst and the relative moment of the increase of activity, the three \ac{EB} entities subclusters (the bottom three plots) differ mainly in the moment of the increased activity, i.e., the ``bump'' in the plot.

One outlier exists in the \ac{LB} entities cluster: \acl{LB} 2a (second plot from the top). 
Its cluster signature shows a high and sudden burst of activity just before the entity is incorporated into Wikipedia, seemingly omitting the period of gradually increasing attention between the initial and final bursts seen at the other two subclusters. 
Its emergence features in Table~\ref{tab:subsubclusters} show that with an average duration of 146 days before an entity is added to the KB, it is almost twice as fast as the global average. Both the cluster signature and the average number of bursts (1.6 on average) show that the entities exhibit a single, steep, high volume burst before being incorporated into the KB. 
Compared to the other two subclusters in the same \ac{LB} group, the number of bursts is substantially lower, and the average burst values are substantially higher. 
This suggests that the entities in the \ac{LB} 2a cluster comprise abruptly or suddenly emerging entities, like events, and indeed, upon closer inspection, events such as the World Music Festival Chicago, the 2013 British Grand Prix, the 2012 Volvo World Match Play Championship, and the 2012 Sundance Film Festival belong to this cluster. 

\begin{table}[t]
\small
\centering
\caption{Comparison of the entity cluster statistics for clusters at Level~2 in Figure~\ref{fig:dendrogram}.}
\label{tab:subsubclusters}
\begin{tabular}{l r r@{ $\pm$ }r r r@{ $\pm$ }r r r@{ $\pm$ }r r}
\toprule
& proportion & \multicolumn{3}{c}{duration (\#days)} & \multicolumn{3}{c}{volume (\#docs)} & \multicolumn{3}{l}{velocity (docs/day)} \\
& & mean & std & med. & mean & std & med. & mean & std & med. \\
\midrule
\ac{EB} 1  & 0.21 & 238 & 148 & 223 & 145 & 868 & 28 & 0.83 & 8.53 & 0.17 \\
\ac{EB} 2a & 0.05 & 220 & 116 & 191 & 74 & 1,270 & 13 & 0.31 & 3.63 & 0.08 \\
\ac{EB} 2b & 0.14 & 204 & 150 & 164 & 94 & 363 & 19 & 0.65 & 2.37 & 0.16 \\
\ac{LB} 1   & 0.27 & 255 & 148 & 230 & 242 & 1,660 & 43 & 1.01 & 4.94 & 0.22 \\
\ac{LB} 2a  & 0.07 & 146 & 121 & 110 & 155 & 557 & 29 & 1.07 & 3.12 & 0.33 \\
\ac{LB} 2b  & 0.26 & 294 & 158 & 297 & 227 & 1,183 & 45 & 0.95 & 5.36 & 0.20 \\
\midrule
& & \multicolumn{3}{c}{n\_bursts} & \multicolumn{3}{c}{burst durations} & \multicolumn{3}{c}{burst values} \\
\midrule
\ac{EB} 1 & 0.21& 3.65 & 2.21 & 3 & 0.03 & 0.03 & 0.02 & 0.04 & 0.07 & 0.02 \\
\ac{EB} 2a & 0.05 & 2.26 & 1.27 & 2 & 0.02 & 0.02 & 0.02 & 0.10 & 0.22 & 0.02 \\
\ac{EB} 2b & 0.14 & 3.22 & 2.32 & 3 & 0.03 & 0.03 & 0.02 & 0.06 & 0.11 & 0.03 \\
\ac{LB} 1 & 0.27 & 4.18 & 2.62 & 4 & 0.03 & 0.03 & 0.02 & 0.02 & 0.05 & 0.01 \\
\ac{LB} 2a & 0.07& 1.60 & 0.88 & 1 & 0.04 & 0.03 & 0.04 & 0.05 & 0.05 & 0.05 \\
\ac{LB} 2b & 0.26& 4.72 & 2.98 & 4 & 0.02 & 0.02 & 0.02 & 0.03 & 0.06 & 0.01 \\
\bottomrule
\end{tabular}
\end{table}

With the exception of this outlier, the entities in the \ac{LB} subclusters show longer emergence durations and substantially higher emergence volumes on average, further supporting our distinction of slowly emerging entities. 
In particular, \acl{LB}~1 and \acl{LB}~2b (the third and first plot from the top, respectively), with 242 and 227 documents on average, with relatively long emergence durations (255 and 294 days on average, respectively), suggest that these clusters contain slowly emerging niche entities, that are not as widely supported by ``the masses'' as, e.g., the entities that are part of the \ac{LB} 2a cluster. 
Manual inspection of the subclusters confirms this, revealing entities such as Summerland Secondary School, but also a long-tail of person entities (e.g., AFL athlete Rodney Filer, jazz musician Kjetil M{\o}ster) in \ac{LB} 1, whereas entities that represent more ``substantive'' events are in \ac{LB} 2a, e.g., the 2012 Benghazi attack, the Greek withdrawal from the Eurozone, and the Incarceration of Daniel Chong.

\subsubsection{Summary}
In this section, we answer our first research question: ``\acl{rq:clusters}''

We performed hierarchical clustering using a burst similarity-metric of the emerging entity time series. 
We discovered two distinct emergence patterns: 
\acl{EB} entities and \acl{LB} entities. 
Both the visual inspection of the cluster signatures, as the analysis of the descriptive statistics of the time series in the clusters support the same findings: 
the \ac{EB} entities are characterized by fewer but higher bursts, with shorter emergence durations and lower emergence volumes.  
The \ac{LB} entities on the other hand, seem to emerge more slowly, with a more gradual increase of exposure in the online text streams, before reaching the point at which the entity is added to the KB. 
This can be seen both in the cluster signature, as in the descriptive statistics, which show longer durations, higher volumes, and a larger number of bursts. 
 
\subsection{Entity Emergence Patterns in Social Media and News}
\label{sec:rq2}

In this section, we answer our second research question: ``\acl{rq:substreams}''
The news or social media document streams represent content from different sources: 
the news stream consists of traditional online news sources, where the content is mostly written by professional journalists;
the social media stream contains mostly user-generated content, and consists of, e.g., forums and blog posts, but also content that was shared through other social media platforms such as Twitter and Facebook (through the bit.ly URL-shortener service). 
Studying whether different sources surface different entities, and exhibit different emergence patterns, can provide insights into how the nature of the source may affect the discovery process. 

In the previous section, we have shown that 79,482 entities emerge in the combined news and social media streams. 
Taking a closer look by splitting out these entities by stream, we find 51,095 of these entities emerge in the news stream (i.e., are mentioned in the news stream), similar to the number of entities that emerge in the social media stream, at 51,356. 
Finally, 30,148 of the emerging entities are mentioned in both streams before being incorporated into Wikipedia. 

In order to answer our second research question, we consider
(i)~the emergence patterns of entities that emerge in social media and news streams, and 
(ii)~entities that appear in both streams vs.\
(iii)~entities that appear in only one of the two streams.

\subsubsection{Global: News vs.\ Social}

First, we compare the emergence patterns of entities in news and social streams. 
We apply the same hierarchical clustering method as we have done in Section~\ref{sec:rq1} on the two subsets of entities that emerge in news and social media streams (where $n_{news}$ = 51,095 and $n_{social}$ = 51,356). 

Unsurprisingly, the emergence patterns are largely the same in the two streams and highly similar to the global patterns studied in Section~\ref{sec:rq1}. 
Both the news and social streams exhibit the same general global emergence pattern, witnessed by the largely similar clusterings we yield.
Both streams exhibit groups that are similar to the \acl{EB} and \acl{LB} entities discovered in the previous section (shown in Figure~\ref{fig:clusters}). \\

\begin{figure}[t]
\centering
\includegraphics[width=.85\linewidth]{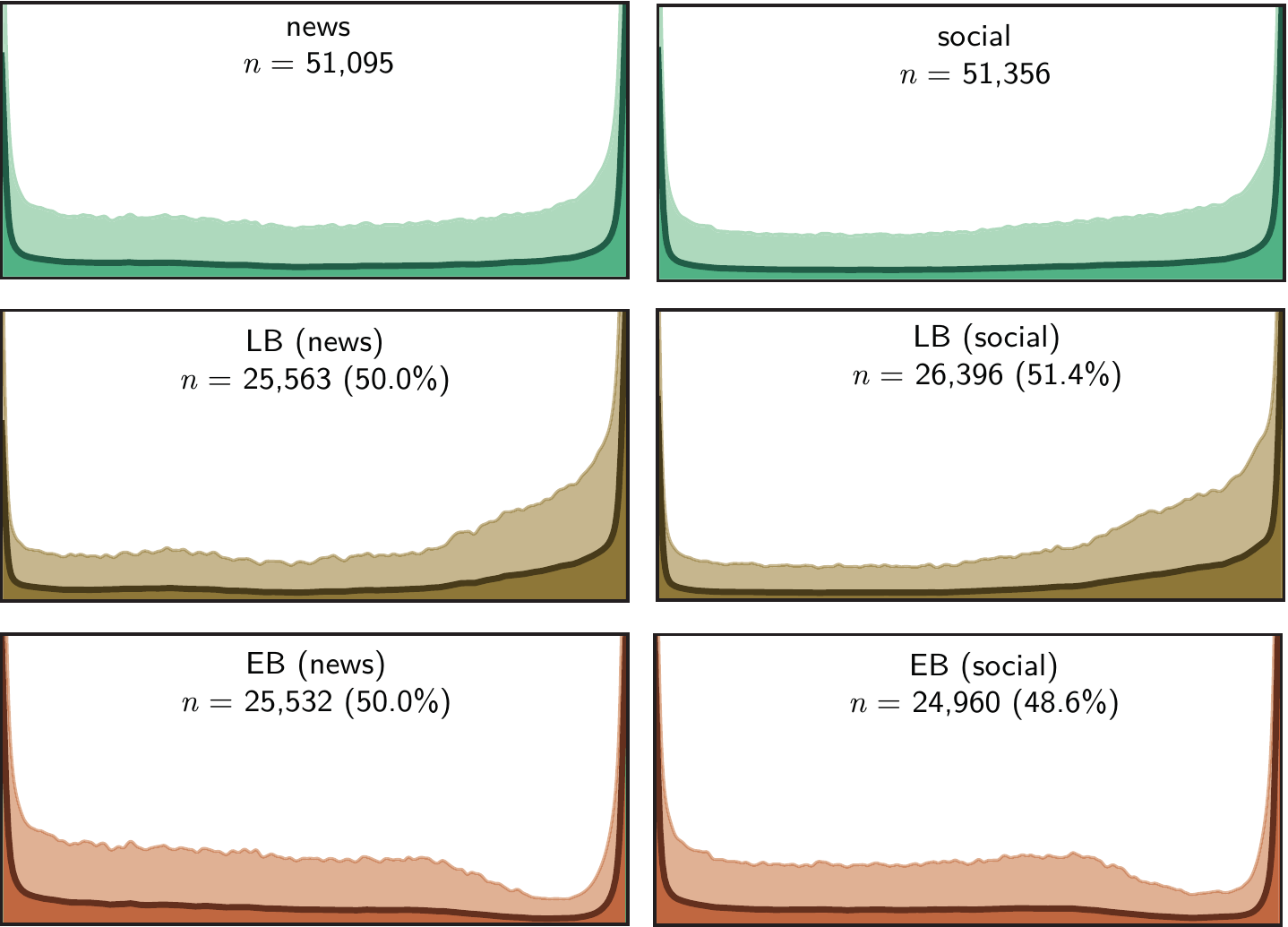} 
\caption{News vs.\ Social stream cluster signatures. 
The top row shows the global cluster signature of the news (left) and social (right) streams. 
The bottom two rows show the signatures of the \acl{LB} and \acl{EB} entity clusters for each stream (news left, social right).}
\label{fig:substreamclustercomparison}
\end{figure}

\noindent
Figure~\ref{fig:substreamclustercomparison} first shows the global emergence patterns (top row, in green), and then the \ac{LB} and \ac{EB} entity clusters in the news (left) and social media (right) streams in the bottom two rows. 
Looking more closely at the global signatures of the news and social media streams, we note that entities that emerge in the news stream have slightly more of their emergence volume mass after the initial burst (i.e., the left-hand side of the plot), compared to the global pattern of the social media stream, which exhibits more of a gradual increase of emergence volume towards the final burst (i.e., the right-hand side of the plot). 
This may be attributed to the slightly higher proportion of \acl{EB} entities in the news stream, which has 50.0\% of its entities falling in this cluster, while the social media stream has 48.6\%. 
The two main clusters (\acl{LB} and \acl{EB}) show broadly similar distributions and patterns, with the \ac{LB} news cluster showing slightly more mass after the initial burst. 
This may be attributed to the ``blind spot'' of the news stream in our dataset (shown in Figure~\ref{fig:substreamvolume}).

\begin{table}
\centering
\caption{Global time series and bursts statistics per type of text stream. }
\label{tab:substreamglobal}
\begin{tabular}{l r@{ $\pm$ }r r r@{ $\pm$ }r r r@{ $\pm$ }r r}
\toprule
stream & \multicolumn{3}{c}{duration (\#days)} & \multicolumn{3}{c}{volume (\#docs)} & \multicolumn{3}{c}{velocity (docs/day)} \\
& mean & std & med. & mean & std & med. & mean & std & med. \\
\midrule
news &   216 & 147 & 170 & 123 & 579 & 28 & 0.75 & 4.28 & 0.19 \\
social & 234 & 153 & 211 &  65 & 239 & 19 & 0.48 & 3.33 & 0.12 \\
\midrule
& \multicolumn{3}{c}{n\_bursts} & \multicolumn{3}{c}{burst durations} & \multicolumn{3}{c}{burst values} \\
\midrule
news &   3.14 & 2.10  & 3 & 0.03 & 0.03 & 0.02 & 0.04 & 0.07 & 0.02 \\
social & 3.75 & 2.54 & 3 & 0.03 & 0.03 & 0.02 & 0.03 & 0.07 & 0.02 \\
\bottomrule
\end{tabular}
\end{table}

Table~\ref{tab:substreamglobal} shows the emergence and burst descriptive statistics of the streams. 
Statistical testing (using the same Kruskal-Wallis test followed by Dunn's multiple comparison test with Holm-Bonferroni adjusted $p$-values used in Section~\ref{subsec:mainclusters}) shows that the differences in the descriptive statistics between both streams are statistically significant at the $\alpha \leq 0.05$ level.
With 216 days on average, entities emerge in news streams more quickly than entities in the social media stream. 
These shorter emergence durations are seen with higher emergence volumes on average: an entity that emerges in news is mentioned on average in 123 documents between their initial and final mention, nearly double the number of documents in social media (65). 
Recall that the total number of documents in the social media stream is larger, at 5.3M documents (versus 1.8M documents for the news stream~---~see also Figure~\ref{fig:substreamvolume}). 
The higher number of documents with comparatively shorter emergence durations further supports the observation that emerging entities in news are picked up quicker than those that emerge in social media. 

Furthermore, in the previous section, we have seen how \acl{EB} entities exhibit fewer but higher bursts, and reasoned they represent more ``urgent'' or timely entities. 
The emergence features of entities emerging in news supports this notion of timeliness or urgency: they exhibit higher and fewer bursts on average. 

In summary, we have shown that entities emerging in news and social media show broadly similar patterns, with both the cluster signatures and descriptive statistics being similar to the emergence patterns in the combined text streams in Section~\ref{sec:rq1}. 
We have also shown that news streams seem to surface entities more quickly than social media streams, which we attribute to the different nature of the streams (professional and authoritative versus unedited and user-generated). 
In Section~\ref{sec:rq3} we revisit this hypothesis by studying how the types of emerging entities differ between streams.

\subsubsection{Who's First?}
Of the 79,482 entities that emerge in the 18 month period our dataset spans, 30,148 appear in both the news and social media stream;  
20,947 entities are mentioned exclusively in the news stream, never appearing in social media (\texttt{news-only}) between surfacing in online text streams and being incorporated into the KB. 
Finally, 21,208 appear only in the social media stream (\texttt{social-only}). 
See also Table~\ref{tab:emergewhere}. 

Of the 30,148 entities that emerge in both streams, the majority appears in the social media stream before they appear in the news stream. 
This may be explained by the nature of the publishing cycles of the two streams; whereas traditional newswire has a more thorough publishing cycle, where stories need to be checked and edited before being published, social media~---~in particular forums and blog posts~---~follows a more unedited and direct publishing cycle. 

The entities that appear in a social media stream first, which we denote \texttt{social-first}, cover 62.9\% ($n$ = 18,967) of the entities that emerge in both streams. 
The opposite pattern, where entities appear in news before they appear in social media (\texttt{news-first}), comprises 29.1\% of the entities that emerge in both streams ($n$ = 8,794). 
Entities that emerge in news first, subsequently appear in social media streams faster than vice versa: 
it takes a \texttt{news-first} entity on average 66 days to appear in the social media stream after surfacing in the news stream, whereas the other way around takes 49 days. 
The remaining comparatively small number of entities are mentioned in both streams on the same day (\texttt{same-time}): 7.9\% ($n$ = 2,387). 
The latter group of entities are expected to emerge more quickly, by virtue of appearing more widely in public discourse, and hence being more urgent and central. 

\begin{table}
\centering
\caption{Time series and burst statistics for entities that emerge in either or both streams (first). Burst and emergence features for our five groups of entities: 
entities that emerge in both streams, but are first mentioned in the news stream (\texttt{news-first}), 
entities that emerge in both streams, but are first mentioned in the social media stream (\texttt{social-first}), 
entities that emerge in both streams, and appear in both on the same day (\texttt{same-time}), 
entities that emerge only in the news stream (\texttt{news-only}), and finally, 
entities that emerge only in the social media stream (\texttt{social-only}).
}
\label{tab:emergewhere}
\begin{tabular}{l r@{ $\pm$ }r r r@{ $\pm$ }r r r@{ $\pm$ }r r}
\toprule
stream & \multicolumn{3}{c}{duration (\#days)} & \multicolumn{3}{c}{volume (\#docs)} & \multicolumn{3}{c}{velocity (docs/day)} \\
& mean & std & med. & mean & std & med. & mean & std & med. \\
\midrule
news first   &  298 & 139 & 305 & 123 & 291 & 53 & 0.58 & 1.59 & 0.21 \\
social first &  281 & 157 & 276 & 182 & 445 & 74 & 0.95 & 3.22 & 0.32 \\
same time    &  197 & 147 & 163 & 192 & 662 & 67 & 2.87 & 23.59 & 0.51 \\
only news    &  250 & 152 & 216 & 415 & 2,215 & 65 & 1.45 & 6.45 & 0.35 \\
only social  &  214 & 148 & 190 & 33 & 134 & 12 & 0.41 & 2.60 & 0.08 \\
\midrule
& \multicolumn{3}{c}{n\_bursts} & \multicolumn{3}{c}{burst durations} & \multicolumn{3}{c}{burst values} \\
\midrule
news first   &  3.92 & 2.50 & 3 & 0.02 & 0.02 & 0.02 & 0.04 & 0.08 & 0.02 \\
social first &  4.10 & 2.86 & 3 & 0.03 & 0.03 & 0.02 & 0.04 & 0.08 & 0.02 \\
same time    & 4.25 & 2.86 & 4 & 0.02 & 0.02 & 0.02 & 0.04 & 0.08 & 0.01 \\
only news    & 3.40 & 2.32 & 3 & 0.03 & 0.03 & 0.02 & 0.03 & 0.08 & 0.01 \\
only social  &  2.99 & 2.08 & 2 & 0.03 & 0.03 & 0.02 & 0.04 & 0.08 & 0.02 \\
\bottomrule
\end{tabular}
\end{table}

The burst and emergence descriptive statistics, as recorded in Table~\ref{tab:emergewhere}, support this view: the \texttt{same-time} entities show comparatively short emergence durations. 
With an average 197 days between being first mentioned and being added to Wikipedia, \texttt{same-time} entities emerge substantially quicker than entities that appear in both streams at different times (at 298 days and 281 days on average for \texttt{news-first} and \texttt{social-first} entities respectively). 
Furthermore, the \texttt{same-time} entities exhibit higher emergence volumes too, accounting for the highest overall velocity at 2.87 documents per day, further supporting the hypothesis that these entities are more urgent and central. 

Entities that emerge only in one of the two media streams show shorter emergence durations than those that appear in both but at different times (at 250 and 214 for news and social media respectively). 
These observations can be explained by the fact that longer duration means that it is more likely for an entity to cross from one stream to the other. 
Similarly, the longer average emergence durations of the \texttt{news-first} and \texttt{social-first} entities are paired with a larger number of bursts (around 4 on average, versus 3 on average for entities that emerge in either the news or social media stream). 
Much like what we saw in the previous section between \acl{LB} and \acl{EB} entities, the shorter durations and smaller number of bursts suggests that entities that emerge in a single stream are more central or important in public discourse.

\subsubsection{Summary}
News and social media streams show broadly similar emergence patterns for entities. 
However, while the patterns may be similar, the population and the behavior of entities emerging in news and social differ significantly. 
More specifically, entities emerging in the social media stream seem to do so slower on average than in news. 
Looking in more detail at the interactions between the two streams, we notice that entities that appear in both streams on the same day are the fastest to be incorporated into the KB. 
Furthermore, we find that entities that first emerge in social media are more quickly picked up in news streams than vice versa. 
 
\subsection{Emergence Patterns of Different Entity Types}
\label{sec:rq3}

In this section, we answer our third research question: ``\acl{rq:entities}''
First, we analyze the descriptive statistics of each entity type in our dataset, to assert whether different types of entities exhibit different behavior in terms of how they emerge in online text streams. 
Next, we study whether entity types are distributed differently over the news and social media text streams.

\subsubsection{Entity Types: Temporal Patterns}
First, we study each entity type in isolation, i.e., we study the descriptive statistics per entity type. Table~\ref{tab:entitytypes} shows all entity types with a frequency of $\geq 400$ in our dataset. 

\begin{figure}[t]
\centering
\includegraphics[width=.95\linewidth]{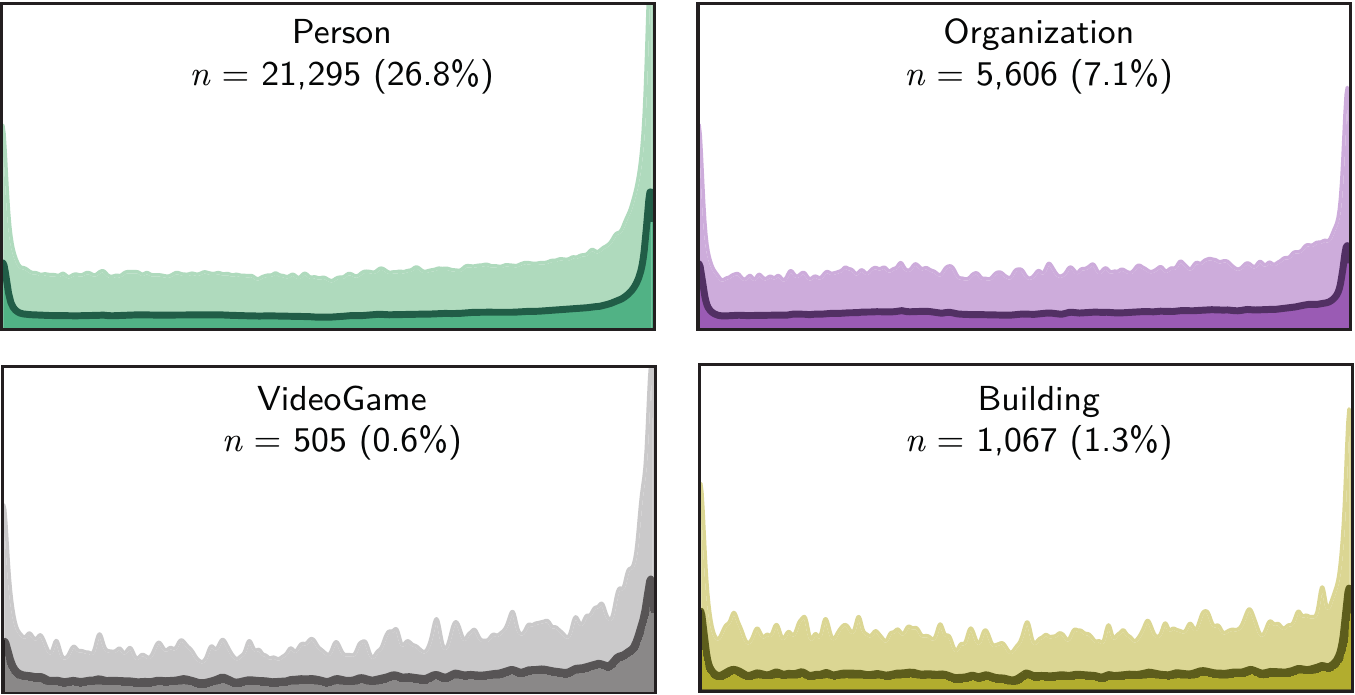} 
\caption{Type signatures of the Person, Organization, VideoGame and Building types. Even though the number of entities of that type differ substantially, the signatures show roughly similar patterns.}
\label{fig:classplots}
\end{figure}

\noindent
We find that the entity type signatures (i.e., the average over all time series for each entity type) are highly similar to the global pattern (visualized in Figure~\ref{fig:global}), suggesting that the time series of mentions are highly variable within an entity type. 
To illustrate this, see Figure~\ref{fig:classplots} for an example of two common entity types (top row) and two less frequently emerging types (bottom row).
Whereas the signature becomes smoother as the number of entities increase, the overall pattern is highly similar across the four types. 

\begin{table}[t]
\small
\centering
\caption{Descriptive statistics per entity type (for types that occur $\geq 400$ in our dataset). }
\label{tab:entitytypes}
\begin{tabular}{l@{}r @{~} r@{ $\pm$ }r @{~} r @{~} r@{ $\pm$ }r @{~} r @{~} r@{ $\pm$ }r @{~} r}
\toprule
stream & n\_samples & \multicolumn{3}{c}{duration (\#days)} & \multicolumn{3}{c}{volume (\#docs)} & \multicolumn{3}{c}{velocity (docs/day)} \\
& & mean & std & med.\ & mean & std & med.\ & mean & std & med.\ \\
\midrule
Person & 21,295 & 270 & 151 & 254 & 243 & 692 & 71 & 1.03 & 3.32 & 0.32 \\
Athlete & 8,018 & 260 & 150 & 235 & 264 & 674 & 76 & 1.05 & 2.45 & 0.37 \\
InformationEntity & 7,847 & 242 & 154 & 210 & 294 & 1,923 & 90 & 1.42 & 6.53 & 0.51 \\
CreativeWork & 7,795 & 243 & 154 & 211 & 294 & 1,928 & 90 & 1.42 & 6.54 & 0.52 \\
Organization & 5,606 & 279 & 153 & 270 & 335 & 1,812 & 71 & 1.40 & 14.44 & 0.31 \\
Place & 3,689 & 274 & 149 & 273 & 122 & 448 & 33 & 0.48 & 1.61 & 0.16 \\
Company & 2,536 & 284 & 156 & 275 & 462 & 1,964 & 108 & 1.98 & 20.88 & 0.47 \\
MusicalWork & 2,474 & 218 & 148 & 181 & 170 & 533 & 78 & 1.13 & 2.23 & 0.49 \\
Movie & 2,033 & 267 & 154 & 247 & 279 & 1,322 & 87 & 1.20 & 6.57 & 0.42 \\
OfficeHolder & 1,929 & 287 & 158 & 284 & 210 & 476 & 73 & 0.85 & 1.88 & 0.31 \\
MusicGroup & 1,649 & 293 & 150 & 289 & 221 & 393 & 86 & 0.95 & 1.97 & 0.34 \\
Artist & 1,624 & 299 & 152 & 302 & 240 & 564 & 75 & 0.95 & 2.19 & 0.30 \\
ArchitecturalStructure & 1,591 & 279 & 149 & 284 & 133 & 436 & 36 & 0.47 & 1.14 & 0.18 \\
PopulatedPlace & 1,521 & 262 & 145 & 244 & 119 & 481 & 31 & 0.53 & 2.15 & 0.15 \\
Building & 1,067 & 281 & 150 & 290 & 125 & 374 & 34 & 0.43 & 0.98 & 0.17 \\
TelevisionShow & 1,043 & 229 & 158 & 193 & 228 & 503 & 87 & 1.33 & 3.05 & 0.57 \\
WrittenWork & 959 & 267 & 147 & 245 & 307 & 864 & 88 & 1.26 & 2.99 & 0.44 \\
EducationalInstitution & 915 & 290 & 144 & 308 & 108 & 564 & 30 & 0.41 & 2.21 & 0.12 \\
Software & 769 & 250 & 156 & 221 & 732 & 5,413 & 192 & 3.04 & 16.33 & 0.92 \\
School & 554 & 280 & 142 & 305 & 56 & 158 & 25 & 0.29 & 2.24 & 0.11 \\
Book & 524 & 272 & 147 & 263 & 286 & 827 & 93 & 1.17 & 3.03 & 0.44 \\
VideoGame & 505 & 229 & 150 & 198 & 381 & 657 & 189 & 2.08 & 3.24 & 1.03 \\
DesignedArtifact & 409 & 217 & 149 & 187 & 1,420 & 7,142 & 214 & 7.04 & 20.42 & 1.39 \\
Infrastructure & 403 & 271 & 146 & 269 & 127 & 560 & 34 & 0.47 & 1.42 & 0.17 \\
\texttt{null} & 39,807 & 225 & 151 & 196 & 98 & 861 & 15 & 0.58 & 3.70 & 0.10 \\
\bottomrule
\end{tabular}
\end{table}

In contrast, the descriptive statistics per entity type does yield clear patterns, as we will describe next.
First, the \texttt{null} class, i.e., the entities that are not assigned an entity type in DBpedia exhibit very low emergence volumes, with an average of 98 documents over 225 days. 
Which may be explained by their nature, as they are not assigned a class in the DBpedia ontology, they are likely to be very long-tail, or unpopular entities. 

Second, we note a group of ``fast'' emerging entity types, i.e., those with short emergence durations and/or high emergence velocities, e.g., \texttt{DesignedArtifact}, \texttt{CreativeWork}, \texttt{Musi\-calWork}, and \texttt{VideoGame}. 
In particular, the \texttt{Designed\-Artifacts} type shows high emergence velocities: it takes entities of this type on average 217 days to be incorporated into the KB, with an average volume of over 7 documents a day (versus 0.87 for the global average). 
The \texttt{DesignedArtifact} type includes entities such as devices and products, e.g., smartphones, tablets, and laptops. 
The relatively fast transition of the entities of this type may be explained by their nature: they have short ``life-cycles'' and may be superseded or replaced at high frequencies. 
Consider, e.g., the release or announcement of a new smartphone: this event typically generates a lot of attention in a short timeframe, which may result in a fast emergence. 
Similar to devices of the \texttt{Designed\-Artifact}-type, creative works (\texttt{Creative\-Work}, including \texttt{Musical\-Work}, \texttt{Written\-Work}, \texttt{Movie}, etc.) share this characteristic: they play a central but short-lived role in public discourse. 

Third, the ``slower'' entities, i.e., those with longer emergence durations and lower emergence volumes, are largely person types such as writers (\texttt{Writer}), artists (\texttt{Artist}), and political figures (\texttt{Office\-Holder}), but also schools (\texttt{School} and \texttt{Educational\-Institution}), and geographical entities (e.g., \texttt{Building}, \texttt{Archi\-tec\-tur\-al\-Struc\-tu\-re},  \texttt{Place}, and \texttt{Popula\-ted\-Place}). 
These entities by their nature may have longer life-cycles, with a more gradual ``rise to fame'' (politicians, artists), and play a less central role in public discourse (schools, buildings). 
The opening of a new school may appear briefly in regional and local news sources, but is unlikely to be globally and widely reported. 
Politicians generally have a long and gradual career, surfacing e.g., in regional media, and do not suddenly ``burst'' into existence. 

To better understand the difference between ``fast'' and ``slow'' entities, we examine the popularity of entities. Table~\ref{tab:typepopularity} lists the average number of pageviews received per entity in 2015, grouped by type.
Entity types that exhibit short emergence durations and high velocities are all in the top 10 (ranks 3, 4, and 9, for \texttt{Video\-Game}, \texttt{Creative\-Work}, and \texttt{Designed\-Artifact}, respectively), whereas the slower entity types all reside towards the lower ranks of the table, e.g., rank 19, 22, and 24 for \texttt{Building}, \texttt{Educational\-Institution} and \texttt{School}, respectively. 
This suggests that entity types that emerge more quickly remain more popular over time. 

\begin{table}[t]
\centering
\caption{``Popularity,'' i.e., average total number of pageviews in 2015 of each entity in our dataset, aggregated per entity type. Ranked in descending order. }
\label{tab:typepopularity}
\begin{tabular}{rlr@{ $\pm$ }rr}
\toprule
Rank & Type & \multicolumn{2}{c}{Mean $\pm$ std} & Median \\
\midrule
1 & Movie & 98,387 & 322,166 & 14,352 \\
2 & TelevisionShow & 97,098 & 309,172 & 9,765 \\
3 & VideoGame & 51,236 & 166,802 & 11,852 \\
4 & CreativeWork & 50,024 & 213,490 & 5,716 \\
5 & InformationEntity & 49,704 & 212,816 & 5,634 \\
6 & Software & 43,657 & 149,499 & 9,582 \\
7 & MusicGroup & 38,883 & 133,336 & 5,400 \\
8 & Artist & 35,607 & 122,032 & 4,116 \\
9 & DesignedArtifact & 29,830 & 82,081 & 7,191 \\
10 & Book & 18,248 & 109,400 & 3,126 \\
11 & WrittenWork & 14,227 & 86,637 & 1,801 \\
12 & Person & 13,772 & 77,791 & 1,568 \\
13 & MusicalWork & 10,443 & 25,009 & 3,523 \\
14 & Athlete & 9,415 & 41,887 & 1,545 \\
15 & Organization & 9,003 & 45,140 & 1,816 \\
16 & Company & 7,624 & 21,371 & 2,566 \\
17 & OfficeHolder & 3,763 & 16,167 & 958 \\
18 & ArchitecturalStructure & 3,189 & 16,978 & 1,042 \\
19 & Building & 3,180 & 20,106 & 987 \\
20 & Infrastructure & 2,813 & 6,769 & 1,085 \\
21 & Place & 2,339 & 12,649 & 827 \\
22 & EducationalInstitution & 1,799 & 3,031 & 862 \\
23 & PopulatedPlace & 1,743 & 9,081 & 694 \\
24 & School & 1,137 & 1,426 & 747 \\
\bottomrule
\end{tabular}
\end{table}

\subsubsection{Entity Types per Stream}
To determine whether the entity types observed in news and social media streams differ significantly, we apply Pearson's chi-squared test. 
This allows us to identify entity types that are observed more than expected (w.r.t.\ the global distribution) in the news and social media streams. 
See Table~\ref{tab:streamchigrams}. 

\begin{table}[t]
\centering
\caption{Entity types ranked by exceeded observed frequency w.r.t. expected frequency using chi-grams. The popularity rank from Table~\ref{tab:typepopularity} is shown in brackets. }
\label{tab:streamchigrams}
\begin{tabular}{ll}
\toprule
More in news than social & More in social than news \\
\midrule
Person (12) & DesignedArtifact (9) \\
Athlete (14) & VideoGame (3) \\
Organization (15) & Infrastructure (20) \\
Place (21) & Book (10) \\
InformationEntity (5) & TelevisionShow (2) \\
CreativeWork (4) & Software (6) \\
Company (16) & School (24) \\
PopulatedPlace (23) & MusicalWork (13) \\
OfficeHolder (17) & WrittenWork (11) \\
ArchitecturalStructure (18) & Movie (1) \\
\bottomrule
\end{tabular}
\end{table}

The majority of the entity types observed more frequently in social media streams (right column of Table~\ref{tab:streamchigrams}) are the ones we identified to be comparatively fast in emerging in the previous section, e.g., \texttt{DesignedArtifact}, \texttt{VideoGame}, \texttt{TelevisionShow}, \texttt{Software}. 
The (average) popularity rankings from Table~\ref{tab:typepopularity} show how 6 out of 10 of the entity types that are observed more frequently in social media are in the top 10 most popular entity types of Table~\ref{tab:typepopularity}. 
The average popularity rank of the social media entity types is 9.9.  At the other end, the types of entities seen more frequently in news are both slower in transition on average (e.g., \texttt{Person}, \texttt{Organization}, and \texttt{Company}), and are characterized by being more ``general'' or less niche types compared to the entity types seen more frequently in the social media stream. 
Videogame or smartphones are likely to see more exposure in social media streams in, e.g., blog and forum posts, but more general entity types such as people, places, and organizations are more natural subjects for (traditional) news media. 
Merely 2 entity types that are more often seen in news streams are in the top 10 most popular entity types in Table~\ref{tab:typepopularity}. 
The average popularity rank of news-specific entity types is 14.5. 
These observations suggest that entities that remain popular over time are more likely to emerge in the social media stream. 

\subsubsection{Summary}
In summary, we have shown that different entity types exhibit substantially different emergence patterns, but entities that belong to a particular type show broadly similar emergence patterns. 
Furthermore, we have shown that different entity types are distributed distinctly over different online text streams, which can be intuitively explained by looking at both the nature of the entity types and the nature of the streams. 
Next, we have seen that entities that emerge fast are more likely to remain popular over time.
Finally, we have seen that in social media streams entities emerge faster, and more remain popular over time in comparison to news streams. 
  
\section{Conclusion}
\label{sec:plos-conclusion}

In this chapter, we studied entities as they emerge in online text streams. We did so by studying a large set of time series of mentions of entities in online news streams before they are added to Wikipedia. 
We studied implicit groups of similarly emerging entities by applying a burst-based agglomerative hierarchical clustering method and explicit groups by isolating entities by whether they emerge in news or social media streams. 
Next, we summarize the findings, implications and limitations of our study into the nature of emerging entities in online text streams. 

\subsection{Findings}
In Section~\ref{sec:rq1}, we applied a clustering method to the time series of mentions of emerging entities to find implicit groups of similarly emerging entities. 
We found that, globally, entities have a long time span between surfacing in online text streams and being incorporated into the KB.  
During this time span, an emerging entity is associated with multiple bursts (i.e., resurfaces into public discourse), however both the emerging entities' introduction into public discourse and subsequent incorporated into Wikipedia occur in the largest document bursts. 
Emergence durations and volumes show large standard deviations, indicating that they differ substantially between entities. 
For this reason, we turned to time series clustering to uncover distinct groups of entities.
We discovered two distinct emergence patterns: \acf{EB} entities and \acf{LB} entities. 
Analysis suggests that \ac{EB} entities comprise mostly ``head'' or popular entities; they exhibit fewer and higher bursts, with shorter emergence durations and lower emergence volumes. 
The \ac{LB} entities emerge more slowly on average, and witness a more gradual increase of exposure in online text streams. 
The emergence patterns we visualized differ substantially from the global average and from, e.g., the type signatures studied in Section~\ref{sec:rq3}, suggesting that the entities in each of the underlying clusters exhibit substantially different and distinct emergence patterns from entities in the other clusters. 

In Section~\ref{sec:rq2}, we showed that entities emerging in news and social media streams display very similar emergence patterns, but that on average, entities that emerge in social media have a longer period between surfacing and being incorporated into the KB. 
We hypothesize that this can be attributed to the nature of the underlying sources.
Traditional news media is more mainstream and professional, with a larger audience and reach, and more authority than social media streams. 
Our findings are in line with those of \citet{enlighten82566}, who compare breaking news on traditional media with that on social media. 
Their findings suggest reported events overlap largely between both media, however, social media exhibits in addition a long tail of minor events, which may explain the longer uptake on average. 
\citet{Leskovec:2009:MDN:1557019.1557077} find that the ``attention span'' for news events on social media both increases and decays at a slower rate than for traditional news sources, which may explain the comparatively slower uptake on social media. 

Finally, we studied entity types in Section~\ref{sec:rq3}. 
We showed that different entity types exhibit substantially different patterns, but entities of a similar type show similar patterns. 
Some entity types, e.g., devices or creative works, on average emerge faster than entities such as buildings, locations, and people. 
At the same time, the former ``faster'' entity types remain more popular over time (as seen through their pageview counts). 
One aspect that distinguishes between ``fast'' and ``slow'' entity types, is that the former are more likely to appear in so-called ``soft news'' (i.e., news that covers sensational or human-interest events and topics, e.g., news related to celebrities and cultural artifacts), whereas the slower entity types are more likely to be associated with more substantive ``hard news'' (i.e., news that encompasses more pressing or urgent events and topics, e.g., reports related to political elections)~\cite{10.2307/2776752}. 
\citet{36915} studied the differences in ``attention span'' of the public (as measured through search engine query volumes) and the traditional news media (as measured through coverage volume) for ``hard'' and ``soft news,'' 
and found that, in line with the findings of \citet{Leskovec:2009:MDN:1557019.1557077}, hard news is associated with a relatively short period of attention from the public (as measured by query volume). 
Soft news exhibits a slower decrease of the public's attention (as seen through slower declines in query volumes), which supports our finding that faster entity types~---~entities more likely to be associated with soft news~---~tend to remain more popular over time. 
Furthermore, the relatively longer attention for soft news may explain the quicker uptake of the ``fast'' entity types, e.g., ``cultural'' artifacts (e.g., movies, TV shows, artists) may emerge more quickly, as they are more widely supported, followed, and more strongly represented in our online public discourse. 

Finally, we showed how entity types are distributed differently over news and social media streams. 
This difference in entity types may be explained by the nature of the streams.
Partly because of the open, democratic, and user-generated nature of the Web 2.0, and blogging in particular~\cite{JCC4:JCC41458}, blogs no longer simply pick up news stories from the traditional media.
The agenda setting power of traditional media is diluting~\cite{doi:10.1177/107769901108800110}.
Moreover, there are situations in which breaking news emerges on social media, e.g., the death of Osama Bin Laden~\cite{Hu:2012:BNT:2207676.2208672}. 
And finally, as mentioned before, the events reported on social media and traditional news overlap, but social media has a long tail of its ``own'' events~\cite{enlighten82566}, which may account for the different distribution of emerging entity types in the social media stream. 

Taking a step back, our findings can be summarized to the observation that emerging entities are not ``born equal,'' i.e., the patterns and features under which an entity emerges differ depending on source and type.

\subsection{Implications}
The findings in this chapter have implications for designing systems to detect emerging entities, and more generally for studying and understanding how entities emerge in public discourse before they are deemed important enough to be incorporated in the KB. 
Here, we list observations and findings that have implications for the discovery process of emerging entities. 
First, we have shown that entities are likely to resurface multiple times in public discourse (i.e., online text streams) before being incorporated into the KB. 
This means that, on average, it takes multiple moments of ``exposure'' for an entity to emerge. 
This suggests that monitoring bursts of newly emerging entities could serve as an effective method for predicting when an entity is about to be incorporated into the KB, i.e., after observing the \emph{initial burst}, it is not too late. 
Furthermore, we have shown that the type of stream in which entities emerge (i.e., news and social media) carries different signals that allow us to model the process of the process of entity emergence. 
More specifically, we have shown how entities that emerge, e.g., in both streams at the same time tend to be added to the KB comparatively quicker than, e.g., those entities that never move from one stream to another. 
These findings suggest that taking different streams into account separately can be beneficial for detecting emerging entities. 
Finally, we have shown that the different types of online text streams also surface different types of entities. 
This provides insights into the difference between ``mainstream media'' and users of the world-wide web, in terms of preferences and interests through the ``agenda setting.''

\subsection{Limitations}
The work presented in this chapter also knows several limitations.

\paragraph{Clusters.}
Part of our findings are derived from the clusters that serve as a starting point for discovering common patterns, similarities, and differences in emergence patterns. 
Clustering and studying cluster signatures is by design a subjective matter~\cite{von2012clustering}. 
Applying unsupervised clustering, entailing different hyperparameters, is by definition a hard task to ``evaluate,'' i.e., to decide whether the resulting clusters show meaningful differences. 
The large (standard) deviations seen in the descriptive statistics within clusters may suggest there exists a wider variety of different entities. 
However, the linear interpolation step we took to cluster time series with variable lengths and different relative timespans (i.e., not temporally aligned) will result in varied time series. 
In our defense, the cluster signatures that result from the clustering method yielded visually discernible and statistically different patterns between clusters, which was not the case for the signatures of the groups of time series from Sections~\ref{sec:rq2} and \ref{sec:rq3} (see, e.g., Figure~\ref{fig:classplots}). 
Finally, the clustering's resulting dendrogram suggests there are distinct and meaningful groups of substantially different time series, as the structure of the dendrogram shows symmetry and clear separations. 

\paragraph{Data.}
Next, the fragmented nature of the dataset that serves as the starting point of our study, the TREC-KBA StreamCorpus 2014, means the coverage, and hence representativeness and comprehensiveness of the data cannot be guaranteed. 
As could be seen in Figure~\ref{fig:substreamvolume}, the dataset contains a blind spot around May 2012, which may affect the time series of all emerging entities. 
To minimize the adverse effects, we normalize all entity document mention time series by the total document volume. 
Furthermore, the choice of underlying sources (that represent the social media and news streams) is limited, e.g., popular social media channels such as Tumblr, Twitter and Facebook are not part of the dataset. 
There may be sampling bias in the types of sources, resulting in a similar bias in the entities. 
That is to say, with another set of sources, we may have had different findings. 
This is unavoidable. 

The entity annotations that represent the starting point of our study into emerging entities cannot be assumed to be 100\% accurate. 
So-called ``cascading errors''~\cite{Finkel:2006:SPC:1610075.1610162} cause the overall accuracy to suffer, by using imperfect tagging by SERIF as input for imperfect FAKBA1 annotations. 
The latter annotations are estimated (from manual inspection) to contain around 9\% incorrectly linked Freebase entities, with around 8\% of SERIF mentions being wrongfully not linked. 
Even more so, the ``difficult'' entity links are long-tail entities, ones that are likely to be included in our filtered set, meaning the accuracy may be relatively worse in our subset of entities. 
However, manually correcting the annotations was beyond the scope of this study, and the large scale of the dataset makes it less likely that wrongfully linked entities are a major issue. 

\begin{table}[t]
\centering
\caption{Top 10 most frequently occurring entities in the FAKBA1 dataset.}
\label{tab:top10entities}
\begin{tabular}{r l r}
\toprule
1 & \texttt{United States} & 236,705,559 \\
2 & \texttt{United Kingdom} & 63,759,238 \\
3 & \texttt{Barack Obama} & 61,412,413 \\
4 & \texttt{China} & 57,273,919 \\
5 & \texttt{Yahoo!} & 49,971,781 \\
6 & \texttt{Facebook} & 45,602,997 \\
7 & \texttt{New York City} & 40,043,377 \\
8 & \texttt{India} & 39,359,865 \\
9 & \texttt{Europe} & 36,342,113 \\
10 & \texttt{Canada} & 29,501,709  \\
\bottomrule
\end{tabular}
\end{table}

Finally, there may be a cultural bias inherent in our choice of datasets: we used English language news sources and social media as well as the English version of Wikipedia. 
As an illustration to the cultural bias inherent to the data sources, see Table~\ref{tab:top10entities} for the top 10 most frequently mentioned entities in the FAKBA1 dataset. 
Hence, one could claim that we studied the emergence patterns of entities for the \emph{English speaking} part of the world. 
Different datasets are likely to yield different findings. 
However, it is unfortunate that the English speaking part of the world is disproportionately represented in our field of research, as witnessed by the biggest constraint in conducting this study: dataset availability.  
We encourage the community to create suitable datasets in other languages and/or reflecting cultural practices in other parts of the planet so as to enable comparative studies.

\subsection{What's Next?}
Following our study into the nature of entities as they emerge in online text streams, and understanding that knowledge bases are dynamic in nature, in the next two chapters we propose automated methods for discovering and searching emerging entities.
More speicfically, in the next chapter we address the task of predicting emerging entities in social streams (Chapter~\ref{ch:ecir}).
Next, we turn to improving their retrieval effectiveness, by constructing dynamic entity representations. More specifically, the dynamic representations aim to capture the different and dynamic ways in which people may refer to or search for entities (Chapter~\ref{ch:wsdm}). 
 \clearpage{}

\clearpage{}
\addtocontents{toc}{\protect\pagebreak[4]}
\chapter{\acl{ch:ecir}}
\label{ch:ecir}

\begin{flushright}
\rightskip=1.8cm``Nec scire fas est omnia.'' \\ 
\vspace{.2em}
\rightskip=.8cm\textit{---Horace, Carmina, IV.}
\end{flushright}

\section{Introduction}
\label{sec:intro}
Entity linking owes a large part of its success to the extensive coverage of today's knowledge bases; they span the majority of popular and well established entities and concepts. 
For most domains this broad coverage is sufficient. 
However, it does not provide a solid basis in domains that refer to ``long-tail'' entities, such as the E-Discovery domain, or in domains where new entities are constantly born, e.g., in the news and social media domain. 
As we have seen in the previous chapter, entities may emerge (and sometimes disappear) before editors of a knowledge base reach consensus on whether an entity should be included in the knowledge base. 
Knowledge bases are never complete: new entities may emerge as events unfold, but at the same time, long-tail, relatively unknown entities may also be added to a KB. 

As a follow-up to Chapter~\ref{ch:plos}, in this chapter we turn our attention to predicting newly emerging entities, i.e., to discover entities that appear in social media streams before they are incorporated into the KB. 
Identifying newly emerging entities that will be incorporated in a knowledge base is important for knowledge base construction, population, and acceleration, and finds applications in complex filtering tasks and search scenarios, where users are not just interested in finding any entity, but also in entity attributes like impact, salience, or importance. 
Identifying emerging entities is closely related to named-entity recognition and classification (NERC), and named-entity normalization (NEN) or disambiguation, with the additional constraint that an entity should have ``impact'' or be important enough to be incorporated in the knowledge base. 
Although impact or importance are hard to model because they depend on the context of a task or domain, we argue that entities that are included in a knowledge base are more important than those that are not, and use this signal for modeling the importance of an entity. 
Through this approach, our method leverages prior knowledge (as encoded in the knowledge base) for discovering newly emerging entities.

Named-entity recognition is a natural approach for identifying these newly emerging entities that are not in the knowledge base. 
However, current models fall short as they do not account for the aforementioned importance or impact of the entity. 
In this chapter, we present an unsupervised method for generating pseudo-ground truth for training a named-entity recognizer to steer its predictions towards newly emerging entities. 
Our method is applicable to any trainable model for named-entity recognition. 
In addition, our method is not restricted to a particular class of entities, but can be trained to predict any type of entity that is in the knowledge base. 

The challenge of discovering newly emerging entities is two-fold:
\begin{inparaenum}[(i)]
\item the first challenge is how to model the attribute of importance, and 
\item the second is the streaming and dynamic nature of the task, where both the input data (social media) and the updates in the knowledge base (i.e., when new entities are added) come in a stream; content that is eligible for addition in Wikipedia today, may no longer be in the future~\cite{10.1371/journal.pone.0038869}, which renders static training annotations unusable. 
\end{inparaenum}

Our approach to discovering newly emerging entities in social streams answers both challenges. 
For the first challenge, we carefully craft the training set of a named-entity recognizer to steer it towards identifying newly emerging entities. 
That is, we leverage prior knowledge of the entities already in the knowledge base, to identify new entities that are likely to share the same attributes, and thus be candidates for inclusion in the knowledge base. 
Just as a named-entity recognizer trained solely on English person-type entities will recognize only such entities, a named-entity recognizer trained on knowledge base entities can be expected to recognize only this type of entity. 
For the second challenge, we provide an unsupervised method for generating pseudo-ground truth from the input stream. 
With this automated method we are not dependent on human annotations that are necessarily limited and domain and language specific, and newly added knowledge will be automatically included.
We focus on social media streams because of the fast paced evolution of content and its unedited nature, which make it a challenging setting for predicting which entities will feature in a knowledge base. 
The main research question we seek to answer in this chapter is: 
\begin{description}[labelwidth=1cm,leftmargin=!]
\item[\ref{rq:ecir}] Can we leverage prior knowledge of entities of interest to bootstrap the discovery of new entities of interest?
\end{description}

\noindent
To answer this question, we propose our novel method, and formulate two sub questions. 
The first sub question aims to study the challenge of the noisy, unedited nature of social media streams. 
In order to generate high quality training data for our named-entity recognizer, we propose and experiment with two sampling methods, and answer the following subquestion: 

\begin{enumerate}[label=\ref{rq:ecir}.\arabic*,labelwidth=1cm,leftmargin=!,align=left]
\item What is the utility of our sampling methods for generating pseudo-ground truth for a named-entity recognizer?\label{rq:ecir1}
\end{enumerate}

\noindent
We measure utility within the task of predicting new knowledge base entities from social streams as the prediction effectiveness of a named-entity recognizer trained using our method.
Next, we study the impact of the amount of prior knowledge on the effectiveness of identifying newly emerging entities.
Our second subquestion is: 

\begin{enumerate}[label=\ref{rq:ecir}.\arabic*,resume,labelwidth=1cm,leftmargin=!,align=left]
\item What is the impact of the size of prior knowledge on predicting new knowledge base entities?\label{rq:ecir2}
\end{enumerate}
 
\section{Approach}
\label{sec:method}

\begin{figure}
  \centering
  \includegraphics[width=\textwidth]{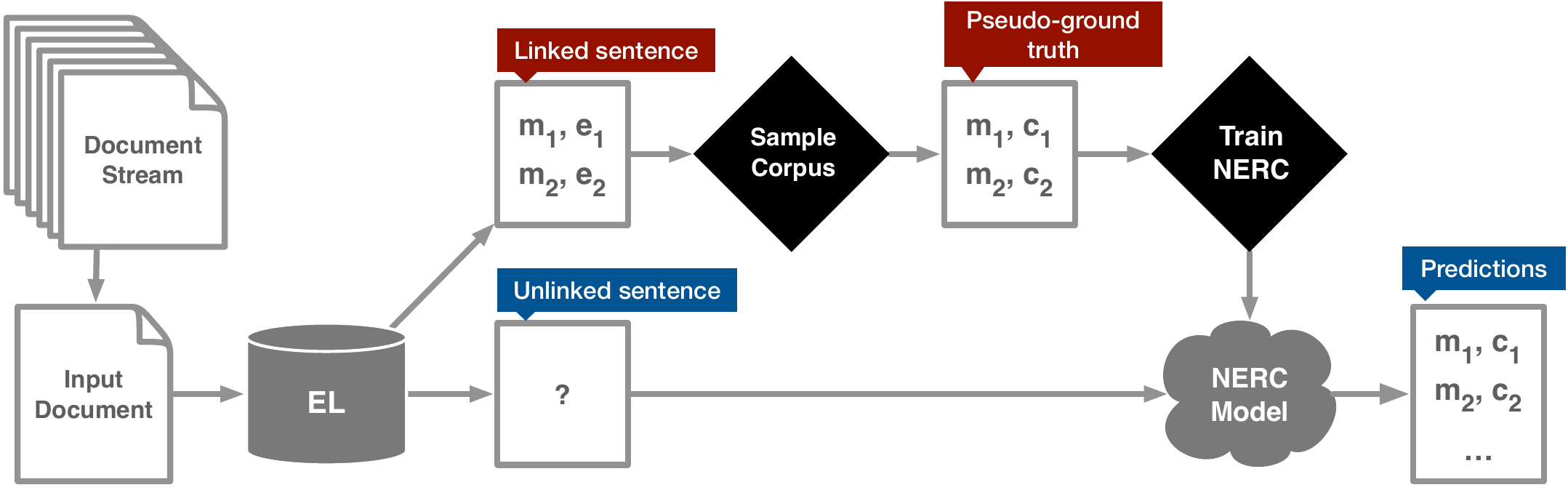}
  \caption{Our approach for \emph{generating pseudo-ground truth}
  for training a NERC method, and for \emph{predicting emerging entities}. 
  }
  \label{fig:uned}
\end{figure}

We view the task of identifying new KB entities in social streams as a combination of an \emph{entity linking} (EL) problem and a \emph{named-entity recognition and classification} (NERC) problem. 
We visualize our method in Figure~\ref{fig:uned}. 
Starting from a document (social media post) in a document stream (social media stream), we extract sentences, and use an EL system to identify referent KB entities in each sentence. 
If any is identified, the sentence is pooled as a candidate training example for our NERC method (we refer to this type of sentence as a \emph{linkable sentence}), otherwise it is routed to our NERC method for identifying newly emerging entities (\emph{unlinkable sentences}): an underlying assumption behind our method is that the first place to look for emerging entities is the set of unlinkable sentences.

Most of our attention in this chapter is devoted to training the NERC method. Two ideas are important here. 
First, we extend the distributional hypothesis~\cite{Firth:1957} 
(i.e., words that occur in the same contexts tend to have similar meaning) from words to entities; we hypothesize that emerging entities that should be included in the knowledge base occur in similar contexts as current knowledge base entities. 
Second, we apply EL on the input stream and transform its output into pseudo-ground truth for NERC; this results in an unsupervised way of generating pseudo-ground truth, with the flexibility of choosing any type of entity or concept described in the KB.
 
\section{Unsupervised Generation of Pseudo-ground Truth}
\label{sec:training}

We start with the output of an entity linking method\footnote{\url{http://semanticize.uva.nl}} \cite{meij2012,Odijk:2013aa}, on a sentence of a document in the stream. 
This output is our source for generating training material. 
The output consists of tuples of \emph{entity mentions} and \emph{entities} ($m, e$ pairs in Figure~\ref{fig:uned}). 

Since we are allowed to use generic corpora from any domain, e.g., news, microblog posts, emails, chat logs, we may expect to have noise in our pseudo-ground truth. 
We apply various sampling methods to select sentences that make up a high quality training corpus. 
These sampling methods are described in Section~\ref{sec:sampling}.

\noindent
After sampling, we convert the remaining sentences in a format suitable for input for our NERC method. 
This format consists of the \emph{entity span}; the sequence of tokens that refers to an entity, i.e. the 
entity mention, and \emph{entity class} for each linked entity ($m, c$ pairs in Figure~\ref{fig:uned}). 
To denote the entity span, we apply the BIO-tagging scheme~\cite{rams:text95}, where each token is tagged with whether it is the 
\textbf{B}eginning of an entity mention, 
\textbf{I}nside an entity mention, or 
\textbf{O}utside of an entity mention, so that a document like:
\begin{quote}
Kendrick Lamar and A\$AP Rocky. That's when I started listening again. Thanks to Brendan.
\end{quote}
\noindent
becomes:
\begin{quote}
Kendrick$_{B}$ Lamar$_{I}$ and$_{O}$ A\$AP$_{B}$ Rocky$_{I}$. That's$_{O}$ when$_{O}$ I$_{O}$ started$_{O}$ listening$_{O}$ again$_{O}$. Thanks$_{O}$ to$_{O}$ Brendan$_{B}$.
\end{quote}
\noindent
The final step is to assign a class label to an entity. 
As not all knowledge bases associate classes to their entities, we use DBpedia for looking up the entity and extracting the entity's DBpedia ontology class, if any; see Section~\ref{sec:exp} for details. 
Our example then becomes:
\begin{quote}
Kendrick$_{B\text{-}PER}$ Lamar$_{I\text{-}PER}$ and$_{O}$ 
A\$AP$_{B\text{-}PER}$ Rocky$_{I\text{-}PER}$. That's$_{O}$ when$_{O}$ I$_{O}$ 
started$_{O}$ listening$_{O}$ again$_{O}$. Thanks$_{O}$ to$_{O}$ 
Brendan$_{B\text{-}PER}$.
\end{quote}
\noindent
Now we can proceed and train our NERC method with our generated pseudo-ground truth. 
We do so using a two-stage approach~\cite{Buitinck:2012:TNR:2368172.2368192}, where both the first stage of recognizing the entity span, and the second stage of classifying the entity type, are implemented using the fast structured perceptron algorithm~\cite{collins2002discriminative},\footnote{\url{https://github.com/larsmans/seqlearn}} which treats both stages as a sequence labeling problem. 
 
\section{Sampling Pseudo-ground Truth}
\label{sec:sampling}

To craft a high quality set of pseudo-ground truth for training our NERC method, in this section, we present two sampling methods: 
\begin{inparaenum}[(a)]
\item sampling based on the entity linking system's confidence score for a linked entity, and 
\item sampling based on the textual quality of an input document.
\end{inparaenum}

\subsection{Sampling Based on Entity Linker's Confidence Score}
\label{par:confidence}
Typically, entity linking systems provide confidence scores for each entity mention ($n$-gram) they are able to link to a knowledge base entity. 
These confidence scores can be used to rank possible entities for a mention, but also for pruning out mention-entity pairs for which the linker is not confident. 
Although the scale of the confidence score depends on the model behind the entity linking system, the scores can be normalized over the candidates for an entity mention, e.g., using linear or z-score normalization. 
We use the $``Sense Probability''$ metric introduced by \citet{Odijk:2013aa} as our confidence score.

\noindent
SenseProbability is calculated by estimating the probability of an $n$-gram being used as an anchor text pointing to a specific entity $e$, with the prior probability of the $n$-gram $n$ being used as an anchor at all (als known as the $\mathit{commonness}$-score~\cite{milne08:learn}.
See equation~\ref{eq:senseprob}.
\begin{equation}
\label{eq:senseprob}
SenseProbability(n,e) =\frac{|L_{n, e}|}{\sum_{e' \in KB}c(n, e')},
\end{equation}
\noindent
where $L_{n,e}$ denotes the set of all links with $n$-gram $n$ and target entity $e$, 
$n$ denotes the $n$-gram to be linked (i.e., anchor text), $e$ denotes the candidate entity, 
$c(n, e)$ denotes the number of times that $n$-gram $n$ links to candidate entity $e$. 

\subsection{Sampling Based on Textual Quality}
\label{par:quality}
Taking the textual quality of content into account has proved helpful in a range of tasks. 
Based on \cite{Wouter-blogs,Manos-tweets}, we consider nine features indicative of textual quality; see Table~\ref{tab:tweet-in-corpus}. 
While not exhaustive, our feature set is primarily aimed at social streams as our target document stream (see Section~\ref{sec:exp}) and suffices for providing evidence on whether this type of sampling is helpful for our purposes. 
Based on these features, we compute a final score for each document $d$ as
\begin{align}
    \mathrm{score}(d) = \frac{1}{|F|}\sum_{f \in F} \frac{f(d)}{\max_f},
\end{align}
\noindent
where $F$ is our set of feature functions (Table~\ref{tab:tweet-in-corpus}) and $\max_f$ is the maximum value of $f$ we have seen so far in the stream of documents. 
Since all features are normalized in $[0,1]$, $\mathrm{score}(d)$ has this same range. 
As a qualitative check, we rank documents from the MSM2013~\cite{cano.ea:2013} dataset using our quality sampling method and list the top-5 and bottom-5 scoring documents in Table~\ref{table:top5-tweets}. 
Top scoring documents are longer and denser in information than low scoring documents. 
We assume that these documents are better examples for training a NERC system.
\begin{table}[t]
  \caption{Features used for sampling documents from which we train a NERC system.}
  \centering
  \begin{tabular}{l p{3.3cm} l p{3.3cm}}
  \toprule
  \textbf{Feature} & \textbf{Description} & \textbf{Feature} & \textbf{Description} \\
  \midrule
  n\_mentions & No. of usernames (@) & avg\_token\_len & Average token length \\
  n\_hashtags & No. of hashtags (\#) & tweet\_len & Length of tweet (char) \\
  n\_urls & Number of URLs & density & Density as in \cite{Hu:2008:CDS:1390334.1390385} \\
  ratio\_upper & Percentage of uppercased chars & personal & Contains personal pronouns (I, me, we, etc.) \\ 
  ratio\_nonalpha & Percentage of non-alphanumeric chars & &  \\
  \bottomrule
  \end{tabular}
\label{tab:tweet-in-corpus}
\end{table}

In the next section, we follow a linear search approach to sampling training examples as input for NERC. 
First, we find an optimal threshold for confidence scores, and fix it. 
For sampling based on textual quality, we turn to the MSM2013 dataset to determine sampling thresholds. 
We calculate the scores for each tweet, and scale them to fall between [0,1]. 
We then plot the distribution of scores, and bin this distribution in three parts: tweets that fall within a single standard deviation of the mean are considered \emph{normal}, 
tweets to the left of this bin are considered \emph{noisy}, 
whilst the remaining tweets to the right of the distribution are considered \emph{nice}. 
We repeat this process for our tweet corpus, using the bin thresholds gleaned from the MSM2013 set.

\begin{table}[t]
  \caption{Ranking of documents in the MSM2013 dataset based on our quality sampling method. Top ranking documents appear longer and denser in information than low ranking documents.}
  \begin{tabularx}{\linewidth}{X}
  \toprule
  \textbf{Top-5 quality documents} \\
  \midrule
  \textit{``Watching the History channel, Hitler's Family. Hitler hid his true family heritage, while others had to measure up to Aryan purity.''}\\
  \midrule
  \textit{``When you sense yourself becoming negative, stop and consider what it would mean to apply that negative energy in the opposite direction.''}\\
  \midrule
  \textit{``So. After school tomorrow, french revision class. Tuesday, Drama rehearsal and then at 8, cricket training. Wednesday, Drama. Thursday ... (c)''}\\
  \midrule
  \textit{These late spectacles were about as representative of the real West as porn movies are of the pizza delivery business Que LOL}\\
  \midrule
  \textit{Sudan's split and emergence of an independent nation has politico-strategic significance. No African watcher should ignore this.}\\
  \midrule
  \textbf{Bottom-5 quality documents} \\
  \midrule
  \textit{Toni Braxton \textasciitilde~ He Wasnt Man Enough for Me \_HASHTAG\_ \_HASHTAG\_? \_URL\_ RT \_Mention\_} \\
  \midrule
  \textit{``tell me what u think The GetMore Girls, Part One \_URL\_''} \\
  \midrule
  \textit{this girl better not go off on me rt} \\
  \midrule
  \textit{``you done know its funky! -- Bill Withers'' ``Kissing My Love'''' \_URL\_ via \_Mention\_''}\\
  \midrule
  \textit{This is great: \_URL\_ via \_URL\_}\\
\bottomrule
\end{tabularx}
\label{table:top5-tweets}
\end{table}
 
\section{Experimental Setup}
\label{sec:exp}

In addressing the problem of predicting emerging entities in social media streams, we concentrate on developing an unsupervised method for generating pseudo-ground truth for a NERC method, and predicting newly emerging entities. 
In particular, we want to know the effectiveness of our unsupervised pseudo-ground truth (UPGT) method over a random baseline and a lexical matching baseline, and the impact on effectiveness of our two sampling methods. 
To answer these questions, we conduct both optimization and prediction experiments.

\subsection{Dataset}
As our document stream, we use tweets from the TREC 2011 Microblog dataset~\cite{mbtoverview:2012}, a collection of 4,832,838 unique English Twitter posts. 
This choice is motivated by the unedited and noisy nature of tweets, which can be challenging for prediction. 
Our knowledge base (KB) is a subset of Wikipedia from January~4, 2012, restricted to entities that correspond to the NERC classes person (PER), location (LOC), or organization (ORG).
We use DBpedia to perform selection, and map the DBpedia classes 
\textit{Organisation}, \textit{Company}, and \textit{Non-ProfitOrganisation} to ORG,
\textit{Place}, \textit{PopulatedPlace}, \textit{City}, and \textit{Country} to LOC, and 
\textit{Person} to PER.\footnote{\url{http://mappings.dbpedia.org/server/ontology/classes/}}
Our final KB contains 1,530,501 entities.

\subsection{Experiment I: Sampling Pseudo-ground Truth} 
To study the utility of our sampling methods, we turn to the impact of setting a threshold on the entity linking system's confidence score (Experiment Ia.) and the effectiveness of our textual quality sampling (Experiment Ib.). 

In Experiment Ia., we perform a sweep over thresholds between 0 and 1, in steps of 0.1, using the same threshold for both the generation of pseudo-ground truth and evaluating the prediction effectiveness of emerging entities. 
Lower thresholds allow low confidence entities in the pseudo-ground truth, and likely generate more data at the expense of noisy output. 
We emphasize that we are not interested in the correlation between noise and confidence score, but rather in the performance of finding emerging entities given the entity linking system's configuration. 
In Experiment Ib., we compare our methods performance with differently sampled pseudo-ground truths, i.e., by sampling tweets that belong to either the \emph{nice}, \emph{normal}, or \emph{noisy} bins, following our textual quality sampling method (cf. Section~\ref{par:quality}).

\subsection{Experiment II: Prediction Experiments} 
\label{subsec:ecirexp2}
To answer our second research question, and study the impact of prior knowledge on detecting emerging entities, we compare the performance of our method (UPGT) to two baselines: 
a \emph{random baseline} (RB) that extracts all $n$-grams from test tweets and considers them emerging entities, and 
a \emph{lexical-matching baseline} (NB) that follows our approach, but generates pseudo-ground truth by applying lexical matching of entity titles, instead of an EL system, and refrains from sampling based on textual quality. 
For this experiment, we use the optimal sampling parameters for generating pseudo-ground truth from our previous experiment, i.e., include linked entities with a confidence score higher than 0.7, and use only normal tweets. 
As we will show in Section~\ref{sec:results}, the threshold of 0.7 balances performance with high recall of entities in the pseudo-ground truth.

\subsection{Evaluation}
We evaluate the quality of the generated pseudo-ground truth on the 
effectiveness of a NERC system trained to predict newly emerging entities. 
As measuring the addition of new entities to the knowledge base is non-trivial, we consider a retrospective scenario: 
Given our KB, we randomly sample entities to yield a smaller KB (KB$_{\text{s}}$). 
This KB$_{\text{s}}$ simulates the available knowledge at the present point in time, whilst the full KB represents the future state. 
By measuring how many entities we are able to detect in our corpus that feature in KB, but not KB$_{\text{s}}$, we can approximate the newly emerging entity prediction task. 
We create KB$_{\text{s}}$ by taking random samples of 20--90\% the size of KB (measured in number of entities), in steps of 10\%. 
We repeat each sampling step ten times to avoid bias. 

For each KB$_{\text{s}}$, we generate pseudo-ground truth for training and test sets for evaluation. 
We use KB$_{\text{s}}$ to link the corpus of tweets, and yield two sets of tweets: 
(a) tweets with linked entities, 
(b) tweets that may contain emerging entities,
analog to the \emph{linked} and \emph{unlinked sentences} from Figure~\ref{fig:uned}. 
The size of these two sets depends on the size of KB$_{\text{s}}$: a smaller KB will yield a larger set of unlinked tweets.
This makes comparisons of results across different KB$_{\text{s}}$ difficult. 
We cater for this bias by randomly sampling 10,000 tweets from both the test set and the pseudo-ground truth, and repeating our experiments ten times.
Using the smallest KB$_{\text{s}}$ (20\%) results in about 15,000 tweets in the pseudo-ground truth. 
Ground truth is assembled by linking the corpus of tweets using the full KB. 
The ground truth consists of 82,305 tweets, with 12,488 unique entities. 

We evaluate the effectiveness of our method in two ways: 
(a)~the ability of NERC to generalize from our pseudo-ground truth, and 
(b)~the accuracy of our predictions. 

For the first, we compare the predicted \emph{entity mentions} to those in our ground truth, akin to traditional NERC evaluation. 
For the second we take the set of correctly predicted entity mentions (true positives), and link them to their referent entities in the ground truth. 
This allows us to measure what we are actually interested in: the fraction of newly discovered entities. 
For both types of evaluation we report on average precision and recall over 100 runs per KB$_{\text{s}}$. 
Statistical significance is tested using a two-tailed paired t-test and is marked as \dubbelop\ for significant differences for $\alpha=.01$.
 
\section{Results}
\label{sec:results}

\subsection{Experiment I: Sampling Pseudo-ground Truth}
\label{par:optimization}

Our first experiment aims to answer~\ref{rq:ecir1}: 

\begin{description}[labelwidth=1cm,leftmargin=!]
\item[\ref{rq:ecir1}] What is the utility of our sampling methods for generating pseudo-ground truth for a named entity recognizer?
\end{description}

\noindent
For this experiment, we fix the size of the KB$_{\text{s}}$ at 50\%. 
We start by looking at the ability of NERC to generalize from our pseudo-ground truth, measured on two aspects: 
(i) effectiveness for identifying mentions of newly emerging entities (i.e., entity \emph{mentions}), and 
(ii) predicting newly emerging entities (i.e., \emph{entities}). 

\begin{figure}[t]
  \centering
  \includegraphics[width=.49\textwidth]{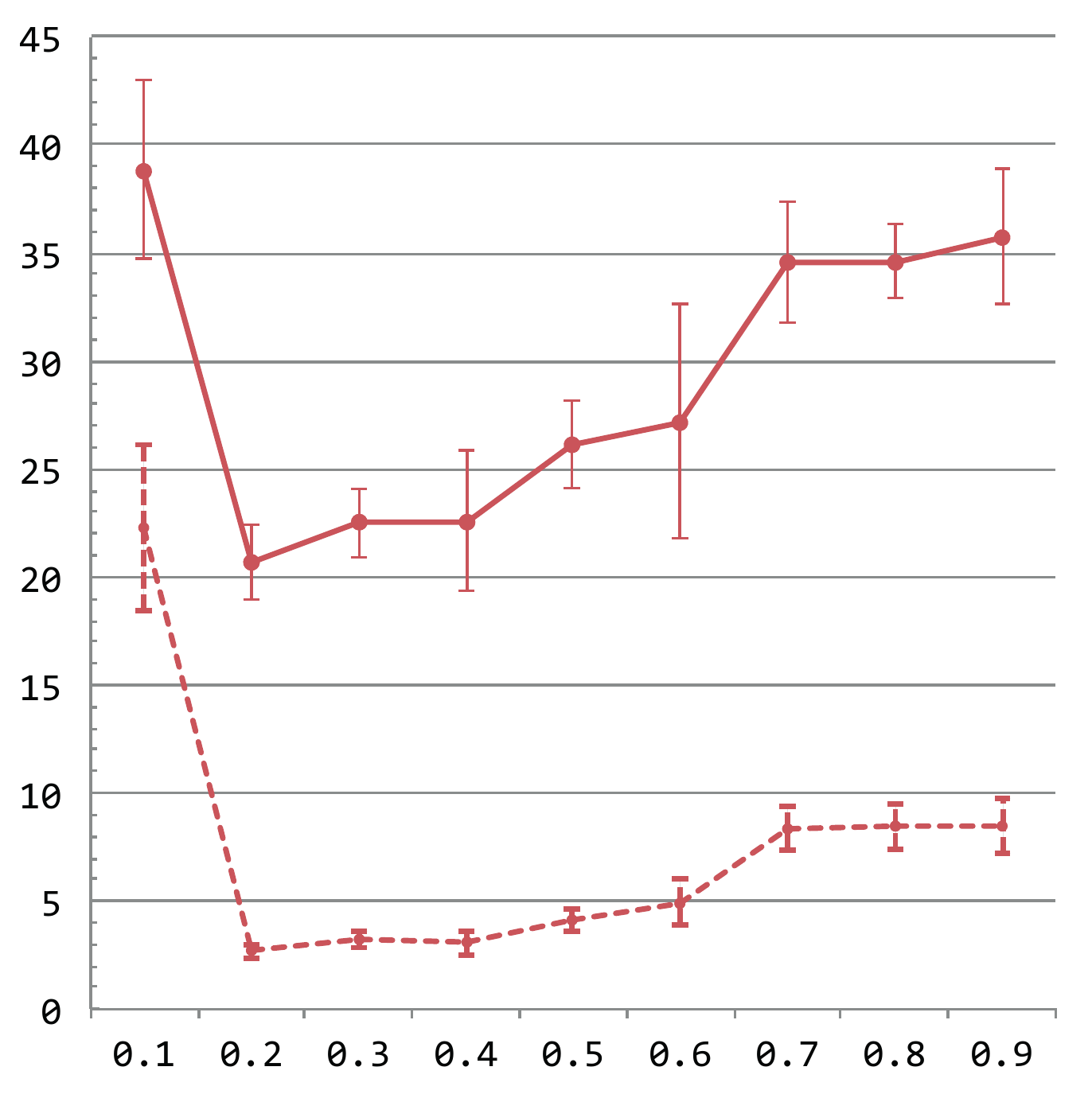}
  \includegraphics[width=.49\textwidth]{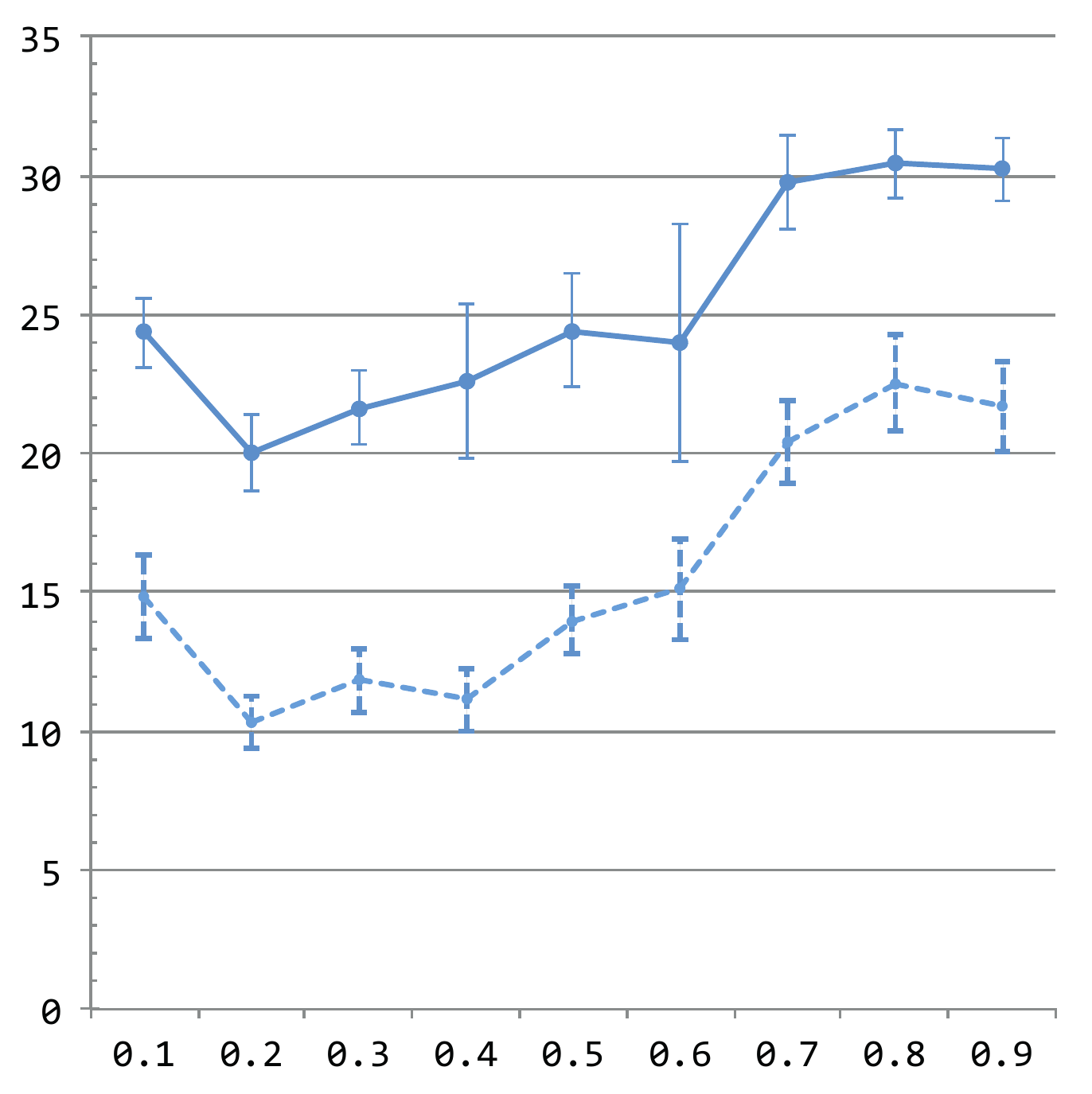}
  \caption{\textbf{Experiment Ia.} Impact of confidence score on UPGT. Effectiveness of identifying entity \emph{mentions} is shown in the left plot, effectiveness of identifying newly emerging \emph{entities} in the right. Threshold set on the confidence score on the x-axis. Precision (solid line) and recall (dotted) are shown on the y-axis.}
  \label{fig:optimization}
\end{figure}

In Experiment Ia., we look at our confidence-based sampling method. 
For identifying mentions of newly emerging entities, we find that effectiveness peaks at 0.1 confidence threshold with a precision of 38.84\%, dips at 0.2 and slowly picks up to 35.75\% as threshold increases (Figure~\ref{fig:optimization}, left). 
For identifying the newly emerging entities that are referred to by the mentions, effectiveness positively correlates with the threshold. 
Effectiveness peaks at the 0.8 confidence threshold, statistically significantly different from 0.7 but not from 0.9 (Figure~\ref{fig:optimization}, right).

\noindent
Interestingly, besides precision, recall also shows a positive correlation with thresholds. 
This suggests that in identifying newly emerging entities, missing training labels are likely to have less impact on performance than generating incorrect, noisy labels. 
This is an interesting finding as it sets the emerging entity prediction-task apart from traditional NERC, where low recall due to incomplete labeling is a well-understood challenge. 

Next, we turn to the characteristics of the pseudo-ground truth that results for each of these thresholds, and provide an analysis of their potential impact on effectiveness.
We find that more data through a larger pseudo-ground truth allows NERC to better generalize and predict a larger number of emerging entities. 
This claim is supported by the number of predicted entity mentions per threshold in Table~\ref{tab:predictions}. 
We find a similar trend as in the precision and recall graph above: the number of predicted emerging entities peaks for the threshold at 0.1 (6,653 entity mentions), and drops between 0.2 and 0.4, and picks up again from 0.5 reaching another local maximum at 0.8. 
The increasing number of predicted emerging entity mentions with stricter thresholds indicates that the NERC model is more successful in learning patterns for separating entities from noisy labels. 
This may be due to the entity linking system linking only those entities it is most confident about, providing a clearer training signal for NERC.

\begin{table}[t]
\small
\caption{Number of predicted emerging entity mentions (P) per threshold on the confidence score (T). GT = Ground truth.}
\label{tab:predictions}
\begin{tabularx}{\hsize}{X rrrrrrrrr}
  \toprule
  \textbf{T} & \textbf{0.1} & \textbf{0.2} & \textbf{0.3} & \textbf{0.4} & \textbf{0.5} & \textbf{0.6} & \textbf{0.7} & \textbf{0.8} & \textbf{0.9} \\
  \textbf{P} & 6,653  & 1,500 & 1,618 & 1,512 & 1,738 & 2,025 & 2,662 & 2,713 & 2,614 \\ 
  \textbf{GT} & 11,429 & 11,533 & 11,291 & 11,078 & 10,955 & 10,935 & 10,799 & 10,881 & 10,855 \\
  \bottomrule
\end{tabularx}
\end{table}

For the rest of our experiments we use a threshold of 0.7 on confidence score because it is deemed optimal in terms of trade-off between performance and quantity of entities in pseudo-ground truth.

In Experiment Ib., we study three different textual quality-based sampling strategies: we consider only tweets that fall in the \emph{normal} bin (i), tweets that fall in the \emph{nice} bin (ii), or tweets that fall in both the \emph{normal} and \emph{nice} bins (iii); see Section~\ref{sec:sampling}. 
For reference, we also report on the performance achieved when no textual quality sampling is used. 
We keep KB$_{\text{S}}$ fixed at 50\%, and use 0.7 for confidence threshold. 

\begin{table}[t]
  \caption{\textbf{Experiment Ib.} Precision and recall for three sampling strategies based on textual quality of documents: nice, normal, normal+nice (mixed). We also report on effectiveness of not using sampling by textual quality for reference. Boldface indicates best performance. Statistical significance is tested against the previous sampling method, e.g., nice to normal.}
  \centering
  \begin{tabularx}{\hsize}{l X X X X}
  \toprule
  & \multicolumn{2}{c}{\textbf{Mention}} & \multicolumn{2}{c}{\textbf{Entity}}  \\
  \cmidrule(r){2-3}\cmidrule{4-5}
  \textbf{Sampling } & \textbf{Precision} & \textbf{Recall} & \textbf{Precision} & \textbf{Recall} \\
  \midrule
  No sampling & 34.26$\pm$2.65 & 8.21$\pm$0.83 & 29.63$\pm$1.67 & 20.25$\pm$1.56  \\
  Normal+nice & 45.50$\pm$4.71\dubbelop & 12.97$\pm$2.03\dubbelop & 36.22$\pm$2.07\dubbelop & 24.86$\pm$1.69\dubbelop \\
  Normal & 66.09$\pm$3.86\dubbelop &  30.94$\pm$3.46\dubbelop & 44.62$\pm$1.51\dubbelop & \textbf{32.20$\pm$1.67}\dubbelop \\
  Nice & \textbf{70.36$\pm$3.07}\dubbelop & \textbf{30.98$\pm$3.25} & \textbf{45.99$\pm$1.34}\dubbelop & 29.69$\pm$1.79 \\
  \bottomrule
  \end{tabularx}
\label{tab:noise-sampling}
\end{table}

Textual quality-based sampling turns out to be twice as effective as no sampling on both identifying mentions of emerging entities, and identifying newly emerging entities. 
Among our sampling strategies, nice proves to be the most effective with a precision of 70.36\% for entity mention identification. 
In emerging entity prediction, the performance of nice and normal strategies hovers around the same levels. 
In terms of recall, nice and normal methods are on par, outperforming both other strategies. 
The success of nice and normal sampling methods can be attributed to the fact that a more coherent and homogeneous training corpus allows the NERC model to more easily learn patterns.

\subsection{Experiment II: Impact of Prior Knowledge}
\label{par:prediction}
Next, we seek to study the impact of the size of the prior knowledge that is available to our emerging entity prediction method.
We do so by answering~\ref{rq:ecir2}: 
\begin{description}[labelwidth=1cm,leftmargin=!]
\item[\ref{rq:ecir2}] What is the impact of the size of prior knowledge on predicting newly emerging entities?
\end{description}
\noindent
We use the optimal combination of our sampling methods from the previous experiments, i.e., a confidence threshold of 0.7, and the normal textual quality sampling. 
We again look at the effectiveness of our methods in both identifying emerging mentions, and emerging entities. 

\begin{figure}[h]
  \centering
  \includegraphics[width=.49\textwidth]{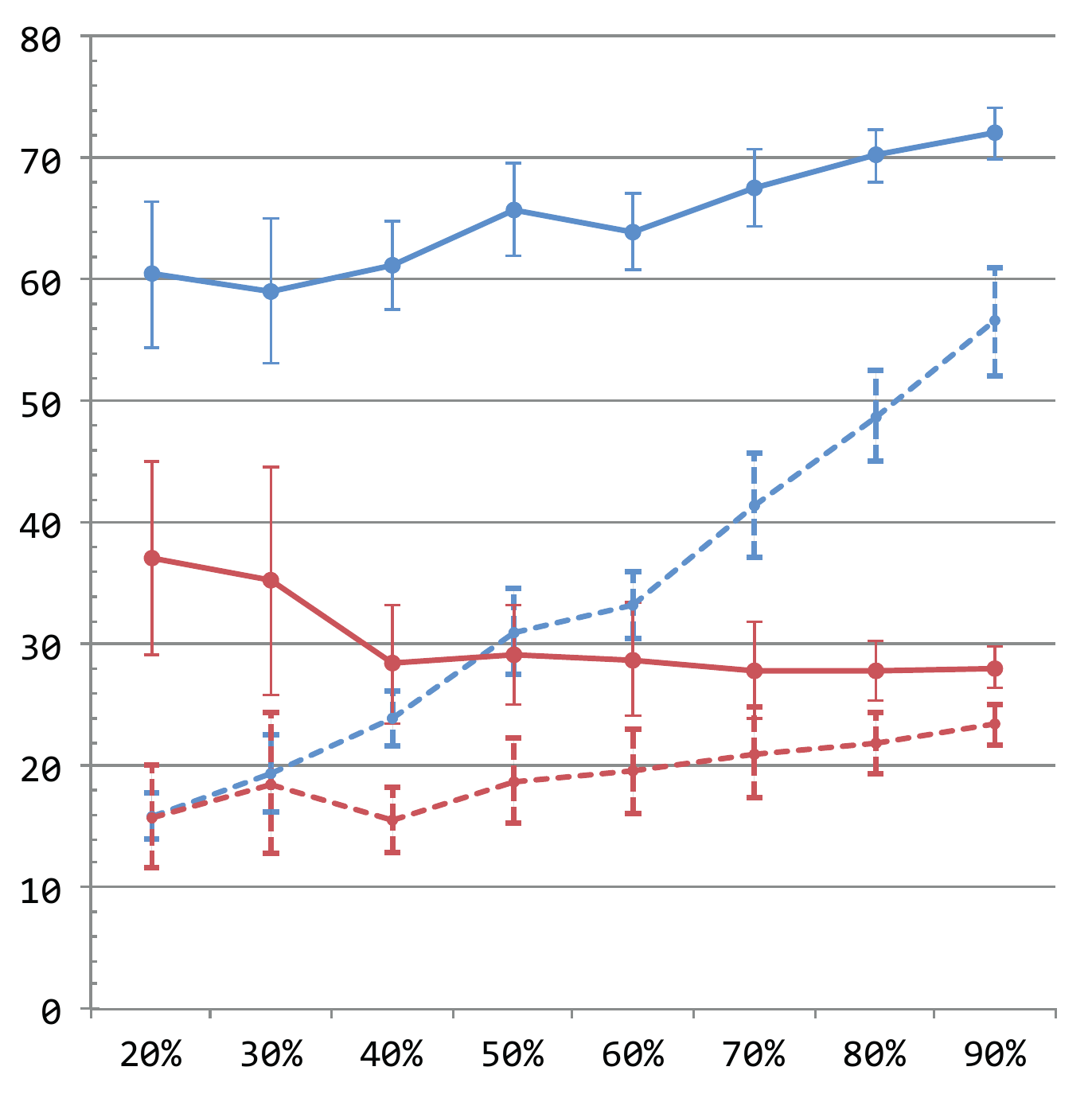}
  \includegraphics[width=.49\textwidth]{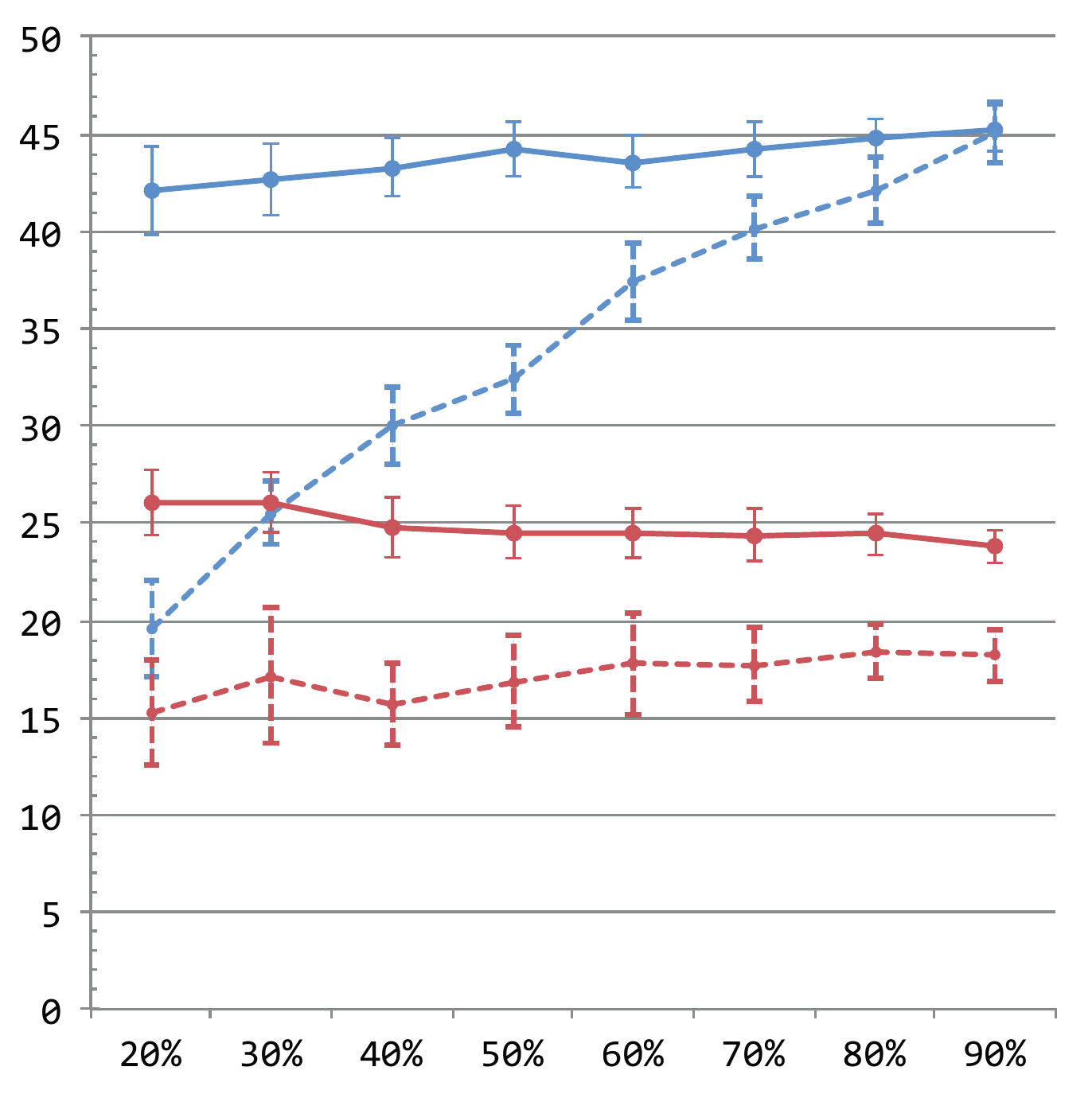}
  \caption{Our method (UPGT, blue line) versus the lexical baseline (red) for both identifying mentions of emerging entities (left) and newly emerging entity detection (right). Knowledge base size is on the $x$-axis, and precision (solid lines) and recall (dotted lines) are marked on the $y$-axis. }
  \label{fig:kb_size}
\end{figure}

Figure~\ref{fig:kb_size} shows the effectiveness of our methods as a function of the size of the knowledge base (i.e., size of KB$_s$). 
For identifying mentions of emerging entities, our method (UPGT, blue line) constantly and statistically significantly outperforms both lexical (red line) and random baselines (not shown). 
In terms of recall, the lexical baseline is on par with UPGT for KB sizes up to 30\%. 
The random baseline shows very low precision for both identifying mentions of emerging entities, and predicting emerging entities over all KB$_{\text{s}}$, but almost perfect recall---which is expected given that it assigns all possible $n$-grams as emerging entity mentions and newly emerging entities (0.69\% precision and 65\% recall for entity mention identification, 1.82\% precision and 94.95\% recall for emerging entity prediction). 

Next, we take a closer look at the results. 
The lexical baseline's recall increases slightly when more prior knowledge is added to the knowledge base. 
This is expected behavior because, as we saw in our previous experiment, the pseudo-ground truth gets more diverse labels, and helps the NERC to generalize. 
The number of unique entities in the pseudo-ground truth shows that the number increases with the size of KB$_{\text{s}}$. 
UPGT assigns labels to 2,500 unique entities at 20\% KB$_{\text{s}}$, which tops at 11,000 unique entities at 90\%. 
These numbers are lower for the lexical baseline (1,800 at 20\% KB$_{\text{s}}$, and 7,000 at 90\% KB$_{\text{s}}$). 
However, for both methods, the number of entities in the ground truth stays around the same. 
The gradual improvement in precision and recall of UPGT for increasing KB$_{\text{s}}$ can be attributed to a broader coverage for labeling (observed through looking at the prior entities), and the main distinction between the lexical baseline and UPGT: stricter labeling through leveraging the entity linking system's confidence score.

Finally, to better understand the performance of our method, we have looked at the set of correct predictions (emerging entities), and false positives, or incorrectly identified emerging entities. 
Our analysis revealed examples of actual emerging entities, i.e., entities that were not included in the initial KB, but did exist in the Wikipedia dump. 
This may be attributed due to, e.g., missing DBpedia class labels, which highlights the challenging setting of evaluating the task. 

On the whole, however, our method is able to deal with missing labels and incomplete data, as observed through its consistent and stable precision, justifying our assumption that data is incomplete by design.

\section{Conclusion}
\label{sec:conclusion}
In this chapter, we tackled the problem of predicting emerging entities in social streams, and presented an effective method of leveraging prior knowledge to bootstrap the discovery of new entities of interest. 
We set out to answer the following question:

\begin{description}[leftmargin=!,labelwidth=1cm]
\item[\ref{rq:ecir}] Can we leverage prior knowledge of entities of interest to bootstrap the discovery of new entities of interest?
\end{description}

\noindent
We presented an unsupervised method for generating pseudo-ground truth using an EL method with a reference KB. 
The pseudo-ground truth serves as training data for a NERC method for detecting emerging entities that are likely to be incorporated in a KB. 
To answer \ref{rq:ecir} we formulated and answered two subquestions.
In Section~\ref{par:optimization} we answer our first subquestion :

\begin{description}[leftmargin=!,labelwidth=1cm]
\item[\ref{rq:ecir1}] What is the utility of our sampling methods for generating pseudo-ground truth for a named-entity recognizer?
\end{description}

\noindent
We do so by introducing and studying two different sampling methods.
The first sampling method is based on the entity linking system's confidence score, where we hypothesize that keeping only high-confidence links will yield less training data but of higher quality, resulting in better predictions. 
The second sampling method is based on the textual quality of the input documents. 
We find that sampling by textual quality improves performance of NERC and consequently our method's performance in predicting emerging entities. 
As setting a higher threshold on the entity linking system's confidence score for generating pseudo-ground truth results in fewer labels but better performance, we show that the NERC is better able to separate noise from entities that are worth including in a KB. 
The entity linker's confidence score is an effective signal for this separation. 
Both our sampling methods significantly improve emerging entity prediction. 

Next, in Section~\ref{par:prediction}, we answer our second subquestion:

\begin{description}[leftmargin=!,labelwidth=1cm]
\item[\ref{rq:ecir2}] What is the impact of the size of prior knowledge on predicting new KB entities?
\end{description}

\noindent
We do so by studying the impact of differently sized (seed) knowledge bases, i.e., different amounts of prior knowledge on the entities of interest. 
We find that in the case of a small amount of prior knowledge, i.e., limited size of the available initial knowledge, our method is able to cope with missing labels and incomplete data, as observed through its consistent and stable precision.
This finding justifies our proposed method that assumes incomplete data by design. 
Furthermore, this finding suggests the scenario of an increasing rate of emerging entity prediction, as more data is fed back to the KB. 
Additionally, we found that a larger number of entities in the KB allows for setting a desirable stricter threshold on the confidence scores, and leads to improvements in both precision and recall. 
This finding suggests an adaptive threshold that takes prior knowledge into account could prove effective.

\noindent
In summary, we have shown that we can effectively leverage prior knowledge (through a reference KB) to detect similar entities that are not in the KB. 
An implication of this finding in the discovery context is that this method allows to effectively support the exploratory search process, as it successfully identifies entities that are similar to a set of seed entities of interest. 

Our work has several limitations. 
We studied the emerging entity prediction task in a streaming scenario.
However, due to limitations of the availability of suitable datasets, we employ a retrospective scenario in Experiment II (Section~\ref{subsec:ecirexp2}), where we randomly sample entities from the KB for measuring the impact of the amount of prior knowledge on prediction performance. 
Taking this retrospective scenario, makes it impossible to measure the impact of \emph{novelty} of entities as they emerge. 
By repeating each experiment ten-fold, we alleviate the problem of keeping or removing popular head entities that impact performance more than tail entities, but a more realistic scenario, e.g., using a comprehensive archive of tweets, spanning multiple months, would allow us to sample KB by time. 
This scenario, similar to the scenario we explored in Chapter~\ref{ch:plos} would be able to provide more insights into the relation between social media dynamics and prediction accuracy of emerging entities. 

A natural extension to the work presented in this chapter would be to include a subsequent entity clustering or disambiguation step, and ultimately to ``feed back'' emerging entities to the KB. 
One approach could be to adapt the work of creating keyphrase-based representation as seed representations~\cite{Hoffart:2014:DEE:2566486.2568003}. 
However, as we have seen in Chapter~\ref{ch:plos}, entities are not static, and may appear in sudden bursts of documents, e.g., through a peak in interest as real-world events unfold. 
In the next chapter, we address this dynamic nature of entities, in the context of entity retrieval. 
We propose a method for dynamically constructing entity representations by leveraging the way in which people refer to entities online, to improve the retrieval effectiveness of entities. 
 \clearpage{}

\clearpage{}
\chapter{\acl{ch:wsdm}}
\label{ch:wsdm}

\begin{flushright}
\rightskip=1.8cm``En wat zijn mijn woorden waard als ik ze niet meer weeg?'' \\
\vspace{.2em}
\rightskip=.8cm---\textit{Sticks, Waar Wacht Je Op?}
\end{flushright}

\section{Introduction}
\label{sec:wsdmintro}

In the third and final chapter of this first part of the thesis we focus on \emph{retrieving} entities of interest from the knowledge base. 
As entities play a central role in exploratory search and answering the 5 W's~\cite{Ahn2010383}, ranking entities to user-issued search queries is an important, but challenging task. 
In this chapter, we study \emph{entity ranking}, where the goal is to position a relevant entity from the knowledge base at the top of the ranking for a given query.

We have seen in Chapter~\ref{ch:plos} that entities may suddenly appear in public discourse, e.g., through events as they unfold in the real world. 
Due to the dynamic nature of entities in public discourse, the way in which people address or refer to entities may suddenly change, providing challenges for traditional retrieval methods that rely on static representations. 
Here, we address the dynamic nature of entities in the general web search scenario, in which a searcher enters a query that can be satisfied by returning an entity from Wikipedia. 
We propose a novel method that collects information from external sources and dynamically constructs entity representations, by optimally combining different entity descriptions from external sources into a single entity representation. 
The method learns directly from users' past interactions (i.e., searches) to adapt the entity representation towards improved retrieval effectiveness of entities. 

The general web search engine scenario is motivated by the fact that many queries issued to general web search engines are related to entities~\cite{Kumar:2010:COB:1772690.1772748}. 
Entity ranking is therefore becoming an ever more important task~\cite{inex2007,inex2008,inex2009,trec2010,trec2011}. 
Entity ranking is inherently difficult due to the potential mismatch between the entity's description in a knowledge base and the way people refer to the same entity when searching for it. 
When we look at how entities are described, two aspects, context and time, are of particular interest and pose challenges to any solution to entity ranking. 
Here we explain both aspects.

\begin{description}
 \item[Context dependency:] 
 Consider the entity \texttt{Germany}. 
 A history student could expect this entity to show up when searching for entities related to World War II. 
 In contrast, a sports fan searching for World Cup 2014 soccer results is also expecting to find the same entity. 
 The challenge then becomes how to capture these different contexts for one single entity. 
 \item[Time dependency:] Entities are not static in how they are perceived. 
 Consider \texttt{Ferguson, Missouri}, which had a fairly standard city description before the shooting of Michael Brown happened in August 2014. 
 After this event, the entity description of Ferguson changed substantially, reflecting people's interest in the event, its aftermath and impact on the city.
\end{description}

\begin{figure}[t]
	\centering
	\includegraphics[trim = 0mm 0mm 0mm 0mm, clip=true, width=\linewidth]{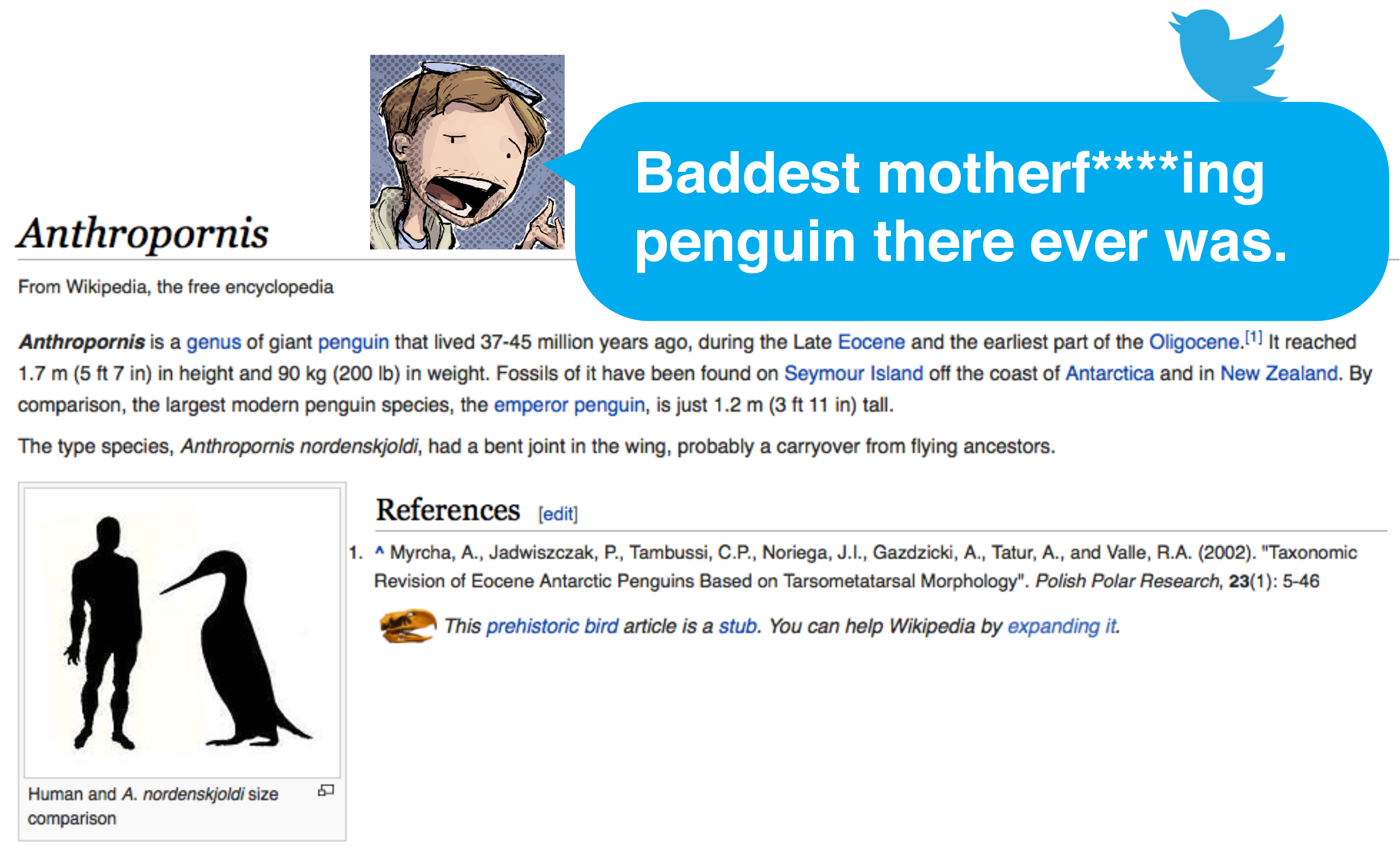}
	\caption{Entity description of \texttt{Anthropornis} in Wikipedia and a tweet with an alternative description of the same entity.}
	\label{fig:penguin}
\end{figure}

\noindent
We propose a method that addresses both challenges raised above. 
First, we use the collective intelligence as offered by a wide range of entity ``description sources'' (e.g., tweets and tags that mention entities), and we combine these into a ``collective entity representation,'' i.e., a representation that encapsulates different ways of how people refer to or talk about the entity. 
Consider the example in Figure~\ref{fig:penguin} in which a tweet offers a very different way to refer to the entity \texttt{Anthropornis} than the original knowledge base description does. 

Second, our method takes care of the time dependency by incorporating dynamic entity description sources, which in turn affect the entity descriptions in near real time. 
Dynamics is part of our method in two ways: 
(i)~we leverage dynamic description sources to expand entity representations, and 
(ii)~we learn how to combine the different entity descriptions for optimal retrieval at specific time intervals. 
The resulting \emph{dynamic collective entity representations} capture both the different contexts of an entity and its changes over time. 
We refer to the collection of descriptions from different sources that are associated with an entity as the entity's \emph{representation}. 
Our method is meant to construct the optimal representation for retrieval, by assigning weights to the descriptions from different sources. 

Collecting terms associated with documents and adding them to the document (document expansion) is in itself not a novel idea. 
Previous work has shown that it improves retrieval effectiveness in a variety of retrieval tasks such as speech retrieval~\cite{singhal1999} and ad-hoc web search~\cite{westerveld2001,wu2012,metzler2009,eiron2003}. 
However, while information retrieval on the open web is inherently dynamic, document expansion for search has mainly been studied in static, context-independent settings, in which expansion terms from one source are aggregated and added to the collection before indexing~\cite{mishne2012,metzler2009,lee2013,westerveld2001}. 
In contrast, our method leverages different dynamic description sources (e.g., queries, social media, web pages), and in this way, uses collective intelligence to bridge the gap between the terms used in a knowledge base's entity descriptions and the terms that people use to refer to entities. 

To achieve this, we represent entities as fielded documents~\cite{macdonald2013}, where each field contains content that comes from a single description source. 
In a dynamic setting such as our entity ranking setting, where new entity descriptions come in as a stream, learning weights for the various fields in batch is not optimal. 
Ideally, the ranker continuously updates its ranking model to successfully rank the entities and incorporate newly incoming descriptions. 
Hence, constructing a dynamic entity representation for optimal retrieval effectiveness boils down to dynamically learning to optimally weight the entity's fields that hold content from the different description sources. 
To this end we exploit implicit user feedback (i.e., clicks) to retrain our model and continually adjust the weights associated to the entity's fields, much like online learning to rank~\cite{hofmann2011}.

Our dynamic collective entity representations generate one additional challenge, which is related to the heterogeneity that exists among entities and among description sources. 
Popular head entities are likely to receive a larger number of external descriptions than tail entities. 
At the same time the description sources differ along several dimensions (e.g., volume, quality, novelty). 
Given this heterogeneity, linearly combining retrieval scores (as is commonly done in structured retrieval models) proves to be suboptimal. 
We therefore extend our method to include features that enable the ranker to distinguish between different types of entities and stages of entity representations. 
The main research question we seek to answer in this chapter is:

\begin{enumerate}[label=\textbf{\ref{rq:wsdm}},labelwidth=1cm,leftmargin=!,align=left]
\item Can we leverage collective intelligence to construct entity representations for increased retrieval effectiveness of entities of interest?
\end{enumerate}

\noindent
To answer this question, we formulate and seek to answer the following three subquestions. 
First, we check the underlying assumption of our method, and answer our first research question: 

\begin{enumerate}[label=\ref{rq:wsdm}.\arabic*,labelwidth=1cm,leftmargin=!,align=left]
\item Does entity ranking effectiveness increase by using dynamic collective entity representations?\label{rq:wsdm1}
\end{enumerate}

\noindent
To answer this question, we compare a baseline entity ranking method based on information in the knowledge base only to our method that incorporates additional description sources (web anchors, queries, tags, and tweets). 

Next, we extend our method, and study the contribution of informing the entity ranking of the entity's description state. 
We seek to answer the second subquestion: 

\begin{enumerate}[label=\ref{rq:wsdm}.\arabic*,resume,labelwidth=1cm,leftmargin=!,align=left]
\item Does entity ranking effectiveness increase when employing additional field and entity importance features?\label{rq:wsdm2}
\end{enumerate}

\noindent
We answer this question by incorporating field and entity importance features to our knowledge base-only baseline ranker and our proposed method, and compare their performance. 
Finally, we answer this chapter's third research question, by studying the dynamic aspect of entity ranking: 
 
\begin{enumerate}[label=\ref{rq:wsdm}.\arabic*,resume,labelwidth=1cm,leftmargin=!,align=left]
\item Does entity ranking effectiveness increase when we continuously learn the optimal way to combine the content from different description sources?\label{rq:wsdm3}
\end{enumerate}

\noindent
We compare a static entity ranking baseline that is trained once at the start, to our proposed method that is retrained at regular intervals. 

The main contribution of this chapter is a novel approach to constructing dynamic collective entity representations, which takes the temporal and contextual dependencies of entity descriptions into account.  
We show that dynamic collective entity representations better capture how people search for entities than their original knowledge base descriptions. 
In addition, we show how field importance features better inform the ranker, thus increasing retrieval effectiveness. 
Furthermore, we show how continuously updating the ranker enables higher ranking effectiveness. 
Finally, we perform extensive analyses of our results and show that incorporating dynamic signals into the dynamic collective entity representation enables a better matching of users' queries to entities.
 
\section{Dynamic Collective Entity Representations}

\subsection{Problem Statement}
The problem of entity ranking is: 
given a query $q$ and a knowledge base $\mathit{KB}$ populated with entities $e \in KB$, find the best matching $e$ that satisfies $q$. 
Both $e$ and $q$ are represented in some (individual or joint) feature space that captures a range of dimensions, which characterize them individually (e.g., content, quality) and jointly (e.g., relationships through click logs). 

The entity ranking problem itself is a standard information retrieval problem, where the system needs to bridge the gap between the vocabulary used by users in queries, and the vocabulary of the entity descriptions. 
This is a long-standing but still open problem that has been tackled from many perspectives. 
One is to design better similarity functions, another is to develop methods for enhancing the feature spaces. 
Our method shares characteristics with both perspectives, as we will now explain.

\subsection{Approach}
\label{subsec:approach}
Our approach to the entity ranking problem consists of two interleaved steps. 
First, we use external description sources (described in Section~\ref{subsec:wsdmdata}) to expand entity representations and reduce the vocabulary gap between queries and entities. 
We do so by representing entities as fielded documents, where each field corresponds to content that comes from one description source. 
Second, we train a classification-based entity ranker, that employs different types of features to learn to weight and combine the content from each field of the entity for optimal retrieval (Section~\ref{subsec:adaptive}). 

External description sources may continually update and change the content in the entity's fields, through user feedback (i.e., clicks) following an issued query, or when users generate content on external description sources that is linked to a KB entity (e.g., Twitter, Delicious). 
Consequently, the feature values that represent the entities change, which may invalidate previously learned optimal feature weights and asks for continuously updating the ranking model.

\subsection{Description Sources}
\label{subsec:wsdmdata}

To construct dynamic collective entity representations we use two types of description sources: 
(i)~a knowledge base ($\mathit{KB}$) from which we extract the initial entity representations, and 
(ii)~a set of external description sources that we use to expand the aforementioned entity representation. 

\begin{figure}[t!]
	\centering
	\includegraphics[trim = 0mm 0mm 0mm 0mm, clip=true, width=\linewidth]{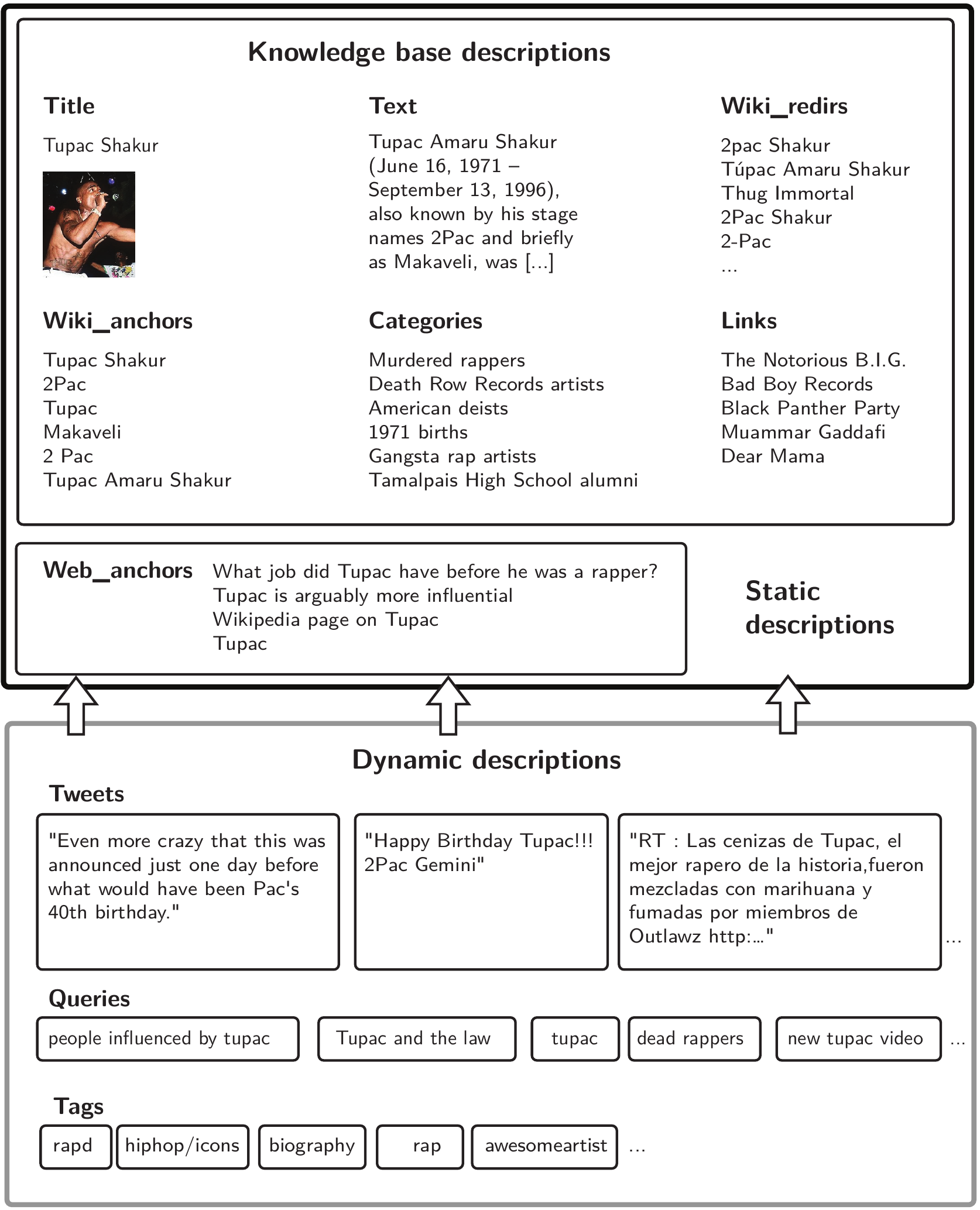}
	\caption{Example expansions for the entity \texttt{Tupac Shakur}.}
	\label{fig:tupac}
\end{figure}

Although in theory all types of external sources are allowed in the model, choosing sources that provide short text descriptions for an entity is favorable as they inherently ``summarize'' entities to a few keywords and do not require an additional step for filtering salient keywords. 
Figure~\ref{fig:tupac} provides an example of expanding the representation of the entity Tupac Shakur\footnote{\url{https://en.wikipedia.org/wiki/Tupac_Shakur}} from several sources, each contributing somewhat different keywords. 

We differentiate between external description sources that are non-timestamped (\emph{static}) and ones that are  timestamped (\emph{dynamic}). 
Non-timestamped sources are those where no time information is available, and sources that are not inherently dynamic, e.g., web archives and aggregates over archives like web anchors. 
Timestamped external description sources are sources whose content is associated with a time\-stamp, and where the nature of the source is inherently dynamic or time-dependent, e.g., tweets or query logs. We describe each type of external description source below, the number of total descriptions, and affected entities, and we provide a summary in Table~\ref{tab:datasummary}.

\begin{table}
  \centering
  \caption{Summary of the nine description sources we consider: Knowledge Base entity descriptions (KB), KB anchors, KB redirects, KB category titles, KB inter-hyperlinks, queries, web anchors, tweets, and tags from Delicious.}
  \label{tab:datasummary}
  \begin{tabular}{lrrr}
    \toprule
    Data source     & Size        & Period     & Affected entities \\
    \midrule
    \multicolumn{4}{l}{\emph{Static expansion sources}}\\
    KB              &  4,898,356   & August 2014 & --                \\ 
    KB anchors      & 15,485,915   & August 2014 & 4,361,608         \\ 
    KB redirects    &  6,256,912   & August 2014 & N/A               \\ 
    KB categories   &  1,100,723   & August 2014 & N/A               \\ 
    KB inter-links  & 28,825,849   & August 2014 & 4,322,703         \\ 
    \midrule
    \multicolumn{4}{l}{\emph{Dynamic expansion sources}}\\
    Queries         & 47,002      & May 2006    & 18,724            \\
    Web anchors     & 9,818,004   & 2012        & 876,063           \\ 
    Twitter         & 52,631      & 2011--2014  & 38,269            \\ 
    Delicious       & 4,429,692   & 2003--2011  & 289,015           \\ 
    \bottomrule
  \end{tabular}
\end{table}

\subsubsection{Initial Entity Representation}

\begin{description}
\item[Knowledge base.]
The knowledge base that we use as our initial index of entities, and which we use to construct the initial entity representations, is a snapshot of Wikipedia from August 3, 2014 with 14,753,852 pages. 
We filter out non-entity pages (``special'' pages such as category, file, and discussion pages), yielding 4,898,356 unique entities. 
The initial entity representations consist of the title and body (i.e., article content) of the Wikipedia page. 
\end{description}

\subsubsection{Static Description Sources}

\begin{description}
\item[Knowledge base.] Knowledge base entities have rich
metadata that can be leveraged for improving retrieval~\cite{shen2014,scholer2004,balog2010}. 
We consider four types of metadata to construct the KB entity representations: (i)~anchor text of inter-knowledge base hyperlinks, (ii)~redirects, (iii)~category titles, and (iv)~titles of entities that are linked from and to each entity. 
Editorial conventions and Wikipedia's quality control ensures these expansions to be of high quality. 

\item[Web anchors.] Moving away from the knowledge base itself, the web provides rich information on how people refer to entities leading to tangible improvements in retrieval~\cite{westerveld2001}. 
We extract anchor texts of links to Wikipedia pages from the Google Wikilinks corpus.\footnote{\url{https://code.google.com/p/wiki-links/}} 
We collect 9,818,004 anchor texts for 876,063 entities. Web anchors differ from KB anchors as they can be of lower quality (due to the absence of editorial conventions) but also of much larger volume. 
While in theory web anchors could be associated with timestamps, in a typical scenario they are aggregated over large archives, where extracting timestamps for diverse web-pages is non-trivial.
\end{description}

\subsubsection{Dynamic Description Sources}

\begin{description}
\item[Twitter.] \citet{mishne2012} show how leveraging terms from 
tweets that do not exist in the pages linked to from tweets can improve retrieval effectiveness of those pages. 
We follow a similar approach and mine all English tweets that contain links to Wikipedia pages that represent the entities in our KB. 
These are extracted from an archive of Twitter's sample stream, spanning four years (2011--2014), resulting in 52,631 tweets for 38,269 entities.

\item[Delicious.] Social tags are concise references to entities and have shown to outperform anchors in several retrieval tasks~\cite{noll2008}. 
We extract tags associated with Wikipedia pages from SocialBM0311\footnote{\url{http://www.zubiaga.org/datasets/socialbm0311/}} resulting in 4,429,692 timestamped tags for 289,015 entities.

\item[Queries.] We use a publicly available query log from MSN sampled between May 1 and May 31, 2006, consisting of 15M queries and their metadata: timestamps and URLs of clicked documents. 
We keep only queries that result in clicks on Wikipedia pages that exist in our snapshot of Wikipedia, resulting in 47,002 queries associated with 18,724 uniquely clicked entities. 
We hold out 30\% of the queries for development (e.g., parameter tuning, feature engineering; 14,101 queries) and use 70\% (32,901 queries) for testing. 
Our use of queries is twofold: 
(i)~queries are used as input to evaluate the performance of our entity ranking approach, and also 
(ii)~as external description source, to expand the entity description with the terms from a query that yield a click on an entity. 
While this dual role of queries may promote head entities that are often searched, we note that ``headness'' of entities differs across description sources, and even tail entities may benefit from external descriptions (as illustrated by the \texttt{Anthropornis} example in Fig.~\ref{fig:penguin}). 
\end{description}

\subsection{Adaptive Entity Ranking}
\label{subsec:adaptive}
The second step in our method is to employ a supervised entity ranker that learns to weight the fields that hold content from the different description sources for optimal retrieval effectiveness. 
Two challenges arise in constructing collective dynamic entity representations:  

\begin{description}
\item[Heterogeneity.] External description sources exhibit different dynamics in terms of volume and quality of content~\cite{li2014}, and differences in number and type of entities to which they link (see, e.g., Table~\ref{tab:datasummary}). 
This heterogeneity causes issues both within entities, since different description sources contribute different amounts and types of content to the entity's fields, and between entities, since popular entities may receive overall more content from external description sources than tail entities. 
\item[Dynamicness.] Dynamic external description sources cause the entity's descriptions to change in near real-time. 
Consequently, a static ranking model cannot capture the evolving and continually changing index that follows from our dynamic scenario, hence we employ an adaptive ranking model that is continuously updated. 
\end{description}
 
\newcommand{\dgvector}[1]{\overline{#1}}

\section{Model}
In the following section, we describe our supervised ranking approach, by first explaining the entity representation,
the set of features we employ for learning an optimal representation for retrieval, and finally the supervised method.

\subsection{Entity Representation}
To deal with dynamic entity representations, which are composed of content from different external description sources, we model entities as fielded documents: 
\begin{align}
	e = \{ \dgvector{f}^e_{title}, \dgvector{f}^e_{text}, \dgvector{f}^e_{anchors}, \ldots, \dgvector{f}^e_{query} \}.
	\label{eq:first}
\end{align}
\noindent
Here, $\dgvector{f}^e$ corresponds to the field term vector that represents $e$'s content from a single source (denoted in subscript).
We refer to this collection of field term vectors as the entity's \emph{representation}. 

The fields with content from dynamic description sources may change over time. 
We refer to the process of adding an external description source's term vector to the entity's corresponding field term vector as an \emph{update}. 
To model these dynamically changing fields, we discretize time (i.e., $T = \{t_1, t_2, t_3, \ldots, t_n\}$), and define \emph{updating} fields as: 
\begin{eqnarray}
    \dgvector{f}^{e}_{query}(t_i) &=& \dgvector{f}^{e}_{query}(t_{i-1})~+
	\begin{cases}
		\dgvector{q}, & \text{if } e_{clicked} \\
		\label{eq:user}
    	0,            & \text{otherwise}
	\end{cases} \\ 
	\dgvector{f}^{e}_{tweets}(t_i) &=& \dgvector{f}^{e}_{tweets}(t_{i-1}) + \dgvector{tweet}_{e}
	\label{eq:tweets}
\\ 
	\dgvector{f}^{e}_{tags}(t_i) &=& \dgvector{f}^{e}_{tags}(t_{i-1}) + \dgvector{tag}_{e}.
	\label{eq:tags}
\end{eqnarray}
\noindent
In equation~\ref{eq:user}, $\dgvector{q}$ represents the term vector
 of query $q$ that is summed element-wise to $e$'s query field term vector ($\dgvector{f}^{e}_{query}$) at time $t_i$, if $e$ is clicked by a user that issues $q$. 
In equation~\ref{eq:tweets},
$\dgvector{tweet}_{e}$ represents the field term vector of a tweet that contains a link to the Wikipedia page of $e$, which also gets added element-wise to the corresponding field ($\dgvector{f}^{e}_{tweets}$). 
Finally, in equation~\ref{eq:tags}, $\dgvector{tag}_{e}$ is the term vector of a tag that a user assigns to the Wikipedia page of $e$. 

To estimate $e$'s relevance to a query $q$ given the above-described representation one could, e.g., linearly combine retrieval scores between $q$'s term vector  $\dgvector{q}$ and each $\dgvector{f} \in e$. 
However, due to the heterogeneity that exists both between the different fields that make up the entity representation, and between different entities (described in Section~\ref{subsec:adaptive}), linearly combining similarity scores may be sub-optimal~\cite{robertson2004}, and hence we 
employ a supervised single-field weighting model~\cite{macdonald2013}.
Here, each field's contribution towards the final score is individually weighted, through learned field weights from implicit user feedback. 

\subsection{Features}
\label{subsec:features}
To learn the optimal entity representation for retrieval, we employ three types of feature that express field and entity importance: 
first, 
\emph{field similarity} features are computed per field and boil down to query--field similarity scores. Next, 
\emph{field importance} features, likewise computed per field, aim to inform the ranker of the status of the field at that point in time (i.e., to favor fields with more and novel content). Finally, we employ 
\emph{entity importance} features, which operate at the entity level and aim to favor recently updated entities. 

\subsubsection{Field Similarity}
\label{subsubsec:similarity}
The first set of features model the similarity between a query and a field, which we denote as $\phi_{sim}$. 
For query--field similarity, we compute TF$\times$IDF cosine similarity. We define
\begin{align}
	\phi_{sim}(q, \dgvector{f}, t_i) = \sum_{w \in q} n(w, \dgvector{f}({t_i})) \cdot \log \frac{|C^{t_i}|}{|\{\dgvector{f}({t_i}) \in C^{t_i} : w \in \dgvector{f}({t_i}) \}|},
	\label{eq:sim}
\end{align}
where $w$ corresponds to a query term, $n(w,\dgvector{f}({t_i}))$ is the frequency of term $w$ in field $f$ at time $t_i$.
$C^{t_i}$ is the collection of fields at time $t_i$, and $|\cdot|$ indicates set cardinality.
More elaborate similarity functions can be used, e.g.,
BM25(F), however, we choose a parameter-less similarity function that requires
no tuning. 
This allows us to directly compare the contribution of the
different expansion fields without having additional factors play a role, such
as length normalization parameters, which affect different fields in
non-trivial ways.

\subsubsection{Field Importance}
The next set of features is also computed per field, $\phi_p$ is meant to capture a field's \emph{importance} at time $t_i$: 
\begin{align}
	\phi_{p}(\dgvector{f}(t_i), e) = S(\dgvector{f}(t_i), e).
\end{align}
\noindent
We instantiate four different field importance features. 
First, we consider two ways to compute a field's length, either in terms~\eqref{eq:len1} or in characters~\eqref{eq:len2}:
\begin{align}
	\phi_{p_{1}}(\dgvector{f}(t_i), e) &= |\dgvector{f}(t_i)|
	\label{eq:len1}
	\\
	\phi_{p_{2}}(\dgvector{f}(t_i), e) &= \sum_{w \in \dgvector{f}(t_i)}|w|
	\label{eq:len2}
\end{align}
\noindent
The third field importance scoring function captures a field's novelty at time $t_i$, to favor fields that have been updated with previously unseen, newly associated terms to the entity (i.e., terms that were not in the original entity representation at $t_0$):
\begin{align}
	\phi_{p_{3}}(\dgvector{f}(t_i), e) & = | \{ w \in \dgvector{f}(t_i) : w \notin \dgvector{f}(t_0)\}|.
	\label{eq:nov}
\end{align}
\noindent
The fourth field importance scoring function expresses whether a field has undergone an update at time $t_i$:
\begin{align}
	\phi_{p_{4}}(\dgvector{f}, t_i, e) = \sum_{j=0}^{i}~+=
\begin{cases}
	1, & \text{if } update(\dgvector{f}(t_j)) \\
    0, & \text{otherwise},
\end{cases}
\end{align}
\noindent
where $update(\dgvector{f}(t_j))$ is a Boolean function indicating whether field $f$ was updated at time $j$, i.e., we sum from $t_0$ through $t_i$ and accumulate updates to the fields.

\subsubsection{Entity Importance}
The feature $\phi_I$ models the entity's importance. 
We compute the time since the entity last received an update to favor recently updated entities:
\begin{align}
	\phi_{I}(e, t_i) = t_i - \max_{\dgvector{f} \in e} time_{\dgvector{f}},
\end{align}
\noindent
here, $t_i$ is the current time, and $time_{\dgvector{f}}$ is the update time of field $\dgvector{f}$. $\max_{\dgvector{f}}$ corresponds to the timestamp of the entity's field that was most recently updated.

\subsection{Machine Learning}
Given the features explained in the previous section, we employ a supervised ranker to learn the optimal feature weights for retrieval: 
\begin{align}
	\Omega = ( \omega_{e}, \omega_{\dgvector{f}_{title}}, \omega_{p1_{title}}, \omega_{p2_{title}}, 	\ldots, \omega_{\dgvector{f}_{text}}, \omega_{p1_{text}}, \ldots ).
\end{align}
\noindent
Here, $\Omega$ corresponds to the weight vector, which is composed of individual weights ($\omega$) for each of the field features (similarity and importance) and the entity importance feature. 

To find the optimal $\Omega$, we train a classification-based re-ranking model, and learn from user interactions (i.e., clicks). 
The model employs the features detailed in the previous section, and the classification's confidence score is used as a ranking signal. 
As input, the ranker receives a feature vector ($x$), extracted for each entity-query pair. 
The associated label $y$ is positive (1) for the entities that were clicked by a user who issued query $q$. 
We define $x$ as
\begin{align}
	x = \{\phi_{{sim}_1}, \phi_{{p1}_1}, \phi_{{p2}_1}, \phi_{{p3}_1}, \phi_{{p4}_1}, 
	\ldots, \notag\\
	\phi_{{sim}_{|e|}}, \phi_{{p1}_{|e|}}, \phi_{{p2}_{|e|}}, \phi_{{p3}_{|e|}}, \phi_{{p4}_{|e|}}, 
	\phi_I \},
\end{align}
\noindent
where $|e|$ corresponds to the number of fields that make up the entity representation ($\dgvector{f} \in e$).

See Algorithm~\ref{alg:dg} for an overview of our machine learning method in pseudo-code. 
As input, our supervised classifier is given a set of $\langle q,e,L\rangle$-tuples (see l.~\ref{alg:line1}). 
These tuples consist of a query ($q$), a candidate entity ($e$), and a (binary) label ($L$): positive (1) or negative (0). 
Given an incoming query,  we first perform top-$k$ retrieval (see Section~\ref{subsec:ML} for details) to yield our initial set of candidate entities: $\mathcal{E}_{candidate}$ (l.~\ref{alg:line2}). 
For each entity, we extract the features which are detailed in Section~\ref{subsec:features} (l.~\ref{alg:line3}) and have the classification-based ranker $R$ output a confidence score for $e$ belonging to the positive class, which is used to rank the candidate entities (l.~\ref{alg:line4}). 
Labels are acquired through user interactions, i.e., entities that are in the set of candidate entities and clicked after issuing a query are labeled as positive instances (l.~\ref{alg:line5}--\ref{alg:line6}), used to retrain $R$ (l.~\ref{alg:line7}).
Finally, after each query, we allow entities to be updated by dynamic description sources: tweets and tags (l.~\ref{alg:line8}), we provide more details in Section~\ref{subsec:expansion}.

\begin{algorithm}[th]
\caption{Pseudo-algorithm for learning optimal $\Omega^{T}$.}
\label{alg:dg}
\begin{algorithmic}[1]
\Require Ranker $R$, Knowledge Base $KB$, Entities $\mathcal{E}$\\
$\mathcal{E} \longleftarrow \{e_1,e_2,\ldots,e_{|KB|}\}$ \\
$e \longleftarrow \{\dgvector{f}_{title},\dgvector{f}_{anchors},\ldots,\dgvector{f}_{text}\}$ \\ \\
$\mathcal{L} = \{\langle q_1,e_1,\{0,1\}\rangle, \langle q_1,e_2,\{0,1\}\rangle, \ldots, \langle q_n,e_m,\{0,1\}\rangle\}$\label{alg:line1} \\
$R \longleftarrow $ Train$(\mathcal{L})$ \\
\While{$q$}
 \State $\mathcal{E}_{candidate}  \longleftarrow $Top-$k$ retrieval$(q)$\label{alg:line2}
 \State $\mathcal{E}_{ranked} \longleftarrow []$
 \\

 \For{$e \in \mathcal{E}_{candidate}$}
  \State $\mathcal{\phi}_e \longleftarrow$ Extract features$(e)$ \label{alg:line3}
  \State $\mathcal{E}_{ranked} \longleftarrow $ Classify$(R, \mathcal{\phi}_e)$\label{alg:line4}
 \EndFor
 \\
 
 \State $e_{clicked} \longleftarrow $ Observe click$(q,\mathcal{E})$ \label{alg:line5}
 \If {$e_{clicked} \in \mathcal{E}_{candidate}$}
   \State $\mathcal{L} \longleftarrow \mathcal{L} \ \cup \{\langle q,e_{clicked},1\rangle\}$
   \State $e_{clicked} \longleftarrow  e_{clicked} \ \cup \{text_{q}\}$
 \Else
   \State $\mathcal{L} \longleftarrow \mathcal{L} \ \cup \{\langle q,e_{clicked},0\rangle\}$
 \EndIf \label{alg:line6} \\
 
 \State $R \longleftarrow $ Train$(\mathcal{L})$\label{alg:line7}
 
 \For{$e \in \mathcal{E}$}
   \State $e \longleftarrow  e \ \cup \{\dgvector{f}_{e,tweet_1},\dgvector{f}_{e,tag_1},\ldots,\dgvector{f}_{e,tweet_i},\dgvector{f}_{e,tag_j}\}$\label{alg:line8}
 \EndFor
\EndWhile
\end{algorithmic}
\end{algorithm}
 
\section{Experimental setup}

In this section, we start by describing our experiments and how they allow us to answer the three subquestions raised in Section~\ref{sec:wsdmintro}, and then, we present our baselines, machine learning setting, and describe our evaluation methodology. 

\subsection{Experiments}
\begin{description}
\item[Experiment 1.]
To answer our first subquestion (\ref{rq:wsdm1}): \emph{does entity ranking effectiveness increase using dynamic collective entity representations?} 
we compare our proposed dynamic collective entity ranking method to a baseline that only incorporates KB fields for entity ranking: \texttt{KBER} (Knowledge Base Entity Representations).
We restrict both the baseline and our Dynamic Collective Entity Representation (\texttt{DCER}) method to the set of field similarity features (Section~\ref{subsubsec:similarity}), which we denote as \texttt{KBER$_{sim}$} and \texttt{DCER$_{sim}$}. 
This allows us to provide clear insights into the contribution of the fields' content in ranking effectiveness. 
In addition, we perform an ablation study and compare the similarity-baseline to several approaches that incorporate content from a single external description source (denoted \texttt{KB+source}$_{sim}$). 

\item[Experiment 2.]
We address \ref{rq:wsdm2}, \emph{does entity ranking effectiveness increase when employing field and entity features}, by comparing the \texttt{KBER}$_{sim}$ baseline that only incorporates field similarity features, to the \texttt{KBER} baseline that incorporates the entity and field importance features, and to our \texttt{DCER} method. 

\item[Experiment 3.]
Finally, to answer \ref{rq:wsdm3}, \emph{does entity ranking effectiveness increase when we continuously learn the optimal entity representations?} 
we compare our proposed entity ranking method \texttt{DCER} to its non-adaptive counterpart (\texttt{DCER$_{na}$}), that we do not periodically retrain. Contrasting the performance of this non-adaptive system with our adaptive system allows us to tease apart the effects of adding more training data, and the effect of the additional content that comes from dynamic external description sources.
Here too, we include an ablation study, and compare to non-adaptive approaches that incorporate content from a single external description source (denoted \texttt{KB+source}$_{na}$).
\end{description} 

\subsection{Baselines}
Due to our focus on dynamic entity representations and adaptive rankers, running our method on datasets from seemingly related evaluation campaigns such as those in TREC~\cite{trec2010,trec2011} and INEX~\cite{inex2007,inex2008,inex2009} is not feasible. 
We are constrained by the size of datasets, i.e., we need datasets that are sufficiently large (thousands of queries) and time-stamped, which excludes the aforementioned evaluation campaigns, and hence direct comparison to results obtained there. 
In a scenario with dynamic fields and continually changing content, 
our method is required to reweigh fields continuously to reflect changes in the fields' content. 
Continually (re-)tuning parameters of existing fielded retrieval methods such as BM25F~\cite{robertson2004}, when documents in the index change is exceedingly expensive, rendering these methods unsuitable for this scenario. 

For these reasons we consider the following supervised baselines in our experiments; 
\texttt{KBER}$_{sim}$ is an online learning classification-based entity ranker, that employs field similarity features on entity representations composed of KB description sources (i.e., title, text, categories, anchors, redirects and links fields). 
\texttt{KBER} is the same baseline system, extended with the full set of features (described in Section~\ref{subsec:features}). 
Finally, \texttt{DCER$_{na}$} is a non-adaptive baseline: 
it incorporates all external description sources, and all features as our proposed \texttt{DCER} method, but does not periodically retrain. \\

\subsection{Data Alignment}
\label{subsec:expansion}
In our experiments we update the fields that jointly represent an entity with external descriptions that come in a streaming manner, from a range of heterogeneous external description sources. 
This assumes that all data sources run in parallel in terms of time.
In a real-world setting this assumption may not always hold as systems need to integrate historical data sources that span different time periods and are of different size for bootstrapping. 
We simulate this scenario by choosing data sources that do not originate from the same time period nor span the same amount of time (see also Table~\ref{tab:datasummary}). 
To remedy this, we introduce a method for time-aligning all data sources (which range from 2009 to 2014) to the timeline of the query log (which dates from 2006). 
We apply a \emph{source-time} transformation to mitigate the dependence of content popularity on time and when it was created
\cite{szabo2010,tsagkias2010}. 
Each query is treated as a time unit, and we distribute the expansions from the different sources over the queries, as opposed to obeying the misaligned timestamps from the query log and expansion sources.

To illustrate: given $n$ queries and a total of $k$ items for a given expansion source, after each query we update the entity representations with $\frac{n}{k}$ expansions of that particular expansion source.
In our dataset we have 32,901 queries, 52,631 tweets, and 4,429,692 tags. 
After each query we distribute 1 query, 2 tweets, and 135 tags over the entities in the index. 
Mapping real time to source time evenly spreads the content of each data source within the timespan of the query log. 
This smooths out bursts but retains the same distribution of content over entities (in terms of, e.g., entity popularity). 
The diminishing effect on burstiness is desirable in the case of misaligned corpora, as bursts are informative of news events, which would not co-occur in the different data sources. 
Although our re-aligned corpora cannot match the quality of real parallel corpora, our method offers a robust lower bound to understand the utility of collective dynamic entity representations of our method. 
We reiterate that our goal in this chapter is to study the effect of dynamic entity representations and adapting rankers, not to leverage temporal features for improving entity retrieval.

\subsection{Machine Learning}
\label{subsec:ML}
We apply machine learning for learning to weight the different fields that make up an entity representation for optimal retrieval effectiveness. 
In response to a query, we first generate an initial set of candidate entities by retrieving the top-$k$ entities to limit the required computational resources for extracting all features for all documents in the collection~\cite{liu2009}.
Our top-$k$ retrieval method involves ranking all entities using the similarity function described in Section~\ref{subsubsec:similarity}, where we collapse the fielded entity representation into a single document.
We set $k=20$ as it has shown a fair tradeoff between high recall (80.1\% on our development set) and low computational expense.
We choose Random Forests as our machine learning algorithm because it has proven robust in a range of diverse tasks (e.g.,~\cite{mohan2011}), and can produce confidence scores that we employ as ranking signal.
In our experiments, we set the number of trees to 500 and the number of features each decision tree considers for the best split to $\sqrt{|\Omega|}$.

\subsection{Evaluation}
For evaluating our method's adaptivity and performance over time, we create a set of incremental time-based train/test splits as in~\cite{bekkerman2004}. 
We first split the query log into $K$ chunks of size $N$: $\{C_{1}, C_{2}, C_{3}, \ldots, C_{K}\}$.
We then allocate the first chunk ($C_{1}$) for training the classifier and start iteratively evaluating each succeeding query. 
Once the second chunk of queries ($C_{2}$) has been evaluated, we expand the training set with it and retrain the classifier. 
We then continue evaluating the next chunk of queries ($C_{3}$). 
This procedure is repeated, continually expanding the training set and retraining the classifier with $N$ queries (we set $N$=500 in our experiments). 
In this scenario, users' clicks are treated as ground truth and the classifier's goal is to rank clicked entities at position 1. We do not distinguish between clicks (e.g., satisfied clicks and non-satisfied clicks), and we leave more advanced user models, e.g., which incorporate skips, as future work. 

To show the robustness of our method we apply five-fold cross-validation over each run, i.e., we generate five alternatively ordered query logs by shuffling the queries. 
We keep the order of the dynamic description sources fixed to avoid conflating the effect of queries' order with that of the description sources. 

Since we are interested in how our method behaves over time, we plot the MAP at each query over the five-folds, as opposed to averaging the scores over all chunks across folds (as in~\cite{bekkerman2004}) and losing this temporal dimension. 
In addition to reporting MAP, we report on P@1, as there is only a single relevant entity per query in our experimental setup. 
We test for statistical significance using a two-tailed paired t-test. 
Significant differences are marked~\upup~for~$\alpha=0.01$.
 
\section{Results and Analysis}
\label{sec:wsdmresults}

We report on the experimental results for each of our three experiments in turn, and provide an analysis of the results to better understand the behavior of our method.

\subsection{Dynamic Collective Entity Representations}
\label{subsec:rq1results}

In our first experiment, we explore the impact of the description sources we use for constructing dynamic collective entity representations, and aim to answer~\ref{rq:wsdm1}. 
We compare the \texttt{KBER}$_{sim}$ baseline, which incorporates field similarity on KB descriptions, to our \texttt{DCER}$_{sim}$ method, which incorporates field similarity features on all description sources (web anchors, tweets, tags, and queries). 
Table~\ref{tab:improvement-summary} shows the performance in terms of MAP and P@1 of the baseline (\texttt{KBER}$_{sim}$) and \texttt{DCER}$_{sim}$ after observing all queries in our dataset. 
We include an oracle run as a performance upper bound given our top-$k$ retrieval scenario. 

\begin{table}[t]
\centering
\caption{Performance of field similarity-based entity ranking methods using KB entity representations (\texttt{KBER}$_{sim}$) and dynamic collective entity representations (\texttt{DCER$_{sim}$}) in terms of MAP and P@1. Significance tested against \texttt{KBER$_{sim}$}. Oracle marks upper bound performance given our top-$k$ scenario.}
\begin{tabular}{l l l}
\toprule
\textbf{Run} & \textbf{MAP} & \textbf{P@1}  \\
\midrule
KBER$_{sim}$ &  0.5579 & 0.4967 \\
DCER$_{sim}$ & \textbf{0.5971\upup} & \textbf{0.5573\upup} \\
\midrule
Oracle & 0.6653 & 0.6653 \\
\bottomrule
\end{tabular}
\label{tab:improvement-summary}
\end{table}

The results show that the dynamic collective entity representations manage to significantly outperform the KB entity representations for both metrics, and that \texttt{DCER$_{sim}$} presents the correct entity at the top of the ranking for over 55\% of the queries. 

\begin{figure}[t!]
	\centering
	\includegraphics[trim = 0mm 4mm 0mm 0mm, clip=true, width=\linewidth]{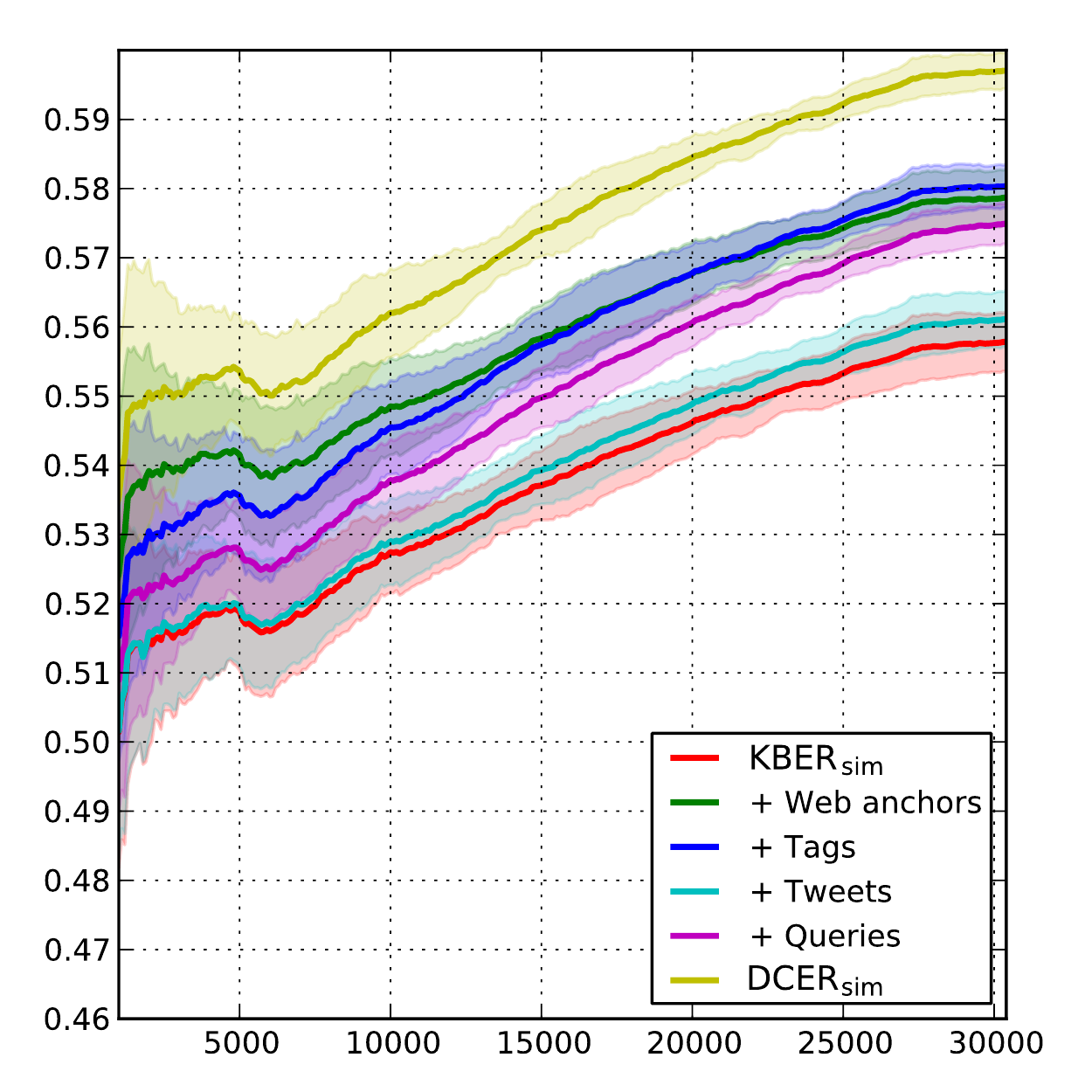}
	\caption{Impact on performance of individual description sources. MAP on the y-axis, number of queries on the x-axis. The line represents MAP, averaged over 5-folds. Standard deviation is shown as a shade around the line. This plot is best viewed in color.} 
	\label{fig:Qablation}
\end{figure}

Next, we look into the impact on performance of individual description sources for our dynamic collective entity representations. 
We add each source individually to the \texttt{KBER$_{sim}$} baseline. 
Figure~\ref{fig:Qablation} shows how each individual description source contributes to produce a more effective ranking, with \texttt{KB+tags} narrowly outperforming \texttt{KB+web} as the best single source. 
Combining all sources into one, yields the best results, outperforming \texttt{KB+tags} by more than 3\%. 
We observe that after about 18,000 queries, \texttt{KB+tags} overtakes the (static) \texttt{KB+web} method, suggesting that newly incoming tags yield higher ranking effectiveness. 
All runs show an upward trend and seem to level out around the 30,000th query. 
This pattern is seen across ranking methods, which indicates that the upward trend can be attributed to the addition of more training data (queries). 

\begin{table}[h]
\caption{Comparison of relative improvement between runs with different field similarity features. We report on MAP and P@1 at query 10,000 and the last query. Rate corresponds to the percentage of improvement between the 10,000th and final query. Significance tested against \texttt{KBER}$_{sim}$.}
\begin{tabularx}{\columnwidth}{lXXXXXX}
\toprule
\textbf{Run} & \textbf{MAP (10k)} & \textbf{MAP (end)} & \textbf{Rate} & \textbf{P@1 (10k)} & \textbf{P@1 (end)} & \textbf{Rate}  \\
\midrule
KBER$_{sim}$ & 0.5274 & 0.5579 & +5.8\% & 0.4648 & 0.4967 & +6.9\% \\
KB+Web$_{sim}$ & 0.5485 & 0.5787\upup & +5.5\% & 0.4965 & 0.5282\upup & +6.4\% \\
\midrule
KB+Tags$_{sim}$ & 0.5455 & 0.5804\upup & +6.4\% & 0.4930 & 0.5317\upup & +7.8\% \\
KB+Tweets$_{sim}$ & 0.5290 & 0.5612\upup & +6.1\% & 0.4673 & 0.5021\upup & +7.5\% \\
KB+Queries$_{sim}$ & 0.5379 & 0.5750\upup & \textbf{+6.9\%} & 0.4813 & 0.5242\upup & \textbf{+8.9\%} \\
\midrule
DCER$_{sim}$ & 0.5620 & \textbf{0.5971\upup} & +6.2\% & 0.5178 & \textbf{0.5573}\upup & +7.6\% \\
\bottomrule
\end{tabularx}
\label{tab:improvement}
\end{table}

\noindent
Table~\ref{tab:improvement} lists the results of all methods along with the improvement rate (relative improvement when going from 10,000 to all queries). 
The runs that incorporate dynamic description sources (i.e., \texttt{KB+tags}, \texttt{KB+tweets}, and \texttt{KB+queries}) show the highest relative improvements (at 6.4\%, 6.1\% and 6.9\% respectively). 
Interestingly, for P@1, the \texttt{KB+queries} method yields a substantial relative improvement (+8.9\%), indicating that the added queries provide a strong signal for ranking the clicked entities at the top.

\noindent
The comparatively lower learning rates of methods that incorporate only static description sources (\texttt{KBER}$_{sim}$ and \texttt{KB+web} yield relative improvements of +5.8\% and +5.5\%, respectively), suggest that the entity content from dynamic description sources effectively contributes to higher ranking performance as the entity representations change and the ranker is able to reach a new optimum. 
\texttt{DCER}$_{sim}$, the method that incorporates all available description sources, shows a comparatively lower relative improvement, which is likely due to it hitting a ceiling, and not much relative improvement can be gained.

\subsubsection{Feature Weights over Time}
\label{subsec:featureweights}
\begin{figure}[t!]
	\centering
	\includegraphics[trim = 0mm 0mm 0mm 0mm, clip=true, width=\linewidth]{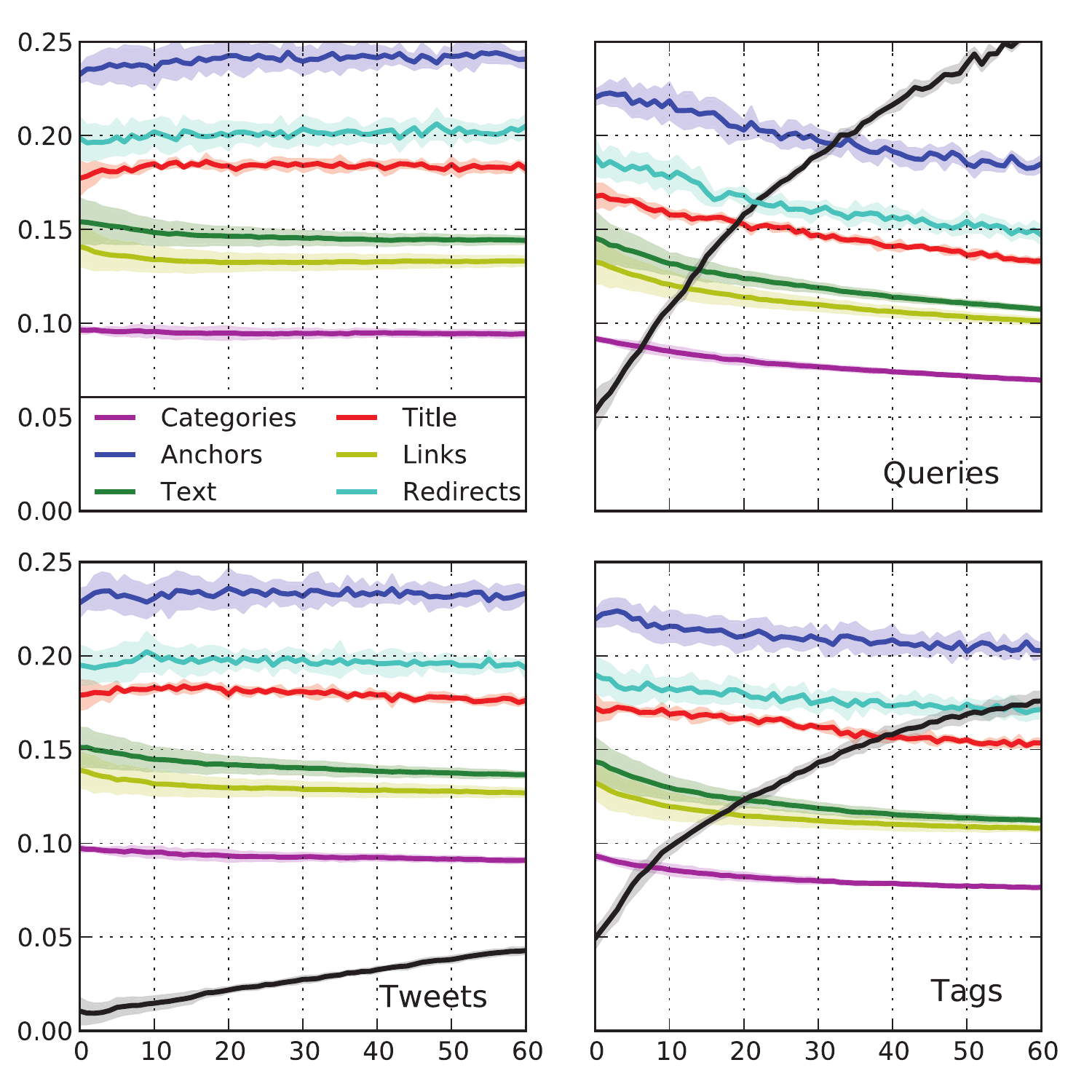}
	\caption{Feature weights over time: y-axis shows (relative) feature weights, x-axis shows each chunk of 500 queries where the ranker is retrained. Starting from top left in a clock-wise direction we show the following runs: KBER$_{sim}$ (baseline), KB+Queries, KB+Tags, KB+Tweets. The black line shows the dynamic description source's weight. This plot is best viewed in color.}
	\label{fig:weights}
\end{figure}

A unique property in our scenario is that, over time, more training data is added to the system, and more descriptions from dynamic sources come in, both of which are expected to improve the performance of our method. 
To better understand our the behavior of our method, we look into the learned feature weights for both a static approach and our dynamic one at each retraining interval. 
We consider as feature importance the average height of a feature when it is used as a split-node in one of the trees that make up the random forest. 
The underlying intuition is that features higher up a decision tree affect a larger fraction of the samples, and hence can be considered more important. 
Figure~\ref{fig:weights} shows the weights for six static fields (categories, title, anchors, links, text, redirects) and three dynamic fields (queries, tweets, tags).

The \texttt{KBER}$_{sim}$ baseline shows little change in the feature weights as more training data is added. 
The anchors' weight increases slightly, while the text and categories fields' weights steadily decline over time. 
The latter two features show a steadily decreasing standard deviation, indicating that the ranker becomes more confident in the assigned weights. 

For \texttt{KB+queries} it is apparent that the queries field weight increases substantially as both the ranking method receives more training data and the entity representations receive more content from the queries description source. 
At the same time, anchors, redirects, and title field weights, which were assigned consistently high weights in \texttt{KBER$_{sim}$}, seem to pay for the increase of the importance of queries. 
Text, category, and links show similar patterns to the baseline run; steadily declining weights, converging in terms of a decreasing standard deviation. 

\noindent
The \texttt{KB+tags} run shows a similar pattern as \texttt{KB+queries}: we observe an increase over time of the weight assigned to the field that holds content from the tag description source, at the cost of original KB fields, resulting in improved ranking effectiveness. 
Looking at \texttt{KB+tweets}, however, we observe a different pattern. 
The tweets field starts out with a very low weight, and although the weight steadily increases, it remains low. 
The ranker here, too, becomes more confident on this source's weight, with the standard deviation dissolving over time. 
Looking at the higher performance of \texttt{KB+tweets} in comparison to the \texttt{KBER}$_{sim}$ baseline, together with the field's weight increasing over time, we conclude that the tweets added to the entity representations provide added value.

\subsection{Modeling Field Importance}
\label{subsec:fieldablation}

In our second experiment, which aims to answer~\ref{rq:wsdm2}, we turn to the contribution of the additional field and entity importance features. 
Table~\ref{tab:improvement2} lists the results of the best performing run with only field similarity features (\texttt{DCER}$_{sim}$), the \texttt{KBER} baseline that incorporates the full set of features, and our proposed \texttt{DCER} method which likewise incorporates the full set of features. 

\begin{table}[t]
\caption{Comparison of relative improvement between the \texttt{KBER} baseline with field importance features for KB fields, and our \texttt{DCER} method with these features for all fields. We also show the best performing run without field and entity importance features. Significance tested against \texttt{KBER}.}
\begin{tabularx}{\columnwidth}{ lXXXXXX }
\toprule
\textbf{Run} & \textbf{MAP (10k)} & \textbf{MAP (end)} & \textbf{Rate} & \textbf{P@1 (10k)} & \textbf{P@1 (end)} & \textbf{Rate} \\
\midrule
Oracle & -- & 0.6653 & -- & -- & 0.6653 & -- \\
DCER$_{sim}$ & 0.5620 & 0.5971 & +6.2\% & 0.5178 & 0.5573 & +7.6\% \\
\midrule
KBER & 0.5853 & 0.6129 & +4.7\% & 0.5559 & 0.5831 & +4.9\% \\
DCER & 0.5923 & \textbf{0.6200\upup} & +4.7\% & 0.5655 & \textbf{0.5925\upup} & +4.8\% \\
\bottomrule
\end{tabularx}
\label{tab:improvement2}
\end{table}

The results show that modeling field and entity importance significantly improves effectiveness of both the \texttt{KBER}$_{sim}$ baseline and the \texttt{DCER}$_{sim}$ runs. 
After running through the entire dataset, the performance of \texttt{DCER} approaches that of the oracle run which also explains why the differences between the two approaches here is rather small: we are very close to the maximum achievable score. 

\begin{figure}[t!]
	\centering
	\includegraphics[trim = 0mm 0mm 0mm 0mm, clip=true, width=\linewidth]{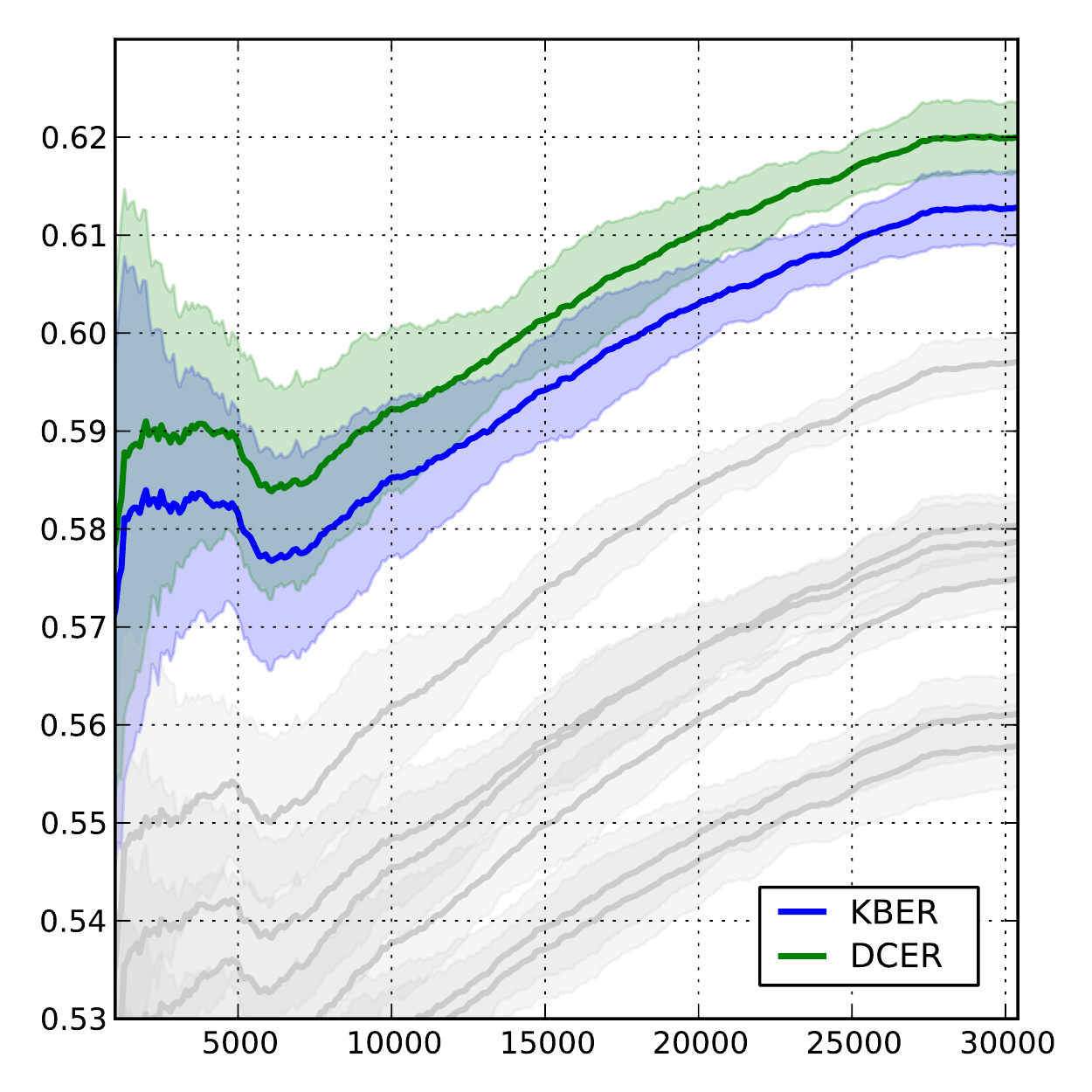}
	\caption{Runs with field similarity, field importance, and entity importance features. MAP on the y-axis, number of queries on the x-axis. The line represents MAP, averaged over 5-folds. Standard deviation is shown as a shade around the line. This plot is best viewed in color.}
	\label{fig:Fablation}
\end{figure}

Figure~\ref{fig:Fablation} shows the performance of the two approaches over time. The pattern is similar to the one in Section~\ref{subsec:rq1results}: both lines show a steady increase, while \texttt{DCER} maintains to be the best performing.

\subsection{Ranker Adaptivity}
\label{subsec:rq2results}

In our third experiment, where we aim to answer~\ref{rq:wsdm3}, we compare our adaptive ranker (\texttt{DCER}), which is continuously retrained, to a non-adaptive baseline (\texttt{DCER$_{na}$}), which is only trained once at the start and is not retrained. 

\begin{table}[h]
\centering
\caption{Comparing relative improvement between runs with and without an adaptive ranker. 
Statistical significance tested between the \texttt{KBER} and \texttt{DCER} approaches and their non-adaptive counterparts.} 
\begin{tabular}{l l l l}
\toprule
\textbf{Run} & \textbf{Adaptive} & \textbf{MAP} & \textbf{P@1} \\
\midrule
KBER$_{na}$ & no & 0.5198 & 0.4392 \\
KBER & yes & 0.5579\upup & 0.4967\upup \\
\midrule
DCER$_{na}$ & no & 0.5872 & 0.5408 \\
DCER & yes & \textbf{0.5971}\upup & \textbf{0.5573}\upup \\
\bottomrule
\end{tabular}
\label{tab:improvement_static-summary}
\end{table}

\noindent
Results in Table~\ref{tab:improvement_static-summary} show that incrementally re-training the ranker is beneficial to entity ranking effectiveness. 
For both \texttt{KBER} and \texttt{DCER} we see a substantial improvement when moving from one single batch training to continuously retraining the ranker. 

\begin{figure}[t!]
	\centering
	\includegraphics[trim = 0mm 0mm 0mm 0mm, clip=true, width=\linewidth]{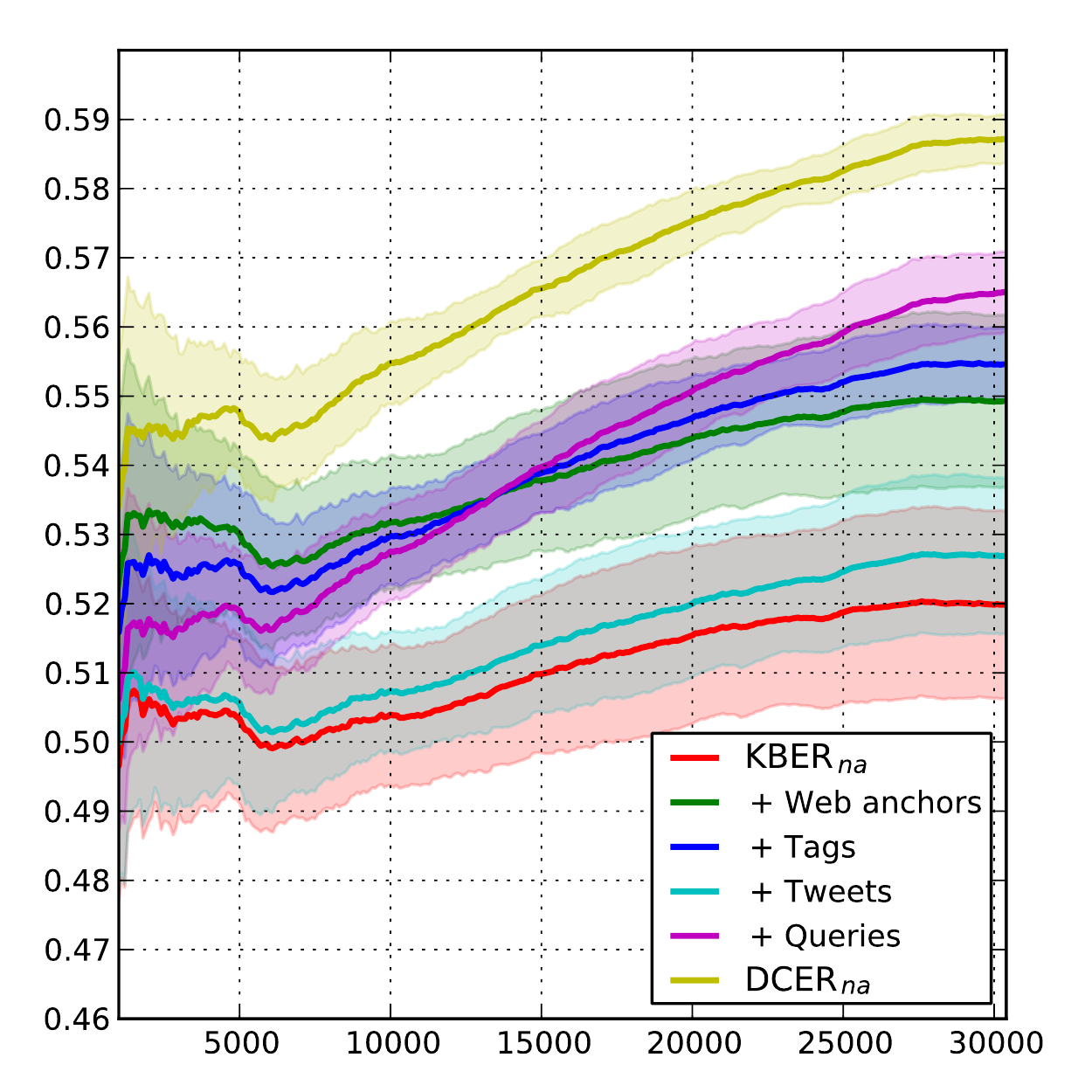}
	\caption{Individual description sources with non-adaptive ranker. MAP on the y-axis, number of queries on the x-axis. The line represents MAP, averaged over 5-folds. Standard deviation is shown as a shade around the line. This plot is best viewed in color.}
	\label{fig:rq2}
\end{figure}

To better understand this behavior, we plot in Figure~\ref{fig:rq2} the performance of all runs we consider over time and at the same time; Table~\ref{tab:improvement_static} provides the detailed scores. 
Broadly speaking we observe similar patterns between adaptive and non-adaptive methods, and we identify three interesting points. 
First, for the non-adaptive methods, the absolute performance is substantially lower across the board. 
Second, for the adaptive methods, the standard deviation (as shown in Figures~\ref{fig:Qablation} and~\ref{fig:rq2}) is substantially lower, which indicates that retraining increases the ranking method's confidence in optimally combining the different descriptions for the entity representation. 
Third, the learning rates in the adaptive setting are substantially higher, reaffirming our observation that learning a single global feature weight vector is not optimal.

\begin{table}[h]
\caption{Comparing relative improvement between non-adaptive runs. Significance tested against \texttt{KBER$_{na}$}.}
\begin{tabularx}{\columnwidth}{lXXXXXX}
\toprule
\textbf{Run} & \textbf{MAP (10k)} & \textbf{MAP (end)} & \textbf{Rate} & \textbf{P@1 (10k)} & \textbf{P@1 (end)} & \textbf{Rate} \\
\midrule
KBER$_{na}$       & 0.5040 & 0.5198 & +3.1\% & 0.4286 & 0.4392 & +2.5\% \\
KB+Web$_{na}$     & 0.5318 & 0.5493\upup & +3.3\% & 0.4698 & 0.4829\upup & +2.8\% \\
\midrule
KB+Tags$_{na}$    & 0.5298 & 0.5546\upup & +4.7\% & 0.4671 & 0.4904\upup & +5.0\% \\
KB+Tweets$_{na}$  & 0.5074 & 0.5269\upup & +3.8\% & 0.4334 & 0.4490\upup & +3.6\% \\
KB+Queries$_{na}$ & 0.5275 & 0.5650\upup & \textbf{+7.1}\% & 0.4659 & 0.5090\upup & +9.2\% \\
\midrule
DCER$_{na}$     & 0.5548 & \textbf{0.5872}\upup & +5.8\% & 0.5063 & 0.5408\upup & +6.8\% \\
\bottomrule
\end{tabularx}
\label{tab:improvement_static}
\end{table}

Table~\ref{tab:improvement_static} shows that the difference in learning rates between methods that only incorporate static description sources (\texttt{KBER}, \texttt{KB+web}) and methods that incorporate dynamic sources is pronounced, in particular for the tags and queries sources. 
This indicates that even with fixed feature weights, the content that comes in from the dynamic description sources yields an improvement in entity ranking effectiveness. 
Finally, the methods that only incorporate static description sources also show lower learning rates than their adaptive counterparts, which indicates that retraining and adapting feature weights is desirable even with static entity representations. 
 
\section{Conclusion}
\label{sec:conclusions}

In this chapter, we addressed the entity ranking task, and set out to improve it by enriching entity representations from entity descriptions collected from a wide variety of sources. 
We set out to answer the following question: \\

\begin{description}[leftmargin=!,labelwidth=1cm]
\item[\ref{rq:wsdm}] Can we leverage collective intelligence to construct entity representations for increased retrieval effectiveness of entities of interest?
\end{description}

\noindent
We have shown an effective way of leveraging collective intelligence to construct entity representations for increased retrieval effectiveness of entities of interest. 

To answer \ref{rq:wsdm} we formulated and answered (in Section~\ref{sec:wsdmresults}) three subquestions.
First, we aim to answer:

\begin{description}[leftmargin=!,labelwidth=1cm]
\item[\ref{rq:wsdm1}] Does entity ranking effectiveness increase by using dynamic collective entity representations?
\end{description}

\noindent 
We demonstrate that incorporating dynamic description sources into dynamic collective entity representations enables a better matching of users' queries to entities, resulting in an increase of entity ranking effectiveness. 
More specifically, we show how each individual description source contributes to produce a more effective ranking (see Figure~\ref{fig:Qablation}).
Looking at the contribution of individual description sources, we see how the highest relative improvements come from dynamic description sources, and social tags prove to be the strongest individual signal. 
In addition, we study how the ranker adapts the individual weights of fields (reflecting the relative importances of expansion sources) in Section~\ref{subsec:featureweights}, and see how dynamic description sources gain higher importances as new descriptions come in. 

Next, we answer subquestion \ref{rq:wsdm2}: 

\begin{description}[leftmargin=!,labelwidth=1cm]
\item[\ref{rq:wsdm2}] Does entity ranking effectiveness increase when employing additional field and entity importance features?
\end{description}

\noindent
We show that informing the ranker on the expansion state of the entities, i.e., employing field and entity importance features, further increases the ranking effectiveness. 

Finally, we answer:

\begin{description}[leftmargin=!,labelwidth=1cm]
\item[\ref{rq:wsdm3}] Does entity ranking effectiveness increase when we continuously learn the optimal way to combine the content from different description sources?
\end{description}

\noindent
We compare an adaptive ranker (i.e., one that we periodically retrain, as descriptions come in) to a static one. 
The results show how retraining the ranker leads to improved ranking effectiveness in dynamic collective entity representations. 
We also show how even the static ranker improves over time, suggesting that even the static ranker benefits from newly incoming descriptions, if not as much as the adaptive method. 

The findings in this work have several implications in the discovery context. 
Namely, we have shown that we can effectively leverage the feedback of searchers (i.e., their queries and clicks) to improve retrieval effectiveness, which may be beneficial in the discovery process. 
Furthermore, we have shown that we can effectively combine heterogeneous information related to an entity of interest into a single representation, which suggests our method can be useful in a scenario where data is available from many different sources, a scenario not uncommon in the context of E-Discovery. 
Consider, e.g., the scenario where emails and attachments in various different formats needs to be considered, a scenario that is not uncommon in E-Discovery~\cite{INR-025}. 
By combining the content that is explicitly linked (e.g., attachments to email addresses), we have shown that regardless of the nature of the data, our ranking method can effectively incorporate the additional data for improved retrieval effectiveness. 

The work presented in this chapter is not without limitations, however. 
Our study of the impact of dynamic collective entity representations was performed in a controlled scenario, where we collect and leverage different data sources. 
One restriction of working with these freely available resources is that it proves hard to find aligned and sizeable datasets. 
In particular, the temporal misalignment between different corpora prevents the analysis of temporal patterns that may span across sources (e.g., queries and tweets showing similar activity around entities when news events unfold). 
In part, these restrictions can be circumvented in future work, e.g., increasing the (comparatively) low number of tweets by enriching them through e.g., entity linking methods~\cite{meij2012}, where entities identified in tweets could be expanded. 
Additional challenges and opportunities may arise when increasing the scale of the data collections. Opportunities may lie in exploiting session or user information for more effective use of user interaction signals. 
Challenges include so-called ``swamping'' \cite{robertson2004} or ``document vector saturation'' \cite{kemp2002}, i.e., entity drift that is more prone to happen when the size of the data collections increase. 

A natural extension to this work is to study this problem in a realistic scenario, with aligned datasets of sufficient size, which would allow us to study the impact of ``swamping'' entities in more detail, as well as provide a more fine-grained analysis of the impact of the importance features.
But more importantly, it would allow to study the link between the temporal patterns of entities (as studied in Chapter~\ref{ch:plos}) and their retrieval effectiveness. 
Novel features that take into account the novelty in fields (e.g., the number of terms that were not in the original entity description), and the diversity could prove useful in adapting for entity drift. 
Furthermore, whilst we consider the changing entity representation now by its impact on retrieval effectiveness, another follow-up may investigate the changes in entity representations on a qualitative level, e.g., by studying per field and/or entity the type of additional descriptions (similar to original vs. novel), the rate of changing representations, etc. 
 \clearpage{}

\part{\acl{pt:2}}
\label{pt:2}

\clearpage{}
\chapter{\acl{ch:sigir}}
\label{ch:sigir}

\begin{flushright}
\rightskip=1.8cm ``I speak to everyone in the same way, \\ whether he is the garbage man or the president of the university.'' \\
\vspace{.2em}
\rightskip=.8cm\textit{---Albert Einstein}
\end{flushright}
\vspace{1em}

\section{Introduction}
\label{sec:intro}
In this first chapter of this part of the thesis, we present our first case study into predicting activities of entities of interest from digital traces.
More specifically, we propose a method for predicting the recipient of an email, by leveraging both the contexts in which the digital traces are created, as their content. 

Knowing the likely recipients of an email can be helpful from a user-perspective (e.g., for pro-actively suggesting recipient), but also in the context of discovery. 
Reliably predicting communication flows may find applications in discovering \emph{entities of interest}, e.g., by comparing actual communication flows to ``expected'' (i.e., predicted) flows, one can identify communication that is ``unexpected,'' which can be valuable signal in the context of digital forensics~\cite{disco2009}. 

Despite the huge increase in the use of social media channels and online collaboration platforms, email remains one of the most popular ways of (online) communication. 
Email traffic has been increasing steadily over the past few years, and recent market studies have projected its continued growth for the years to come~\cite{radicati-email-stats-2016-2020}. 
Recipient recommendation methods aim at providing the sender of an email with the appropriate predicted recipients of the email that is currently being written. 
These methods furthermore allow us to gain a better understanding of communication patterns in enterprises, potentially revealing underlying structures of communication networks. 

In this chapter, we focus on recipient prediction in an enterprise setting, as it allows us to leverage the full content and structure of the communication network, as opposed to strictly local (ego network) approaches, that are restricted to only the sender's direct network. 
Furthermore, enterprise email databases are common objects of study in the E-Discovery scenario, as can be seen by the TREC Legal track which aims to closely resemble a real-world E-Discovery task~\cite{treclegal2009,treclegal2010}. 
\citet{Henseler:2010:NFL:1991676.1991681} likewise studied email in an E-Discovery setting. 
He found that combining keyword search with communication pattern analysis reduces the amount of information reviewers need to process. 

We propose a novel hybrid generative model, which incorporates both the communication graph structure and email content signals for recipient prediction. 
Our model predicts recipients from scratch, i.e., without assuming seed recipients for prediction.
Furthermore, our model can quickly deal with updates in both the communication graph and the profiles of recipients when new emails are sent around. 

Our main research question revolves around the contribution of \emph{context} of digital sources (i.e., social context or organizational structure) and \emph{content} of digital sources (i.e., the actual email content) towards effectiveness of predicting communication. 
More specifically, this chapter aims to answer \ref{rq:sigir}:

\begin{description}
\item[\ref{rq:sigir}] Can we predict email communication through modeling email content and communication graph properties?
\end{description}

\noindent
On top of that we investigate how we can optimally estimate the various components of our model. 
We train our recipient prediction model on one email collection (the Enron email collection) and test it on another (the Avocado email collection). 

Our main finding is that a combination of the communication graph and email content outperforms the individual components. 
We obtain optimal performance when we incorporate the number of emails a recipient has received so far and the number of emails a given sender sent to a recipient at that point in time in our model. 
Other options, like using PageRank as a recipient prior, do not improve performance. 
Our findings show that both the \emph{context} and \emph{content} of digital sources contribute to prediction effectiveness.

\section{Communication Graph}
\label{sec:graph}
To model the \emph{context} in which communication happens, we construct a communication graph from all emails sent by users in our email collections. 
This graph implicitly captures the (social or professional) relations between users in an email network, and hence models an aspect of the context of communication. 

We consider the email traffic as a directed graph $G$, consisting of a set of vertices and arcs (directed edges) $\langle V,A \rangle$. 
Each vertex $v \in V$ represents a distinct email address in the corpus (i.e., a sender $S$ or recipient $R$ in terms of our modeling in Section~\ref{sec:model}), and arcs $a \in A$ that connect them represent the communication between the two corresponding addresses (i.e., emails exchanged), directed from the sender node to the recipient(s) node(s). 
The arcs are weighted by the number of emails sent from one user to the other. 

The communication graph allows us to model the network and interactions by considering several well-established graph-based metrics, that represent contextual signals such as the proximity between emailers (i.e., connection strength), and larger structures within the network of individuals. 
One example is to measure a user's relative importance in the communication graph through her \emph{PageRank}-score. 
The PageRank algorithm measures a user's (node's) relative importance through its connected arcs and their corresponding weights. 
In this model, each arc is a ``vote of support,'' and thus users with a larger number of interactions receive a higher PageRank score~\cite{ilprints422}. 

\noindent
We update the communication graph after each email that is being sent, i.e., each email either adds a new arc to the graph, or updates the weight of an existing one. 
We describe the utility of our communication graph for recipient prediction in Section~\ref{sec:model}.

\section{Modeling}
\label{sec:model}

We propose a generative model that is aimed at calculating the probabilities of recipients given the sender and the content of the email. 
Instead of predicting a single recipient, we cast the task as a ranking problem in which we try to rank the appropriate recipients as high as possible. 

More formally, let $R$ be a candidate recipient, $S$ the sender of an email, and $E$ the email itself. 
Our final ranking will be based on the probability of observing $R$ given $S$ and $E$: $P(R \mid S,E)$. 
We use Bayes' Theorem to rewrite this probability:
\begin{align}
 P(R \mid S,E) = \frac{ P(R) \cdot P(S \mid R) \cdot P(E \mid R,S) }{ P(S) \cdot P(E \mid S) }.
 \label{eq:prse}
\end{align}
\noindent
We can explain equation~\ref{eq:prse} as follows: 
the ``relevance'' of a recipient is determined by 
(i)~her prior probability (how likely is this person to receive email in general, $P(R)$ in equation~\ref{eq:prse}), 
(ii)~the probability of observing the sender with this particular recipient ($P(S \mid R)$ in equation~\ref{eq:prse}), and 
(iii)~the likelihood of this email to be generated from communication between the recipient and the sender ($P(E \mid R,S)$ in equation~\ref{eq:prse}). 
To obtain the final probability, we normalize using the prior probability of the sender, and the likelihood of observing this email given its sender ($P(S) \cdot P(E \mid S)$ in equation~\ref{eq:prse}). 

For ranking purposes we can ignore $P(S)$ and $P(E \mid S)$, which will be the same for all recipients. 
Our final ranking function is displayed in equation~\ref{eq:prse-final}. 
\begin{align}
 P(R \mid S,E) \propto P(E \mid R,S) \cdot P(S \mid R) \cdot P(R).
 \label{eq:prse-final}
\end{align}
\noindent
In the next three sections we explain how we estimate the three components of the model: the email likelihood ($P(E \mid R,S)$), the sender likelihood ($P(S \mid R)$), and the recipient prior ($P(R)$).

\subsection{Email Likelihood}
\begin{figure}[t]
 \centering
 \includegraphics[width=\linewidth]{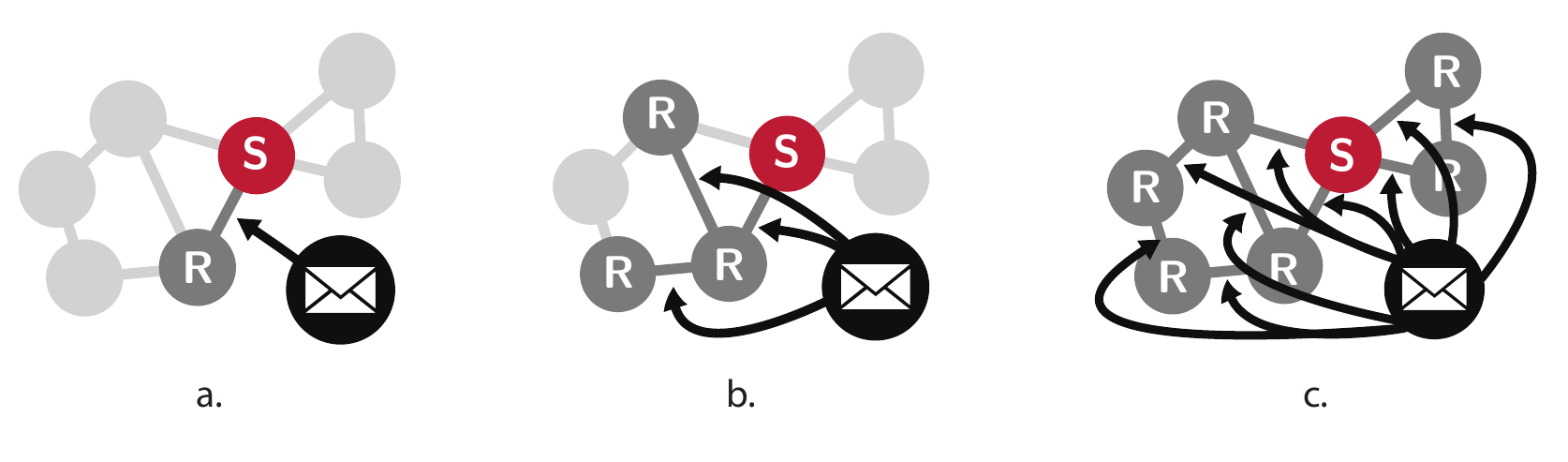}
 \caption{The three unigram LMs which we model from email communication. Here, the red node represents the email sender ($S$), the dark grey node the (candidate) recipient node(s) ($R$), and the light grey nodes the remaining nodes in graph $G$.
 Figure shows which emails are used to estimate $P(E \mid R,S)$ (a.), $P(w \mid R)$ (b.), and $P(w)$ (c.) }
 \label{fig:LMs}
\end{figure}

We have several options when it comes to estimating $P(E \mid R,S)$. 
We could, for example, incorporate individual emails as latent variables. 
However, in this chapter we opt to directly estimate the email likelihood using the terms in the email (viz.\ equation~\ref{eq:pers}).
\begin{align}
 P(E \mid R,S) = \prod_{w \in E} \left[\lambda P(w \mid R,S) + \gamma P(w \mid R) + \beta P(w)\right].
 \label{eq:pers}
\end{align}
\noindent
In this estimation, $P(w \mid R,S)$ indicates the probability of observing a term $w$ in all emails exchanged between $S$ and $R$. 
To prevent zero probabilities, we smooth this probability with the term probability in all emails sent and received by $R$ (i.e., $P(w \mid R)$ in equation~\ref{eq:pers}) and the term probability over the whole collection (i.e., $P(w)$ in equation~\ref{eq:pers}). 
In other words, we apply unigram language modeling for all the emails between $S$ and $R$, which we smooth by the unigram LM of $R$ (i.e., both incoming and outgoing emails), and the unigram LM of the entire background corpus (i.e., all emails in the collection up to that point in time).
See also Figure~\ref{fig:LMs}. 

We introduce three parameters, $\lambda$, $\beta$, and $\gamma$ that represent the relative importance of each of the three components of equation~\ref{eq:pers}, with $\lambda + \gamma + \beta = 1$, to combine the three term probabilities. 

We calculate each of the three term probabilities using the maximum likelihood estimate, i.e., 
$P(w \mid \cdot) = \frac{n(w,\cdot)}{|\cdot|}$, 
the frequency of term $w$ in the set of documents divided by the length of this set in number of terms.

\subsection{Sender Likelihood}
We move to the estimation of $P(S \mid R)$, the likelihood of observing the sender for a given recipient. 
Here, we use the communication graph constructed from email exchanges. 
The closer a recipient is in this graph and the stronger her connection to the sender $S$, the more likely it is that the two ``belong together.'' 
We estimate this connection strength in two ways: by considering 
(i)~the \emph{frequency} ($\mathit{freq}$), or the number of emails $S$ sent to $R$ at that point in time, and 
(ii)~\emph{co-occurrence} ($\mathit{co}$), or the number of times $S$ and $R$ co-occur as addressees in an email. 
More specifically, the frequency-based probability is defined as:
\begin{align}
 P_{\mathit{freq}}(S \mid R) = \frac{ n(e, S \rightarrow R) }{ \sum_{S' \in \mathcal{S}} n(e, S' \rightarrow R) },
 \label{eq:psrfreq}
\end{align}
\noindent
where $n(e, x \rightarrow R)$ indicates the number of emails sent from $x$ to $R$ and $\mathcal{S}$ is the set of all senders in the graph at the current point in time. 
The co-occurrence-based probability is defined as:
\begin{align}
P_{\mathit{co}}(S \mid R) = \frac{n(e, \rightarrow R,S)}{n(e, \rightarrow R) + n(e, \rightarrow S)},
\label{eq:psrco}
\end{align}
\noindent 
where $n(e, \rightarrow R,S)$ corresponds to the number of emails sent to \emph{both} $R$ and $S$ (i.e., number of emails of which $S$ and $R$ are both recipients), and $n(e, \rightarrow X)$ the number of emails sent to $X$. 

\subsection{Recipient Likelihood}
Finally, we introduce a recipient prior, that is, the email independent probability that a recipient will be observed. This probability is unrelated to both the email at hand ($E$) and the sender of that email ($S$) and can be estimated without knowing these two variables. 
Again, we can choose from a variety of ways to estimate this prior probability, but we stick to two obvious choices. 
First, we use the \emph{number of emails received} by $R$, normalized by the total number of emails sent at that point in time ($\mathit{rec}$). 
This estimation indicates how likely it is that any given email would be sent to this recipient. 
Second, we calculate recipient $R$'s \emph{PageRank} score ($\mathit{pr}$) as an indication of how important $R$ is in the communication graph.

\section{Experimental Setup}
\label{sec:exp-setup}
To study the contribution of the context and content at which digital traces are created, we evaluate our recipient recommendation model using a realistic experimental setup. 
Our experiments aim at demonstrating the prediction effectiveness of the individual components of our model, i.e., the communication graph (CG) component and the email content component (EC), and their combination (CG+EC) as in equation~\ref{eq:pers}. 
We use real enterprise email databases to evaluate our model. 
We optimize our models on the Enron email collection~\cite{klimt2004enron}, and test it on the Avocado collection, which consists of email boxes of employees of an IT firm that developed products for the mobile Internet market~\cite{avocado}. 
The two collections are described in Table~\ref{tab:emailsummaries}.

For both collections we follow the same method to select the set of users for evaluation. 
We first split users into three groups based on their email activity: high activity, medium activity, and low activity. 
This way we can study the correlation between a user's level of activity and the model's performance. 
As email networks typically show a long-tailed distribution~\cite{McCallum:2007:TRD:1622637.1622644}, with a small number of users responsible for a large volume of the sent mails, and a large number of users responsible for a small volume, we define a user's activity by taking the logarithm of the number of sent emails. 
We prune users that have less then 100 sent mails, and compute the resulting distribution's mean ($\mu$) and standard deviation ($\sigma$) and split the distribution into three bins: 
first, 
(i)~users with \emph{low} activity (LOW) are those that fall below $\mu-\frac{1}{2}\sigma$, 
(ii)~\emph{medium} active users (MED) are those between $\mu-\frac{1}{2}\sigma$ and $\mu+\frac{1}{2}\sigma$, 
and finally, (iii)~\emph{highly} active users (HIGH) are the ones over $\mu+\frac{1}{2}\sigma$. 
From each bin we randomly sample 50 users, which results in our final evaluation set of 150 users.

\begin{table}[t]
  \caption{Summary of Enron and Avocado email collections. We list
  the time span in months (Period), total number of Emails, total number of employee addresses (Addr.), the average number of emails
  sent ($\overline{S/p}$) and received ($\overline{R/p}$) per
  address.}
  \begin{tabularx}{\linewidth}{Xccccc}
    \toprule
    & Period & Emails & Addr. & $\overline{S/p}$ & $\overline{R/p}$ \\
    \midrule
        Enron   & {45}    & {252,424}    & {6,145}    & {\phantom{0}34}              & {294} \\
        Avocado & {58}    & {607,011}    & {2,068}    & {174}              & {321}  \\
    \bottomrule
  \end{tabularx}
  \label{tab:emailsummaries}
\end{table}

\subsection{Evaluation}
\begin{figure}[t]
 \centering
 \includegraphics[width=\linewidth]{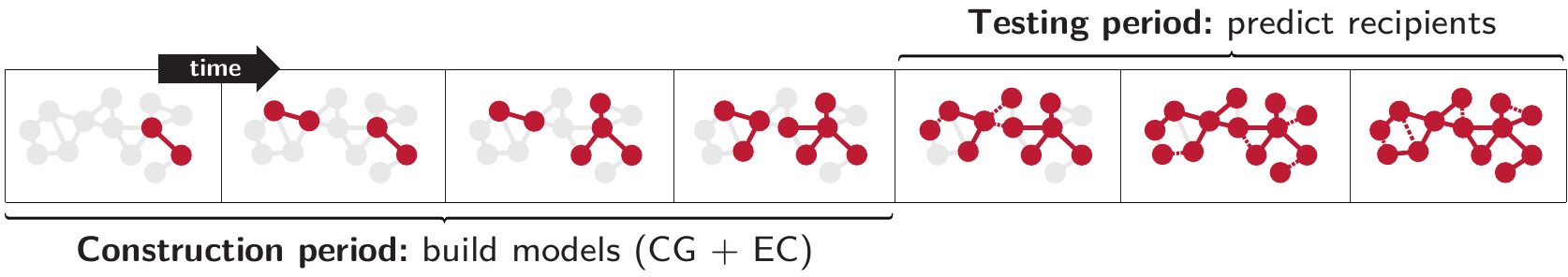}
 \caption{Our evaluation methodology, where we have an initial \emph{construction period} to generate language models and build the communication graph. During the subsequent \emph{testing period} we select emails for which predict recipients (represented by the dotted lines). We continue building the communication graph and updating the language models in the testing period.}
 \label{fig:exp-setup-time}
\end{figure}

Before we start predicting email recipients we allow our model to gather evidence from all email communication up to that point. 
More specifically, we use an initial \emph{construction period} to generate the users' language models and the communication graph. 
We start to predict recipients in the subsequent \emph{testing period}. 
During both periods our model is updated for each sent email. 
We split each user's period of activity (starting from the user's first sent email, up to the last sent mail) into the \emph{construction period}, covering $\frac{2}{3}$ of the emails, and the \emph{testing period}, which is $\frac{1}{3}$ of the emails. 
See also Figure~\ref{fig:exp-setup-time}). 

For each sender in our user evaluation set, we select 10 emails, evenly distributed over the testing period as evaluation points. 
For each of these emails we rank the top 10 recipients and compare to the actual recipients of the email. 
We report on mean average precision (MAP), as it allows us to identify improvements in the ranking of recipients.
We indicate the best performing model in boldface and test statistical significance using a two-tailed paired $t$-test. Significant differences are marked $\blacktriangle$/$\blacktriangledown$ for $\alpha$ = 0.01 and $\vartriangle$/$\triangledown$ for $\alpha$ = 0.05.

\subsection{Parameter Tuning}
We use the Enron collection to tune the parameters $\lambda$, $\gamma$, and $\beta$ in equation~\ref{eq:pers}, and to determine which methods for estimating the sender ($P(S \mid R)$) and recipient ($P(R)$) likelihood work best. 
The results of the parameter tuning are displayed in Tables \ref{tab:snresults} and \ref{tab:paramtuning}.
The final settings we use for testing our model on the Avocado collection are the following:
$\lambda = 0.6$, 
$\gamma = 0.2$, 
$\beta = 0.2$, 
we estimate $P(S \mid R)$ using the number of emails $S$ sent to $R$, $P_{\mathit{freq}}(S \mid R)$, and we estimate $P(R)$ using the number of emails received by $R$, $P_{\mathit{rec}}(R)$. 

\begin{table}[t]
  \centering
  \caption{System performance (MAP) on the Enron dataset over
  different methods for estimating $P(R)$ and $P(S \mid R)$ (Section~\ref{sec:model}).}
  \begin{tabularx}{\linewidth}{XXXXX}
  \toprule
   & \small{$P_{\mathit{pr}}(R)$, $P_{co}(S \mid R)$} & \small{$P_{\mathit{pr}}(R)$, $P_{\mathit{freq}}(S \mid R)$} & \small{$P_{\mathit{rec}}(R)$, $P_{\mathit{co}}(S \mid R)$} & \small{$P_{\mathit{rec}}(R)$, $P_{\mathit{freq}}(S \mid R)$} \\
  \midrule
	\small{LOW} & 0.2207	&	0.1488	&	0.2317	&	\textbf{0.4365} \\
	\small{MED} & 0.1961	&	0.2334	&	0.2116	&	\textbf{0.3857} \\
	\small{HIGH} & 0.1016	&	0.1213	&	0.1169	&	\textbf{0.2060} \\
  \midrule
	\small{ALL} & 0.1755	&	0.1676	&	0.1893	&	\textbf{0.3480} \\
  \bottomrule
  \end{tabularx}
  \label{tab:snresults}
\end{table}

\begin{table}[t]
    \centering
    \caption{System performance (MAP) on the Enron dataset over a
    parameter sweep for the parameters $\lambda$, $\gamma$, and $\beta = 1 - (\lambda+\gamma)$ in
    equation~\ref{eq:pers} with a step size of 0.2. }
    \begin{tabular}{cccc}
        \toprule
        \large{$\nicefrac{\lambda\downarrow}{\gamma\rightarrow}$}             &  0.2     & 0.4       & 0.6        \\
        \midrule
        0.2          & 0.4670   & 0.4699    & 0.4752     \\
        0.4          & 0.5070   & 0.5095    &           \\
        0.6          & \textbf{0.5258}   &          &   \\
        \bottomrule
    \end{tabular}
    \label{tab:paramtuning}
\end{table}

\section{Results and Analysis}
We compare the found optimal settings for the CG and EC components to their combination using our training collection (Enron). 
Table~\ref{tab:enronresults} shows that our hybrid model significantly outperforms either of the single models across all groups of users in the Enron collection, even if the performance increase is modest. 
The highest performance is achieved in the LOW and MED user groups: lower user activity correlates positively with performance. 

\begin{table}[t]
  \centering
  \caption{System performance (MAP) on Enron and Avocado. We test for statistically significant differences of the individual models (CG and EC) against the combined model (CG+EG). }
  \begin{tabular}{@{}l@{~~~} c@{~~}c@{~~}c c@{~~}c@{~~}c@{}}
  \toprule
  & \multicolumn{3}{c}{Enron} & \multicolumn{3}{c}{Avocado} \\
  \cmidrule(r){2-4}\cmidrule{5-7}
  & CG & EC & CG+EC & CG & EC & CG+EC \\
  \midrule
  LOW & 0.4365\downdown & 0.5757\downdown & \textbf{0.5833} & 0.6502\downdown	&	0.6946\downdown	&	\textbf{0.7077} \\
  MED & 0.3857\downdown & 0.5161\downdown & \textbf{0.5325} & 0.6052\down	&	0.6328\downdown	&	\textbf{0.6542} \\
  HIGH & 0.2060\downdown & 0.4779\down & \textbf{0.4853} & \textbf{0.6652}\up	&	0.5739\downdown	&	0.6136\\
  \midrule
  ALL & 0.3480\downdown & 0.5258\downdown & \textbf{0.5362} & 0.6402\phantom{\down}	&	0.6352\downdown	&	\textbf{0.6597} \\
  \bottomrule
  \end{tabular}
  \label{tab:enronresults}
  \label{tab:avoresults}
\end{table}

We present the results of our final experiments on the Avocado set in Table~\ref{tab:avoresults}. 
Compared to the results on the Enron set, which we used for tuning our parameters, our model's performance is higher throughout on the Avocado set, both across the different models, and within each subgroup of users. 
An indication for this difference in absolute performance scores comes from the collection statistics in Table~\ref{tab:emailsummaries}. 
Here we see that the Avocado set contains fewer unique addressees, spans a longer period of time, and contains a larger number of emails per person. 
As a possible factor contributing to our models' higher performance on the Avocado collection, we point to the fact that our model has more data available to leverage for ranking a smaller number of candidates.
While different datasets may need different models, the consistently high scores show that the components work in isolation and in combination over different datasets.

\subsection{Between Groups Comparison}
Similar to what we saw in the experiment on the Enron collection, higher user activity generally seems to result in lower performance on the Avocado dataset (Table~\ref{tab:avoresults}).
The CG model is an exception and outperforms our combined model for highly active users. 
The combined model achieves significant performance improvements over the content model in each subset of users. 

To better understand these patterns, we turn to the characteristics of the different user subgroups. 
We plot the users' numbers of emails (both sent and received), indicating their activity, and juxtapose it to the size of their \emph{egonet}, which corresponds to the set of directly connected neighbors in the communication graph~\cite{Akoglu:2010:OSA:2144032.2144081}. 
This \emph{egonet} represents the users they interact(ed) with, and is indicative of a user's reach or embedding inside the communication graph. 

\begin{figure}[t]
 \centering
 \includegraphics[width=\linewidth]{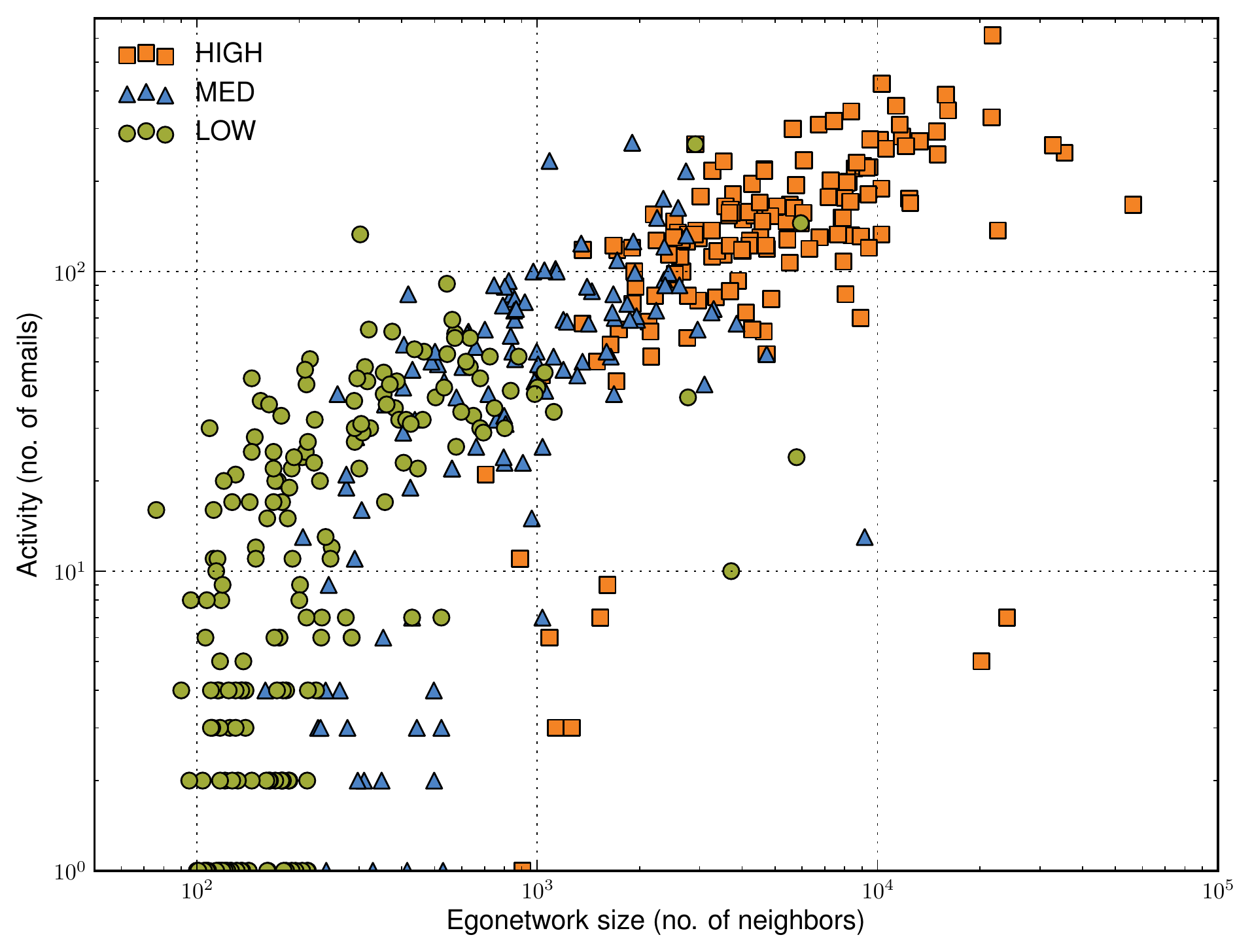}
 \caption{The three user groups in Avocado, showing each user's activity (y-axis) against the size of its egonetwork (x-axis).}
 \label{fig:egonet}
\end{figure}

The resulting plot is shown in Figure~\ref{fig:egonet}. 
There is a clear clustering: highly active users (orange squares) who send and receive a large number of emails, also have a larger number of people they interact with. 
While more textual content allows the generative model to create richer recipient profiles, in turn enabling more informed recipient ranking, there is a catch to a larger egonet too.
The sender-recipient communication smoothing in our generative model results in a larger number of high-scoring candidates for highly active users.
This makes it more difficult for the ranker to discern the true recipient(s) from the larger pool.

\subsection{Performance over Time}
Motivated by the fact that our model is updated at each sent email, we study its performance over time. 
To get an indication of whether our model's performance improves or deteriorates over time, we apply linear regression on the data points of each model, and fit a trend line. 
See Table~\ref{tab:slopes} for the trend line slopes for each individual component, and the combined model. 

\begin{table}[t]
  \centering
  \caption{Trend line slopes after applying linear regression to the individual and combined components data.}
  \begin{tabular}{lr}
  \toprule
  \textbf{Model} & \textbf{Slope} \\
  \midrule
  Email Content (EC) & $1.36\cdot 10^{-4}$ \\
  Communication Graph (CG) & $-1.38\cdot 10^{-4}$ \\
  Combined (EC+CG) & $1.27\cdot 10^{-4}$ \\
  \bottomrule
  \end{tabular}
  \label{tab:slopes}
\end{table}

Both the EC and combined model's trend lines have positive slopes, whilst the CG model has a negative slope. 
This indicates that our language modeling approach benefits from the larger amount of textual content it receives for each recipient over time, allowing the generation of richer recipient profiles for better email likelihood estimations. 
The CG approach on the other hand, deteriorates over time, suffering from the increased size and complexity of the communication graph. 
We note that, as time progresses, the communication graph model ``settles in,'' and becomes less likely to pick up on changing balances or shifting communication patterns in the communication graph. 
For future work we argue for a time-aware model, that is able to adapt to shifts in the communication graph over time, by taking recency into account.

\subsection{Rankers' Correlation}
\begin{figure}[t]
 \centering
 \includegraphics[width=\linewidth]{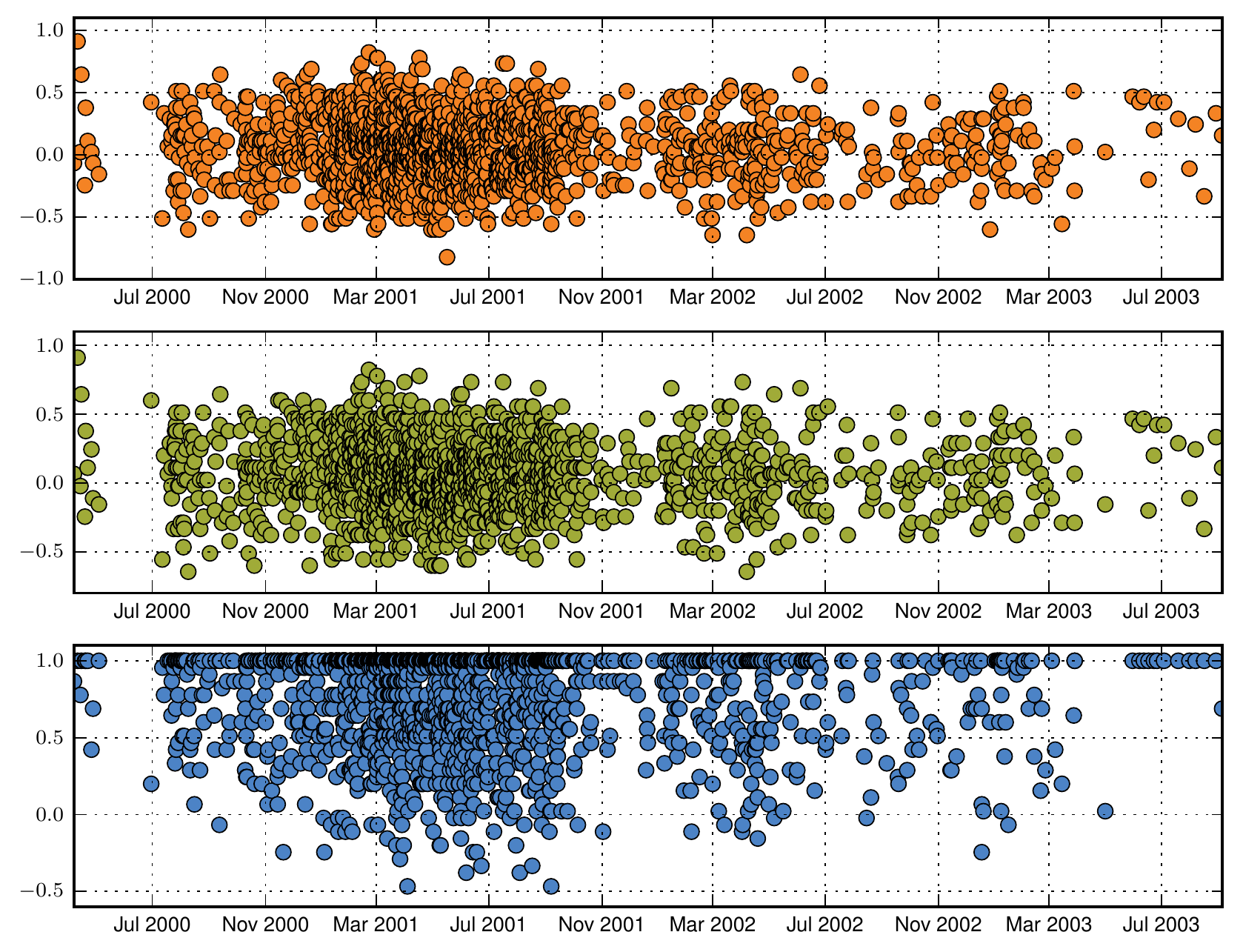}
 \caption{Kendall's $\tau$ over time, between CG and EC (orange), CG and CG+EC (green), and EC and CG+EC (blue).}
 \label{fig:correlation}
\end{figure}

To analyze the performance of our combined model in comparison to the isolated ones, we look at their rankings and compute the Kendall tau rank correlation coefficient ($\tau$) between pairs of models.
The top plot in Figure~\ref{fig:correlation} shows how agreement between CG and EC is relatively low, centering around the 0 mark with an average of 0.0471. 
This pattern largely coincides with that of the second plot, which depicts agreement between CG and our combined model. 
The average correlation coefficient is only slightly higher at 0.0562.
Finally, the agreement is highest between the EC model and our combined model, at on average 0.7425. 
The high agreement offers an explanation for the comparatively low performance of our combined model in the HIGH subset of the Avocado set. 
The EC model's comparatively low performance (as seen in Table~\ref{tab:avoresults}), indicates that the combined model is negatively affected by following EC's incorrect rankings. 

\section{Conclusion}

In this chapter, we presented a novel hybrid model for email recipient prediction that leverages both the information from an email network's communication graph and the textual content of the messages. 
Our model starts from scratch, in that it does not assume or need seed recipients, and it is updated for each email sent. 
The proposed model achieves high performance on two email collections. 
We have answered \ref{rq:sigir}:

\begin{description}
\item[\ref{rq:sigir}] Can we predict email communication through modeling email content and communication graph properties?
\end{description}

\noindent
We show that the communication graph properties and the email content signals both provide a strong baseline for predicting recipients, but are complementary, i.e., combining both signals achieves the highest accuracy for predicting the recipients of email. 

We have shown that the number of received emails is an effective method for estimating the prior probability of observing a recipient, and the number of emails sent between two users is an effective way of estimating the ``connectedness'' between these two users, and proves to be a helpful signal in ranking recipients. 

An implication of the findings in this chapter is that both the \emph{context} in which digital traces are created, and the \emph{content} of the digital traces are valuable signals in predicting real-world behavior, in the case of email communication.
This has implications for both designing systems to predict recipients, as it does for analyzing and understanding communication flows in an enterprise, e.g., in a discovery scenario where one may want to discover ``unexpected'' communication. 

The work presented in this chapter has several limitations, first, from a modeling perspective;
We identified characteristic weaknesses of our individual models' robustness to specific circumstances. 
As witnessed by the decrease in performance for highly active users, our email content model seems unfit to deal with users that exchange a large number of emails with a large number of people. 
The deteriorating performance over time shows that our communication graph models' performance suffers with expanding graphs as they develop. 

To address these issues, one possible direction for future work is to incorporate time into our model. 
We can, for example, use decay functions to weigh the edges between users and promote more recent communication. 
Similarly, we can use time-dependent language models that favor recent documents to incorporate time. 
Whereas in this work we choose to keep a simple, and hence interpretable model, a machine learned method may improve prediction accuracy, e.g., by using the signals we employed as features. 
Finally, for future work, the presented model can be applied to novel tasks, e.g., role or anomaly detection in email corpora, or leak prevention. 

In the next chapter, we take a similar approach to the work presented in this chapter. More specifically, we analyze a novel dataset of interaction logs, and address a prediction task. 
Much like the recipient recommendation task in this chapter, we study the contribution of the context and the content of the digital traces, with the goal of predicting real-world activity of our entities of interest. 
\clearpage{}

\clearpage{}
\chapter{\acl{ch:umap}}
\label{ch:umap}

\begin{flushright}
\rightskip=1.8cm``The advantage of a bad memory is that one enjoys \\ several times the same good things for the first time.'' \\
\vspace{.2em}
\rightskip=.8cm---\textit{Friedrich Nietzsche}
\end{flushright}
\vspace{1em}

\noindent
Our second case-study into the link between digital traces and real-world behavior or activities of entities of interest, revolves around mobile devices. 
Mobile devices are an especially interesting source of digital traces, as the closeness, portability, and popularity of our smartphones means they are part of most of the aspects of our daily lives. 
Mobile device hence take a central role as potential ``containers for evidence'' in the E-Discovery scenario~\cite{Casey:2000:DEC:555668}. 
Smartphones may carry signals of our whereabouts, context, and activities in many different forms, e.g., from emails, (voice) messages, call and SMS logs, to browser data, and other data logs~\cite{Barmpatsalou2013323}. 

A relatively new source of digital traces that can provide rich insights into users' activities and behavior are automated personal assistants. 
These intelligent assistants have been introduced in major online service offerings and mobile devices over the last several years. 
Systems such as Siri, Google Now, Echo, M, and Cortana support a range of reactive and proactive scenarios, ranging from question answering to alerting about plane flights and traffic. 
The embedding of these personal assistants in our day-to-day life makes them a rich potential resource of ``digital evidence,'' i.e., digital traces that can be employed for finding evidence of activity in the real world~\cite{INR-025}. 

Several of these personal assistants provide reminder services aimed at helping people to remember plans for future tasks they may otherwise forget. 
Potentially representing the user's (future) location, activity, or planned tasks, interaction logs with these types of reminder services could serve as a valuable signal in an E-Discovery scenario. 
In this chapter, we study a large-scale log of user-created reminders from Microsoft Cortana, to better understand the context under which digital traces are produced, and to study whether we can predict the (planned) behavior of the \emph{entities of interest} of Part~\ref{pt:2}: the producers of digital traces. 

Motivated by the findings in the previous chapter, in this chapter too we study the impact of the \emph{contexts} in which the traces are created. 
Here, we consider the creation time of a reminder as the context. 
In addition, we study the predictive power of the textual \emph{content} of the digital traces, by incorporating the reminder text for predicting the reminder task's execution time. 

\begin{table}[h]
  \caption{Example interaction sequence for setting a reminder.}
  \centering
  \label{tab:examplereminder}
  \begin{tabular}{lll}
    \toprule
    \textbf{Turn} & \textbf{Who} & \textbf{Text} \\
    \midrule
    1 & User & Remind me to do the laundry. \\
    2 & System & When would you like to be reminded? \\
    3 & User & Sunday at noon. \\
    4 & System & Alright, remind you to do the laundry at 12:00PM on Sunday, is \\ & & that right? \\
    5 & User & Yes. \\
    6 & System & Great, I'll remind you! {success chime} \\
    \bottomrule
  \end{tabular}
  \label{tab:summaries}
\end{table}

\noindent
Table~\ref{tab:examplereminder} presents an example of the types of reminder dialogs recorded in the dataset. 
These logs offer insights into the reminder generation process, including the types of tasks for which people formulate reminders, the task descriptions, the times that reminders of different types are created, and the periods of time between the creation of reminders and notifications. 
Beyond analysis of the nature and timing of reminders, we demonstrate how information about patterns of reminder usage and general trends seen across users can be harnessed to predict when users are likely to execute certain tasks, which can be of interest in an E-Discovery scenario. 
We focus primarily on time-based reminders, i.e., reminders for tasks planned for a future time. 
In this chapter, we aim to answer \ref{rq:umap}: 
\emph{``Can we identify patterns in the times at which people create reminders, and, via notification times, when the associated tasks are to be executed?''}

\section{Reminder Types}
\label{sec:remindertypes}
We first investigate user behavior around reminder creation by studying the common tasks linked to setting reminders. 
In this section, we focus on the question: 
\emph{``Is there a body of common tasks that underlie the reminder creation process?''} 
Understanding the common tasks allows us to better see common usage patterns, which will help in predictive models and in finding unexpected behavior, e.g., through outlier detection. 
To answer this question, we perform a data-driven qualitative analysis of Cortana reminder logs. 
Specifically, we extract reminders that are observed frequently and across multiple users. 
Then, we employ a manual labeling strategy, and categorize the reminders in a task taxonomy to better understand the task types that drive users to create reminders. 

\subsection{Reminder Composition}
Before we analyze the different tasks at the root at reminder setting, we describe the composition of a typical reminder. In the left column of Table~\ref{tab:reminderpredicates}, we present three examples of common reminders. The examples show a structure that is frequently observed in the logged reminders. Reminders are typically composed as predicate sentences. They contain a phrase related to an action that the user would like to perform (typically a verb phrase) and a referenced object that is the target of the action to be performed. 

\begin{table}[ht]
  \centering
  \caption{Example reminders as predicates.}
  \label{tab:reminderpredicates}
  \begin{tabular}{ll}
    \toprule
	\textbf{Reminder} & \textbf{Predicate} \\
    \midrule
	``Remind me to take out the trash'' & Take out (me, the trash) \\
	``Remind me to put my clothes in dryer'' & Put (me, clothes in dryer) \\
	``Remind me to get cash from the bank'' & Get (me, cash from the bank) \\
    \bottomrule
  \end{tabular}
\end{table}

\subsection{Data}
\label{subsec:data}
A session for setting a reminder consists of a dialog, where the user and the intelligent assistant interact in multiple turns. 
Typically, the user starts by issuing the command for setting a reminder, and dictates the reminder. 
Optionally, the user specifies the reminder's notification time. 
Next, the assistant requests to specify the time (if the user has not yet specified it), or provides a summary of the reminder, i.e., the task description and notification time, asking the user to confirm or change the proposed reminder (see Table~\ref{tab:examplereminder}). 

We analyze a sample of two months of Cortana reminder logs, spanning all of January and February 2015. 
We ensure the data used in this chapter preserves user privacy, as we filter the data to patterns commonly observed across multiple users, and study behavior in aggregate, refraining from studying individual patterns. 
In summary, we are not looking at any user's behavior at the individual level, but across a large population, to uncover broad and more general patterns.

We pre-process this set of reminders by including only reminders from the US market (the only market which had Cortana enabled on mobile devices at that time). 
To narrow the scope of our analysis, we focus on time-based reminders and remove location (e.g., \emph{``remind me to do X when I am at Y''}) and person-based reminders (e.g., \emph{``remind me to give X when I see Y''}), which are less common and more challenging to study across users due to their personal nature. 
Finally, we retain only reminders that are confirmed by the user (turn 6 in Table~\ref{tab:examplereminder}). 
The resulting sample contains 576,080 reminders from 92,264 users. 
For each reminder, we extract the reminder task description and notification time from Cortana's summary (turn 4 in Table~\ref{tab:examplereminder}). 
We also extract the creation time based on the local time of the user's device. 
Each reminder is represented by: 

\begin{itemize}
\item[{\bf $r_{task}$}] The textual task description of the reminder, i.e., the phrase which encodes the future task or action to be taken, as dictated by the user. We extract the text from Cortana's final summary response (``do the laundry'' from turn 4 in Table~\ref{tab:examplereminder}).
\item[{\bf $r_{CT}$}] The creation time of the reminder. This represents the time at which the user encodes the reminder. We represent $r_{CT}$ as a discretized time-value; Section~\ref{subsec:method} defines the discretization process. We extract this timestamp from the client's device.
\item[{\bf $r_{NT}$}] The notification time set for the reminder to fire an alert. 
This data represents the time at which the user wishes to be reminded about a future task or action. 
We represent $r_{NT}$ in the same discretized manner as $r_{CT}$. We extract the notification time from Cortana's summary response (turn 4 in Table~\ref{tab:examplereminder}). 
\item[{\bf $r_{\Delta T}$}] Subtracting the creation time from the notification time yields the time-delta, the delay between the reminder's creation and notification time. 
Intuitively, reminders with smaller time-deltas represent short-term or immediate tasks (\emph{``remind me to take the pizza out of the oven''}), whereas reminders with larger time-deltas represent tasks planned further ahead in time (\emph{``remind me to make a doctor's appointment''}).
\end{itemize}

\subsection{Identifying Common Tasks}
To understand the common needs that underlie the reminder creation process, and hence common user behavior, we first identify common reminders, i.e., reminders that are frequently observed across multiple users. 
Studying common reminders can aid analysts in understanding common usage patterns, and hence can help in identifying patterns or reminders that diverge. 
In addition, common usage patterns can aid system designers in creating predictive models. 
We employ a mixed methods approach, comprising data-driven and qualitative methodologies, to extract and identify common task types.

\subsubsection{Frequent Task Description Extraction} 
First, we extract common task descriptions, by leveraging the predicate (verb+object) structure described at the start of this section. 
To ensure the underlying task descriptions represent broad tasks, we filter to retain only descriptions that start with a verb (or a multi-word phrasal verb) that occurs at least 500 times, across at least ten users, with at least five objects. 
This yields a set of 52 frequent verbs, which covers 60.9\% of the reminders in our sample. The relatively small number of verbs that cover the majority of reminders in our log indicates there likely exists a large `head' of common tasks that give rise to reminder creation. 
To analyze the underlying tasks, we include the most common objects, by pruning objects observed less than five times with a verb. 
This yields a set of 2,484 unique task descriptions (i.e., verb+object pairs), covering 17.9\% of our sample log. 

\subsubsection{Manual Labeling} 
Next, we aim to identify the tasks which underlie the frequent task descriptions, and categorize them into a task type taxonomy. 
Specifically, by manual inspection, we identified several key dimensions that separate tasks. 
In particular, we find that dimensions that commonly separate tasks are (i) the cognitive load the task incurs (i.e., how much does the task represent an interruption of the user's activity), (ii) the context in which the task is to be executed (i.e., at home, at work), and (iii) the (expected) duration of the task. 
This enabled us to categorize the extracted frequent task descriptions into one of six broader task types, with several subclasses, which we present in the following section with example reminders. 

\clearpage
\section{Task Type Taxonomy}

\begin{figure}
\centering
\includegraphics[width=\linewidth]{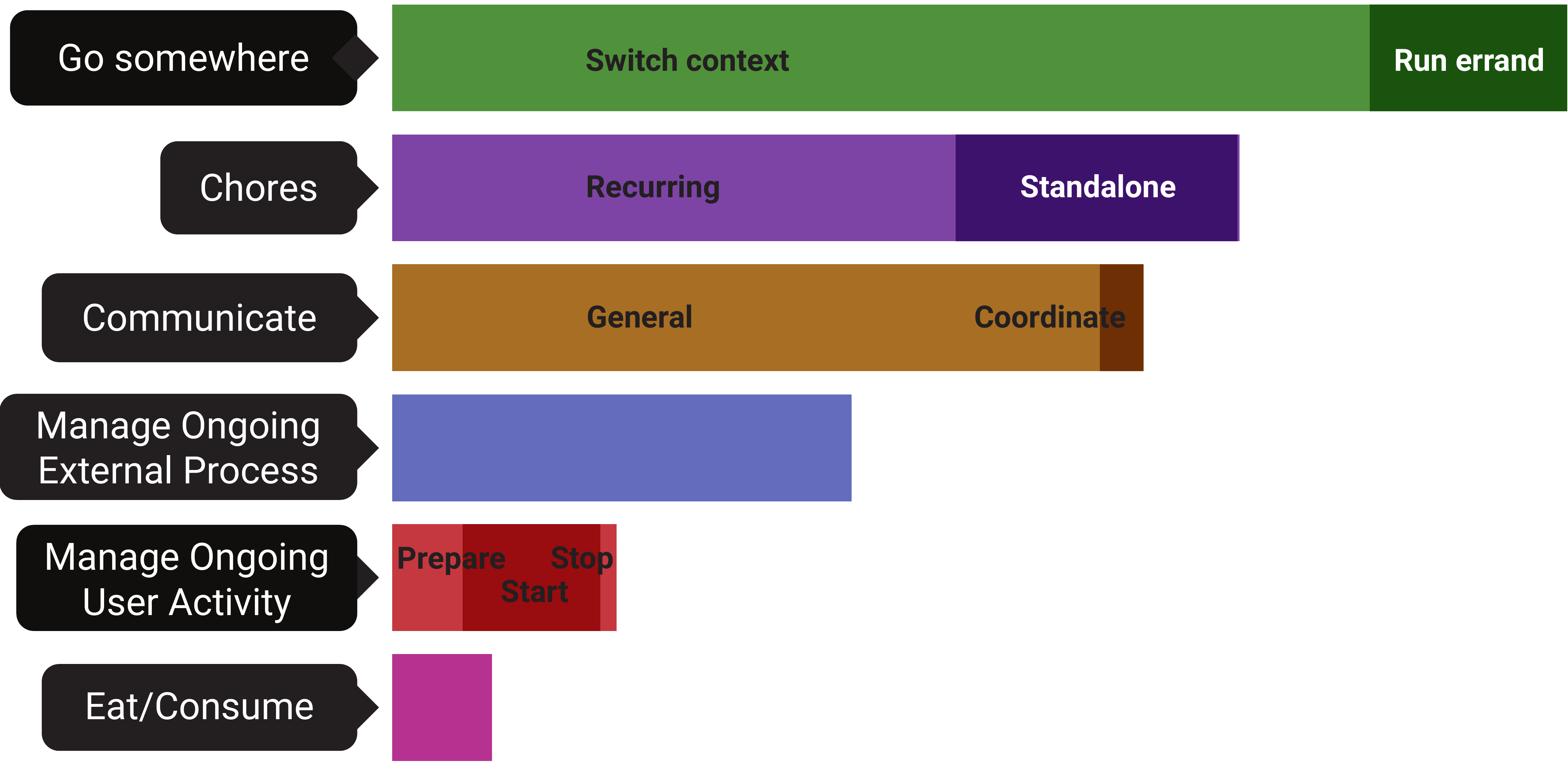}
\caption{Bar plot showing the proposed reminder task-type taxonomy.}
\label{fig:tasktypeoverview}
\end{figure}

In this section, we describe each of the six task types in turn, and provide examples of the associated verb+object patterns. 
First, for an overview of the task types and their relative occurrence in our data, see Figure~\ref{fig:tasktypeoverview}.
The example objects are shown in decreasing order of frequency, starting with the most common. 
Note that verbs are not uniquely associated with a single task type, but the verb+object-pair may determine the task type (compare, e.g., \emph{``set alarm''} to \emph{``set doctor appointment''}).

\clearpage
\subsubsection{1. Go Somewhere (33.0\%)} 
One third of the frequent tasks refer to the user moving from one place to another. 
See Table~\ref{tab:gosomewhere}.
We distinguish between two subtypes: the first subtype is running an errand (83.2\%), where the reminder refers to executing a task at some location (e.g., ``pick up milk''). 
Running an errand represents an interruption of the user's activity, but a task of a relatively small scale, i.e., it represents a task that briefly takes up the user's availability. 
The second subtype is more comprehensive, and represents tasks that are characterized by a switch of context (16.8\%), e.g., moving from one context or activity to another (``go to work'', ``leave for office''), which has a larger impact on the user's availability.

\begin{table}[h]
    \caption{Go Somewhere.}
    \label{tab:gosomewhere}
    \centering
  \begin{tabular}{ll}
    \toprule
    \textbf{Run errand} &  \\
    \midrule
    grab [something] &  laundry, lunch, headphones  \\
    get [something] &  batteries  \\
    pick up [something/someone] &  laundry, person, pizza  \\
    buy [something] &  milk, flowers, coffee, pizza  \\
    bring [something] &  laptop, lunch, phone charger  \\
    drop off [something] &  car, dry cleaning, prescription  \\
    return [something] &  library books  \\
    \midrule
    \textbf{Switch context} &  \\
    \midrule
    leave (for) [some place] &  house, work, airport  \\
    come [somewhere] &  home, back to work, in  \\
    be [somewhere] &  be at work, at home  \\
    go (to) [somewhere] &  gym, work, home, appointment,   \\
    stop by [some place] &  the bank, at Walmart  \\
    have (to) [something] &  work, appointment  \\
    \bottomrule
  \end{tabular}
\end{table}

\clearpage

\subsubsection{2. Chores (23.8\%)} 
The second most common type of reminders represent daily chores. 
See Table~\ref{tab:chores}. 
We distinguish two subtypes: recurring (66.5\%) and standalone chores (33.5\%). 
Both types represent smaller-scale tasks that briefly interrupt the user's activity.

\begin{table}[h]
  \caption{Chores.}
  \label{tab:chores}
  \centering
  \begin{tabular}{ll}
	\toprule
	\textbf{Recurring} & \\
	\midrule
	take out [something] & trash, bins \\
	feed [something] & dogs, meter, cats, baby \\
	clean [something] & room, house, bathroom \\
	wash [something] & clothes, hair, dishes, car \\
	charge [something] & phone, fitbit, batteries\\
	do [something] & laundry, homework, taxes, yoga \\
	pay [something] & pay rent, bills, phone bill \\
	set [something] & alarm, reminder \\
	\midrule
	\textbf{Standalone} & \\
	\midrule
	write [something] & a check, letter, thank you note \\
	change [something] & laundry, oil, air filter \\
	cancel [something] & amazon prime, netflix \\
	order [something] & pizza, flowers \\
	renew [something] & books, driver's license, passport \\
	book [something] & hotel, flight \\
	mail [something] & letter, package, check \\
	submit [something] & timesheet, timecard, expenses \\
	fill out [something] & application, timesheet, form \\
	print [something] & tickets, paper, boarding pass \\
	pack [something] & lunch, gym clothes, clothes \\
    \bottomrule
  \end{tabular}
\end{table}

\clearpage

\subsubsection{3. Communicate (21.1\%)} 
Next, a common task is to remind to contact (\emph{``call,'' ``phone,'' ``text''}) another individual, either a person (e.g., \emph{``mom,'' ``jack,'' ``jane''}), organization/company (\emph{``AT\&T''}), or other (\emph{``hair dresser,'' ``doctor's office''}). 
See Table~\ref{tab:communicate}. 
We identify two subtypes: the majority reflects general, unspecified communication (94.7\%) (e.g., \emph{``call mom''}), and a smaller part (5.3\%) represents coordination or planning tasks (e.g., \emph{``make doctor's appointment''}). 
Both subtypes represent tasks that briefly interrupt the user's activity.

\begin{table}[h]
  \caption{Communicate}
  \label{tab:communicate}
  \centering
  \begin{tabularx}{.75\linewidth}{ll}
	\toprule
	\textbf{General} & \\
	\midrule
	send [something] & email, text, report \\
	email [someone] & dad, mom \\
	text [someone] & mom, dad \\
	call [someone] & mom, dad \\
	tell [someone] [something] & my wife I love her, happy \\
	&  birthday mom \\
	\midrule
	\textbf{Coordinate} & \\
	\midrule
	set [an appointment] & doctors appointment \\
	make [an appointment] & doctors appointment, reservation \\
	schedule [an appointment] & haircut, doctors appointment \\
    \bottomrule
  \end{tabularx}
\end{table}

\subsubsection{4. Manage Ongoing External Process (12.9\%)} 
These reminders represent manipulation of an ongoing, external process, i.e., tasks where the user monitors or interacts with something, e.g., the laundry or oven. 
See Table~\ref{tab:externalprocess}.
These tasks briefly interrupt a user's activity.

\begin{table}[h]
  \centering
  \caption{Manage Ongoing External Process}
  \label{tab:externalprocess}
  \begin{tabularx}{.75\linewidth}{ll}
	\toprule
	turn [on/off] [something] & water, oven, stove, heater \\
	check [something] & email, oven, laundry, food \\
	start [something] & dishwasher, laundry \\
	put [something] in [something] & pizza in oven, clothes in dryer \\
	take [something] out & pizza, chicken, laundry \\
    \bottomrule
  \end{tabularx}
\end{table}

\clearpage

\subsubsection{5. Manage Ongoing User Activity (6.3\%)} 
This class of reminders is similar to the previous class, however, as opposed to the user interacting with an external process, these reflect a change in the user's own activity. 
They incur a higher cost on the user's availability and cognitive load. 
See Table~\ref{tab:useractivity}. 
We distinguish three subtypes: preparing (31.4\%), starting (61.4\%), and stopping an activity (7.2\%). 

\begin{table}[h]
  \caption{Manage Ongoing User Activity}
  \label{tab:useractivity}
  \centering
  \begin{tabularx}{.7\linewidth}{ll}
	\toprule
	\textbf{Activity/Prepare} & \\
	\midrule
	get ready [to/for] & work, home \\
	\midrule
	\textbf{Activity/Start} & \\
	\midrule
	start [some activity] & dinner, cooking, studying \\
	make [something] & food, breakfast, grocery list \\
	take [something] & a shower, a break \\
	play [something] & game, xbox, basketball \\
	watch [something] & tv, the walking dead, seahawks game \\
	\midrule
	\textbf{Activity/Stop} & \\
	\midrule
	stop [some activity] & reading, playing \\
	finish [something] & homework, laundry, taxes \\
    \bottomrule
  \end{tabularx}
\end{table}

\subsubsection{6. Eat/Consume (2.8\%)} 
Another frequent reminder type refers to consuming something, most typically food (\emph{``have lunch''}) or medicine (\emph{``take medicine''}). 
These are also small and range from brief interruptions (\emph{``take pills''}) to longer interruptions (\emph{``have dinner''}).
See Table~\ref{tab:eatconsume}. 

\begin{table}[h]
  \centering
  \caption{Eat/Consume}
  \label{tab:eatconsume}
  \begin{tabularx}{.7\linewidth}{ll}
	\toprule
	take [something] & Medicine \\
	eat [something] & lunch, dinner, breakfast, pizza \\
	have [something] & lunch, a snack, breakfast \\
    \bottomrule
  \end{tabularx}
\end{table}

\clearpage

\section{Reminder Patterns}
\label{sec:reminderpatterns}
Next, we study the temporal patterns of reminders. 
We seek to understand when people create reminders, when reminders are set to notify users, and the average delay between creation and notification time for different reminders. 
Such knowledge could prove useful in providing likelihoods about when certain tasks tend to happen or predicting (follow-up) tasks. 
In this section, we focus on the research question: 
\emph{``Can we identify patterns in the times at which people create reminders, and, via notification times, when the associated tasks are to be executed?''}

We study reminders on several levels of granularity. 
In Section~\ref{subsec:globalpatterns} we look at global patterns and trends across all reminders. 
Next, we study temporal patterns per task type in Section~\ref{subsec:tasktype}. 
In Section~\ref{subsec:terms}, we perform a temporal analysis of task description terms. 
Finally, we study the relation between reminder creation and notification times in Section~\ref{subsec:time}. 
First, we explain how we represent the reminder's creation and notification time to enable our analyses.

\subsection{Method}
\label{subsec:method}
To study common temporal patterns of reminders, we discretize time by dividing each day into the following six four-hour buckets: 
(i) late night [00:00-04:00), 
(ii) early morning [04:00-08:00), 
(iii) morning [08:00-12:00), 
(iv) afternoon [12:00-16:00), 
(v) evening [16:00-20:00), and 
(vi) night [20:00-00:00). 
By combining this time-of-day division with the days of week we yield a 7 by 6 matrix $M$, whose columns represent days, and rows times. 
Each $r_{CT}$ and $r_{NT}$ can be represented as a cell in matrix $M$, i.e., $M_{i,j}$ where $i$ corresponds to the day of week, and $j$ to the time of day. 
Furthermore, we distinguish between $M^{CT}$ and $M^{NT}$, the matrices whose cells contain reminders that are created, or respectively set to notify, at a particular day and time. 
We represent each reminder as an object, with the attributes described in Section~\ref{subsec:data}: the reminder's task description ($r_{task}$), creation time ($r_{CT}$), notification time ($r_{NT}$), and time-delta ($r_{\Delta T}$). 
To study the temporal patterns, we look at the number of reminders that are created, or whose notifications are set, per cell. 
We compute conditional probabilities over the cells in $M^{CT}$ and $M^{NT}$, where the reminder? creation or notification time is conditioned on the task type, time, or the terms in the task description.
\begin{equation}
\label{eq:ctgivenw}
P(r_{CT} \mid w) = \frac{|\{r \in R : w \in r_{task} \wedge r_{CT} = X \}|}{|\{ w \in r_{task}, r \in M^{CT} \}|}
\end{equation}
\begin{equation}
\label{eq:ntgivenw}
P(r_{NT} \mid w) = \frac{|\{r \in R : w \in r_{task} \wedge r_{NT} = X \}|}{|\{ w \in r_{task}, r \in M^{NT} \}|}
\end{equation}
\noindent
To estimate the conditional probability of a notification or creation time, given a term from the task description, we take the set of reminders containing term $w$ that are created or whose notification is set at time $X$, over the total number of reminders that contain the word (see equation~\ref{eq:ctgivenw} and equation~\ref{eq:ntgivenw}). 
By computing this probability for each cell in $M^{NT}$ or $M^{CT}$, (i.e., $\Sigma_{i,j \in M}P(r_{NT} = i,j \mid w)$) we generate a probability distribution over matrix $M$. 

\begin{equation}
\label{eq:ntgivenct}
P(r_{NT} = X \mid r_{CT} = i,j) = \sum_{i,j \in M^{CT}} \frac{|\{ r \in M^{CT}_{i,j} : r_{NT} = X \}|}{| M^{CT}_{i,j} |}
\end{equation}
\noindent
To study the common patterns of the periods of time between the creation of reminders and notifications, we estimate a probability distribution for a reminder's notification time given a creation time (see equation~\ref{eq:ntgivenct}). 
We compute this probability by taking the reminders in each cell of $M^{CT}$ that have their notifications set to fire at time $X$, over all the reminders in that cell. 

Finally, we study the delays between setting and executing reminders, by collecting counts and plotting histograms of r$_{\Delta T}$ of reminders for a given subset, e.g., $\{ r_{CT} \in R : w \in r \}$ or $\{ r_{CT} \in R : r_{CT} = X \}$.

\subsection{Global Patterns}
\label{subsec:globalpatterns}
We now describe broad patterns of usage, and answer the following questions: 
\emph{``At which times during the day do people plan (i.e., create reminders), and at which times do they execute tasks (i.e., reminder notification trigger)?''} and 
\emph{``How far in advance do people plan tasks?''} 
To answer these questions, we examine the temporal patterns in our log data over the aggregate of all reminders in the two-month sample (576,080 reminders). 

\begin{figure}[t]
\centering
\includegraphics[width=\linewidth]{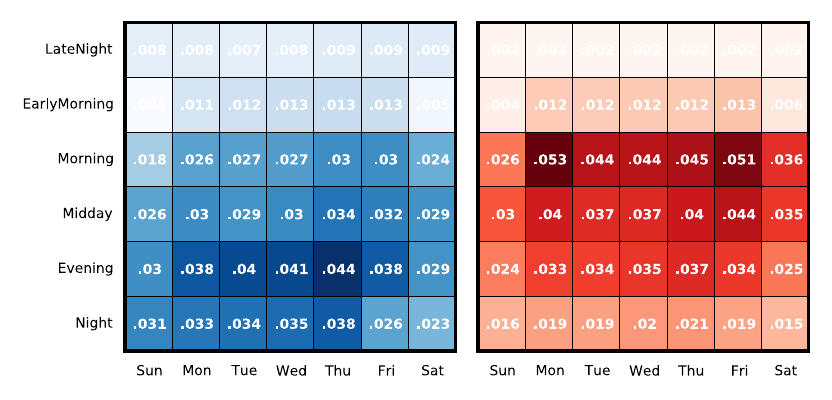}
\caption{Distribution of reminder creation times (left plot) and reminder notification times (right plot) for all reminders in two-month sample ($n$ = 576,080)}
\label{fig:all}
\end{figure}

Figure~\ref{fig:all} shows the prior probability of a reminder's creation time, $P(r_{CT})$, and notification time, $P(r_{NT})$, in each cell in $M^{CT}$ and $M^{NT}$. 
Looking at Figure~\ref{fig:all}, we see that in our sample, planning (reminder creation) most frequently happens later in the day, more so than during office hours (morning and midday). 
This observation could be explained by the user's availability; users may have more time to interact with their mobile devices in the evenings. 
Additionally, the end of the day is a natural moment for ``wrapping up the day,'' i.e., looking backward and forward to completed and future tasks. 

Turning our attention to notification times, the right plot of Figure~\ref{fig:all} shows a slightly different pattern: people execute tasks (i.e. notifications trigger) throughout the day, from morning to evening. 
This shows that users want to be reminded of tasks throughout the day, in different contexts (e.g., both at home and at work). 
This is reflected in our task-type taxonomy, where tasks are related to both contexts. 
We also note how slightly more notifications trigger on weekdays than in weekends, and more notifications trigger at the start and end of the workweek than midweek and in weekends. 
This observation may be attributed to the same phenomenon for reminder creation; users may tend to employ reminders for activities that switch between week and weekend contexts. 
Finally, comparing the two plots shows the notification times are slightly less uniformly distributed than creation times, e.g., users create reminders late at night, when it is relatively unlikely for notifications to fire.

\begin{figure}[t!]
\centering
\includegraphics[width=.9\linewidth]{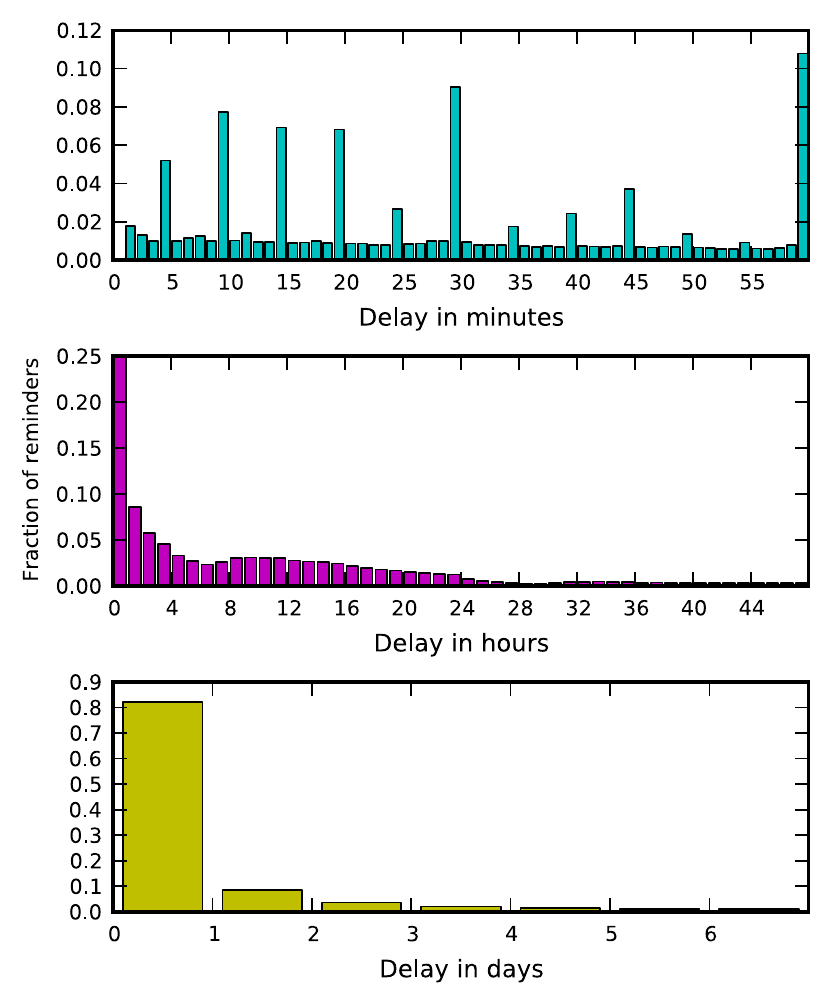}
\caption{Histograms of delays (in minutes, hours, and days from top to bottom) between reminder creation and notification times.}
\label{fig:alldelays}
\end{figure}

Next, to determine how far in advance users typically plan, we look at the delays between reminder creation and notification in Figure~\ref{fig:alldelays}. 
The top plot shows distinct spikes around five-minute intervals, which are due to reminders with a relative time indication (e.g., \emph{``remind me to take out the pizza in 5 minutes''}). 
These intervals are more likely to come to mind than more obscure time horizons (e.g., \emph{``remind me to take out the pizza at 6.34pm''}). 
The second and third plots clearly illustrate that the majority of reminders have a short delay: around 25\% of the reminders are set to notify within the same hour (second plot), and around 80\% of the reminders are set to notify within 24 hours (third plot). 
Interestingly, there is a small hump around 8--9 hours in the second plot, which may be explained by reminders that span a night, e.g., created at the end of the day, to notify early the next day), or a `working day' (reminder creation in the morning, notification at the end of the day). 

In summary, we have shown that on average people tend to set plans in the evening, and execute them throughout the day. 
Furthermore, most tasks that drive reminder setting are for short-term tasks to be executed in the next 24 hours.

\subsection{Temporal Patterns of Different Task Types}
\label{subsec:tasktype}
In this section, we explore whether different task types are characterized by distinct temporal patterns that differ from the global patterns seen in the previous section. 
To do so, we take the subset of the 2,484 unique frequent reminders in our two-month sample of reminders. 
We extract all reminders that match the exact reminder task descriptions from our full dataset, and yield a subset of 125,376 reminders with task type-labels, that we use for analysis. 
We aim to answer the same questions raised in the previous section, but at the level of task type, as opposed to a characterization of the global aggregate. 

\begin{figure}[t!]
\centering
\includegraphics[width=6.6cm,trim=0 4mm 37mm 0,clip]{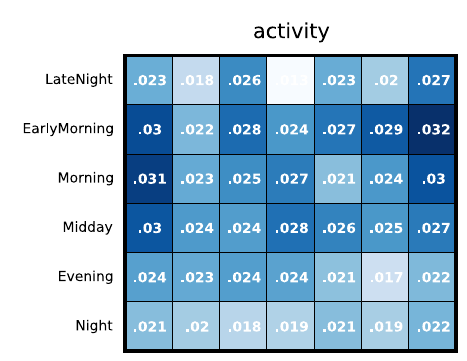}
\includegraphics[width=4.925cm,trim=12mm 4mm 37mm 0,clip]{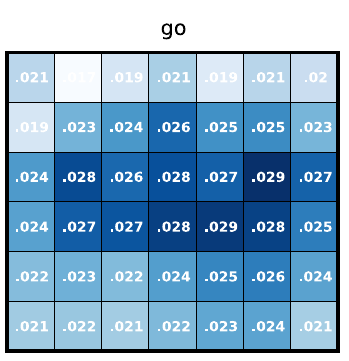}
\includegraphics[width=6.6cm,trim=0 0 37mm 2mm,clip]{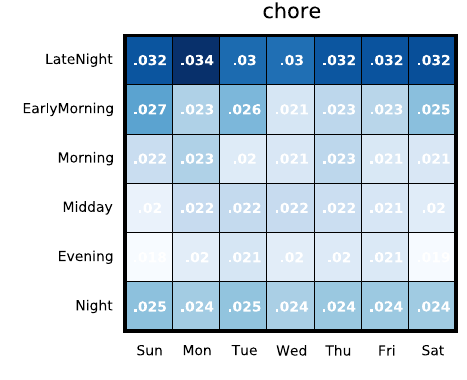}
\includegraphics[width=4.925cm,trim=12mm 0 37mm 2mm,clip]{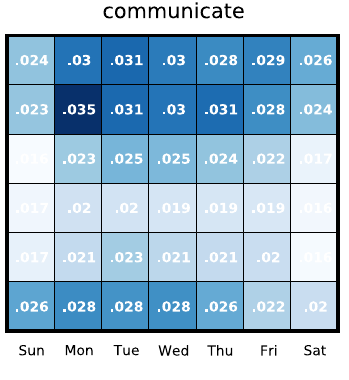}
\caption{Reminder creation time probability distributions over time, for different task types. }
\label{fig:tasktypescreation}
\end{figure}

\paragraph{Creation and notification times.} 
First, we look at the probability distribution of reminder creation times per task type, i.e., $P(r_{CT} \mid tasktype)$. 
Looking at the distributions for each task type, we discover two broader groups: per task type, reminders are either created mostly in morning and midday blocks (roughly corresponding to office hours), or outside these blocks. 
Figure~\ref{fig:tasktypescreation} shows examples of both types: ``Activity'' and ``Going somewhere'' reminders are mostly created during office hours, while e.g., ``Communicate'' and ``Chore'' reminders are more prone to be created in evenings. 
Another interesting observation is that activity-related reminders are comparatively frequent on weekends. 

\begin{figure}[t]
\centering
\includegraphics[height=5cm,trim=0 4mm 73mm 0,clip]{08-umap16/img/task_communicate}
\includegraphics[width=5cm,trim=48mm 4mm 0 0,clip]{08-umap16/img/task_communicate}
\includegraphics[width=5cm,trim=48mm 4mm 0 0,clip]{08-umap16/img/task_go}
\includegraphics[height=4.925cm,trim=0 0 73mm 4mm,clip]{08-umap16/img/task_communicate}
\includegraphics[width=4.925cm,trim=48mm 0 0 2mm,clip]{08-umap16/img/task_chore}
\includegraphics[width=4.925cm,trim=48mm 0 0 2mm,clip]{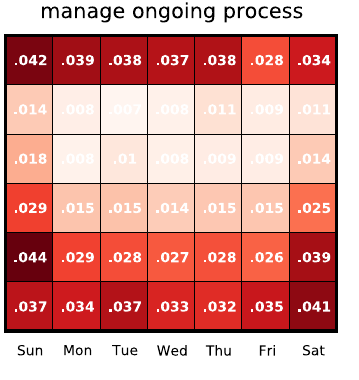}
\caption{Reminder notification time probability distributions over time, per task type.}
\label{fig:tasktypenotification}
\end{figure}

\noindent
Next, we study reminder notification times per task type, i.e., $P(r_{NT} \mid tasktype)$. 
Here, a similar pattern emerges. 
Broadly speaking, there are two types of tasks: those set to notify during office hours, and those that trigger outside these hours. 
See Figure~\ref{fig:tasktypenotification} for examples. 
``Communicate'' and ``Go'' fall under the former type, whereas ``Chore'' and ``Manage ongoing process'' fall under the latter. 
The nature of the tasks explains this distinction: the former relate to work-related tasks (communication, work-related errands), whilst the majority of the latter represent activities that are more common in a home setting (cooking, cleaning). 

\begin{figure}[t]
\centering
\includegraphics[height=5cm,trim=0 0 73mm 4mm,clip]{08-umap16/img/task_communicate}
\includegraphics[width=4.95cm,trim=48mm 0mm 0 0,clip]{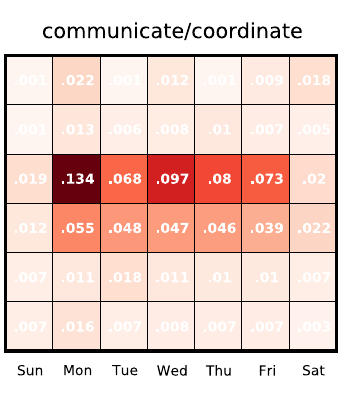}
\includegraphics[width=4.95cm,trim=48mm 0mm 0 0,clip]{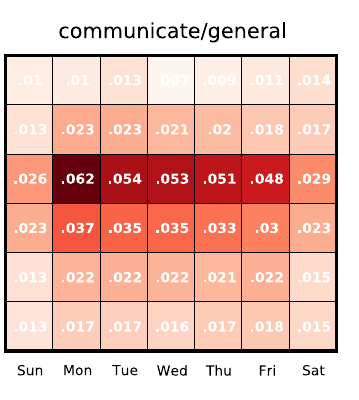}
\caption{``Communicate'' subclass notification time probability distributions over time.}
\label{fig:tasktypenotification-b}
\end{figure}

Taking a closer look at the Communicate task subclasses in Figure~\ref{fig:tasktypenotification-b}, we show how ``Communicate/General'' and ``Communicate/Coordinate'' differ: the former is more uniformly distributed, whilst the latter is denser around office hours. 
The general subtask too has comparatively more reminders that trigger in weekends, whereas coordinate is more strongly centered on weekdays. The distinct patterns suggest these subclasses indeed represent different types of tasks. 

\begin{figure}[t]
\centering
\includegraphics[width=.9\linewidth]{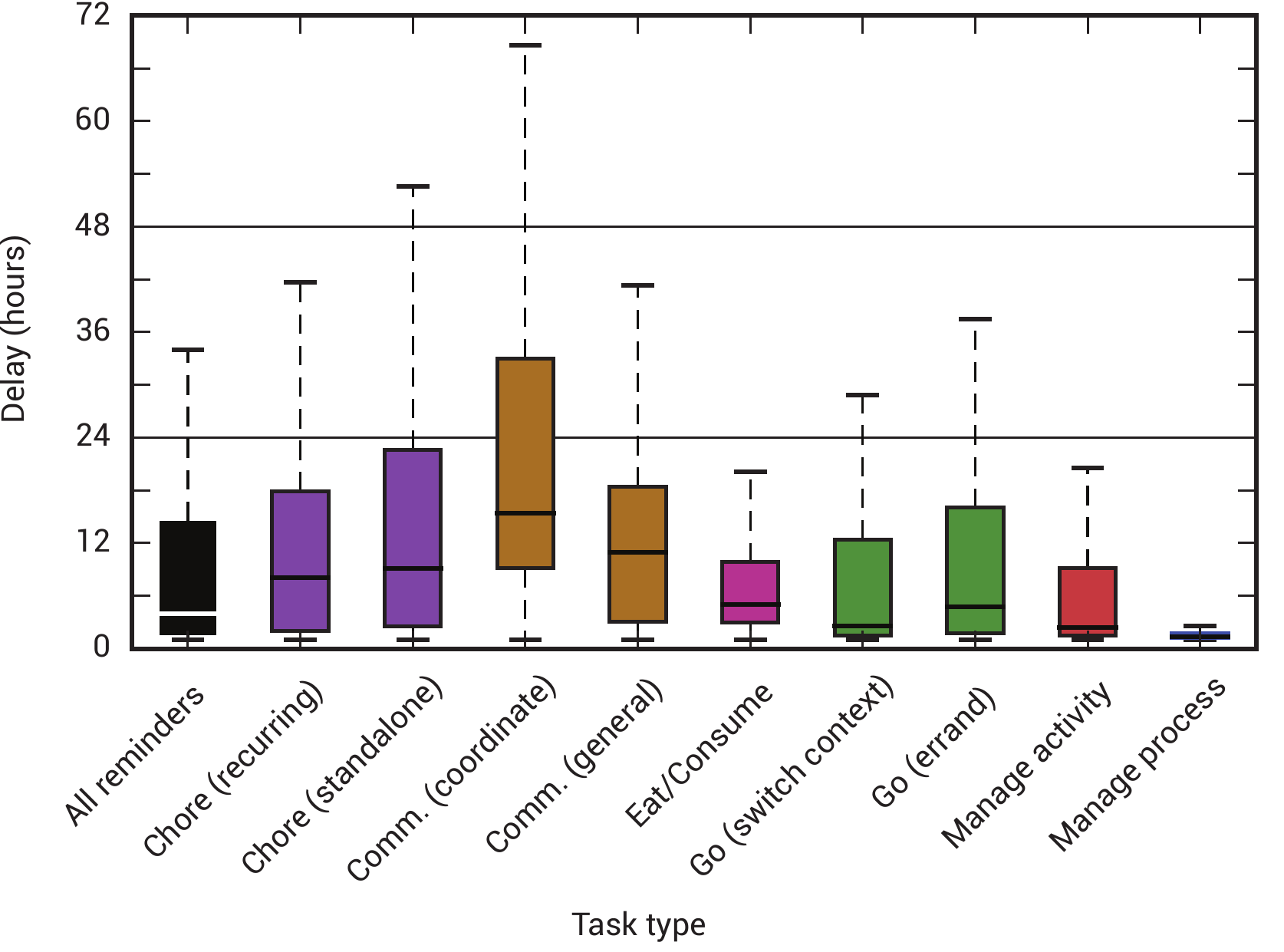}
\caption{Boxplot showing delay between reminder creation and notification times ($n$ = 125,376).}
\label{fig:tasktype_delay_boxplot}
\end{figure}

\paragraph{Reminder creation and notification delay.}
To better understand differences in the lead times between reminder creation and notification, we present an overview of the distribution of reminder delays per task type in Figure~\ref{fig:tasktype_delay_boxplot}. 
In general, the lower the boxplot lies on the $y$-axis, the lower the lead time, i.e., the shorter the delay between creating the reminder and executing the task. 
It is worth comparing, e.g., the plot of ``Manage ongoing process,'' to both ``Go'' or ``Communicate'' task types: 
execution of managing ongoing processes tasks seems to be planned with a much shorter lead time than the other types of task. 
Considering the nature of the tasks, where ongoing processes often represent the close monitoring or checking of a process (e.g., cooking or cleaning tasks), it is understandable that the delays are on the order of a few minutes, rather than hours. 
``Communicate/Coordinate'' has the largest delay on average, i.e., it is the task type people plan furthest in advance.

\begin{figure}[t]
\centering
\includegraphics[width=.9\linewidth,trim=0 41mm 0 35mm,clip]{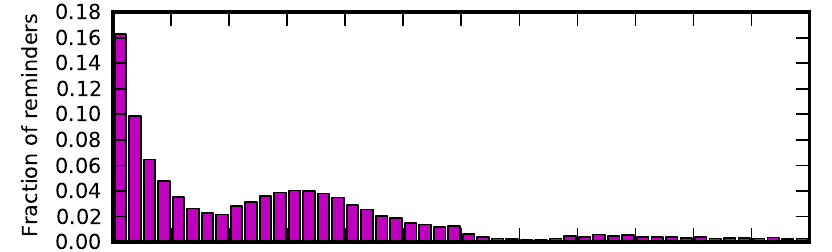}
\includegraphics[width=.9\linewidth,trim=0 34mm 0 35mm,clip]{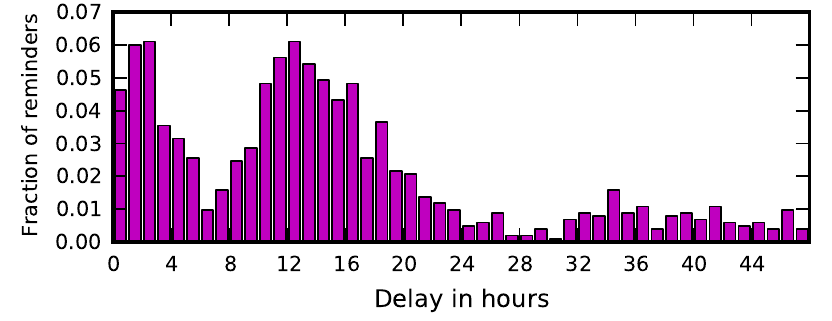}
\caption{Delay (lead time) between reminder creation and notification for ``Communicate'' subtasks. 
Showing ``Communicate/General'' in the top plot, ``Communicate/Coordinate'' in the bottom.}
\label{fig:tasktype_delay_comm}
\end{figure}

A more detailed examination of the differences between the ``Communicate'' subtasks, illustrated in Figure~\ref{fig:tasktype_delay_comm}, reveals that ``Communicate/General'' subtasks are more likely to be executed with lower lead time, as noted by the peak at hour 0 in the top plot. 
The ``Communicate/Coordinate'' subtask is about as likely to be executed the next day, as seen by the high peak around the 12 hour mark in the bottom plot. 
Much like the observations made in the previous section, the difference in the patterns between both ``Communicate'' subtasks suggests that the distinction between the subtypes is meaningful. 
Differences are not only found on a semantic level through our qualitative analysis, but also in temporal patterns.

In summary, we have shown how task type-specific temporal patterns differ from the aggregate patterns in Section~\ref{subsec:globalpatterns}. 

\subsection{Temporal Patterns of Terms in Reminder Task Descriptions}
\label{subsec:terms}

One can hypothesize that the terms in task descriptions show distinct temporal patterns, i.e., reminders that contain the term ``pizza'' are likely to trigger around dinner time. 
Presence of these temporal patterns may be leveraged for reminder creation or notification time prediction. 
To study this, we manually inspected the temporal distribution of task descriptions' terms of the 500 most frequent terms. 
More specifically, we compute conditional probabilities for a cell in $M^{CT}$ or $M^{NT}$ given a term $w$ (see equation~\ref{eq:ctgivenw} and Eq.~\ref{eq:ntgivenw}). 
We found several intuitive patterns, which we illustrate below with examples. 
These are simply presented to motivate the intuition behind the term features used in our predictive model in Section~\ref{sec:predicting}.

\begin{figure}[t]
\centering
\includegraphics[height=4.925cm,trim=0 4mm 73mm 0,clip]{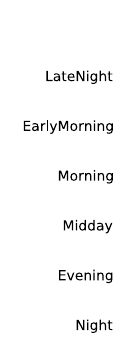}
\includegraphics[width=4.925cm,trim=12mm 4mm 37mm 0,clip]{08-umap16/img/term_church}
\includegraphics[width=4.925cm,trim=48mm 4mm 1mm 0,clip]{08-umap16/img/term_church}
\includegraphics[height=4.925cm,trim=0 2mm 73mm 0mm,clip]{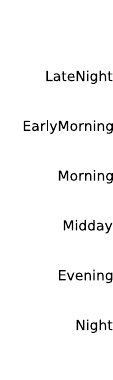}
\includegraphics[width=4.925cm,trim=12mm 0mm 37mm 0,clip]{08-umap16/img/term_appointment}
\includegraphics[width=4.925cm,trim=48mm 0mm 1mm 0,clip]{08-umap16/img/term_appointment}
\caption{Creation and notification times for reminders with the terms ``church'' (top row) and ``appointment'' (bottom row).}
\label{fig:terms}
\end{figure}

Figure~\ref{fig:terms} shows creation and notification times of task descriptions that contain the terms ``church'' or ``appointment.'' 
The ``appointment'' plot shows a strong pattern around the morning and midday blocks, representing office hours. 
Reminders that contain ``church'' show a clear and intuitive pattern too; they are largely created from Saturday night through Sunday morning, and are set to notify on Sunday early morning and mornings. When we examine the delays between reminder creation and notification, clear patterns emerge. 

\begin{figure}[t!]
\centering
\includegraphics[width=.515\linewidth,trim=0 0 0 0,clip]{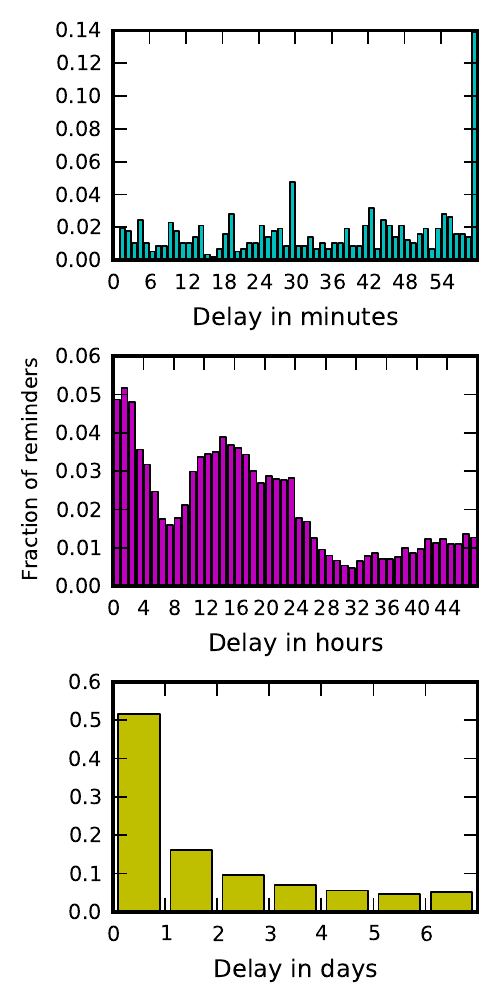}
\includegraphics[width=.425\linewidth,trim=9mm 0 0 0,clip]{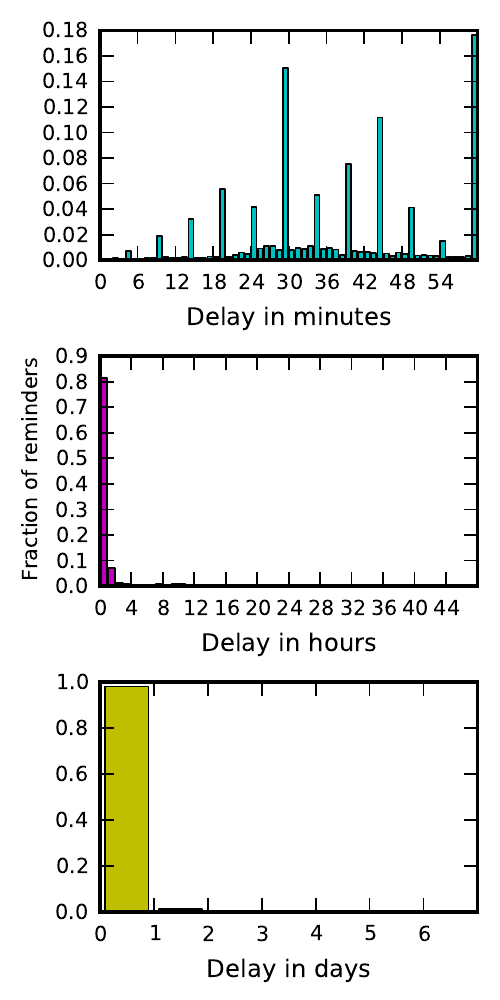}
\caption{Delays between reminder creation and notification for reminder task descriptions containing the terms ``appointment'' (left column) and ``laundry'' (right column).}
\label{fig:termdelays}
\end{figure}

In Figure~\ref{fig:termdelays}, we compare the average delays of reminders containing the term ``appointment'' to ``laundry.'' 
Clearly, on average, ``appointment'' reminders have longer delays, reflected by the nature of the task (which may involve other individuals and hence require more planning), whereas ``laundry'' reminders are more likely to reflect short-term tasks (which may be performed individually). 

In summary, we see distinct temporal patterns in task descriptions' terms. 
In Section~\ref{sec:predicting}, we study the generalizability of these patterns. 

\subsection{Temporal Patterns of Notification Times}
\label{subsec:time}
Finally, we look at correlations between reminders' creation and notification times. 
Motivated by the observation that most reminders are set to notify shortly after they are created, we study the probability of a reminder's notification time given its creation time, $P(r_{NT} \mid r_{CT})$. 

\begin{figure}[t!]
\centering
\includegraphics[width=\linewidth]{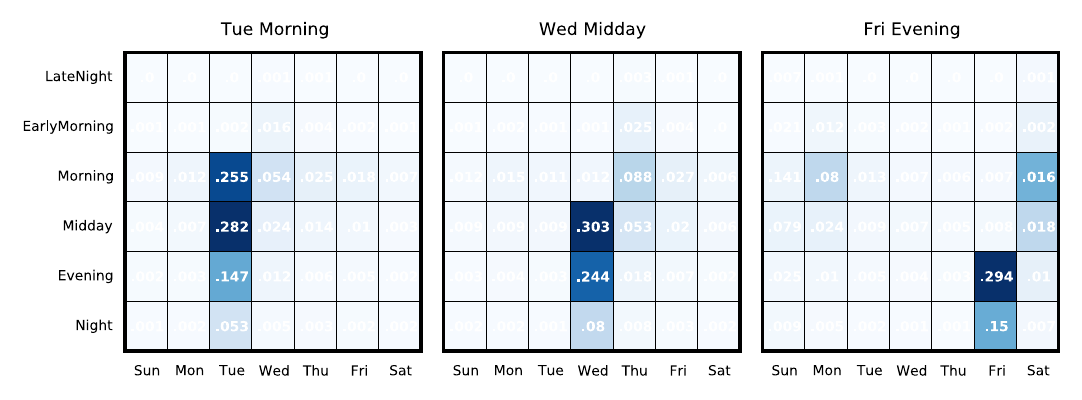}
\caption{Reminder notification time probability distributions over time (i.e., $P(r_{NT} \mid r_{CT})$), for three different $r_{CT}$.}
\label{fig:ntgivenct}
\end{figure}

See Figure~\ref{fig:ntgivenct} for examples. 
Looking at the plots in detail, we see how reminders across different creation times appear similar: they are most likely to have their notification fire within the same cell or the next, confirming earlier observations that the majority of reminders are short-term (i.e., same cell). 
However, upon closer inspection, we see that as the reminders' creation time moves towards later during the day, reminders are more likely to be set to notify the next day. 
Furthermore, in the third plot from the left, we see how reminders created on Friday evenings have a small but substantial probability of having their notification fire on Monday morning (i.e., the reminder spans the weekend). 
These patterns show how delay between reminder creation and notification time is low on average, but the length of delay is not independent from the creation time. 

In summary, we have shown distinct temporal patterns of reminders of different task types, and of the terms in task descriptions. 
Finally, we have shown that a reminder's notification time is most likely set shortly after creation time, but the later in the day a reminder is created, the more likely the notification time is further in the future. 
 
\section{Predicting Notification Time}
\label{sec:predicting}
In the previous section, we have shown temporal patterns in reminder creation and notification time of four types: aggregate patterns, task type-related, term-based, and time-based. 
To study whether these patterns can be effectively harnessed, we address a prediction task. 
Specifically, we turn to the task of predicting the day of the week in which a task is most likely to happen (i.e., predicting $r_{NT}$). 
Motivated by our observation that the majority of the reminders are set to trigger soon after being set (Section~\ref{subsec:globalpatterns} and \ref{subsec:time}), and by the patterns we observed of the task descriptions' terms (Section~\ref{subsec:terms}), we aim to answer the following research questions: 
\emph{``Is the reminder's creation time indicative of its notification time?''} and 
\emph{``Do term-based features yield an increase in predictive power?''} 
The aim of our experiments is not to develop complex or novel predictive models, but instead to study whether the patterns discussed earlier generalize so as to contribute to overall predictive performance. 

We cast the task of predicting the day of week a reminder is set to notify as a multiclass classification task, where each day of the week corresponds to a class. 
The input to our predictive model is the reminder's task description ($r_{task}$), creation time ($r_{CT}$), and the target class is the notification time ($r_{NT}$) day of week. 
We measure the predictive power of the patterns identified in the previous sections via term-based and (creation) time-based features. 
Specifically, for term-based features, we extract bag of word features (unigrams), and our time-based features correspond to $R_{CT}$'s time of day (row) and day of week (column), and the minutes since the start of week. 
See Table~\ref{tab:umapfeatures} for an overview. 

\begin{table}[ht]
  \centering
  \caption{Features used for prediction.}
  \begin{tabular}{ll}
    \toprule
    \textbf{Features} & \textbf{Description} \\
    \midrule
    Term Features & Unigram Bag of Word features \\
    Time features & $r_{CT}$ Time of day \\
                  & $r_{CT}$ Day of week \\
                  & $r_{CT}$ Minutes since start of week \\
    \bottomrule
  \end{tabular}
  \label{tab:umapfeatures}
\end{table}

\subsection{Experimental Setup}
We use Gradient Boosted Decision Trees for classification. 
This method has proven to be robust and efficient in large-scale learning problems~\cite{Friedman:2002:SGB:635939.635941}. 
The ability of this method to deal with non-linearity in the feature space and heterogeneous features makes it a natural choice. 
To address the multiclass nature of our problem, we employ a one vs. all classification strategy, where we train seven binary classifiers and output the prediction with the highest confidence as final prediction. 
We compare the accuracy to randomly picking a day of the week (with accuracy of $\frac{1}{7} = 0.1429$) and to a more competitive baseline that predicts the notification will be for the same day that the notification was created (\texttt{BL-SameDay}).

For the experiments, we sample six months of data (January through June 2015).?All data were filtered according to the process described in Section~\ref{subsec:data}, resulting in a total of 1,509,340 reminders.?We split this data sequentially: the first 70\% (approx. January 1 to May 7) forms the training set and the last 30\% (approx. May 8 to June 30) forms the test set.?We use the first two months of the training set for the analysis described in Sections~\ref{sec:remindertypes} and \ref{sec:reminderpatterns} as well as for parameter tuning before we retrained on the entire training set.?
In the next section, we report predictive performance on the held out test set, specifically, we report macro and micro-averaged accuracy over the classes (\texttt{Macro} and \texttt{Micro}, respectively). 
We compare three approaches: one that leverages time features based on the reminder's creation time (\texttt{Time only}), one with term features (\texttt{Terms only}), and finally a model that leverages both types of features (\texttt{Full model}). 
We test for statistical significance using $t$-tests, comparing our predictive models against \texttt{BL-SameDay}. 
The symbols \dubbelop~and \dubbelneer~denote statistically significant differences (greater than the baseline and worse than the baseline, respectively) at $\alpha$ = 0.01.

\subsection{Results}
Table~\ref{tab:results} shows the results of our prediction task. 
First, we note that the baseline of predicting the notification time to be the same day as the creation time, at 0.5147 micro-averaged accuracy, performs much better than random (at 0.1429). 
This indicates users mostly set reminders to plan for events in the short-term. 
Next, we see that the \texttt{Time only} model with a micro-averaged accuracy of 0.6279 significantly improves over the baseline, indicating that the reminder creation time helps further improve prediction accuracy. 
As noted earlier, tasks planned late at night are more likely to be executed on a different day, and the use of creation time helps leverage this and more general patterns. 
Finally, the model that uses only features based on the task description (\texttt{Terms only}) performs better than random, but does not outperform the baseline. 
However, when combined with the time model (\texttt{Full model}) we see an increase of 8.2\% relative to the time only model. 
We conclude that the creation time provides the most information for predicting when a task will be performed, but the task description provides significant additional information. 
Framed differently, we show that both the \emph{context} in which digital traces are produced (i.e., the time of day), and the \emph{content} of the digital traces (i.e., the textual description of the task) provide signals for better predicting real-world behavior of the producers of digital traces. 

\begin{table}[ht]
  \centering
  \caption{Average accuracy of day-of-week prediction task. Statistical significance tested against \texttt{BL-SameDay}.}
  \begin{tabular}{llll}
    \toprule
    \textbf{Run} & \textbf{Micro} & \textbf{Macro} & \textbf{Error reduction} \\
    \midrule
    \texttt{Full model} & 0.6788\dubbelop & 0.6761\dubbelop & +0.3381 \\
    \texttt{Time only}  & 0.6279\dubbelop & 0.6258\dubbelop & +0.2333 \\
    \texttt{Terms only} & 0.1777\dubbelneer & 0.1772\dubbelneer & -0.6944 \\
    \midrule
    \texttt{BL-SameDay} & 0.5147 & 0.5165 & \\
    \bottomrule
  \end{tabular}
  \label{tab:results}
\end{table}

\section{Conclusion}
In this chapter, we have performed a large-scale analysis of reminder data from user activity in a natural setting to answer the following research question:

\begin{description}
\item[\ref{rq:umap}] Can we identify patterns in the times at which people create reminders, and, via notification times, when the associated tasks are to be executed?
\end{description}

\noindent
We employed a mixed methods approach, comprising data-driven and qualitative methodologies, to extract and identify common task types from reminders that are frequently observed across multiple users. 
To answer \ref{rq:umap}, we answer several subquestions in this chapter, the findings of which we summarize below. 

\paragraph{Task type taxonomy.} 
Through our log analyses we have shown common reminders and made an attempt at identifying and categorizing the types of tasks that underlie them. 
We identified that the majority of reminders in our sample refer to either daily household chores, running errands, or switching contexts. 

\paragraph{Temporal patterns.} 
Next, we show how reminders display different temporal patterns depending on the task type they represent, the reminder's creation time, and the terms in the task description. 
Most notably, we have shown that people mostly plan for tasks in the evening hours, and most tasks are set to execute in mornings and throughout the day. 
On average, people plan in advance for the short-term, i.e., 80\% of the reminders in our sample log are set to trigger on the same day as reminder creation. 

\paragraph{Prediction task.} Finally, we demonstrated that we can leverage these patterns to predict the day of the week that a reminder is most likely to trigger, i.e., the day the task is most likely to be executed. 
Specifically, we confirm that the reminder's creation time is a strong indicator of notification time, but that including the task description further improves accuracy over the strongest baseline, with a 33\% reduction in error. \\

\noindent
In line with our findings in the previous chapter, here we confirm that both the \emph{context} in which digital traces are produced (the time of day), and the (textual) \emph{content} of the digital traces (the reminder description) provide valuable signals in predicting the real-world behavior and activity of our entities of interest: the producers of digital traces. 
Furthermore, the findings in this chapter have implications for designing systems to help with task completion, and more generally for developing technology to reduce prospective memory failures. 
The analysis and prediction task show that we can leverage large-scale interaction logs to predict a users' (planned) activities. 

There are several limitations in our log analyses. 
First, we performed this analysis on a specific subset of reminders: reminders from one geographic locale and for a single type of reminder: time-based. 
There are opportunities to understand cultural and linguistic factors in reminder creation by considering reminders from multiple regions. 
We additionally seek to investigate other types of reminders, such as those involving people, places, and events. 
Second, it is difficult to quantify the comprehensiveness of the task type taxonomy, which covers common reminders. 
The ontology may therefore not cover more intricate, personal, specific, or complex reminders, the nature of which needs to be better understood. 
Finally, our approach and analysis is entirely log based. 
The taxonomy's categories were manually labeled, and we make inferences and assumptions about the tasks that people are engaged in. 
User studies are needed to better understand the reminder process, including the generation and value of reminders, including how people behave when they are notified. 

Future work includes developing more sophisticated models (e.g., considering personalized signals) to improve prediction performance. 
Besides the value of the insights gained in this chapter from the discovery perspective, i.e., from understanding the link between people's interaction logs with intelligent assistants and their (planned) activities and tasks, the insights and predictions about tasks and the use of reminders can also prove valuable with the development of systems endowed with the ability to proactively reserve time, manage conflicts, remind people about tasks they might forget, and, more generally, help people achieve their goals. \clearpage{}

\bookmarksetup{startatroot} 
\addtocontents{toc}{\bigskip} 
\clearpage{}
\chapter{Conclusions}
\label{ch:conclusions}

\begin{flushright}
\rightskip=1.8cm``Si finis bonus est, totum bonum erit.'' \\
\vspace{.2em}
\rightskip=.8cm---\textit{Gesta Romanorum, Tale LXVII}
\end{flushright}
\vspace{1em}

\noindent
In this thesis, we have studied \emph{discovery} in digital traces. 
We have done so in two parts:
in Part~\ref{pt:1} we have studied the \emph{content} of (textual) digital traces, and in Part~\ref{pt:2} we have turned to the \emph{contexts} in which digital traces are created, with the goal of leveraging digital traces for predicting real-world activity. 

Part~\ref{pt:1} revolves around semantic search. 
Semantic search is a search paradigm that uses structured knowledge to increase retrieval effectiveness~\cite{vandijk2011}, and allows us to better support the exploratory search process that is inherent to the discovery process~\cite{Ahn2010383}. 
More specifically, in this part our entities of interest are the real-world entities that occur (i.e., are mentioned) in digital traces. 
We leverage publicly available knowledge bases such as Wikipedia and Freebase, and study \emph{emerging entities}: entities that are not (yet) described in the knowledge base. 
This is motivated by the fact that in typical discovery scenarios (e.g., investigative journalism or digital forensics), the entities of interest may not be known a priori. 
In three chapters, we 
analyze the temporal patterns of emerging entities in online text streams (Chapter~\ref{ch:plos}), 
predict them in social media streams (Chapter~\ref{ch:ecir}),
and enrich their representations by leveraging ``collective intelligence'' (i.e., how people refer to and search for the entities on the web), for increased retrieval effectiveness (Chapter~\ref{ch:wsdm}). 

The second part of this thesis revolves around the contexts in which digital traces are created.
Here, we shift our focus, and our entities of interest are the producers of digital traces, i.e., the people who leave behind digital traces. 
Our goal is to predict real-world activity from digital traces, a core task in E-Discovery~\cite{INR-025}.
We present two case studies:
in the first, we study enterprise email collections, and we analyze aspects that govern communication between employees and aim to predict email communication by leveraging the communication graph and email content (Chapter~\ref{ch:sigir}). 
Next, we study user interaction logs with a personal intelligent assistant (Microsoft Cortana), that represents the users' planned activities and tasks (Chapter~\ref{ch:umap}). 
In this chapter, we show common usage patterns of the reminder service, and show how we can leverage the uncovered patterns for predicting the most likely day at which a task will be executed. 

In this final chapter of the thesis, we re-iterate the research questions we answered, and summarize our methods and findings in Section~\ref{subsec:findings}. In Section~\ref{subsec:futurework} we reflect on future work. 

\section{Main Findings, Limitations, and Implications}
\label{subsec:findings}
Here, we summarize the methods, findings, and their implications. 
We do so by answering the research questions raised in Chapter~\ref{ch:introduction}. 

\subsection*{Part~\ref{pt:1}.~\acl{pt:1}}

In the first content chapter of this thesis (Chapter~\ref{ch:plos}), we start our exploration of real-world entities in digital traces with a large-scale analysis of how entities emerge. 
In Chapter~\ref{ch:plos} we answer our first research question:

\begin{description}
\item[\ref{rq:plos}] Are there common temporal patterns in how entities of interest emerge in online text streams?
\end{description}

\noindent
To answer this question, we analyze a large collection of entity time series, i.e., time series of entity mentions in the lead time between an entity's first mention in online text streams and its subsequent incorporation into the knowledge base. 

We apply an unsupervised hierarchical clustering method to uncover groups of entities that exhibit different emergence patterns. 
We discover two main emergence patterns. 
First, roughly half of the entities we study emerge in a ``bursty'' fashion, i.e., they surface in online text streams without a precedent, and blast into activity before they are incorporated in the KB. 
Other entities display a ``delayed'' pattern, where they appear in public discourse, experience a period of inactivity, and then resurface before being incorporated into the KB. 
In addition to these emergence patterns, we found distinct differences in the emergence patterns of different types of entities, suggesting that different entities of interest exhibit different emergence patterns. 

The work presented in this chapter has several implications. 
Most notably, the analyses have implications for designing systems to detect emerging entities before they are part of the KB. 
We have shown how entities tend to resurface multiple times before being added to Wikipedia.
This suggests that burst detection may be an effective way of ``catching'' emerging entity before they are part of the KB. 
Furthermore, we have shown that the different streams in which the entities emerge are indicative of how fast the entity may be added to the KB. 

The work presented in this chapter also know several limitations.
First, evaluating unsupervised clustering without labels is not trivial: \citet{von2012clustering} aptly question whether clustering is an ``art or science.'' 
However, we do have indications to believe the clusters we generated are meaningful. 
First, the cluster signatures show distinct and statistically significantly different patterns, whereas different groupings of time series (e.g., by stream or by entity types) did not yield any discernible patterns. 
Furthermore, the structure of the dendrogram (i.e., the hierarchical cluster tree) suggests there are clear subgroups in the data. 

Another limitation to our study is related to the dataset used. 
First, we rely on automatically generated annotations (FAKBA1 dataset), which means we cannot assume a 100\% accurate set of annotations. 
What's more, the set of long-tail entities that are likely in our subset of ``emerging entities,'' are known to be comparatively more difficult to link, as most entity linking methods have a strong reliance on ``popularity''-based features, such as, commonness~\cite{Mihalcea:2007:WLD:1321440.1321475}.
Next to the quality of the annotations, the coverage or selection of sources in the data may be a concern for how well our findings generalize to other domains or datasets. 
Sampling bias, and e.g., the absence of popular social media platforms (such as Facebook, Twitter, or Tumblr) means the findings may be different with other datasets. 
Finally, the cultural bias that is inherent to the dataset selection.
By studying only English sources, we effectively studied how entities emerge in the English speaking world.

\bigskip\noindent
The analysis of emerging entities in Chapter~\ref{ch:plos} shows that knowledge bases are never complete: new entities may emerge as events unfold, but at the same time, long-tail, relatively unknown entities may also be added to a KB. 
As a follow-up to this chapter, we turn our attention to predicting newly emerging entities in Chapter~\ref{ch:ecir}. 
More specifically, we focus on predicting mentions of emerging KB entities in social media streams, motivated by the high pace and noisy nature that characterizes social media streams. 
Our method leverages an entity linking system to automatically generate training data for a named-entity recognizer to learn to recognize KB entities. 
We answer:

\begin{description}
	\item[\ref{rq:ecir}] Can we leverage prior knowledge of entities of interest to bootstrap the discovery of new entities of interest?
\end{description}

\noindent
To answer this question, we propose a novel unsupervised method for generating pseudo-training data to train a named-entity recognizer and classifier (NERC) for predicting mentions of emerging entities that are likely to be added to a KB. 
The method includes two sampling methods for selecting high quality training samples, the first selects samples based on the textual quality of the sample, and the second leverages the confidence score of the entity linking system to select samples of which the entity linker is more confident, in an attempt to reduce noisy samples. 
We measure the effectiveness of these sampling methods by their impact on the accuracy of predicting newly emerging entities. 
Furthermore, we perform an additional analysis, where we study the impact of the amount of prior knowledge.
We approximate this quantity of prior knowledge by randomly sampling different (size) reference knowledge bases, i.e., we leave out a part of the knowledge base, and study the prediction effectiveness. 

We find that sampling by textual quality improves the performance of our NERC method and consequently our method's performance in predicting emerging entities. 
Furthermore, we show that setting a higher threshold on the entity linking system's confidence score for generating pseudo-ground truth results in fewer labels but better performance. 
We show that the NERC is better able to separate noise from entities that are worth including in a knowledge base. 
Finally, we show that in the case of a small amount of prior knowledge, i.e., limited size of the available initial knowledge, our method is able to cope with missing labels and incomplete data, as observed through its consistent and stable precision. 
This finding justifies our proposed method that assumes incomplete data by design. 

The implication of our findings is that we our approach can effectively support the exploratory search process, by retrieving similar entities to those in a reference KB, that are not yet part of the KB. 

A limitation of our work is the retrospective scenario we employ in our evaluation.
For this reason, we cannot measure the impact of the novelty of entities of interest as they emerge. 

\bigskip\noindent
Finally, in Chapter~\ref{ch:wsdm} we address the follow-up task of enriching entity representations of emerging entities with descriptions coming from a variety of sources, in a real-time, streaming manner. 
Collecting entity descriptions from different sources, and combining them in a single representation for improved retrieval effectiveness, is beneficial in scenarios where information about the same entities is spread over multiple sources, e.g., in a discovery scenario, where multiple sources that represent the activity of, e.g., people in a company (e.g., email collections, collaboration platforms, social media posts) can be used to paint the full picture. 
We answer: 

\begin{description}
  \item[\ref{rq:wsdm}] Can we leverage collective intelligence to construct entity representations for increased retrieval effectiveness of entities of interest?
\end{description}

\noindent
We answer this question by proposing an effective way of leveraging collective intelligence (i.e., entity descriptions from multiple sources) to construct entity representations for optimal retrieval effectiveness of entities of interest. 
More specifically, we collect descriptions from knowledge bases, social media, and the web, to increase retrieval effectiveness of entities. 
We do this in a dynamic scenario, i.e., we distinguish between static description sources, where descriptions are aggregated and added to entity representations, and dynamic description sources, where descriptions come in a real-time, streaming manner. 

The main challenge with leveraging collective intelligence is the heterogeneity between entities: 
some entities may receive many descriptions (i.e., head entities), while others may receive few (tail entities).
At the same time, different description sources represent very different content, compare, e.g., the structured and clean entity description from a knowledge base, to the contexts in which people refer to entities on social media. 

First, we demonstrate that incorporating dynamic description sources into dynamic collective entity representations enables a better matching of users' queries to entities, resulting in an increase of entity ranking effectiveness. 
In addition, we show that informing the ranker on the expansion state of the entities, further increases the ranking effectiveness. 
Finally, we show how retraining the ranker leads to improved ranking effectiveness in dynamic collective entity representations, however, even a static (not periodically retrained) ranker's performance improves over time, suggesting that even the static ranker benefits from newly incoming descriptions. 

The findings in this chapter imply that information from different types of sources can be effectively combined to improve retrieval effectiveness of entities of interest.
The scenario of different (heterogeneous) data sources that may be related, is not uncommon in the E-Discovery scenario, e.g., when an email box and associated attachments or files can all be traced to a single entity. 

One limitation in the work presented in this chapter was related to the experimental design, which was constrained by the lack of temporally aligned and sizeable datasets. 
In particular, the temporal misalignment between different corpora prevents the analysis of temporal patterns that affect entities in unforeseen ways.

\subsection*{Part~\ref{pt:2}.~\acl{pt:2}}

In the first chapter of this part, the digital traces under study are enterprise email, and the entities of interest are emailers. 

\noindent
Our aim in Chapter~\ref{ch:sigir} is to provide insights in the aspects that guide communication between people. 
Specifically, we study the impact of aspects of the communication graph (e.g., the strength of ties between emailers, or the proximity of emailers in the enterprise network), and the impact of email content (e.g., the similarity of email content between two emailers).
We apply these signals for predicting enterprise email communication, which may find application in detecting unexpected communication. 
We answer:

\begin{description}
	\item[\ref{rq:sigir}] Can we predict email communication through modeling email content and communication graph properties?
\end{description}

\noindent
We present a hybrid model for email recipient prediction that leverages both the information from the email network's communication graph, and the Language Models of the emailers, estimated with the content of the emails that are sent by each user. 
Our model starts from scratch, in that it does not assume or need seed recipients, and it is updated for each email sent. 

We show that the communication graph properties and the email content signals both provide a strong baseline for predicting recipients, but are complementary, i.e., combining both signals achieves the highest accuracy for predicting the recipients of email. 
Furthermore, we show that the number of received emails is an effective method for estimating the prior probability of observing a recipient, and we show that the number of emails sent between two users is an effective way of estimating the ``connectedness'' between the two users. 

The implication of our findings is that both the \emph{context} in which digital traces are created (i.e., the position of an emailer in the communication graph, her ties with surrounding emailers), and the \emph{content} (i.e., the email itself) are important signals in predicting real-world behavior (i.e., email communication).

\bigskip\noindent
Next, we study interaction logs with an intelligent personal assistant in Chapter~\ref{ch:umap}. 
Intelligent assistants are proliferating across (both mobile and desktop) devices, they have a close proximity to and embedding in our day-to-day life (in part thanks to their conversational and ``personal'' nature).
For these reasons, interaction logs with personal assistants are a rich resource for digital evidence.
In this chapter we take a similar approach to the previous, and analyze a novel dataset of interaction logs, and address a prediction task to study how digital traces can be employed to predict real-world activity of our entities of interest. 

User interaction logs of intelligent personal assistants may contain many rich contextual clues, and represent the user's (planned) activities or geographic location at any given time. 
We are particularly interested in the reminder service of one such personal assistant (Microsoft Cortana), as this service is a heavily used feature by its users, and reminders represent (planned) tasks of users, and can hence be used to gain insights into their (planned) activities. 
In combination with other data, these logs could be a rich signal for inferring real-world activity, a core task in discovery. 
We answer:

\begin{description}
	\item[\ref{rq:umap}] Can we identify patterns in the times at which people create reminders, and, via notification times, when the associated tasks are to be executed?
\end{description}

\noindent
We analyze a large-scale user log, comprising over 500,000 time-based reminders of over 90,000 users. 
We identify a body of common tasks types that give rise to the reminders across a large number of users. 
We subsequently arrange these tasks into a taxonomy.
Finally, we study their temporal patterns, and we address a prediction task, where we aim to predict when a task reminder is set to trigger, i.e., the user aims to execute the task, given the reminder task description and creation time. 

We show how reminders display different temporal patterns depending on the task type they represent, the reminder's creation time, and the terms in the task description. 
We show that the time at which a user creates a reminder is a strong indication of when the task is scheduled to be executed. 
Furthermore, we show that including the text of the task reminder further improves prediction accuracy. 
Much like in Chapter~\ref{ch:sigir}, we confirm that combining the \emph{content} of the digital traces (i.e., the textual description of the reminder), and the \emph{context} in which they are created (i.e., the time of day), achieves the highest prediction accuracy, suggesting that both signals are complimentary. 

As in the previous chapter, the findings in this chapter imply that both the \emph{context} (the time of day), and the \emph{content} (the reminder description) of digital traces provide signals that help in predicting real-world behavior of our entities of interest: the producers of digital traces. 

This chapter is not without limitations either. 
First, we make assumptions about the behavior and activities of people, based on interaction logs. 
Without conducting user studies or using other ways to infer the actual activities of people --- beyond what they say to Cortana, we cannot make any statements about people's activities. 

\section{Future Research Directions}
\label{subsec:futurework}
In this section, we summarize some of the limitations to the work presented in this thesis, and identify some areas for potential follow-up work. 

\subsection*{Part~\ref{pt:1}.~\acl{pt:1}}

Chapter~\ref{ch:plos} is a starting point of studying how entities emerge in public discourse, and what happens in the lead time between the entity first surfaces, and is subsequently deemed ``important enough'' to be added to the KB. 
As a next step, we should take a closer look at the detailed circumstances under which entities emerge, by not only considering the number of documents they appear in over time, but also in which contexts, e.g., by looking at the content of the articles themselves. 
Another interesting aspect of emerging entities, that falls out of the scope of the present work, is the notion of when entities are ``deemed important enough,'' i.e., when the collective reaches consensus. 
For example, one could study the emergence patterns of entities that are removed from the KB. 
Finally, the observations made in this chapter could be explored in a prediction task, where, e.g., given a partial entity time series, the task would be to predict the point at which the entity will be incorporated in the KB.

The method we present in Chapter~\ref{ch:ecir} addresses entity mention detection for emerging entities, i.e., we leverage a knowledge base to label entity mentions, so a named-entity recognizer can identify similar but unlinked mentions in social media posts. 
A natural follow-up to would be to ``close the loop,'' i.e., feed back the emerging entities to  populate the knowledge base. 
Closing the loop would open the door to a fully automated knowledge base population process, the scenario where the knowledge base is automatically populated with entity mentions and representations, which in turn means the pseudo-training data generated by the entity linking system remains up-to-date at all times. 

To achieve this, an additional step of constructing seed entity representations is necessary.
One approach is to mine keyphrases~\cite{Hoffart:2014:DEE:2566486.2568003} from the tweets in which the emerging entities appear. 
To effectively construct these representations, an entity clustering or disambiguation step would be beneficial, to collect enough content to extract keyphrases from~\cite{Davis:2012:NED:2390524.2390639,Rao:2010:SCD:1944566.1944687}. 

Another challenge that lies in this direction is that of supervision; letting the system roam free may introduce noise (i.e., false positives) into the knowledge base, which may over time corrupt training data and hence derail the system. 
Studying ways of incorporating a ``human-in-the-loop,'' when evaluating and judging predictions, using e.g., crowdsourcing finds applications in any task in which a self-learning, automated, system runs freely~\cite{Finin:2010:ANE:1866696.1866709,Demartini:2012:ZLP:2187836.2187900}. 
The human-in-the-loop paradigm for evaluating and tracking algorithmic predictions is of particular interest in the E-Discovery domain, where machine-generated predictions can have severe impact on the outcomes of, e.g., legal cases~\cite{INR-025}.

A natural extension to the work presented in Chapter~\ref{ch:wsdm} is to study our method in a real life scenario, with real-time, streaming data, and to provide a more fine-grained analysis of the impact of the importance features and to to study the link between the temporal patterns of entities (as studied in Chapter~\ref{ch:plos}) and their retrieval effectiveness. 

Larger and temporally aligned data collections would also increase the potential challenge of ``swamping''~\cite{robertson2004} or ``document vector saturation''~\cite{kemp2002}, where description sources may completely overtake an entity representation. 
This swamping phenomenon could be addressed by incorporating a temporal decay on the terms that make up the representation, or imposing a size limit on the fields. 
Furthermore, additional feature engineering, e.g. informing the ranker of the diversity of terms in a field, or the novelty of the terms compared to the original entity representation, may also prove beneficial when the number and volume of descriptions increase. 

In addition, we focused on a concrete end-to-end task in this chapter, we evaluate the quality of the representations by measuring the retrieval effectiveness, given the added entity descriptions. 
Another direction would be to study the shifting entity representations from a more analytic perspective. 
One could study the entity drift, or whether entity representations change over time in meaningful ways, e.g., by studying which words are more strongly associated to entities at different points in time. 
Similar to studying how word use may shift over time~\cite{Kenter:2015:AHM:2806416.2806474}, studying entity representation drift may provide insights into the type of entities or events that are prone to change more or less intensely. 

\subsection*{Part~\ref{pt:2}.~\acl{pt:2}}
Our study in Chapter~\ref{ch:sigir} focused on the (relative) contribution of two different aspects; communication graph and email content properties. 
Engineering-wise, however, prediction accuracy could be improved by a number of approaches:
first, employing the scores of our generative model as features in a machine learning model, instead of using them directly for ranking, means multiple signals could be easily combined, and prediction accuracy may increase. 
Next, taking a more fine-grained or local approach to the communication graph could further boost prediction accuracy, by incorporating (implicit) organizational structures of the enterprise. 
Finally, the analysis of our experimental results highlighted that dealing with growing networks, and handling time in general, proved a challenge. 
Incorporating time-awareness into both components of the model may prove beneficial for performance over time, 
and studying the effect of communication patterns in evolving networks~\cite{Aggarwal:2014:ENA:2620784.2601412,Tylenda:2009:TTL:1731011.1731020}, may alleviate these weaknesses, and further improve our model's prediction accuracy. 

Another opportunity for the work presented in this chapter is to adapt the presented method to novel tasks. 
One could, e.g., look at role detection or inferring a company's hierarchical structure through leveraging both social network aspects and email content (e.g., common topics amongst employees).
Furthermore, a predictive model with accurate recipient predictions could be employed for rating the ``likelihood'' of historic interactions, and thus be applied for outlier or anomaly detection, to discover communication in an enterprise that is ``unexpected,'' and therefore may be of interest to, e.g., a digital forensic analyst.

The prediction task studied in Chapter~\ref{ch:umap} served as a starting point for further research into predicting (future) tasks and activities given historic signals. 
Future work includes developing more sophisticated models.
Whereas the current work focused on global patterns, seen frequently and across many users, considering more personalized signals may increase prediction accuracy; it is likely that users exhibit distinct patterns and usage of the reminder service. 
Furthermore, incorporating more signals into the task, e.g., leveraging geographical signals and location services to better infer the current context of the user, to be able to distinguish between 
being at work, or at home, or driving/in transit, 
or even a more fine-grained context, by looking up the type of location the user currently is at (e.g., a restaurant, grocery store). 

Furthermore, more sophisticated approaches to understanding the language of the reminder descriptions may prove beneficial in clustering similar task reminders, to battle sparsity and yield stronger (temporal) signals for different task types.
Using topic modeling, distributional or dimension reduction techniques (through e.g., word or reminder embeddings) could prove beneficial in clustering similar task descriptions. 

Finally, the insights and predictions about tasks and the use of a personal assistant's reminder service could also prove valuable in a more practical or commercial setting, e.g., with developing systems endowed with the ability to proactively reserve time for users, automatically manage their scheduling and planning conflicts, remind people about tasks they might forget, and, more generally, to help people achieve their goals. 
\clearpage{}

\backmatter
\clearpage{}

\renewcommand{\bibsection}{\chapter{Bibliography}}
\renewcommand{\bibname}{Bibliography}
\markboth{Bibliography}{Bibliography}
\renewcommand{\bibfont}{\footnotesize}
\setlength{\bibsep}{0pt}
\bibliographystyle{abbrvnatnourl}
\bibliography{thesis-min}
\clearpage{}

\clearpage{}
\chapter*{SIKS Dissertation Series}
\phantomsection

\rhead[SIKS Dissertation Series]{}
\lhead[]{SIKS Dissertation Series}

\setstretch{1}
\newcommand{\siksitem}[3]{\item\footnotesize{ #1 (#2) \textit{#3}}}
\newcounter{sikscounter}
\newenvironment{sikslist}[1]{
	\setcounter{sikscounter}{0}
	\vspace{2mm}\noindent{\small\textbf{#1}}\vspace{2mm}
	\begin{list}{\footnotesize\arabic{sikscounter}}
  	{   \usecounter{sikscounter}
		\setlength{\topsep}{0pt}
		\setlength{\partopsep}{0pt}
		\setlength{\listparindent}{0pt}
		\setlength{\labelsep}{5pt}
		\setlength{\itemsep}{-3pt}
		\setlength{\labelwidth}{10pt}
		\setlength{\itemindent}{0pt}
		\setlength{\leftmargin}{8pt}}} 
{
    \end{list}
}

\raggedcolumns
\setlength{\multicolsep}{1em}
\setlength{\columnsep}{2em}
\begin{multicols}{2}

\begin{sikslist}{1998}

\siksitem{Johan van den Akker}{CWI}{DEGAS: An Active, Temporal Database of Autonomous Objects}

\siksitem{Floris Wiesman}{UM}{Information Retrieval by Graphically Browsing Meta-Information}

\siksitem{Ans Steuten}{TUD}{A Contribution to the Linguistic Analysis of Business Conversations}

\siksitem{Dennis Breuker}{UM}{Memory versus Search in Games}

\siksitem{E. W. Oskamp}{RUL}{Computerondersteuning bij Straftoemeting}

\end{sikslist}
\begin{sikslist}{1999}

\siksitem{Mark Sloof}{VUA}{Physiology of Quality Change Modelling: Automated modelling of}

\siksitem{Rob Potharst}{EUR}{Classification using decision trees and neural nets}

\siksitem{Don Beal}{UM}{The Nature of Minimax Search}

\siksitem{Jacques Penders}{UM}{The practical Art of Moving Physical Objects}

\siksitem{Aldo de Moor}{KUB}{Empowering Communities: A Method for the Legitimate User-Driven}

\siksitem{Niek J. E. Wijngaards}{VUA}{Re-design of compositional systems}

\siksitem{David Spelt}{UT}{Verification support for object database design}

\siksitem{Jacques H. J. Lenting}{UM}{Informed Gambling: Conception and Analysis of a Multi-Agent Mechanism}

\end{sikslist}
\begin{sikslist}{2000}

\siksitem{Frank Niessink}{VUA}{Perspectives on Improving Software Maintenance}

\siksitem{Koen Holtman}{TUe}{Prototyping of CMS Storage Management}

\siksitem{Carolien M. T. Metselaar}{UvA}{Sociaal-organisatorische gevolgen van kennistechnologie}

\siksitem{Geert de Haan}{VUA}{ETAG, A Formal Model of Competence Knowledge for User Interface}

\siksitem{Ruud van der Pol}{UM}{Knowledge-based Query Formulation in Information Retrieval}

\siksitem{Rogier van Eijk}{UU}{Programming Languages for Agent Communication}

\siksitem{Niels Peek}{UU}{Decision-theoretic Planning of Clinical Patient Management}

\siksitem{Veerle Coup\'e}{EUR}{Sensitivity Analyis of Decision-Theoretic Networks}

\siksitem{Florian Waas}{CWI}{Principles of Probabilistic Query Optimization}

\siksitem{Niels Nes}{CWI}{Image Database Management System Design Considerations, Algorithms and Architecture}

\siksitem{Jonas Karlsson}{CWI}{Scalable Distributed Data Structures for Database Management}

\end{sikslist}
\begin{sikslist}{2001}

\siksitem{Silja Renooij}{UU}{Qualitative Approaches to Quantifying Probabilistic Networks}

\siksitem{Koen Hindriks}{UU}{Agent Programming Languages: Programming with Mental Models}

\siksitem{Maarten van Someren}{UvA}{Learning as problem solving}

\siksitem{Evgueni Smirnov}{UM}{Conjunctive and Disjunctive Version Spaces with Instance-Based Boundary Sets}

\siksitem{Jacco van Ossenbruggen}{VUA}{Processing Structured Hypermedia: A Matter of Style}

\siksitem{Martijn van Welie}{VUA}{Task-based User Interface Design}

\siksitem{Bastiaan Schonhage}{VUA}{Diva: Architectural Perspectives on Information Visualization}

\siksitem{Pascal van Eck}{VUA}{A Compositional Semantic Structure for Multi-Agent Systems Dynamics}

\siksitem{Pieter Jan 't Hoen}{RUL}{Towards Distributed Development of Large Object-Oriented Models}

\siksitem{Maarten Sierhuis}{UvA}{Modeling and Simulating Work Practice}

\siksitem{Tom M. van Engers}{VUA}{Knowledge Management}

\end{sikslist}
\begin{sikslist}{2002}

\siksitem{Nico Lassing}{VUA}{Architecture-Level Modifiability Analysis}

\siksitem{Roelof van Zwol}{UT}{Modelling and searching web-based document collections}

\siksitem{Henk Ernst Blok}{UT}{Database Optimization Aspects for Information Retrieval}

\siksitem{Juan Roberto Castelo Valdueza}{UU}{The Discrete Acyclic Digraph Markov Model in Data Mining}

\siksitem{Radu Serban}{VUA}{The Private Cyberspace Modeling Electronic}

\siksitem{Laurens Mommers}{UL}{Applied legal epistemology: Building a knowledge-based ontology of}

\siksitem{Peter Boncz}{CWI}{Monet: A Next-Generation DBMS Kernel For Query-Intensive}

\siksitem{Jaap Gordijn}{VUA}{Value Based Requirements Engineering: Exploring Innovative}

\siksitem{Willem-Jan van den Heuvel}{KUB}{Integrating Modern Business Applications with Objectified Legacy}

\siksitem{Brian Sheppard}{UM}{Towards Perfect Play of Scrabble}

\siksitem{Wouter C. A. Wijngaards}{VUA}{Agent Based Modelling of Dynamics: Biological and Organisational Applications}

\siksitem{Albrecht Schmidt}{UvA}{Processing XML in Database Systems}

\siksitem{Hongjing Wu}{TUe}{A Reference Architecture for Adaptive Hypermedia Applications}

\siksitem{Wieke de Vries}{UU}{Agent Interaction: Abstract Approaches to Modelling, Programming and Verifying Multi-Agent Systems}

\siksitem{Rik Eshuis}{UT}{Semantics and Verification of UML Activity Diagrams for Workflow Modelling}

\siksitem{Pieter van Langen}{VUA}{The Anatomy of Design: Foundations, Models and Applications}

\siksitem{Stefan Manegold}{UvA}{Understanding, Modeling, and Improving Main-Memory Database Performance}

\end{sikslist}
\begin{sikslist}{2003}

\siksitem{Heiner Stuckenschmidt}{VUA}{Ontology-Based Information Sharing in Weakly Structured Environments}

\siksitem{Jan Broersen}{VUA}{Modal Action Logics for Reasoning About Reactive Systems}

\siksitem{Martijn Schuemie}{TUD}{Human-Computer Interaction and Presence in Virtual Reality Exposure Therapy}

\siksitem{Milan Petkovic}{UT}{Content-Based Video Retrieval Supported by Database Technology}

\siksitem{Jos Lehmann}{UvA}{Causation in Artificial Intelligence and Law: A modelling approach}

\siksitem{Boris van Schooten}{UT}{Development and specification of virtual environments}

\siksitem{Machiel Jansen}{UvA}{Formal Explorations of Knowledge Intensive Tasks}

\siksitem{Yongping Ran}{UM}{Repair Based Scheduling}

\siksitem{Rens Kortmann}{UM}{The resolution of visually guided behaviour}

\siksitem{Andreas Lincke}{UvT}{Electronic Business Negotiation: Some experimental studies on the interaction between medium, innovation context and culture}

\siksitem{Simon Keizer}{UT}{Reasoning under Uncertainty in Natural Language Dialogue using Bayesian Networks}

\siksitem{Roeland Ordelman}{UT}{Dutch speech recognition in multimedia information retrieval}

\siksitem{Jeroen Donkers}{UM}{Nosce Hostem: Searching with Opponent Models}

\siksitem{Stijn Hoppenbrouwers}{KUN}{Freezing Language: Conceptualisation Processes across ICT-Supported Organisations}

\siksitem{Mathijs de Weerdt}{TUD}{Plan Merging in Multi-Agent Systems}

\siksitem{Menzo Windhouwer}{CWI}{Feature Grammar Systems: Incremental Maintenance of Indexes to Digital Media Warehouses}

\siksitem{David Jansen}{UT}{Extensions of Statecharts with Probability, Time, and Stochastic Timing}

\siksitem{Levente Kocsis}{UM}{Learning Search Decisions}

\end{sikslist}
\begin{sikslist}{2004}

\siksitem{Virginia Dignum}{UU}{A Model for Organizational Interaction: Based on Agents, Founded in Logic}

\siksitem{Lai Xu}{UvT}{Monitoring Multi-party Contracts for E-business}

\siksitem{Perry Groot}{VUA}{A Theoretical and Empirical Analysis of Approximation in Symbolic Problem Solving}

\siksitem{Chris van Aart}{UvA}{Organizational Principles for Multi-Agent Architectures}

\siksitem{Viara Popova}{EUR}{Knowledge discovery and monotonicity}

\siksitem{Bart-Jan Hommes}{TUD}{The Evaluation of Business Process Modeling Techniques}

\siksitem{Elise Boltjes}{UM}{Voorbeeldig onderwijs: voorbeeldgestuurd onderwijs, een opstap naar abstract denken, vooral voor meisjes}

\siksitem{Joop Verbeek}{UM}{Politie en de Nieuwe Internationale Informatiemarkt, Grensregionale politi\"ele gegevensuitwisseling en digitale expertise}

\siksitem{Martin Caminada}{VUA}{For the Sake of the Argument: explorations into argument-based reasoning}

\siksitem{Suzanne Kabel}{UvA}{Knowledge-rich indexing of learning-objects}

\siksitem{Michel Klein}{VUA}{Change Management for Distributed Ontologies}

\siksitem{The Duy Bui}{UT}{Creating emotions and facial expressions for embodied agents}

\siksitem{Wojciech Jamroga}{UT}{Using Multiple Models of Reality: On Agents who Know how to Play}

\siksitem{Paul Harrenstein}{UU}{Logic in Conflict. Logical Explorations in Strategic Equilibrium}

\siksitem{Arno Knobbe}{UU}{Multi-Relational Data Mining}

\siksitem{Federico Divina}{VUA}{Hybrid Genetic Relational Search for Inductive Learning}

\siksitem{Mark Winands}{UM}{Informed Search in Complex Games}

\siksitem{Vania Bessa Machado}{UvA}{Supporting the Construction of Qualitative Knowledge Models}

\siksitem{Thijs Westerveld}{UT}{Using generative probabilistic models for multimedia retrieval}

\siksitem{Madelon Evers}{Nyenrode}{Learning from Design: facilitating multidisciplinary design teams}

\end{sikslist}
\begin{sikslist}{2005}

\siksitem{Floor Verdenius}{UvA}{Methodological Aspects of Designing Induction-Based Applications}

\siksitem{Erik van der Werf}{UM}{AI techniques for the game of Go}

\siksitem{Franc Grootjen}{RUN}{A Pragmatic Approach to the Conceptualisation of Language}

\siksitem{Nirvana Meratnia}{UT}{Towards Database Support for Moving Object data}

\siksitem{Gabriel Infante-Lopez}{UvA}{Two-Level Probabilistic Grammars for Natural Language Parsing}

\siksitem{Pieter Spronck}{UM}{Adaptive Game AI}

\siksitem{Flavius Frasincar}{TUe}{Hypermedia Presentation Generation for Semantic Web Information Systems}

\siksitem{Richard Vdovjak}{TUe}{A Model-driven Approach for Building Distributed Ontology-based Web Applications}

\siksitem{Jeen Broekstra}{VUA}{Storage, Querying and Inferencing for Semantic Web Languages}

\siksitem{Anders Bouwer}{UvA}{Explaining Behaviour: Using Qualitative Simulation in Interactive Learning Environments}

\siksitem{Elth Ogston}{VUA}{Agent Based Matchmaking and Clustering: A Decentralized Approach to Search}

\siksitem{Csaba Boer}{EUR}{Distributed Simulation in Industry}

\siksitem{Fred Hamburg}{UL}{Een Computermodel voor het Ondersteunen van Euthanasiebeslissingen}

\siksitem{Borys Omelayenko}{VUA}{Web-Service configuration on the Semantic Web: Exploring how semantics meets pragmatics}

\siksitem{Tibor Bosse}{VUA}{Analysis of the Dynamics of Cognitive Processes}

\siksitem{Joris Graaumans}{UU}{Usability of XML Query Languages}

\siksitem{Boris Shishkov}{TUD}{Software Specification Based on Re-usable Business Components}

\siksitem{Danielle Sent}{UU}{Test-selection strategies for probabilistic networks}

\siksitem{Michel van Dartel}{UM}{Situated Representation}

\siksitem{Cristina Coteanu}{UL}{Cyber Consumer Law, State of the Art and Perspectives}

\siksitem{Wijnand Derks}{UT}{Improving Concurrency and Recovery in Database Systems by Exploiting Application Semantics}

\end{sikslist}
\begin{sikslist}{2006}

\siksitem{Samuil Angelov}{TUe}{Foundations of B2B Electronic Contracting}

\siksitem{Cristina Chisalita}{VUA}{Contextual issues in the design and use of information technology in organizations}

\siksitem{Noor Christoph}{UvA}{The role of metacognitive skills in learning to solve problems}

\siksitem{Marta Sabou}{VUA}{Building Web Service Ontologies}

\siksitem{Cees Pierik}{UU}{Validation Techniques for Object-Oriented Proof Outlines}

\siksitem{Ziv Baida}{VUA}{Software-aided Service Bundling: Intelligent Methods \& Tools for Graphical Service Modeling}

\siksitem{Marko Smiljanic}{UT}{XML schema matching: balancing efficiency and effectiveness by means of clustering}

\siksitem{Eelco Herder}{UT}{Forward, Back and Home Again: Analyzing User Behavior on the Web}

\siksitem{Mohamed Wahdan}{UM}{Automatic Formulation of the Auditor's Opinion}

\siksitem{Ronny Siebes}{VUA}{Semantic Routing in Peer-to-Peer Systems}

\siksitem{Joeri van Ruth}{UT}{Flattening Queries over Nested Data Types}

\siksitem{Bert Bongers}{VUA}{Interactivation: Towards an e-cology of people, our technological environment, and the arts}

\siksitem{Henk-Jan Lebbink}{UU}{Dialogue and Decision Games for Information Exchanging Agents}

\siksitem{Johan Hoorn}{VUA}{Software Requirements: Update, Upgrade, Redesign - towards a Theory of Requirements Change}

\siksitem{Rainer Malik}{UU}{CONAN: Text Mining in the Biomedical Domain}

\siksitem{Carsten Riggelsen}{UU}{Approximation Methods for Efficient Learning of Bayesian Networks}

\siksitem{Stacey Nagata}{UU}{User Assistance for Multitasking with Interruptions on a Mobile Device}

\siksitem{Valentin Zhizhkun}{UvA}{Graph transformation for Natural Language Processing}

\siksitem{Birna van Riemsdijk}{UU}{Cognitive Agent Programming: A Semantic Approach}

\siksitem{Marina Velikova}{UvT}{Monotone models for prediction in data mining}

\siksitem{Bas van Gils}{RUN}{Aptness on the Web}

\siksitem{Paul de Vrieze}{RUN}{Fundaments of Adaptive Personalisation}

\siksitem{Ion Juvina}{UU}{Development of Cognitive Model for Navigating on the Web}

\siksitem{Laura Hollink}{VUA}{Semantic Annotation for Retrieval of Visual Resources}

\siksitem{Madalina Drugan}{UU}{Conditional log-likelihood MDL and Evolutionary MCMC}

\siksitem{Vojkan Mihajlovic}{UT}{Score Region Algebra: A Flexible Framework for Structured Information Retrieval}

\siksitem{Stefano Bocconi}{CWI}{Vox Populi: generating video documentaries from semantically annotated media repositories}

\siksitem{Borkur Sigurbjornsson}{UvA}{Focused Information Access using XML Element Retrieval}

\end{sikslist}
\begin{sikslist}{2007}

\siksitem{Kees Leune}{UvT}{Access Control and Service-Oriented Architectures}

\siksitem{Wouter Teepe}{RUG}{Reconciling Information Exchange and Confidentiality: A Formal Approach}

\siksitem{Peter Mika}{VUA}{Social Networks and the Semantic Web}

\siksitem{Jurriaan van Diggelen}{UU}{Achieving Semantic Interoperability in Multi-agent Systems: a dialogue-based approach}

\siksitem{Bart Schermer}{UL}{Software Agents, Surveillance, and the Right to Privacy: a Legislative Framework for Agent-enabled Surveillance}

\siksitem{Gilad Mishne}{UvA}{Applied Text Analytics for Blogs}

\siksitem{Natasa Jovanovic'}{UT}{To Whom It May Concern: Addressee Identification in Face-to-Face Meetings}

\siksitem{Mark Hoogendoorn}{VUA}{Modeling of Change in Multi-Agent Organizations}

\siksitem{David Mobach}{VUA}{Agent-Based Mediated Service Negotiation}

\siksitem{Huib Aldewereld}{UU}{Autonomy vs. Conformity: an Institutional Perspective on Norms and Protocols}

\siksitem{Natalia Stash}{TUe}{Incorporating Cognitive/Learning Styles in a General-Purpose Adaptive Hypermedia System}

\siksitem{Marcel van Gerven}{RUN}{Bayesian Networks for Clinical Decision Support: A Rational Approach to Dynamic Decision-Making under Uncertainty}

\siksitem{Rutger Rienks}{UT}{Meetings in Smart Environments: Implications of Progressing Technology}

\siksitem{Niek Bergboer}{UM}{Context-Based Image Analysis}

\siksitem{Joyca Lacroix}{UM}{NIM: a Situated Computational Memory Model}

\siksitem{Davide Grossi}{UU}{Designing Invisible Handcuffs. Formal investigations in Institutions and Organizations for Multi-agent Systems}

\siksitem{Theodore Charitos}{UU}{Reasoning with Dynamic Networks in Practice}

\siksitem{Bart Orriens}{UvT}{On the development an management of adaptive business collaborations}

\siksitem{David Levy}{UM}{Intimate relationships with artificial partners}

\siksitem{Slinger Jansen}{UU}{Customer Configuration Updating in a Software Supply Network}

\siksitem{Karianne Vermaas}{UU}{Fast diffusion and broadening use: A research on residential adoption and usage of broadband internet in the Netherlands between 2001 and 2005}

\siksitem{Zlatko Zlatev}{UT}{Goal-oriented design of value and process models from patterns}

\siksitem{Peter Barna}{TUe}{Specification of Application Logic in Web Information Systems}

\siksitem{Georgina Ram\'\i rez Camps}{CWI}{Structural Features in XML Retrieval}

\siksitem{Joost Schalken}{VUA}{Empirical Investigations in Software Process Improvement}

\end{sikslist}
\begin{sikslist}{2008}

\siksitem{Katalin Boer-Sorb\'an}{EUR}{Agent-Based Simulation of Financial Markets: A modular, continuous-time approach}

\siksitem{Alexei Sharpanskykh}{VUA}{On Computer-Aided Methods for Modeling and Analysis of Organizations}

\siksitem{Vera Hollink}{UvA}{Optimizing hierarchical menus: a usage-based approach}

\siksitem{Ander de Keijzer}{UT}{Management of Uncertain Data: towards unattended integration}

\siksitem{Bela Mutschler}{UT}{Modeling and simulating causal dependencies on process-aware information systems from a cost perspective}

\siksitem{Arjen Hommersom}{RUN}{On the Application of Formal Methods to Clinical Guidelines, an Artificial Intelligence Perspective}

\siksitem{Peter van Rosmalen}{OU}{Supporting the tutor in the design and support of adaptive e-learning}

\siksitem{Janneke Bolt}{UU}{Bayesian Networks: Aspects of Approximate Inference}

\siksitem{Christof van Nimwegen}{UU}{The paradox of the guided user: assistance can be counter-effective}

\siksitem{Wauter Bosma}{UT}{Discourse oriented summarization}

\siksitem{Vera Kartseva}{VUA}{Designing Controls for Network Organizations: A Value-Based Approach}

\siksitem{Jozsef Farkas}{RUN}{A Semiotically Oriented Cognitive Model of Knowledge Representation}

\siksitem{Caterina Carraciolo}{UvA}{Topic Driven Access to Scientific Handbooks}

\siksitem{Arthur van Bunningen}{UT}{Context-Aware Querying: Better Answers with Less Effort}

\siksitem{Martijn van Otterlo}{UT}{The Logic of Adaptive Behavior: Knowledge Representation and Algorithms for the Markov Decision Process Framework in First-Order Domains}

\siksitem{Henriette van Vugt}{VUA}{Embodied agents from a user's perspective}

\siksitem{Martin Op 't Land}{TUD}{Applying Architecture and Ontology to the Splitting and Allying of Enterprises}

\siksitem{Guido de Croon}{UM}{Adaptive Active Vision}

\siksitem{Henning Rode}{UT}{From Document to Entity Retrieval: Improving Precision and Performance of Focused Text Search}

\siksitem{Rex Arendsen}{UvA}{Geen bericht, goed bericht. Een onderzoek naar de effecten van de introductie van elektronisch berichtenverkeer met de overheid op de administratieve lasten van bedrijven}

\siksitem{Krisztian Balog}{UvA}{People Search in the Enterprise}

\siksitem{Henk Koning}{UU}{Communication of IT-Architecture}

\siksitem{Stefan Visscher}{UU}{Bayesian network models for the management of ventilator-associated pneumonia}

\siksitem{Zharko Aleksovski}{VUA}{Using background knowledge in ontology matching}

\siksitem{Geert Jonker}{UU}{Efficient and Equitable Exchange in Air Traffic Management Plan Repair using Spender-signed Currency}

\siksitem{Marijn Huijbregts}{UT}{Segmentation, Diarization and Speech Transcription: Surprise Data Unraveled}

\siksitem{Hubert Vogten}{OU}{Design and Implementation Strategies for IMS Learning Design}

\siksitem{Ildiko Flesch}{RUN}{On the Use of Independence Relations in Bayesian Networks}

\siksitem{Dennis Reidsma}{UT}{Annotations and Subjective Machines: Of Annotators, Embodied Agents, Users, and Other Humans}

\siksitem{Wouter van Atteveldt}{VUA}{Semantic Network Analysis: Techniques for Extracting, Representing and Querying Media Content}

\siksitem{Loes Braun}{UM}{Pro-Active Medical Information Retrieval}

\siksitem{Trung H. Bui}{UT}{Toward Affective Dialogue Management using Partially Observable Markov Decision Processes}

\siksitem{Frank Terpstra}{UvA}{Scientific Workflow Design: theoretical and practical issues}

\siksitem{Jeroen de Knijf}{UU}{Studies in Frequent Tree Mining}

\siksitem{Ben Torben Nielsen}{UvT}{Dendritic morphologies: function shapes structure}

\end{sikslist}
\begin{sikslist}{2009}

\siksitem{Rasa Jurgelenaite}{RUN}{Symmetric Causal Independence Models}

\siksitem{Willem Robert van Hage}{VUA}{Evaluating Ontology-Alignment Techniques}

\siksitem{Hans Stol}{UvT}{A Framework for Evidence-based Policy Making Using IT}

\siksitem{Josephine Nabukenya}{RUN}{Improving the Quality of Organisational Policy Making using Collaboration Engineering}

\siksitem{Sietse Overbeek}{RUN}{Bridging Supply and Demand for Knowledge Intensive Tasks: Based on Knowledge, Cognition, and Quality}

\siksitem{Muhammad Subianto}{UU}{Understanding Classification}

\siksitem{Ronald Poppe}{UT}{Discriminative Vision-Based Recovery and Recognition of Human Motion}

\siksitem{Volker Nannen}{VUA}{Evolutionary Agent-Based Policy Analysis in Dynamic Environments}

\siksitem{Benjamin Kanagwa}{RUN}{Design, Discovery and Construction of Service-oriented Systems}

\siksitem{Jan Wielemaker}{UvA}{Logic programming for knowledge-intensive interactive applications}

\siksitem{Alexander Boer}{UvA}{Legal Theory, Sources of Law \& the Semantic Web}

\siksitem{Peter Massuthe}{TUE, Humboldt-Universitaet zu Berlin}{Operating Guidelines for Services}

\siksitem{Steven de Jong}{UM}{Fairness in Multi-Agent Systems}

\siksitem{Maksym Korotkiy}{VUA}{From ontology-enabled services to service-enabled ontologies (making ontologies work in e-science with ONTO-SOA)}

\siksitem{Rinke Hoekstra}{UvA}{Ontology Representation: Design Patterns and Ontologies that Make Sense}

\siksitem{Fritz Reul}{UvT}{New Architectures in Computer Chess}

\siksitem{Laurens van der Maaten}{UvT}{Feature Extraction from Visual Data}

\siksitem{Fabian Groffen}{CWI}{Armada, An Evolving Database System}

\siksitem{Valentin Robu}{CWI}{Modeling Preferences, Strategic Reasoning and Collaboration in Agent-Mediated Electronic Markets}

\siksitem{Bob van der Vecht}{UU}{Adjustable Autonomy: Controling Influences on Decision Making}

\siksitem{Stijn Vanderlooy}{UM}{Ranking and Reliable Classification}

\siksitem{Pavel Serdyukov}{UT}{Search For Expertise: Going beyond direct evidence}

\siksitem{Peter Hofgesang}{VUA}{Modelling Web Usage in a Changing Environment}

\siksitem{Annerieke Heuvelink}{VUA}{Cognitive Models for Training Simulations}

\siksitem{Alex van Ballegooij}{CWI}{RAM: Array Database Management through Relational Mapping}

\siksitem{Fernando Koch}{UU}{An Agent-Based Model for the Development of Intelligent Mobile Services}

\siksitem{Christian Glahn}{OU}{Contextual Support of social Engagement and Reflection on the Web}

\siksitem{Sander Evers}{UT}{Sensor Data Management with Probabilistic Models}

\siksitem{Stanislav Pokraev}{UT}{Model-Driven Semantic Integration of Service-Oriented Applications}

\siksitem{Marcin Zukowski}{CWI}{Balancing vectorized query execution with bandwidth-optimized storage}

\siksitem{Sofiya Katrenko}{UvA}{A Closer Look at Learning Relations from Text}

\siksitem{Rik Farenhorst}{VUA}{Architectural Knowledge Management: Supporting Architects and Auditors}

\siksitem{Khiet Truong}{UT}{How Does Real Affect Affect Affect Recognition In Speech?}

\siksitem{Inge van de Weerd}{UU}{Advancing in Software Product Management: An Incremental Method Engineering Approach}

\siksitem{Wouter Koelewijn}{UL}{Privacy en Politiegegevens: Over geautomatiseerde normatieve informatie-uitwisseling}

\siksitem{Marco Kalz}{OUN}{Placement Support for Learners in Learning Networks}

\siksitem{Hendrik Drachsler}{OUN}{Navigation Support for Learners in Informal Learning Networks}

\siksitem{Riina Vuorikari}{OU}{Tags and self-organisation: a metadata ecology for learning resources in a multilingual context}

\siksitem{Christian Stahl}{TUE, Humboldt-Universitaet zu Berlin}{Service Substitution: A Behavioral Approach Based on Petri Nets}

\siksitem{Stephan Raaijmakers}{UvT}{Multinomial Language Learning: Investigations into the Geometry of Language}

\siksitem{Igor Berezhnyy}{UvT}{Digital Analysis of Paintings}

\siksitem{Toine Bogers}{UvT}{Recommender Systems for Social Bookmarking}

\siksitem{Virginia Nunes Leal Franqueira}{UT}{Finding Multi-step Attacks in Computer Networks using Heuristic Search and Mobile Ambients}

\siksitem{Roberto Santana Tapia}{UT}{Assessing Business-IT Alignment in Networked Organizations}

\siksitem{Jilles Vreeken}{UU}{Making Pattern Mining Useful}

\siksitem{Loredana Afanasiev}{UvA}{Querying XML: Benchmarks and Recursion}

\end{sikslist}
\begin{sikslist}{2010}

\siksitem{Matthijs van Leeuwen}{UU}{Patterns that Matter}

\siksitem{Ingo Wassink}{UT}{Work flows in Life Science}

\siksitem{Joost Geurts}{CWI}{A Document Engineering Model and Processing Framework for Multimedia documents}

\siksitem{Olga Kulyk}{UT}{Do You Know What I Know? Situational Awareness of Co-located Teams in Multidisplay Environments}

\siksitem{Claudia Hauff}{UT}{Predicting the Effectiveness of Queries and Retrieval Systems}

\siksitem{Sander Bakkes}{UvT}{Rapid Adaptation of Video Game AI}

\siksitem{Wim Fikkert}{UT}{Gesture interaction at a Distance}

\siksitem{Krzysztof Siewicz}{UL}{Towards an Improved Regulatory Framework of Free Software. Protecting user freedoms in a world of software communities and eGovernments}

\siksitem{Hugo Kielman}{UL}{A Politiele gegevensverwerking en Privacy, Naar een effectieve waarborging}

\siksitem{Rebecca Ong}{UL}{Mobile Communication and Protection of Children}

\siksitem{Adriaan Ter Mors}{TUD}{The world according to MARP: Multi-Agent Route Planning}

\siksitem{Susan van den Braak}{UU}{Sensemaking software for crime analysis}

\siksitem{Gianluigi Folino}{RUN}{High Performance Data Mining using Bio-inspired techniques}

\siksitem{Sander van Splunter}{VUA}{Automated Web Service Reconfiguration}

\siksitem{Lianne Bodenstaff}{UT}{Managing Dependency Relations in Inter-Organizational Models}

\siksitem{Sicco Verwer}{TUD}{Efficient Identification of Timed Automata, theory and practice}

\siksitem{Spyros Kotoulas}{VUA}{Scalable Discovery of Networked Resources: Algorithms, Infrastructure, Applications}

\siksitem{Charlotte Gerritsen}{VUA}{Caught in the Act: Investigating Crime by Agent-Based Simulation}

\siksitem{Henriette Cramer}{UvA}{People's Responses to Autonomous and Adaptive Systems}

\siksitem{Ivo Swartjes}{UT}{Whose Story Is It Anyway? How Improv Informs Agency and Authorship of Emergent Narrative}

\siksitem{Harold van Heerde}{UT}{Privacy-aware data management by means of data degradation}

\siksitem{Michiel Hildebrand}{CWI}{End-user Support for Access to\\ Heterogeneous Linked Data}

\siksitem{Bas Steunebrink}{UU}{The Logical Structure of Emotions}

\siksitem{Zulfiqar Ali Memon}{VUA}{Modelling Human-Awareness for Ambient Agents: A Human Mindreading Perspective}

\siksitem{Ying Zhang}{CWI}{XRPC: Efficient Distributed Query Processing on Heterogeneous XQuery Engines}

\siksitem{Marten Voulon}{UL}{Automatisch contracteren}

\siksitem{Arne Koopman}{UU}{Characteristic Relational Patterns}

\siksitem{Stratos Idreos}{CWI}{Database Cracking: Towards Auto-tuning Database Kernels}

\siksitem{Marieke van Erp}{UvT}{Accessing Natural History: Discoveries in data cleaning, structuring, and retrieval}

\siksitem{Victor de Boer}{UvA}{Ontology Enrichment from Heterogeneous Sources on the Web}

\siksitem{Marcel Hiel}{UvT}{An Adaptive Service Oriented Architecture: Automatically solving Interoperability Problems}

\siksitem{Robin Aly}{UT}{Modeling Representation Uncertainty in Concept-Based Multimedia Retrieval}

\siksitem{Teduh Dirgahayu}{UT}{Interaction Design in Service Compositions}

\siksitem{Dolf Trieschnigg}{UT}{Proof of Concept: Concept-based Biomedical Information Retrieval}

\siksitem{Jose Janssen}{OU}{Paving the Way for Lifelong Learning: Facilitating competence development through a learning path specification}

\siksitem{Niels Lohmann}{TUe}{Correctness of services and their composition}

\siksitem{Dirk Fahland}{TUe}{From Scenarios to components}

\siksitem{Ghazanfar Farooq Siddiqui}{VUA}{Integrative modeling of emotions in virtual agents}

\siksitem{Mark van Assem}{VUA}{Converting and Integrating Vocabularies for the Semantic Web}

\siksitem{Guillaume Chaslot}{UM}{Monte-Carlo Tree Search}

\siksitem{Sybren de Kinderen}{VUA}{Needs-driven service bundling in a multi-supplier setting: the computational e3-service approach}

\siksitem{Peter van Kranenburg}{UU}{A Computational Approach to Content-Based Retrieval of Folk Song Melodies}

\siksitem{Pieter Bellekens}{TUe}{An Approach towards Context-sensitive and User-adapted Access to Heterogeneous Data Sources, Illustrated in the Television Domain}

\siksitem{Vasilios Andrikopoulos}{UvT}{A theory and model for the evolution of software services}

\siksitem{Vincent Pijpers}{VUA}{e3alignment: Exploring Inter-Organizational Business-ICT Alignment}

\siksitem{Chen Li}{UT}{Mining Process Model Variants: Challenges, Techniques, Examples}

\siksitem{Jahn-Takeshi Saito}{UM}{Solving difficult game positions}

\siksitem{Bouke Huurnink}{UvA}{Search in Audiovisual Broadcast Archives}

\siksitem{Alia Khairia Amin}{CWI}{Understanding and supporting information seeking tasks in multiple sources}

\siksitem{Peter-Paul van Maanen}{VUA}{Adaptive Support for Human-Computer Teams: Exploring the Use of Cognitive Models of Trust and Attention}

\siksitem{Edgar Meij}{UvA}{Combining Concepts and Language Models for Information Access}

\end{sikslist}
\begin{sikslist}{2011}

\siksitem{Botond Cseke}{RUN}{Variational Algorithms for Bayesian Inference in Latent Gaussian Models}

\siksitem{Nick Tinnemeier}{UU}{Organizing Agent Organizations. Syntax and Operational Semantics of an Organization-Oriented Programming Language}

\siksitem{Jan Martijn van der Werf}{TUe}{Compositional Design and Verification of Component-Based Information Systems}

\siksitem{Hado van Hasselt}{UU}{Insights in Reinforcement Learning: Formal analysis and empirical evaluation of temporal-difference}

\siksitem{Base van der Raadt}{VUA}{Enterprise Architecture Coming of Age: Increasing the Performance of an Emerging Discipline}

\siksitem{Yiwen Wang}{TUe}{Semantically-Enhanced Recommendations in Cultural Heritage}

\siksitem{Yujia Cao}{UT}{Multimodal Information Presentation for High Load Human Computer Interaction}

\siksitem{Nieske Vergunst}{UU}{BDI-based Generation of Robust Task-Oriented Dialogues}

\siksitem{Tim de Jong}{OU}{Contextualised Mobile Media for Learning}

\siksitem{Bart Bogaert}{UvT}{Cloud Content Contention}

\siksitem{Dhaval Vyas}{UT}{Designing for Awareness: An Experience-focused HCI Perspective}

\siksitem{Carmen Bratosin}{TUe}{Grid Architecture for Distributed Process Mining}

\siksitem{Xiaoyu Mao}{UvT}{Airport under Control. Multiagent Scheduling for Airport Ground Handling}

\siksitem{Milan Lovric}{EUR}{Behavioral Finance and Agent-Based Artificial Markets}

\siksitem{Marijn Koolen}{UvA}{The Meaning of Structure: the Value of Link Evidence for Information Retrieval}

\siksitem{Maarten Schadd}{UM}{Selective Search in Games of Different Complexity}

\siksitem{Jiyin He}{UvA}{Exploring Topic Structure: Coherence, Diversity and Relatedness}

\siksitem{Mark Ponsen}{UM}{Strategic Decision-Making in complex games}

\siksitem{Ellen Rusman}{OU}{The Mind's Eye on Personal Profiles}

\siksitem{Qing Gu}{VUA}{Guiding service-oriented software engineering: A view-based approach}

\siksitem{Linda Terlouw}{TUD}{Modularization and Specification of Service-Oriented Systems}

\siksitem{Junte Zhang}{UvA}{System Evaluation of Archival Description and Access}

\siksitem{Wouter Weerkamp}{UvA}{Finding People and their Utterances in Social Media}

\siksitem{Herwin van Welbergen}{UT}{Behavior Generation for Interpersonal Coordination with Virtual Humans On Specifying, Scheduling and Realizing Multimodal Virtual Human Behavior}

\siksitem{Syed Waqar ul Qounain Jaffry}{VUA}{Analysis and Validation of Models for Trust Dynamics}

\siksitem{Matthijs Aart Pontier}{VUA}{Virtual Agents for Human Communication: Emotion Regulation and Involvement-Distance Trade-Offs in Embodied Conversational Agents and Robots}

\siksitem{Aniel Bhulai}{VUA}{Dynamic website optimization through autonomous management of design patterns}

\siksitem{Rianne Kaptein}{UvA}{Effective Focused Retrieval by Exploiting Query Context and Document Structure}

\siksitem{Faisal Kamiran}{TUe}{Discrimination-aware Classification}

\siksitem{Egon van den Broek}{UT}{Affective Signal Processing (ASP): Unraveling the mystery of emotions}

\siksitem{Ludo Waltman}{EUR}{Computational and Game-Theoretic Approaches for Modeling Bounded Rationality}

\siksitem{Nees-Jan van Eck}{EUR}{Methodological Advances in Bibliometric Mapping of Science}

\siksitem{Tom van der Weide}{UU}{Arguing to Motivate Decisions}

\siksitem{Paolo Turrini}{UU}{Strategic Reasoning in Interdependence: Logical and Game-theoretical Investigations}

\siksitem{Maaike Harbers}{UU}{Explaining Agent Behavior in Virtual Training}

\siksitem{Erik van der Spek}{UU}{Experiments in serious game design: a cognitive approach}

\siksitem{Adriana Burlutiu}{RUN}{Machine Learning for Pairwise Data, Applications for Preference Learning and Supervised Network Inference}

\siksitem{Nyree Lemmens}{UM}{Bee-inspired Distributed Optimization}

\siksitem{Joost Westra}{UU}{Organizing Adaptation using Agents in Serious Games}

\siksitem{Viktor Clerc}{VUA}{Architectural Knowledge Management in Global Software Development}

\siksitem{Luan Ibraimi}{UT}{Cryptographically Enforced Distributed Data Access Control}

\siksitem{Michal Sindlar}{UU}{Explaining Behavior through Mental State Attribution}

\siksitem{Henk van der Schuur}{UU}{Process Improvement through Software Operation Knowledge}

\siksitem{Boris Reuderink}{UT}{Robust Brain-Computer Interfaces}

\siksitem{Herman Stehouwer}{UvT}{Statistical Language Models for Alternative Sequence Selection}

\siksitem{Beibei Hu}{TUD}{Towards Contextualized Information Delivery: A Rule-based Architecture for the Domain of Mobile Police Work}

\siksitem{Azizi Bin Ab Aziz}{VUA}{Exploring Computational Models for Intelligent Support of Persons with Depression}

\siksitem{Mark Ter Maat}{UT}{Response Selection and Turn-taking for a Sensitive Artificial Listening Agent}

\siksitem{Andreea Niculescu}{UT}{Conversational interfaces for task-oriented spoken dialogues: design aspects influencing interaction quality}

\end{sikslist}
\begin{sikslist}{2012}

\siksitem{Terry Kakeeto}{UvT}{Relationship Marketing for SMEs in Uganda}

\siksitem{Muhammad Umair}{VUA}{Adaptivity, emotion, and Rationality in Human and Ambient Agent Models}

\siksitem{Adam Vanya}{VUA}{Supporting Architecture Evolution by Mining Software Repositories}

\siksitem{Jurriaan Souer}{UU}{Development of Content Management System-based Web Applications}

\siksitem{Marijn Plomp}{UU}{Maturing Interorganisational Information Systems}

\siksitem{Wolfgang Reinhardt}{OU}{Awareness Support for Knowledge Workers in Research Networks}

\siksitem{Rianne van Lambalgen}{VUA}{When the Going Gets Tough: Exploring Agent-based Models of Human Performance under Demanding Conditions}

\siksitem{Gerben de Vries}{UvA}{Kernel Methods for Vessel Trajectories}

\siksitem{Ricardo Neisse}{UT}{Trust and Privacy Management Support for Context-Aware Service Platforms}

\siksitem{David Smits}{TUe}{Towards a Generic Distributed Adaptive Hypermedia Environment}

\siksitem{J. C. B. Rantham Prabhakara}{TUe}{Process Mining in the Large: Preprocessing, Discovery, and Diagnostics}

\siksitem{Kees van der Sluijs}{TUe}{Model Driven Design and Data Integration in Semantic Web Information Systems}

\siksitem{Suleman Shahid}{UvT}{Fun and Face: Exploring non-verbal expressions of emotion during playful interactions}

\siksitem{Evgeny Knutov}{TUe}{Generic Adaptation Framework for Unifying Adaptive Web-based Systems}

\siksitem{Natalie van der Wal}{VUA}{Social Agents. Agent-Based Modelling of Integrated Internal and Social Dynamics of Cognitive and Affective Processes}

\siksitem{Fiemke Both}{VUA}{Helping people by understanding them: Ambient Agents supporting task execution and depression treatment}

\siksitem{Amal Elgammal}{UvT}{Towards a Comprehensive Framework for Business Process Compliance}

\siksitem{Eltjo Poort}{VUA}{Improving Solution Architecting Practices}

\siksitem{Helen Schonenberg}{TUe}{What's Next? Operational Support for Business Process Execution}

\siksitem{Ali Bahramisharif}{RUN}{Covert Visual Spatial Attention, a Robust Paradigm for Brain-Computer Interfacing}

\siksitem{Roberto Cornacchia}{TUD}{Querying Sparse Matrices for Information Retrieval}

\siksitem{Thijs Vis}{UvT}{Intelligence, politie en veiligheidsdienst: verenigbare grootheden?}

\siksitem{Christian Muehl}{UT}{Toward Affective Brain-Computer Interfaces: Exploring the Neurophysiology of Affect during Human Media Interaction}

\siksitem{Laurens van der Werff}{UT}{Evaluation of Noisy Transcripts for Spoken Document Retrieval}

\siksitem{Silja Eckartz}{UT}{Managing the Business Case Development in Inter-Organizational IT Projects: A Methodology and its Application}

\siksitem{Emile de Maat}{UvA}{Making Sense of Legal Text}

\siksitem{Hayrettin Gurkok}{UT}{Mind the Sheep! User Experience Evaluation \& Brain-Computer Interface Games}

\siksitem{Nancy Pascall}{UvT}{Engendering Technology Empowering Women}

\siksitem{Almer Tigelaar}{UT}{Peer-to-Peer Information Retrieval}

\siksitem{Alina Pommeranz}{TUD}{Designing Human-Centered Systems for Reflective Decision Making}

\siksitem{Emily Bagarukayo}{RUN}{A Learning by Construction Approach for Higher Order Cognitive Skills Improvement, Building Capacity and Infrastructure}

\siksitem{Wietske Visser}{TUD}{Qualitative multi-criteria preference representation and reasoning}

\siksitem{Rory Sie}{OUN}{Coalitions in Cooperation Networks (COCOON)}

\siksitem{Pavol Jancura}{RUN}{Evolutionary analysis in PPI networks and applications}

\siksitem{Evert Haasdijk}{VUA}{Never Too Old To Learn: On-line Evolution of Controllers in Swarm- and Modular Robotics}

\siksitem{Denis Ssebugwawo}{RUN}{Analysis and Evaluation of Collaborative Modeling Processes}

\siksitem{Agnes Nakakawa}{RUN}{A Collaboration Process for Enterprise Architecture Creation}

\siksitem{Selmar Smit}{VUA}{Parameter Tuning and Scientific Testing in Evolutionary Algorithms}

\siksitem{Hassan Fatemi}{UT}{Risk-aware design of value and coordination networks}

\siksitem{Agus Gunawan}{UvT}{Information Access for SMEs in Indonesia}

\siksitem{Sebastian Kelle}{OU}{Game Design Patterns for Learning}

\siksitem{Dominique Verpoorten}{OU}{Reflection Amplifiers in self-regulated Learning}

\siksitem{Anna Tordai}{VUA}{On Combining Alignment Techniques}

\siksitem{Benedikt Kratz}{UvT}{A Model and Language for Business-aware Transactions}

\siksitem{Simon Carter}{UvA}{Exploration and Exploitation of Multilingual Data for Statistical Machine Translation}

\siksitem{Manos Tsagkias}{UvA}{Mining Social Media: Tracking Content and Predicting Behavior}

\siksitem{Jorn Bakker}{TUe}{Handling Abrupt Changes in Evolving Time-series Data}

\siksitem{Michael Kaisers}{UM}{Learning against Learning: Evolutionary dynamics of reinforcement learning algorithms in strategic interactions}

\siksitem{Steven van Kervel}{TUD}{Ontologogy driven Enterprise Information Systems Engineering}

\siksitem{Jeroen de Jong}{TUD}{Heuristics in Dynamic Sceduling: a practical framework with a case study in elevator dispatching}

\end{sikslist}
\begin{sikslist}{2013}

\siksitem{Viorel Milea}{EUR}{News Analytics for Financial Decision Support}

\siksitem{Erietta Liarou}{CWI}{MonetDB/DataCell: Leveraging the Column-store Database Technology for Efficient and Scalable Stream Processing}

\siksitem{Szymon Klarman}{VUA}{Reasoning with Contexts in Description Logics}

\siksitem{Chetan Yadati}{TUD}{Coordinating autonomous planning and scheduling}

\siksitem{Dulce Pumareja}{UT}{Groupware Requirements Evolutions Patterns}

\siksitem{Romulo Goncalves}{CWI}{The Data Cyclotron: Juggling Data and Queries for a Data Warehouse Audience}

\siksitem{Giel van Lankveld}{UvT}{Quantifying Individual Player Differences}

\siksitem{Robbert-Jan Merk}{VUA}{Making enemies: cognitive modeling for opponent agents in fighter pilot simulators}

\siksitem{Fabio Gori}{RUN}{Metagenomic Data Analysis: Computational Methods and Applications}

\siksitem{Jeewanie Jayasinghe Arachchige}{UvT}{A Unified Modeling Framework for Service Design}

\siksitem{Evangelos Pournaras}{TUD}{Multi-level Reconfigurable Self-organization in Overlay Services}

\siksitem{Marian Razavian}{VUA}{Knowledge-driven Migration to Services}

\siksitem{Mohammad Safiri}{UT}{Service Tailoring: User-centric creation of integrated IT-based homecare services to support independent living of elderly}

\siksitem{Jafar Tanha}{UvA}{Ensemble Approaches to Semi-Supervised Learning Learning}

\siksitem{Daniel Hennes}{UM}{Multiagent Learning: Dynamic Games and Applications}

\siksitem{Eric Kok}{UU}{Exploring the practical benefits of argumentation in multi-agent deliberation}

\siksitem{Koen Kok}{VUA}{The PowerMatcher: Smart Coordination for the Smart Electricity Grid}

\siksitem{Jeroen Janssens}{UvT}{Outlier Selection and One-Class Classification}

\siksitem{Renze Steenhuizen}{TUD}{Coordinated Multi-Agent Planning and Scheduling}

\siksitem{Katja Hofmann}{UvA}{Fast and Reliable Online Learning to Rank for Information Retrieval}

\siksitem{Sander Wubben}{UvT}{Text-to-text generation by monolingual machine translation}

\siksitem{Tom Claassen}{RUN}{Causal Discovery and Logic}

\siksitem{Patricio de Alencar Silva}{UvT}{Value Activity Monitoring}

\siksitem{Haitham Bou Ammar}{UM}{Automated Transfer in Reinforcement Learning}

\siksitem{Agnieszka Anna Latoszek-Berendsen}{UM}{Intention-based Decision Support. A new way of representing and implementing clinical guidelines in a Decision Support System}

\siksitem{Alireza Zarghami}{UT}{Architectural Support for Dynamic Homecare Service Provisioning}

\siksitem{Mohammad Huq}{UT}{Inference-based Framework Managing Data Provenance}

\siksitem{Frans van der Sluis}{UT}{When Complexity becomes Interesting: An Inquiry into the Information eXperience}

\siksitem{Iwan de Kok}{UT}{Listening Heads}

\siksitem{Joyce Nakatumba}{TUe}{Resource-Aware Business Process Management: Analysis and Support}

\siksitem{Dinh Khoa Nguyen}{UvT}{Blueprint Model and Language for Engineering Cloud Applications}

\siksitem{Kamakshi Rajagopal}{OUN}{Networking For Learning: The role of Networking in a Lifelong Learner's Professional Development}

\siksitem{Qi Gao}{TUD}{User Modeling and Personalization in the Microblogging Sphere}

\siksitem{Kien Tjin-Kam-Jet}{UT}{Distributed Deep Web Search}

\siksitem{Abdallah El Ali}{UvA}{Minimal Mobile Human Computer Interaction}

\siksitem{Than Lam Hoang}{TUe}{Pattern Mining in Data Streams}

\siksitem{Dirk B\"orner}{OUN}{Ambient Learning Displays}

\siksitem{Eelco den Heijer}{VUA}{Autonomous Evolutionary Art}

\siksitem{Joop de Jong}{TUD}{A Method for Enterprise Ontology based Design of Enterprise Information Systems}

\siksitem{Pim Nijssen}{UM}{Monte-Carlo Tree Search for Multi-Player Games}

\siksitem{Jochem Liem}{UvA}{Supporting the Conceptual Modelling of Dynamic Systems: A Knowledge Engineering Perspective on Qualitative Reasoning}

\siksitem{L\'eon Planken}{TUD}{Algorithms for Simple Temporal Reasoning}

\siksitem{Marc Bron}{UvA}{Exploration and Contextualization through Interaction and Concepts}

\end{sikslist}
\begin{sikslist}{2014}

\siksitem{Nicola Barile}{UU}{Studies in Learning Monotone Models from Data}

\siksitem{Fiona Tuliyano}{RUN}{Combining System Dynamics with a Domain Modeling Method}

\siksitem{Sergio Raul Duarte Torres}{UT}{Information Retrieval for Children: Search Behavior and Solutions}

\siksitem{Hanna Jochmann-Mannak}{UT}{Websites for children: search strategies and interface design - Three studies on children's search performance and evaluation}

\siksitem{Jurriaan van Reijsen}{UU}{Knowledge Perspectives on Advancing Dynamic Capability}

\siksitem{Damian Tamburri}{VUA}{Supporting Networked Software Development}

\siksitem{Arya Adriansyah}{TUe}{Aligning Observed and Modeled Behavior}

\siksitem{Samur Araujo}{TUD}{Data Integration over Distributed and Heterogeneous Data Endpoints}

\siksitem{Philip Jackson}{UvT}{Toward Human-Level Artificial Intelligence: Representation and Computation of Meaning in Natural Language}

\siksitem{Ivan Salvador Razo Zapata}{VUA}{Service Value Networks}

\siksitem{Janneke van der Zwaan}{TUD}{An Empathic Virtual Buddy for Social Support}

\siksitem{Willem van Willigen}{VUA}{Look Ma, No Hands: Aspects of Autonomous Vehicle Control}

\siksitem{Arlette van Wissen}{VUA}{Agent-Based Support for Behavior Change: Models and Applications in Health and Safety Domains}

\siksitem{Yangyang Shi}{TUD}{Language Models With Meta-information}

\siksitem{Natalya Mogles}{VUA}{Agent-Based Analysis and Support of Human Functioning in Complex Socio-Technical Systems: Applications in Safety and Healthcare}

\siksitem{Krystyna Milian}{VUA}{Supporting trial recruitment and design by automatically interpreting eligibility criteria}

\siksitem{Kathrin Dentler}{VUA}{Computing healthcare quality indicators automatically: Secondary Use of Patient Data and Semantic Interoperability}

\siksitem{Mattijs Ghijsen}{UvA}{Methods and Models for the Design and Study of Dynamic Agent Organizations}

\siksitem{Vinicius Ramos}{TUe}{Adaptive Hypermedia Courses: Qualitative and Quantitative Evaluation and Tool Support}

\siksitem{Mena Habib}{UT}{Named Entity Extraction and Disambiguation for Informal Text: The Missing Link}

\siksitem{Kassidy Clark}{TUD}{Negotiation and Monitoring in Open Environments}

\siksitem{Marieke Peeters}{UU}{Personalized Educational Games: Developing agent-supported scenario-based training}

\siksitem{Eleftherios Sidirourgos}{UvA/CWI}{Space Efficient Indexes for the Big Data Era}

\siksitem{Davide Ceolin}{VUA}{Trusting Semi-structured Web Data}

\siksitem{Martijn Lappenschaar}{RUN}{New network models for the analysis of disease interaction}

\siksitem{Tim Baarslag}{TUD}{What to Bid and When to Stop}

\siksitem{Rui Jorge Almeida}{EUR}{Conditional Density Models Integrating Fuzzy and Probabilistic Representations of Uncertainty}

\siksitem{Anna Chmielowiec}{VUA}{Decentralized k-Clique Matching}

\siksitem{Jaap Kabbedijk}{UU}{Variability in Multi-Tenant Enterprise Software}

\siksitem{Peter de Cock}{UvT}{Anticipating Criminal Behaviour}

\siksitem{Leo van Moergestel}{UU}{Agent Technology in Agile Multiparallel Manufacturing and Product Support}

\siksitem{Naser Ayat}{UvA}{On Entity Resolution in Probabilistic Data}

\siksitem{Tesfa Tegegne}{RUN}{Service Discovery in eHealth}

\siksitem{Christina Manteli}{VUA}{The Effect of Governance in Global Software Development: Analyzing Transactive Memory Systems}

\siksitem{Joost van Ooijen}{UU}{Cognitive Agents in Virtual Worlds: A Middleware Design Approach}

\siksitem{Joos Buijs}{TUe}{Flexible Evolutionary Algorithms for Mining Structured Process Models}

\siksitem{Maral Dadvar}{UT}{Experts and Machines United Against Cyberbullying}

\siksitem{Danny Plass-Oude Bos}{UT}{Making brain-computer interfaces better: improving usability through post-processing}

\siksitem{Jasmina Maric}{UvT}{Web Communities, Immigration, and Social Capital}

\siksitem{Walter Omona}{RUN}{A Framework for Knowledge Management Using ICT in Higher Education}

\siksitem{Frederic Hogenboom}{EUR}{Automated Detection of Financial Events in News Text}

\siksitem{Carsten Eijckhof}{CWI/TUD}{Contextual Multidimensional Relevance Models}

\siksitem{Kevin Vlaanderen}{UU}{Supporting Process Improvement using Method Increments}

\siksitem{Paulien Meesters}{UvT}{Intelligent Blauw: Intelligence-gestuurde politiezorg in gebiedsgebonden eenheden}

\siksitem{Birgit Schmitz}{OUN}{Mobile Games for Learning: A Pattern-Based Approach}

\siksitem{Ke Tao}{TUD}{Social Web Data Analytics: Relevance, Redundancy, Diversity}

\siksitem{Shangsong Liang}{UvA}{Fusion and Diversification in Information Retrieval}

\end{sikslist}
\begin{sikslist}{2015}

\siksitem{Niels Netten}{UvA}{Machine Learning for Relevance of Information in Crisis Response}

\siksitem{Faiza Bukhsh}{UvT}{Smart auditing: Innovative Compliance Checking in Customs Controls}

\siksitem{Twan van Laarhoven}{RUN}{Machine learning for network data}

\siksitem{Howard Spoelstra}{OUN}{Collaborations in Open Learning Environments}

\siksitem{Christoph B\"osch}{UT}{Cryptographically Enforced Search Pattern Hiding}

\siksitem{Farideh Heidari}{TUD}{Business Process Quality Computation: Computing Non-Functional Requirements to Improve Business Processes}

\siksitem{Maria-Hendrike Peetz}{UvA}{Time-Aware Online Reputation Analysis}

\siksitem{Jie Jiang}{TUD}{Organizational Compliance: An agent-based model for designing and evaluating organizational interactions}

\siksitem{Randy Klaassen}{UT}{HCI Perspectives on Behavior Change Support Systems}

\siksitem{Henry Hermans}{OUN}{OpenU: design of an integrated system to support lifelong learning}

\siksitem{Yongming Luo}{TUe}{Designing algorithms for big graph datasets: A study of computing bisimulation and joins}

\siksitem{Julie M. Birkholz}{VUA}{Modi Operandi of Social Network Dynamics: The Effect of Context on Scientific Collaboration Networks}

\siksitem{Giuseppe Procaccianti}{VUA}{Energy-Efficient Software}

\siksitem{Bart van Straalen}{UT}{A cognitive approach to modeling bad news conversations}

\siksitem{Klaas Andries de Graaf}{VUA}{Ontology-based Software Architecture Documentation}

\siksitem{Changyun Wei}{UT}{Cognitive Coordination for Cooperative Multi-Robot Teamwork}

\siksitem{Andr\'e van Cleeff}{UT}{Physical and Digital Security Mechanisms: Properties, Combinations and Trade-offs}

\siksitem{Holger Pirk}{CWI}{Waste Not, Want Not!: Managing Relational Data in Asymmetric Memories}

\siksitem{Bernardo Tabuenca}{OUN}{Ubiquitous Technology for Lifelong Learners}

\siksitem{Lo\"\i s Vanh\'ee}{UU}{Using Culture and Values to Support Flexible Coordination}

\siksitem{Sibren Fetter}{OUN}{Using Peer-Support to Expand and Stabilize Online Learning}

\siksitem{Zhemin Zhu}{UT}{Co-occurrence Rate Networks}

\siksitem{Luit Gazendam}{VUA}{Cataloguer Support in Cultural Heritage}

\siksitem{Richard Berendsen}{UvA}{Finding People, Papers, and Posts: Vertical Search Algorithms and Evaluation}

\siksitem{Steven Woudenberg}{UU}{Bayesian Tools for Early Disease Detection}

\siksitem{Alexander Hogenboom}{EUR}{Sentiment Analysis of Text Guided by Semantics and Structure}

\siksitem{S\'andor H\'eman}{CWI}{Updating compressed colomn stores}

\siksitem{Janet Bagorogoza}{TiU}{KNOWLEDGE MANAGEMENT AND HIGH PERFORMANCE: The Uganda Financial Institutions Model for HPO}

\siksitem{Hendrik Baier}{UM}{Monte-Carlo Tree Search Enhancements for One-Player and Two-Player Domains}

\siksitem{Kiavash Bahreini}{OU}{Real-time Multimodal Emotion Recognition in E-Learning}

\siksitem{Yakup Ko\c c}{TUD}{On the robustness of Power Grids}

\siksitem{Jerome Gard}{UL}{Corporate Venture Management in SMEs}

\siksitem{Frederik Schadd}{TUD}{Ontology Mapping with Auxiliary Resources}

\siksitem{Victor de Graaf}{UT}{Gesocial Recommender Systems}

\siksitem{Jungxao Xu}{TUD}{Affective Body Language of Humanoid Robots: Perception and Effects in Human Robot Interaction}

\end{sikslist}
\begin{sikslist}{2016}

\siksitem{Syed Saiden Abbas}{RUN}{Recognition of Shapes by Humans and Machines}

\siksitem{Michiel Christiaan Meulendijk}{UU}{Optimizing medication reviews through decision support: prescribing a better pill to swallow}

\siksitem{Maya Sappelli}{RUN}{Knowledge Work in Context: User Centered Knowledge Worker Support}

\siksitem{Laurens Rietveld}{VUA}{Publishing and Consuming Linked Data}

\siksitem{Evgeny Sherkhonov}{UvA}{Expanded Acyclic Queries: Containment and an Application in Explaining Missing Answers}

\siksitem{Michel Wilson}{TUD}{Robust scheduling in an uncertain environment}

\siksitem{Jeroen de Man}{VUA}{Measuring and modeling negative emotions for virtual training}

\siksitem{Matje van de Camp}{TiU}{A Link to the Past: Constructing Historical Social Networks from Unstructured Data}

\siksitem{Archana Nottamkandath}{VUA}{Trusting Crowdsourced Information on Cultural Artefacts}

\siksitem{George Karafotias}{VUA}{Parameter Control for Evolutionary Algorithms}

\siksitem{Anne Schuth}{UvA}{Search Engines that Learn from Their Users}

\siksitem{Max Knobbout}{UU}{Logics for Modelling and Verifying Normative Multi-Agent Systems}

\siksitem{Nana Baah Gyan}{VUA}{The Web, Speech Technologies and Rural Development in West Africa: An ICT4D Approach}

\siksitem{Ravi Khadka}{UU}{Revisiting Legacy Software System Modernization}

\siksitem{Steffen Michels}{RUN}{Hybrid Probabilistic Logics: Theoretical Aspects, Algorithms and Experiments}

\siksitem{Guangliang Li}{UvA}{Socially Intelligent Autonomous Agents that Learn from Human Reward}

\siksitem{Berend Weel}{VUA}{Towards Embodied Evolution of Robot Organisms}

\siksitem{Albert Mero\~no Pe\~nuela}{VUA}{Refining Statistical Data on the Web}

\siksitem{Julia Efremova}{Tu/e}{Mining Social Structures from Genealogical Data}

\siksitem{Daan Odijk}{UvA}{Context \& Semantics in News \& Web Search}

\siksitem{Alejandro Moreno C\'elleri}{UT}{From Traditional to Interactive Playspaces: Automatic Analysis of Player Behavior in the Interactive Tag Playground}

\siksitem{Grace Lewis}{VUA}{Software Architecture Strategies for Cyber-Foraging Systems}

\siksitem{Fei Cai}{UvA}{Query Auto Completion in Information Retrieval}

\siksitem{Brend Wanders}{UT}{Repurposing and Probabilistic Integration of Data: An Iterative and data model independent approach}

\siksitem{Julia Kiseleva}{TU/e}{Using Contextual Information to Understand Searching and Browsing Behavior}

\siksitem{Dilhan Thilakarathne}{VUA}{In or Out of Control: Exploring Computational Models to Study the Role of Human Awareness and Control in Behavioural Choices, with Applications in Aviation and Energy Management Domains}

\siksitem{Wen Li}{TUD}{Understanding Geo-spatial Information on Social Media}

\siksitem{Mingxin Zhang}{TUD}{Large-scale Agent-based Social Simulation: A study on epidemic prediction and control}

\siksitem{Nicolas H\"oning}{TUD}{Peak reduction in decentralised electricity systems -Markets and prices for flexible planning}

\siksitem{Ruud Mattheij}{UvT}{The Eyes Have It}

\siksitem{Mohammad Khelghati}{UT}{Deep web content monitoring}

\siksitem{Eelco Vriezekolk}{UT}{Assessing Telecommunication Service Availability Risks for Crisis Organisations}

\siksitem{Peter Bloem}{UvA}{Single Sample Statistics, exercises in learning from just one example}

\siksitem{Dennis Schunselaar}{TUe}{Configurable Process Trees: Elicitation, Analysis, and Enactment}

\siksitem{Zhaochun Ren}{UvA}{Monitoring Social Media: Summarization, Classification and Recommendation}

\siksitem{Daphne Karreman}{UT}{Beyond R2D2: The design of nonverbal interaction behavior optimized for robot-specific morphologies}

\siksitem{Giovanni Sileno}{UvA}{Aligning Law and Action: a conceptual and computational inquiry}

\siksitem{Andrea Minuto}{UT}{MATERIALS THAT MATTER: Smart Materials meet Art \& Interaction Design}

\siksitem{Merijn Bruijnes}{UT}{Believable Suspect Agents: Response and Interpersonal Style Selection for an Artificial Suspect}

\siksitem{Christian Detweiler}{TUD}{Accounting for Values in Design}

\siksitem{Thomas King}{TUD}{Governing Governance: A Formal Framework for Analysing Institutional Design and Enactment Governance}

\siksitem{Spyros Martzoukos}{UvA}{Combinatorial and Compositional Aspects of Bilingual Aligned Corpora}

\siksitem{Saskia Koldijk}{RUN}{Context-Aware Support for Stress Self-Management: From Theory to Practice}

\siksitem{Thibault Sellam}{UvA}{Automatic Assistants for Database Exploration}

\siksitem{Bram van de Laar}{UT}{Experiencing Brain-Computer Interface Control}

\siksitem{Jorge Gallego Perez}{UT}{Robots to Make you Happy}

\siksitem{Christina Weber}{UL}{Real-time foresight: Preparedness for dynamic innovation networks}

\siksitem{Tanja Buttler}{TUD}{Collecting Lessons Learned}

\siksitem{Gleb Polevoy}{TUD}{Participation and Interaction in Projects. A Game-Theoretic Analysis}

\siksitem{Yan Wang}{UVT}{The Bridge of Dreams: Towards a Method for Operational Performance Alignment in IT-enabled Service Supply Chains}

\end{sikslist}
\begin{sikslist}{2017}

\siksitem{Jan-Jaap Oerlemans}{UL}{Investigating Cybercrime}

\siksitem{Sjoerd Timmer}{UU}{Designing and Understanding Forensic Bayesian Networks using Argumentation}

\siksitem{Dani\"el Harold Telgen}{UU}{Grid Manufacturing: A Cyber-Physical Approach with Autonomous Products and Reconfigurable Manufacturing Machines}

\siksitem{Mrunal Gawade}{CWI}{MULTI-CORE PARALLELISM IN A COLUMN-STORE}

\siksitem{Mahdieh Shadi}{UvA}{Collaboration Behavior}

\siksitem{Damir Vandic}{EUR}{Intelligent Information Systems for Web Product Search}

\siksitem{Roel Bertens}{UU}{Insight in Information: from Abstract to Anomaly}

\siksitem{Rob Konijn}{VUA}{Detecting Interesting Differences:Data Mining in Health Insurance Data using Outlier Detection and Subgroup Discovery}

\siksitem{Dong Nguyen}{UT}{Text as Social and Cultural Data: A Computational Perspective on Variation in Text}

\siksitem{Robby van Delden}{UT}{(Steering) Interactive Play Behavior}

\siksitem{Florian Kunneman}{RUN}{Modelling patterns of time and emotion in Twitter \#anticipointment}

\siksitem{Sander Leemans}{UT}{Robust Process Mining with Guarantees}

\siksitem{Gijs Huisman}{UT}{Social Touch Technology: Extending the reach of social touch through haptic technology}

\siksitem{Shoshannah Tekofsky}{UvT}{You Are Who You Play You Are: Modelling Player Traits from Video Game Behavior}

\siksitem{Peter Berck, Radboud University}{RUN}{Memory-Based Text Correction}

\siksitem{Aleksandr Chuklin}{UvA}{Understanding and Modeling Users of Modern Search Engines}

\siksitem{Daniel Dimov}{VUA}{Crowdsourced Online Dispute Resolution}

\siksitem{Ridho Reinanda}{UvA}{Entity Associations for Search}

\siksitem{Jeroen Vuurens}{TUD}{Proximity of Terms, Texts and Semantic Vectors in Information Retrieval}

\siksitem{Mohammadbashir Sedighi}{TUD}{Fostering Engagement in Knowledge Sharing: The Role of Perceived Benefits, Costs and Visibility}

\siksitem{Jeroen Linssen}{UT}{Meta Matters in Interactive Storytelling and Serious Gaming (A Play on Worlds)}

\siksitem{Sara Magliacane}{VUA}{Logics for causal inference under uncertainty}

\siksitem{David Graus}{UvA}{Entities of Interest --- Discovery in Digital Traces}

\end{sikslist}
 
\end{multicols}
\clearpage{}

\clearpage{}
\chapter{Samenvatting}

Tegenwoordig laten we continu --- en soms zonder het te weten --- digitale sporen achter, door online content te browsen, \emph{sharen}, \emph{liken}, doorzoeken, bekijken, of beluisteren. Geaggregeerd kunnen deze digitale sporen waardevolle inzichten bieden in ons gedrag, onze voorkeuren, bezigheden, en eigenschappen. Terwijl veel mensen zich zorgen maken om de hiermee gepaarde bedreiging van onze privacy, heeft het op grote schaal verzamelen en gebruiken van onze digitale sporen ook veel goeds gebracht. Bijvoorbeeld de toegang tot ongekende hoeveelheden informatie en kennis die de zoekmachines --- al lerende van hun gebruikers --- ons verschaffen, of het ontdekken van nieuwe bijwerkingen van medicijnen door de zoekopdrachten die bij een zoekmachine binnenkomen te analyseren.

Of het nu in online diensten, journalistiek, digitaal sporenonderzoek, of wetenschap is, men richt zich steeds meer op digitale sporen om nieuwe informatie te ontdekken. Neem bijvoorbeeld het Enron schandaal, de controverse rond het emailgebruik van Hillary Clinton, of de Panama Papers: voorbeelden van zaken die draaien rond het analyseren, doorzoeken, onderzoeken, en binnenstebuiten keren van grote hoeveelheden digitale sporen om tot nieuwe inzichten, kennis, en informatie te komen. 

Dit proefschrift gaat over het ontdekkingsproces in grootschalige verzamelingen digitale sporen. Het proefschrift bevindt zich op de kruising van zoekmachinetechnologie, taaltechnologie, en toegepaste \emph{machine learning}, en presenteert computationele methoden die ondersteuning bieden aan het doorzoeken en inzichtelijk maken van grootschalige verzamelingen digitale sporen. We richten ons op tekstuele digitale sporen, zoals emails, en social media. Daarnaast richten we ons op twee aspecten die centraal staan in het ontdekkingsproces. 

Allereerst richten we ons op de inhoud van digitale sporen. Onze onderzoekssubjecten zijn de personen, plaatsen, en organisaties die worden genoemd in tekstuele digitale sporen. Deze zogenaamde \emph{real-world entities} zijn van centraal belang in het doorzoeken en inzichtelijk maken van de inhoud van digitale sporen. In dit deel richten we ons op het analyseren, herkennen, en beter vindbaar maken van nieuwe entiteiten die opkomen in digitale sporen. 

In het tweede deel richten we ons op de producenten van digitale sporen. Niet de inhoud, maar de context waarin de digitale sporen tot stand komen staat hierbij centraal. Ons doel is toekomstige activiteiten te voorspellen door gebruik te maken van digitale sporen. We tonen aan dat we de ontvanger van een email kunnen voorspellen door gebruik te maken van de communicatie netwerken van emailverkeer. Daarnaast tonen we aan dat we kunnen voorspellen wanneer iemand van plan is een bepaalde taak uit te voeren, door gebruik te maken van geaggregeerde interactie-data met een persoonlijke assistent-applicatie. \clearpage{}

\clearpage{}
\makeatletter
\@openrightfalse

\chapter{Summary}

In the era of big data, we continuously --- and at times unknowingly --- leave behind digital traces, by browsing, sharing, posting, liking, searching, watching, and listening to online content. When aggregated, these digital traces can provide powerful insights into the behavior, preferences, activities, and traits of people. While many have raised privacy concerns around the use of aggregated digital traces, it has undisputedly brought us many advances, from the search engines that learn from their users and enable our access to unforeseen amounts of data, knowledge, and information, to, e.g., the discovery of previously unknown adverse drug reactions from search engine logs. 

Whether in online services, journalism, digital forensics, law, or research, we increasingly set out to exploring large amounts of digital traces to discover new information. Consider for instance, the Enron scandal, Hillary Clinton's email controversy, or the Panama papers: cases that revolve around analyzing, searching, investigating, exploring, and turning upside down large amounts of digital traces to gain new insights, knowledge, and information. This \emph{discovery} task is at its core about ``finding evidence of activity in the real world.''

This dissertation revolves around \emph{discovery in digital traces}, and sits at the intersection of Information Retrieval, Natural Language Processing, and applied Machine Learning. We propose computational methods that aim to support the exploration and sense-making process of large collections of digital traces. We focus on textual traces, e.g., emails and social media streams, and address two aspects that are central to discovery in digital traces.

In the first part, we focus on the textual content of digital traces. Here, our entities of interest are the people, places, and organizations that appear in digital traces. These so-called real-world entities are central to enabling exploratory search in digital traces. More specifically, we analyze, predict and show how to improve retrieval of newly \emph{emerging} entities, i.e., previously unknown entities that surface in digital traces.

In the second part, our entities of interest are the people who leave behind the digital traces. Here, we focus on the contexts in which digital traces are created, and aim to predict users' future activity, by leveraging their historic digital traces. We show we can predict email recipients by leveraging the email communication network, and we show we can predict when users plan to execute activities, by leveraging logs of interactions with an intelligent assistant.
\stepcounter{page}
\setcounter{page}{174}\clearpage{}

\newpage{\blankpage}

\includepdf[pages=-]{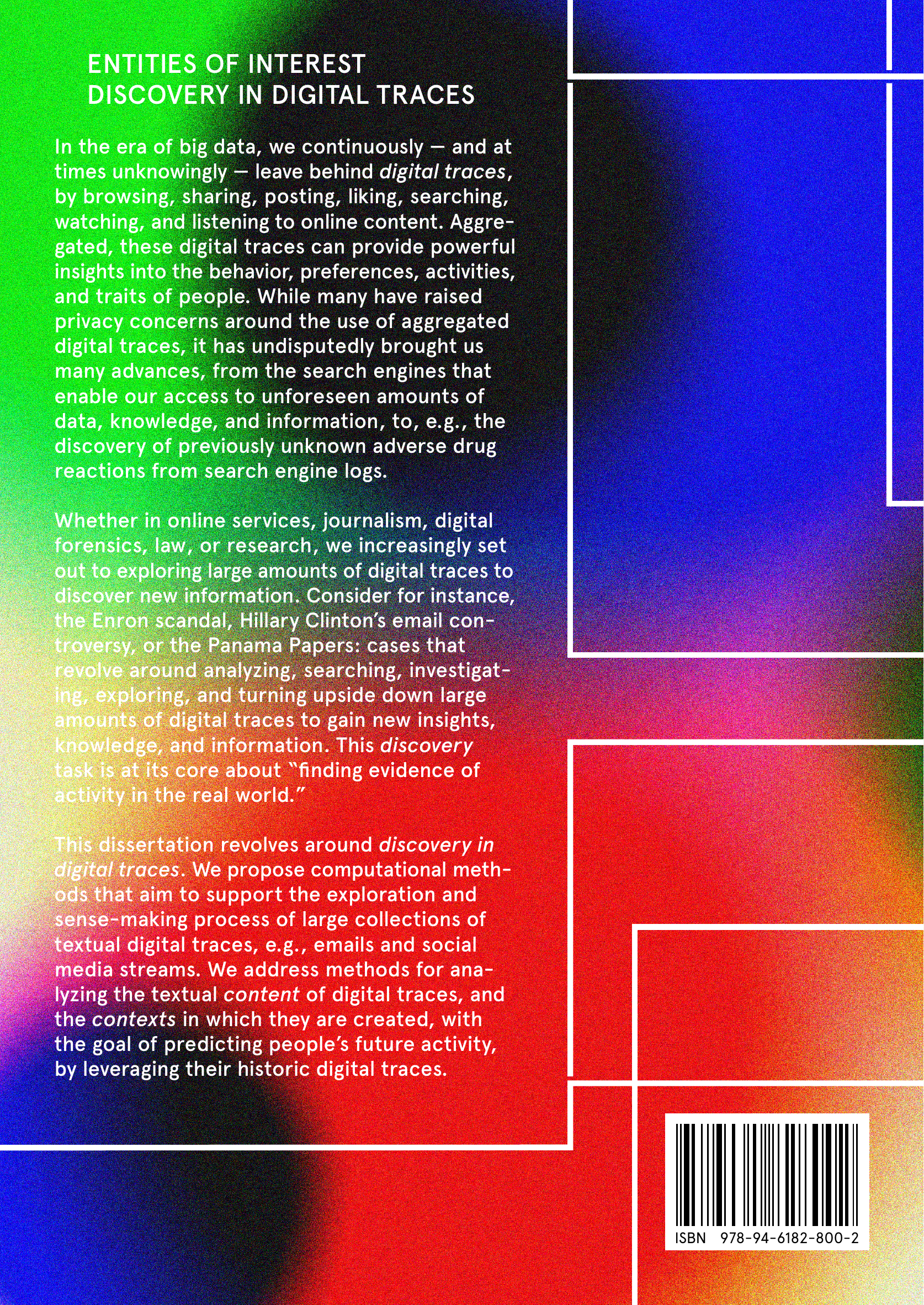}

\end{document}